\documentclass[12pt,oneside]{mitthesis}
\usepackage{lgrind}
\usepackage{cmap}
\usepackage[T1]{fontenc}
\usepackage[utf8]{inputenc}
\usepackage[english]{babel}
\usepackage{libertine}
\usepackage[babel=true]{microtype}
\usepackage{graphicx}
\usepackage{xspace} 
\usepackage{bm} 
\usepackage{array} 
\newcolumntype{L}[1]{>{\raggedright\let\newline\\\arraybackslash\hspace{0pt}}m{#1}}
\newcolumntype{C}[1]{>{\centering\let\newline\\\arraybackslash\hspace{0pt}}m{#1}}
\newcolumntype{R}[1]{>{\raggedleft\let\newline\\\arraybackslash\hspace{0pt}}m{#1}}
\usepackage{subcaption}
\usepackage{hhline}
\usepackage[normalem]{ulem}
\usepackage{booktabs}
\usepackage{makecell} 
\usepackage{titlesec} 
\usepackage{indentfirst} 
\usepackage{float} 
\usepackage{dblfloatfix} 
\usepackage{changepage}

\usepackage{pifont}
\newcommand{\delete}[1]{{\color{red}\sout{}}}

\usepackage{multirow}
\usepackage[usenames,dvipsnames,svgnames,table]{xcolor}

\usepackage[numbers,sort]{natbib}

\newcommand{\todo}[1]{\textbf{\fcolorbox{black}{darkcyan}{\color{white}{TODO}}} \underline{$\overline{\hbox{\textcolor{darkcyan}{\emph{#1}}}}$}}
\definecolor{amber}{rgb}{1.0, 0.49, 0.0}
\definecolor{darkgreen}{rgb}{0.0, 0.2, 0.13}
\definecolor{darkbyzantium}{rgb}{0.36, 0.22, 0.33}
\definecolor{darkseagreen}{rgb}{0.56, 0.74, 0.56}
\definecolor{darkspringgreen}{rgb}{0.09, 0.45, 0.27}
\definecolor{dollarbill}{rgb}{0.52, 0.73, 0.4}

\newcommand{\squeezeme}{ \setlength{\itemsep}{0pt}
     \setlength{\parsep}{3pt}
     \setlength{\topsep}{3pt}
     \setlength{\partopsep}{0pt}
     \setlength{\leftmargin}{1.5em}
     \setlength{\labelwidth}{1em}
     \setlength{\labelsep}{0.5em} }

\usepackage{amsmath}
\usepackage[nolessnomore, italic]{mathastext}

\usepackage{comment}

\newcommand{\specialcell}[2][c]{%
	\begin{tabular}[#1]{@{}r@{}}#2\end{tabular}}
\newcommand{\specialcellll}[2][c]{%
	\begin{tabular}[#1]{@{}c@{}}#2\end{tabular}}
\newcommand{\specialcel}[2][t]{%
	\begin{tabular}[#1]{@{}l@{}}#2\end{tabular}}
\usepackage{tabularx}
\newcolumntype{Z}{>{\raggedleft\let\newline\\\arraybackslash\hspace{0pt}}X}
\newcolumntype{Y}{>{\raggedright\let\newline\\\arraybackslash\hspace{0pt}}X}

\usepackage{mathtools}
\usepackage{titlesec}
\usepackage{microtype} 

\usepackage{array}

\makeatletter
\let\MYcaption\@makecaption
\makeatother

\usepackage{fixltx2e}

\usepackage[usenames,dvipsnames,table]{xcolor}
\usepackage{blindtext}

\usepackage{xspace}
\usepackage{ifthen}

\usepackage{booktabs}
\usepackage[all]{nowidow}
\usepackage{multirow}

\usepackage{comment}
\usepackage{enumitem}
\usepackage{setspace}
\usepackage{booktabs}
\usepackage{multirow}
\usepackage{amssymb}

\usepackage{tikz}
\usetikzlibrary{arrows}
\usetikzlibrary{patterns}
\usetikzlibrary{calc}
\usetikzlibrary{shapes}
\usetikzlibrary{positioning,fit}
\usepackage{float}
\usepackage{lipsum}

\usepackage[nolessnomore, italic]{mathastext}
\usepackage[T1]{fontenc}
\usepackage{hhline}
\usepackage[normalem]{ulem}
\usepackage{indentfirst}
\usepackage{footmisc}

\usepackage[us,12hr]{datetime}
\usepackage{verbatim}
\usepackage{float}

\usepackage{graphicx}
\usepackage{textcomp}

\usepackage{amsmath,amssymb,amsfonts}
\usepackage{algorithm}
\usepackage[noend]{algpseudocode}

\algrenewcommand\alglinenumber[1]{\fontsize{10}{10} #1:}

\usepackage{multirow}

\usepackage{etoolbox}

\usepackage{tikz}
\newcommand{\circlednumber}[1]{\raisebox{0.5pt}{\protect%
  \tikz[baseline=(myanchor.base)]{
  \node[circle,fill=.,inner sep=1pt] (myanchor) {\color{-.}\footnotesize #1};}%
}}
\newcommand*\circlednumberr[1]{\raisebox{0.5pt}{\protect%
  \tikz[baseline=(char.base)]{
  \node[shape=circle,draw,densely dotted,thick,inner sep=1pt] (char) {\footnotesize #1};}%
}}

\definecolor{amber}{rgb}{1.0, 0.49, 0.0}
\definecolor{darkgreen}{rgb}{0.0, 0.2, 0.13}
\definecolor{darkbyzantium}{rgb}{0.36, 0.22, 0.33}
\definecolor{darkseagreen}{rgb}{0.56, 0.74, 0.56}
\definecolor{darkspringgreen}{rgb}{0.09, 0.45, 0.27}
\definecolor{dollarbill}{rgb}{0.52, 0.73, 0.4}

\definecolor{darkcyan}{rgb}{0.0, 0.55, 0.55}
\definecolor{forestgreen}{rgb}{0.0, 0.27, 0.13}
\definecolor{azure}{rgb}{0.0, 0.5, 1.0}
\definecolor{amber}{rgb}{1.0, 0.49, 0.0}

\definecolor{darkpink}{rgb}{0.88, 0.28, 0.54}

\newcommand{\CF}[1]{\textbf{\color{blue}{[#1]}}}

\newcommand{\hl}[1]{{\color{black}#1}}
\newcommand{\rev}[1]{{\color{black}#1}}
\newcommand{\sg}[1]{{\color{black}#1}}

\newcommand{\revonur}[1]{{\color{black}#1}}
\newcommand{\sgrev}[1]{{\color{black}#1}}

\newcommand{\revII}[1]{{\color{black}#1}}
\newcommand{\sgii}[1]{{\color{black}#1}}

\newcommand{\revIII}[1]{{\color{black}#1}}
\newcommand{\comm}[1]{{\color{NavyBlue}#1}}

\newcommand{\revIV}[1]{{\color{black}#1}}
\newcommand{\revV}[1]{{\color{black}#1}}

\newcommand{\revonurcan}[1]{{\color{black}#1}}

\newcommand{\revmicro}[1]{{\color{black}#1}}

\definecolor{mypink}{RGB}{224,8,95}
\definecolor{myblue}{RGB}{31,116,186}
\definecolor{mygreen}{RGB}{35,155,51}
\definecolor{myorange}{RGB}{201,108, 32}
\definecolor{mypurple}{RGB}{122,32,201}
\usepackage{array}
\newcolumntype{P}[1]{>{\centering\arraybackslash}p{#1}}

\usepackage{titlesec}
\titleformat{\subsection}[block]{\normalfont\normalsize\bfseries}{\thesubsection}{1em}{}

\newcommand{\etal}{\textit{et al}. }
\newcommand{\ie}{\textit{i}.\textit{e}., }
\newcommand{\eg}{\textit{e}.\textit{g}., }

\usepackage{url}

\usepackage[flushleft]{threeparttable}
\usepackage{ragged2e}

\renewcommand{\thefigure}{\arabic{chapter}-\arabic{figure}}
\renewcommand{\thetable}{\arabic{chapter}-\arabic{table}}

\definecolor{amber}{rgb}{1.0, 0.49, 0.0}
\definecolor{darkgreen}{rgb}{0.0, 0.2, 0.13}
\definecolor{darkbyzantium}{rgb}{0.36, 0.22, 0.33}
\definecolor{darkseagreen}{rgb}{0.56, 0.74, 0.56}
\definecolor{darkspringgreen}{rgb}{0.09, 0.45, 0.27}
\definecolor{dollarbill}{rgb}{0.52, 0.73, 0.4}

\definecolor{darkcyan}{rgb}{0.0, 0.55, 0.55}
\definecolor{forestgreen}{rgb}{0.0, 0.27, 0.13}
\definecolor{azure}{rgb}{0.0, 0.5, 1.0}
\definecolor{amber}{rgb}{1.0, 0.49, 0.0}

\definecolor{darkpink}{rgb}{0.88, 0.28, 0.54}

\definecolor{dblue}{rgb}{0.00, 0.00, 0.55}
\definecolor{ddblue}{rgb}{0.00, 0.00, 0.90}

\definecolor{magenta}{rgb}{255,0,255}
\definecolor{red}{rgb}{255,0,0}

\newcommand{\ba}{BitAlign\xspace}
\newcommand{\ms}{MinSeed\xspace}
\newcommand{\mech}{{\damla{SeGraM}}\xspace} 

\newcommand{\damla}[1]{{\color{black}#1}}
\newcommand{\damlaI}[1]{{\color{black}#1}}
\newcommand{\damlaII}[1]{{\color{black}#1}}
\newcommand{\damlaIII}[1]{{\color{black}#1}}

\newcommand{\sgh}[1]{{\color{black}#1}}
\newcommand{\sghi}[1]{{\color{black}#1}}
\newcommand{\zulal}[1]{{\color{black}#1}}

\usepackage[binary-units=true]{siunitx}

\usepackage[breaklinks=true,hidelinks]{hyperref}

\usepackage{fancyhdr}
\def\secondpage{\clearpage\null\vfill
\pagestyle{empty}
\begin{minipage}[b]{0.9\textwidth}
\center{Copyright \copyright\ \the\year\ Damla Senol Cali \par All Rights Reserved} 
\end{minipage}
\vspace*{2\baselineskip}

\pagebreak
\hspace{0pt}
\vfill
\begin{center}
\emph{To my parents - Mine and Sinan,\\my sister - Irmak,\\my husband - Tunca.}
\end{center}
\vfill
\hspace{0pt}
\cleardoublepage

\rfoot{\thepage}}

\makeatletter
\g@addto@macro{\maketitle}{\secondpage}
\makeatother

\begin{document}
\pagenumbering{roman}
\title{Accelerating Genome Sequence Analysis via Efficient Hardware/Algorithm Co-Design} 

\author{Damla Senol Cali} 
\department{Department of Electrical Engineering and Computer Science}

\degree{Master of Science in Electrical Engineering and Computer Science}

\degreemonth{June}
\degreeyear{1990}
\thesisdate{May 18, 1990}


\supervisor{William J. Dally}{Associate Professor}

\chairman{Arthur C. Smith}{Chairman, Department Committee on Graduate Theses}

\maketitle



\newpage
\newpage 

\pagestyle{plain}
\section*{Acknowledgments}

\damlaII{
I have many people to thank who have supported me in very different ways during my Ph.D. journey over the last six years. First and foremost, I am very grateful for my advisors, Prof. Onur Mutlu and Prof. Saugata Ghose. Onur has generously provided many opportunities, offered guidance, mentorship, and patience throughout my Ph.D., which have been key to my growth and success. His passion for excellence in all aspects of research has enabled me to push my boundaries and be a part of several top-notch research projects. I have learned greatly from his feedback and insights during the numerous iterations we made for the paper submissions, video recordings, and conference talks. In addition to the high standards he set for research, the collaborative environment he has offered within the SAFARI Research Group and also with our industry partners has helped me to work with a diverse group of people all around the world and exposed me to many different research directions. 

I am very grateful to Saugata for being a great mentor and offering help whenever I need it. During my all ups and downs throughout my Ph.D. journey, he has supported me with great patience and guidance. His tremendous amount of help and support during paper submissions has helped me to stay motivated, positive, and encouraged despite the challenging and stressful deadlines. In addition to the calming and encouraging environment he has provided, his passion for teaching has been inspirational.

I am very grateful to Can Alkan, my dissertation committee member and my mentor since my undergrad. His passion and vision for genomics have been one of the key enablers of all the research I have conducted during my Ph.D. 

I am also grateful to my dissertation committee member, James Hoe, for his valuable feedback and for always letting me think about the big picture of my research. 

This long and challenging journey would have been impossible without the great support and friendship I have had in my research group. Despite the thousands of miles we have between us, I have never felt distant or lonely. First, I want to thank Jeremie Kim, my very first friend at SAFARI, who helped me to survive my first year. I am very grateful for his support and the contributions he made to my research throughout my Ph.D. I am very thankful for Zulal Bingol, an endless source of support and positivity, a friend and colleague whom I can trust and count on whenever I need the most. I am very thankful for dear Nastaran Hajinazar, who always makes me smile and brings joy and happiness to everyone around her. I am very thankful for Can Firtina, who is a great friend since my undergrad and a very helpful colleague during my Ph.D. I want to thank Giray Yaglikci for his friendship and our endless chats in the office. I want to thank Gagandeep Singh for providing a fun working environment and for our TV series discussions. I want to thank Amirali Boroumand for being a great friend and keeping me sane during my Ph.D. I want to thank Konstantinos Kanellopoulos for our long hours of brainstorming and bringing an immense wealth of knowledge. I want to thank Geraldo Francisco Oliveira for his friendship and delicious cake recipes. I want to thank Nour Almadhoun Alserr for her kindness and support during the final year of my Ph.D. I also want to thank Nika Mansourighiasi for her friendship and our helpful research discussions. I am tremendously lucky to have all these invaluable lifelong friendships, and I am immensely grateful to all of them.

I am also very grateful to Lavanya Subramanian, Rachata Ausavarungnirun, Mohammed Alser, and Juan Gómez-Luna for their mentorship. They have provided great insights and support throughout my Ph.D. journey with their technical expertise, many fruitful discussions, and their patience and kindness towards me.

I want to acknowledge all the other members of our research group at CMU and ETH for being both great friends and colleagues: Minesh Patel, Minh Sy Quang Truong, Banu Cavlak, Joel Lindegger, Ziyi Zuo, Jisung Park, Hasan Hassan, Max Rumpf, Haiyu Mao, Aditya Manglik, Sam Cheung, Akanksha Baranwal, Nandita Vijaykumar, Kevin Hsieh, Donghyuk Lee, Hongyi Xin, Yixin Luo, Vivek Seshadri, Kevin Chang, and all other past and current SAFARI, ARCANA, and Bilkent CompGen members for their discussions, feedback, collaboration, and support. 

I would also like to thank my internship mentors and managers, Sree Subramoney and Gurpreet S. Kalsi, who provided a stimulating environment and an industrial perspective on hardware acceleration during my internships at Intel Labs. I sincerely thank Intel for these opportunities.

I gratefully acknowledge the generous support from Intel, Google, Microsoft, Samsung, VMware, and other industrial partners of SAFARI Research Group, the Semiconductor Research Corporation, \damlaII{and} the National Institutes of Health (grant HG006004).

Finally, I want to thank my family and friends for their love and support during this long journey. Especially, I am very grateful to my parents, Mine and Sinan. I cannot thank my mom enough for her endless support, encouragement, unconditional love, and sacrifice. She has taught me how to be a very strong woman. I cannot thank my dad enough for being a wonderful role model with his ambition, determination, high standards for success, and the inspiration to pursue my dreams. I am very grateful for his endless support and love. I am very grateful to my sister, Irmak, for being my best friend, an endless source of joy, laughter, and love. I cannot thank my sister enough for making me smile even on my hardest days and always being by my side. I am very grateful to my husband, Tunca, for always believing in me and providing endless care and love. Despite all my highs and downs, with his unwavering support, understanding, and encouragement, I have always found my way. Without the comfortable, joyful, and loving environment he has provided, I could not have accomplished any of my achievements. It would have been impossible for me to become who I am today without my parents, my sister, and my husband, and I will forever be indebted and grateful to them. This dissertation would in no way be possible without them, so it is dedicated to my loving family: Mine, Sinan, Irmak, and Tunca. 
}

\cleardoublepage

\begin{abstractpage}
\pagestyle{plain}
Genome sequence analysis plays a pivotal role in enabling many medical and scientific advancements in personalized medicine, outbreak tracing, the understanding of evolution, and forensics. Modern genome sequencing machines can rapidly generate massive amounts of genomics data at low cost. However, the analysis of genome sequencing data is currently bottlenecked by the computational power and memory bandwidth limitations of existing systems, as many of the steps in genome sequence analysis must process a large amount of data. 
Moreover,  as  sequencing  technologies  advance,  the growth in the rate that sequencing devices generate genomics data is far outpacing the corresponding growth in computational power, placing greater pressure on these bottlenecks.

\damlaII{Our goals in this dissertation are to (1) understand where the current tools and algorithms do not perform well in order to develop better tools and algorithms, and (2) understand the limitations of existing hardware systems when running these tools and algorithms in order to design efficient customized accelerators. Towards this end, we propose four major works, where we characterize the real-system behavior of the genome sequence analysis pipeline and its associated tools, expose the bottlenecks and tradeoffs of the pipeline and tools, and co-design fast and efficient algorithms along with scalable and energy-efficient customized hardware accelerators for the key pipeline bottlenecks to enable faster genome sequence analysis.}

First, we comprehensively analyze the tools in the genome assembly pipeline for long reads in multiple dimensions (i.e., accuracy, performance, memory usage, and scalability), uncovering bottlenecks and tradeoffs that different combinations of tools and different underlying systems lead to. We show that we need high-performance, memory-efficient, low-power, and scalable designs for genome sequence analysis in order to exploit the advantages that genome sequencing provides. Second, we propose GenASM, an acceleration framework that builds upon bitvector-based approximate string matching (ASM) to accelerate multiple steps of the genome sequence analysis pipeline. We co-design our highly-parallel, scalable and memory-efficient algorithms with low-power and area-efficient hardware accelerators. We evaluate GenASM for three different use cases of ASM in genome sequence analysis and show that GenASM is significantly faster and more power- and area-efficient than state-of-the-art software and hardware tools for each of these use cases. Third, we implement an FPGA-based prototype for GenASM, where state-of-the-art 3D-stacked memory (HBM2) offers high memory bandwidth and FPGA resources offer high parallelism by instantiating multiple copies of the GenASM accelerators. \damla{Fourth, we propose \mech, the first hardware acceleration framework for sequence-to-graph mapping and alignment.} Instead of representing the reference genome as a single linear DNA sequence, genome graphs provide a better representation of the diversity among populations by encoding variations across individuals in a graph data structure, avoiding a bias towards any one reference. \mech enables the efficient mapping of a sequenced genome to a graph-based reference, providing more comprehensive and accurate genome sequence analysis. \damla{For \mech, we co-design algorithms and accelerators for memory-efficient minimizer-based seeding and bitvector-based, highly-parallel sequence-to-graph alignment. Compared to state-of-the-art software tools for sequence-to-graph mapping and alignment, we show that \mech significantly increases throughput and reduces power consumption for both short and long reads.}

Overall, we demonstrate that genome sequence analysis can be accelerated by co-designing scalable and energy-efficient customized accelerators along with efficient algorithms for the key steps of genome sequence analysis. 
\damla{We also hope that this dissertation inspires future work in co-designing algorithms and hardware together to create powerful frameworks that accelerate other genomics workloads and emerging applications.}

\end{abstractpage}


\cleardoublepage



{\setstretch{1.5}
\tableofcontents
\newpage
\phantomsection
\addcontentsline{toc}{chapter}{List of Figures}
\listoffigures
\newpage
\phantomsection
\addcontentsline{toc}{chapter}{List of Tables}
\listoftables
}

\pagenumbering{arabic}
\setcounter{page}{1}
\chapter{Introduction}

\section{Problem and Dissertation Statement} 
\label{sec:introduction} 

Genome sequencing, which determines the DNA sequence of an organism, plays a pivotal role in
enabling many medical and scientific advancements in personalized medicine~\cite{alkan2009personalized,flores2013p4,ginsburg2009genomic,chin2011cancer,Ashley2016}, evolutionary theory~\cite{ellegren2014genome,Prado-Martinez2013,Prohaska2019}, and forensics~\cite{yang2014application,borsting2015next,alvarez2017next}.
Modern genome sequencing machines\revonur{~\cite{cali2017nanopore,minionwebpage,gridionwebpage,promethionwebpage,sequelwebpage,miseqwebpage,nextseqwebpage,novaseqwebpage}} can rapidly generate massive amounts of genomics data at low cost~\cite{Shendure2017,alser2020accelerating,mutlu2019aacbb,alser2020technology}, but are unable to extract \revII{an} organism's \revII{complete} DNA in one piece.  Instead, these machines \revII{extract} 
\revII{smaller random} fragments of the original DNA sequence, known as \emph{reads}.
These reads \revII{then} pass through a computational process known as \emph{genome sequence analysis.}
However, the analysis of genome sequencing data is currently bottlenecked by the computational power and memory bandwidth limitations of existing systems, as many of the steps in genome sequence analysis must process a large amount of data. Moreover,  as  sequencing  technologies  advance,  the growth in the rate that sequencing devices generate genomics data is far outpacing the corresponding growth in computational power, placing greater pressure on these bottlenecks.


Our goals in this dissertation are to (1) understand where the current tools and algorithms do \damlaIII{\emph{not}} perform well in order to develop better tools and algorithms, and (2) understand the limitations of existing hardware systems when running these tools and algorithms in order to design efficient customized accelerators. Towards this end, we propose four major works.
First, we comprehensively analyze the tools in the genome assembly pipeline for long reads in multiple dimensions (i.e., accuracy, performance, memory usage, and scalability), uncovering bottlenecks and tradeoffs that different combinations of tools and different underlying systems lead to. We show that we need high-performance, memory-efficient, low-power, and scalable designs for genome sequence analysis in order to exploit the advantages that genome sequencing provides. Second, we propose GenASM, an acceleration framework that builds upon bitvector-based approximate string matching (ASM) to accelerate multiple steps of the genome sequence analysis pipeline. We co-design our highly-parallel, scalable and memory-efficient algorithms with low-power and area-efficient hardware accelerators. We evaluate GenASM for three different use cases of ASM in genome sequence analysis and show that GenASM is significantly faster and more power- and area-efficient than state-of-the-art software and hardware tools for each of these use cases. Third, we implement an FPGA-based prototype for GenASM, where state-of-the-art 3D-stacked memory (HBM2) offers high memory bandwidth and FPGA resources offer high parallelism by instantiating multiple copies of the GenASM accelerators. \damla{Fourth, we propose \mech, the first hardware acceleration framework for sequence-to-graph mapping and alignment.} Instead of representing the reference genome as a single linear DNA sequence, genome graphs provide a better representation of the diversity among populations by encoding variations across individuals in a graph data structure, avoiding a bias towards any one reference. \mech enables the efficient mapping of a sequenced genome to a graph-based reference, providing more comprehensive and accurate genome sequence analysis. \damla{For \mech, we co-design algorithms and accelerators for memory-efficient minimizer-based seeding and bitvector-based, highly-parallel sequence-to-graph alignment. Compared to state-of-the-art software tools for sequence-to-graph mapping and alignment, we show that \mech significantly increases \damlaIII{throughput} and reduces \damlaIII{power consumption} for both short and long reads.}

\medskip
Our dissertation statement is as follows:
\textbf{\emph{Genome sequence analysis can be accelerated by co-designing fast and efficient algorithms along with scalable and energy-efficient customized hardware accelerators for the key bottleneck steps of the \damlaIII{genome analysis} pipeline.}}

\section{Overview of Our Approach} 
\label{sec:overview} 

\damlaII{In line with our dissertation statement, we present four works, where we \textbf{characterize} the real-system behavior of the genome sequence analysis pipeline and its associated tools, \textbf{expose} the bottlenecks and tradeoffs of the pipeline and tools, and \textbf{co-design} fast and efficient algorithms along with scalable and energy-efficient customized hardware accelerators for the key pipeline bottlenecks to enable faster genome sequence analysis.}

In our first work, we present the first \damlaIII{experimental analysis of} state-of-the-art tools associated with each step of the genome assembly pipeline using long reads. We analyze the tools in multiple dimensions that are important for both developers and users/practitioners: accuracy, performance, memory usage and scalability. We reveal new bottlenecks and tradeoffs that different combinations of tools and different underlying systems lead to, based on our extensive experimental analyses. We also provide guidelines for both practitioners, such that they can determine the appropriate tools and tool combinations that can satisfy their goals, and tool developers, such that they can make design choices to improve current and future tools. 

In our second work, we propose GenASM, the first approximate string matching (ASM) acceleration framework for genome sequence analysis. GenASM performs bitvector-based ASM, which can efficiently accelerate multiple steps of genome sequence analysis. 
We modify the underlying ASM algorithm (Bitap~\damlaIII{\cite{baeza1992new, wu1992fast}}) to significantly increase its parallelism and reduce its memory footprint. 
Using this modified algorithm, we design the first hardware accelerator for Bitap. Our hardware accelerator consists of specialized systolic-array-based compute units and on-chip SRAMs that are designed to match the rate of computation with memory capacity and bandwidth, resulting in an efficient design whose performance scales linearly as we increase the number of compute units working in parallel. We demonstrate that GenASM provides significant performance and power benefits for three different use cases in genome sequence analysis. First, GenASM accelerates read alignment for both long reads and short reads. For long reads, GenASM outperforms state-of-the-art software and hardware accelerators by 116$\times$ and \revII{$3.9\times$}, respectively, while reducing power consumption by $37\times$ and 2.7$\times$. For short reads, GenASM outperforms state-of-the-art software and hardware accelerators by $111\times$ and $1.9\times$. Second, GenASM accelerates pre-alignment filtering for short reads, with $3.7\times$ the performance of a state-of-the-art pre-alignment filter, while \revonur{reducing power consumption by \revonur{$1.7\times$}} and significantly improving the filtering accuracy. Third, GenASM accelerates edit distance calculation, with \revIII{22--12501$\times$} \revonur{and 9.3--400$\times$ speedups over the state-of-the-art \revonur{software} library and FPGA-based accelerator, respectively, while \revV{reducing power consumption} by \revonur{548--582$\times$} and \revIII{$67\times$}.} We also briefly discuss \revonur{four other} use cases that can benefit from GenASM.

In our third work, we propose BitMAc, which is an FPGA-based prototype for GenASM. In BitMAc, we map our GenASM algorithms on Stratix 10 MX FPGA with a state-of-the-art 3D-stacked memory (HBM2), where HBM2 offers high memory bandwidth and FPGA resources offer high parallelism by instantiating multiple copies of the GenASM accelerators. After \damlaIII{modifying} the GenASM algorithms for better mapping to existing FPGA resources, we show that BitMAc provides 64\% logic utilization and 90\% on-chip memory utilization, while having \SI{48.9}{\watt} of total power consumption. We compare BitMAc with state-of-the-art CPU-based and GPU-based read alignment tools. Compared to the alignment steps of the CPU-based read mappers, (1)~for long reads, BitMAc provides $761\times$ and 136$\times$ speedup, while reducing power consumption by $1.9\times$ and $2.0\times$, and (2)~for short reads, BitMAc provides 92$\times$ and 130$\times$ speedup, while reducing power consumption by $2.2\times$ and $2.0\times$. We also show that BitMAc provides significant speedup compared to the GPU-based baseline, while reducing   power consumption.

In our fourth work, we propose \mech, the first hardware acceleration framework for sequence-to-graph mapping \damla{and alignment}. Reference genomes are conventionally represented as a linear sequence. However, this linear representation of the reference genome \damlaIII{results in} ignoring the variations that exist in a population (i.e., genetic diversity) and introducing biases for the downstream analysis. To address these limitations, recently, graph-based representations of the genomes (i.e., \emph{genome graphs}) have gained attention. As shown in many prior \damlaII{works~\cite{alser2017gatekeeper,turakhia2018darwin,cali2020genasm,fujiki2018genax,kim2020geniehd,alser2019sneakysnake,goyal2017ultra,fujiki2020seedex,bingol2021gatekeeper,nag2019gencache,kim2018grim,lavenier2016dna,kaplan2018rassa,kaplan2020bioseal}}, sequence-to-sequence mapping is one of the major bottlenecks of the genome sequence analysis pipeline and \damlaIII{needs} to be accelerated using specialized hardware. Since graph-representation of the genome is much more complex than the linear representation, sequence-to-graph mapping is placing a greater pressure on this bottleneck. Thus, in this work, our goal is to design a high-performance, scalable, power- and area-efficient hardware accelerator for sequence-to-graph mapping that support both short and long reads. We base \mech on a memory-efficient minimizer-based seeding algorithm and a bitvector-based, highly-parallel sequence-to-graph alignment algorithm. We \emph{co-design} both of our algorithms with high-performance, area- and power-efficient hardware accelerators. \damlaII{\mech consists of two components: (1)~\ms, which provides hardware support to execute our minimizer-based seeding algorithm, and (2)~\ba, which provides hardware support to execute our bitvector-based sequence-to-graph alignment algorithm.} \damla{For sequence-to-graph mapping with long reads, we find that \mech achieves $8.8\times$ and $7.3\times$ speedup over 12-thread execution of state-of-the-art sequence-to-graph mapping tools (GraphAligner~\damlaIII{\cite{rautiainen2020graphaligner}} and vg~\damlaIII{\cite{garrison2018variation}}, respectively), while reducing power consumption by $4.9\times$ and $6.5\times$. For sequence-to-graph mapping with short reads, we find that \mech achieves $168\times$ and $726\times$ speedup over 12-thread execution of GraphAligner and vg, respectively, while reducing power consumption by $4.7\times$ and $4.9\times$. For sequence-to-graph alignment, we show that \ba provides $41\times$--$539\times$ speedup over \damlaII{PaSGAL~\damlaIII{\cite{jain2019accelerating}}, a} state-of-the-art sequence-to-graph alignment tool}.

\section{Contributions} 

This dissertation makes the following \textbf{key contributions}: 

\begin{enumerate}
\squeezeme
\item  We present the first work that analyzes state-of-the-art tools associated with each step of the genome assembly pipeline using long reads. For the 5 different steps of the pipeline, we analyze 12 different tools and make 21 observations for these tools.
    \begin{enumerate}
    \squeezeme
    \item We analyze the tools in multiple dimensions that are important for both developers and users/practitioners: accuracy, performance, memory usage and scalability.
    \item We reveal new bottlenecks and tradeoffs that different combinations of tools and different underlying systems lead to, based on our extensive experimental analyses.
    \item We show that basecalling is the most important step of the pipeline to overcome the high error rates of \damlaIII{the} nanopore sequencing technology~\damlaIII{\cite{cali2017nanopore,lu2016oxford,magi2017nanopore,clarke2009continuous,deamer2016three,marx2015nanopores,branton2008potential,laver2015assessing,ip2015minion,kasianowicz1996characterization,jain2018nanopore,quick2014reference}}.
    \item We show that there is a tradeoff between accuracy and performance when choosing the tool for the assembly step. Miniasm~\damlaIII{\cite{li2016minimap}}, coupled with an additional polishing step can lead to faster overall assembly than using Canu~\damlaIII{\cite{koren2017canu}} itself, while producing high-quality assemblies.
    \item We make observations that can guide researchers and practitioners in making conscious and effective choices for each step of the genome assembly pipeline using long reads. Also, with the help of bottlenecks we find, developers can improve the current tools or build new ones that are both accurate and fast, in order to overcome the high error rates of the long read sequencing technologies.
    \item We show that we need high-performance, memory-efficient, low-power, and scalable designs for genome sequence analysis in order to exploit the advantages that genome sequencing provides.
    \end{enumerate} 

\item We present GenASM, a novel approximate string matching acceleration framework for genome sequence analysis. GenASM is a power- and area-efficient hardware implementation of our new Bitap-based algorithms. GenASM is a fast, efficient, and flexible framework for both short and long reads, which \sg{can be used to accelerate} \emph{multiple steps} of the genome sequence analysis pipeline. 
    \begin{enumerate}
    \squeezeme 
    \item \sg{To avoid implementing more complex hardware for the dynamic programming \revII{based} algorithm\revonur{~\cite{Fei2018, kaplan2020bioseal, turakhia2018darwin, gupta2019rapid, Banerjee2019,jiang2007reconfigurable,rucci2018swifold, chen2014accelerating}, we base GenASM upon} the \textit{Bitap} algorithm~\cite{baeza1992new, wu1992fast}. Bitap uses only fast and simple bitwise operations to perform \revonur{approximate} string matching, making it amenable to \revonur{efficient} hardware acceleration. To our knowledge, GenASM is the first work that enhances and accelerates Bitap.}
    \item We modify Bitap to add efficient support for long reads and enable parallelism \revII{within each ASM operation}. We also propose the \emph{first} Bitap-compatible traceback algorithm. \revIV{We open source our software implementations of \revV{the GenASM algorithms}~\cite{genasmgithub}}.
    \item In GenASM, we \emph{co-design} our modified Bitap \sgii{algorithm and} our new Bitap-compatible \revII{\emph{traceback}} algorithm with an area- and power-efficient hardware \sgrev{accelerator.} Our hardware \sgrev{accelerator} \revII{(1)~\sgrev{balances} the compute resources \sgii{with} available memory capacity \revII{and bandwidth} per compute unit \sgrev{to avoid wasting resources}, (2)~\sgrev{achieves} high performance and \revII{power} efficiency \sgrev{by using specialized compute units that we design to exploit} data locality, and (3)~\sgrev{scales linearly in performance with the number of parallel compute units that we add to the system}.}
    \item We show that GenASM can accelerate \emph{three use cases} of approximate string matching (ASM) in genome sequence analysis (\revV{i.e.,} read alignment~\damlaIII{\cite{levenshtein1966binary,smith1981identification,needleman1970general}}, pre-alignment filtering~\damlaIII{\cite{alser2017gatekeeper,alser2019sneakysnake,Alser2019,alser2017magnet,Xin2013, Xin2015}}, edit distance \revIII{calculation~\damlaIII{\cite{levenshtein1966binary,vsovsic2017edlib}}).} 
    \item We \sg{find that GenASM is \revonur{greatly faster and more \revII{power-}efficient} for all three use cases than state-of-the-art software and hardware baselines.}
    \end{enumerate} 

\item We propose BitMAc, where we leverage a modern FPGA with high-bandwidth memory (HBM) for presenting an FPGA-based prototype for our GenASM accelerators. In BitMAc, we map GenASM on Stratix 10 MX FPGA~\damlaIII{\cite{stratix10mx}} with a state-of-the-art 3D-stacked memory (HBM2~\damlaIII{\cite{hbm}}), where HBM2 offers high memory bandwidth. We exploit intra-level parallelism by instantiating multiple processing elements (PEs) for the DC execution, and inter-level parallelism by running multiple independent GenASM executions in parallel.

    \begin{enumerate} 
    \squeezeme 
    \item We implement our DC and TB accelerator datapaths using SystemVerilog and incorporate the \damlaIII{on-chip memory blocks (i.e., M20Ks)} and the HBM2 interface for both top and bottom HBM2 stacks using M20K and HBM2 IPs, \damlaIII{respectively}. After re-modifying the GenASM algorithms for a better mapping to existing FPGA resources, the final and complete BitMAc design has 4 BitMAc accelerators connected to each pseudo-channel (128 in total), where each BitMAc accelerator contains a DC accelerator with 16 PEs, a TB accelerator, an FSM, and 13.2KB of M20Ks. We synthesize \revonur{and place \& route} the complete BitMAc design clocked at 200 MHz.
    \item We show that BitMAc provides 64\% logic utilization and 90\% on-chip memory utilization, while having \SI{48.9}{\watt} of total power consumption.
    \item We compare BitMAc with state-of-the-art CPU-based and GPU-based read alignment tools and show that BitMAc provides significant speedup compared to the the baselines, while reducing the power consumption.
    \item We show that due to the simplicity of the GenASM algorithms, BitMAc is a low-cost and scalable solution for bitvector-based sequence alignment.
    \end{enumerate} 

\item \damla{We propose \mech, a hardware acceleration framework for sequence-to-graph mapping and alignment. \mech targets both the seeding and alignment steps of sequence-to-graph mapping, with support for both short (e.g., Illumina~\damlaIII{\cite{reuter2015high,van2014ten}}) and long (e.g., PacBio~\damlaIII{\cite{english2012mind,roberts2013advantages,rhoads2015pacbio,wenger2019accurate,nakano2017advantages,van2018third,mantere2019long,amarasinghe2020opportunities}}, ONT~\damlaIII{\cite{cali2017nanopore,lu2016oxford,magi2017nanopore,clarke2009continuous,deamer2016three,marx2015nanopores,branton2008potential,laver2015assessing,ip2015minion,kasianowicz1996characterization,jain2018nanopore,quick2014reference}}) read sequencing technologies.
For seeding, we base \mech on a memory-efficient minimizer-based seeding algorithm, and for alignment, we develop a new bitvector-based, highly-parallel sequence-to-graph alignment algorithm. We \emph{co-design} both of our algorithms with high-performance, area- and power-efficient hardware accelerators.
    \begin{enumerate} 
    \squeezeme 
    \item To our knowledge, \mech is the first acceleration framework for sequence-to-graph mapping and alignment. \mech aims to 
    \sgh{alleviate existing performance bottlenecks}
    for both short and long read \sgh{analysis}.
    \item We propose \ms, the first hardware accelerator for minimizer-based seeding. \ms can be used for the seeding steps of both graph-based mapping and traditional \sgh{sequence-to-sequence} mapping.
    \item We propose \ba, the first hardware accelerator for sequence-to-graph alignment. \ba is based upon a new bitvector-based sequence-to-graph alignment algorithm \sgh{that we develop}, and can be also used as a sequence-to-sequence aligner.
    \item We couple \mech with high-bandwidth memory (HBM) to enable more effective data movement, exploit the high internal bandwidth, and improve the overall performance and energy efficiency.
    \item We evaluate \mech using a combination of accelerator synthesis and detailed performance modeling. We find that \mech is significantly more efficient than state-of-the-art sequence-to-graph mapping and sequence-to-graph alignment tools.
    \end{enumerate} 
}
\end{enumerate}

\section{Dissertation Outline}

This dissertation is organized into 8 chapters. 
Chapter 2 describes necessary background on genome sequencing, sequencing technologies, genome sequence analysis, and genome graphs. Chapter 3 presents our experimental study of the genome assembly pipeline using long reads. 
Chapter 4 presents GenASM, a high-performance and low-power approximate string matching acceleration framework for genome sequence analysis. 
Chapter 5 presents BitMAc, an FPGA-based near-memory prototype of the GenASM accelerators.
Chapter 6 presents \mech, the first hardware acceleration framework for sequence-to-graph mapping.
\damla{Chapter 7 presents the \damlaII{expected} long-term impact of the works in this dissertation and more generally, accelerating genome sequence analysis.
Finally, Chapter 8} presents conclusions and future research directions that are enabled by this dissertation.

\chapter{Background} 

We describe the necessary background on genome sequencing, genome sequence analysis, genome assembly and read mapping pipelines, \damla{genome graphs, and sequence-to-graph mapping} to help the reader to understand our observations and proposed designs for accelerating genome sequence analysis. 

\section{Genome Sequencing}\label{sec:background:sequencing}

\revII{Genome sequencing, which determines the DNA sequence of an organism,}
\revonur{plays a pivotal role in
enabling many medical and scientific} advancements in \revonur{personalized medicine~\cite{alkan2009personalized,flores2013p4,ginsburg2009genomic,chin2011cancer,Ashley2016}, evolutionary theory~\cite{ellegren2014genome,Prado-Martinez2013,Prohaska2019}, and forensics~\cite{yang2014application,borsting2015next,alvarez2017next}}.
Modern genome sequencing machines\revonur{~\cite{cali2017nanopore,minionwebpage,gridionwebpage,promethionwebpage,sequelwebpage,miseqwebpage,nextseqwebpage,novaseqwebpage}} can rapidly generate massive amounts of genomics data at low cost~\cite{Shendure2017,alser2020accelerating,mutlu2019aacbb,alser2020technology}, but are unable to extract \revII{an} organism's \revII{complete} DNA in one piece.  Instead, these machines \revII{extract} 
\revII{smaller random} fragments of the original DNA sequence, known as \emph{reads}.


\sg{State-of-the-art sequencing machines produce \revII{broadly} one of two kinds of reads.
\emph{Short reads} (consisting of no more than a few hundred DNA base \revonur{pairs~\cite{chaisson2004fragment,trapnell2009map}}) are generated using \revII{short-read sequencing (SRS)} technologies\revonur{~\cite{reuter2015high,van2014ten}, which have been on the market for more than a decade}. Because each read fragment is so short compared to the entire DNA (e.g., a \sgrev{human's} DNA consists of over 3~billion base \revonur{pairs}\revII{~\cite{venter2001sequence}}), short reads incur a number of reproducibility (e.g., non-deterministic mapping) and computational challenges\revIII{~\cite{firtina2016genomic,Xin2013,Xin2015,alkan2011limitations,treangen2011repetitive,alser2020technology,alser2017gatekeeper,xin2016optimal,mutlu2019aacbb}}.
\emph{Long reads} (\revII{consisting of} \revonur{\revIII{thousands to} millions of DNA base pairs}) are generated using \revII{long-read sequencing (LRS) technologies, of which Oxford Nanopore Technologies’ (ONT) nanopore sequencing~\cite{cali2017nanopore,lu2016oxford,magi2017nanopore,clarke2009continuous,deamer2016three,marx2015nanopores,branton2008potential,laver2015assessing,ip2015minion,kasianowicz1996characterization,jain2018nanopore,quick2014reference} and Pacific Biosciences’ (PacBio) single-molecule real-time (SMRT) sequencing~\cite{english2012mind,roberts2013advantages,rhoads2015pacbio,wenger2019accurate,nakano2017advantages,van2018third,mantere2019long,amarasinghe2020opportunities}} are the most \revII{widely used ones}. 
\revIII{LRS} technologies are relatively \revII{new}, and they avoid} many of the challenges faced by short reads.

\revII{LRS technologies} have \revII{three} key advantages compared to \revII{SRS technologies}.
\revII{First, LRS devices can generate very long reads, which 
\sgii{(1)~reduces the non-deterministic mapping problem faced by short reads, as long reads are significantly more likely to be unique
and therefore have fewer potential mapping locations in the reference genome; and
(2)~span larger parts of the \revIII{repeated or} complex regions of a genome, enabling detection of \revIII{genetic variations} \revIV{that might} exist in these regions~\cite{van2018third}.}
Second, \sgii{LRS \revIII{devices
perform} real-time sequencing, and can enable concurrent sequencing and analysis~\cite{quick2016real,roberts2013advantages,logsdon2020long}.}
Third, ONT's pocket-sized device (MinION~\cite{minionwebpage}) provides portability, making sequencing possible at remote places using laptops or mobile devices.}
This enables a number of new applications, 
such as rapid infection diagnosis and outbreak tracing (e.g., COVID-19, Ebola, Zika, \revIII{swine flu\revonur{~\cite{quick2016real,wu2020new,harcourt2020isolation,james2020lampore,da2020evolution,greninger2015rapid,wang2015minion,faria2016mobile}}).
Unfortunately,} \revII{LRS} devices are much more error-prone in sequencing (with a typical error rate of 10--15\%~\cite{jain2018nanopore, weirather2017comprehensive,ardui2018single,van2018third}) compared to \revII{SRS} \revonur{devices} (typically 0.1\%~\cite{glenn2011field,quail2012tale,goodwin2016coming}), \revonur{which leads to new computational challenges\revIII{~\cite{cali2017nanopore}}.}

\section{Genome Sequence Analysis}\label{sec:background:gsa}

Since the whole genome of most organisms cannot be sequenced all at once, the genome is broken into smaller fragments. After each fragment is sequenced, small pieces of DNA sequences (i.e., \textit{reads}) are generated. The locations of the sample fragments on the whole genome are usually random. Thus, the sequences of DNA fragments (i.e., \textit{reads}) should pass through computational mechanisms to gather meaningful information out of them, which is called \emph{genome sequence analysis}.

There are two types of genome sequence analysis mechanisms: (1)~assemble the reads without a template reference sequence (i.e., \textit{de novo assembly}), and (2)~map the reads with respect to a reference sequence (i.e., \textit{read mapping}).



\section{Genome Assembly Pipeline Using Long Reads}\label{sec:background:assembly}


\begin{figure}[b!]
\centering
\includegraphics[width=\columnwidth,keepaspectratio]{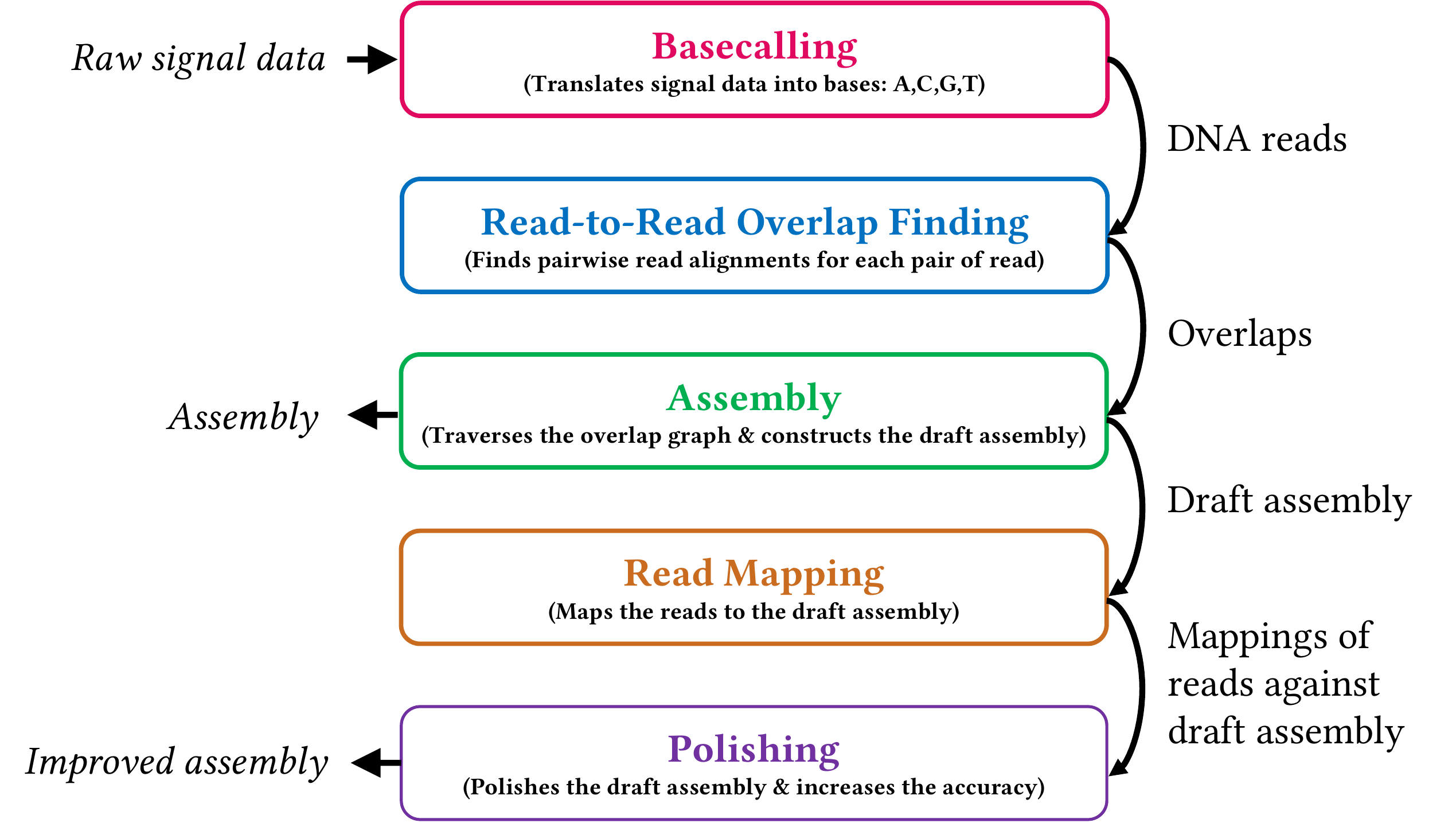}
\caption{Genome assembly pipeline using long reads, with its five steps and the associated tools for each step.}\label{fig:assembly-pipeline}
\end{figure}

Figure~\ref{fig:assembly-pipeline} shows each step of the genome assembly pipeline using long reads. 
The output of nanopore sequencers is raw signal data that represents changes in electric current when a DNA strand passes through nanopore. Thus, the pipeline starts with the raw signal data. The first step, \emph{basecalling}, translates this raw signal output of MinION into bases (A, C, G, T) to generate DNA reads. It is important to note that basecalling is the only step unique to nanopore sequencing, and the rest of the steps are applicable for both of the long read sequencing technologies (ONT and PacBio). The second step computes all pairwise read alignments or suffix-prefix matches between each pair of reads, called \emph{read-to-read overlaps}. Overlap-layout-consensus (OLC) algorithms are used for the assembly of nanopore sequencing reads since OLC-algorithms perform better with longer error-prone reads \cite{pop2009genome}. OLC-based assembly algorithms generate an \emph{overlap graph}, where each node denotes a read and each edge represents the suffix-prefix match between the corresponding two nodes. The third pipeline step, \emph{genome assembly}, traverses this overlap graph, producing the layout of the reads and then constructing a draft assembly. To increase the accuracy of the assembly, further \emph{polishing}, i.e., post-assembly error correction, may be required. The fourth step of the pipeline is mapping the original basecalled reads to the generated draft assembly from the previous step (i.e., read mapping). The fifth and final step of the pipeline is polishing the assembly with the help of mappings from the previous step.

\section{Read Mapping Pipeline}\label{sec:background:mapping}

\begin{figure}[t!]
\centering
\includegraphics[width=\columnwidth,keepaspectratio]{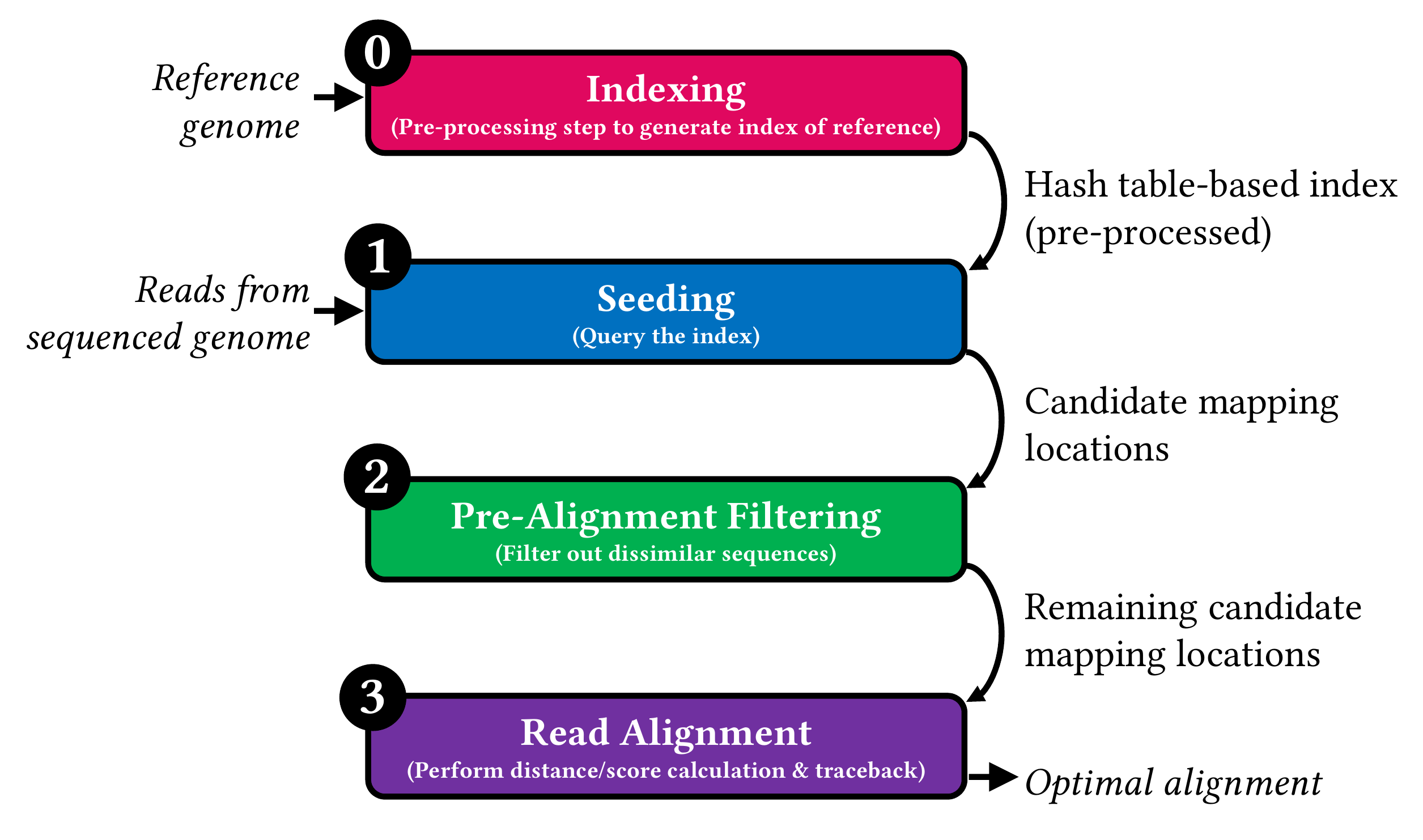}
\caption{Four steps of read mapping.} 
\label{fig:pipeline}
\end{figure}

Another common approach for \revonur{genome sequence analysis} is to perform \revII{\textit{read mapping}, where \sgii{each \textit{read} of an organism's sequenced} genome \sgii{is} matched against the \textit{reference genome for the organism's species} to find \sgii{the read's original location.}}
As Figure~\ref{fig:pipeline} shows, \revonur{typical}
read mapping\revonur{~\cite{li2018minimap2,li2013aligning,alkan2009personalized,Xin2013,langmead2012fast,li2009soap2}} is a four-step process, \damlaII{which is also known as \textit{seed-and-extend strategy}}. 
First, read mapping starts with \textit{indexing}~\circlednumber{0}, \revII{which is an offline pre-processing step performed on a known reference genome}. Second, \sgrev{once a sequencing machine generates reads from a DNA sequence, the} \textit{seeding} \sgrev{process}~\circlednumber{1} queries the index structure \sgii{to determine} \revII{the candidate \revIII{(i.e., potential)} mapping locations} of each read 
in the reference genome using substrings (i.e., \textit{seeds}) from each read. Third, \sgii{for each read,} \textit{pre-alignment filtering}~\circlednumber{2} uses filtering heuristics to examine the similarity between \sgii{a} read and 
\sgii{the portion of the reference genome at each of the read's candidate mapping locations.}
These filtering heuristics aim to eliminate most of the dissimilar \revII{pairs of reads and candidate \sgii{mapping locations}} to decrease the number of required alignments \revIII{in the next step}. Fourth, for all of the \sgii{remaining} candidate \revII{mapping locations}, \textit{read alignment}~\circlednumber{3}
\revII{runs a dynamic programming based algorithm} to determine 
\sgii{which of the candidate mapping locations in the reference matches best with the input read.}
As part of this step, traceback \sgii{is performed} between the reference 
and the \revonur{input \sgii{read to} find the \emph{optimal alignment}, which is the alignment with the highest \sgrev{likelihood of being correct (based on a scoring \revII{function}\revIII{~\cite{gotoh1982improved,miller1988sequence,waterman1984efficient}})}}.
\revonur{\sgrev{The optimal} alignment is defined using a \emph{CIGAR string}\revII{~\cite{li2009sequence}}, which shows the sequence \sgrev{and position of each match, substitution, insertion, and deletion for the read \revII{with respect to the \sgii{selected mapping location}} of the reference.}}

\damla{
\section{Genome Graphs}\label{sec:background:graphs}

\zulal{Genetic variation between individuals is observed by \sgh{comparing the differences between their two genomes.}
These differences, such as single-nucleotide polymorphisms (i.e., SNPs), insertions and deletions (i.e., indels), and structural variations (i.e., SVs), maintain genetic diversity between populations and within communities~\cite{10002015global}. However, the presence of these genomic portions creates limitations for mapping the sequenced reads to a reference genome~\cite{paten2017genome, degner2009effect, brandt2015mapping, gunther2019presence}, since the reference is commonly represented as a linear 
DNA sequence~\cite{schneider2017evaluation}. \sgh{Using a} single reference introduces \emph{reference allele bias}, by emphasizing the \sgh{\emph{alleles} (i.e., gene variants) that} are present in the reference \sgh{individual}~\cite{paten2017genome}. \damlaII{Alternate locus sequences~(i.e., ALT, alternative subsequences for diverging regions of the reference DNA sequence~\cite{alt,jager2016alternate}) produced along with recent linear reference versions} or utilizing pangenome models to include the collection of population genomes~\cite{computational2018computational, paten2017genome} can alleviate the effect of reference allele bias. These factors lead to low read mapping accuracy around the diversity regions~(SNPs, indels and SVs) and eventually cause false detection of SVs~\cite{rakocevic2019fast}. \damlaII{Thus,} the current practices for variant detection techniques mostly depend on complex combinations of alignment patterns~\cite{alkan2011genome}. 

Sequence graphs of \sgh{a genome are better suited} for expressing the \sgh{differences or ambiguities in diversity regions than linear reference sequences}~\cite{garrison2018variation}. Therefore, there is a \sgh{growing trend} towards utilizing genome graphs~\cite{pevzner2001eulerian, nurk2021complete, rautiainen2020graphaligner, garrison2018variation, rakocevic2019fast, kim2019graph, dilthey2015improved} \sgh{to more efficiently and accurately express the reference and its associated} diversity annotations. Genome graphs are also more effective for presenting pangenomes~\cite{paten2017genome} and extending the linear reference with alternate locus sequences to mitigate reference allele bias~\cite{degner2009effect}.}

\damlaII{Genome graphs represent the reference genome \emph{and} known genetic variations in the population as a graph-based data structure. As we show in Figure~\ref{fig:genomegraph-example}, a node represents one or more base pairs, and edges connect the base pairs in a node to all of the possible base pairs that come next in the sequence, with multiple outgoing edges from a node capture genetic variations. Thus, different paths in the graph translate to different sequences.}

\begin{figure}[h!]
\centering
\includegraphics[width=\columnwidth,keepaspectratio]{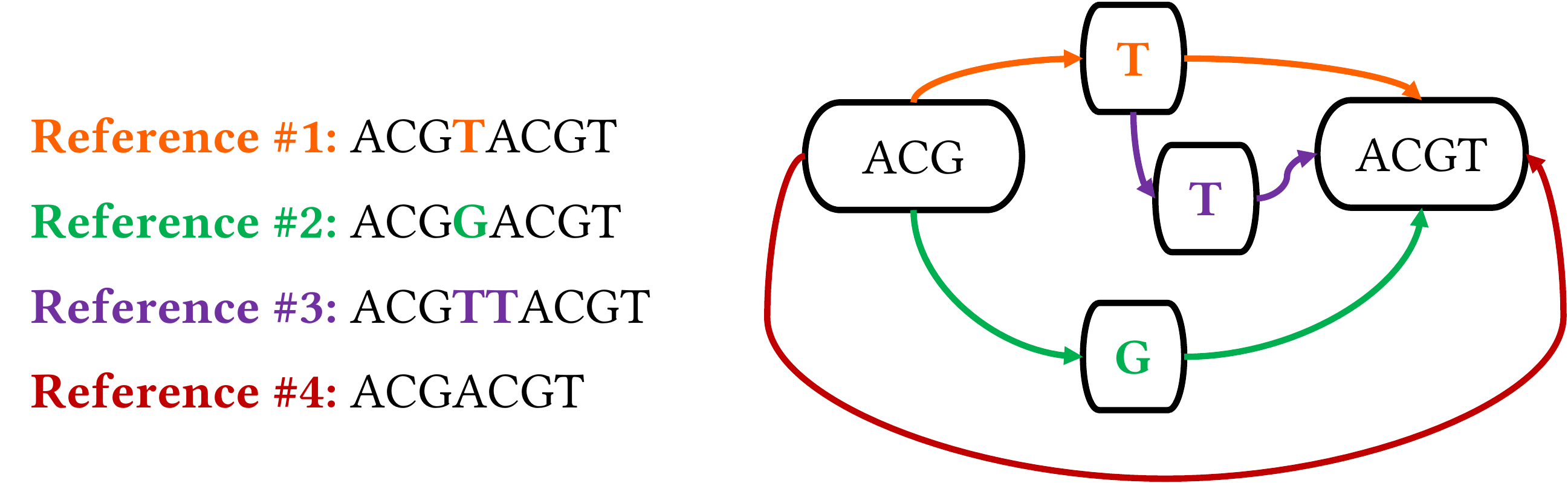}
\caption{Example of a genome graph that represents 4 related but different genomic sequences.} 
\label{fig:genomegraph-example}
\end{figure}

\damlaII{Genome graphs are growing in popularity for a number of applications, such as variation calling~\cite{garrison2018variation}, genome assembly~\cite{compeau2011apply, pevzner2001eulerian, zerbino2008velvet, simpson2009abyss}, error correction~\cite{salmela2014lordec}, and multiple sequence alignment~\cite{paten2011cactus, lee2002multiple}.} With an increasing importance and usage of genome graphs, having efficient tools for \damla{mapping genomic} sequences to these graphs become crucial. 

\section{Sequence-to-Graph Mapping}\label{sec:background-genomemapping}

Similar to conventional sequence-to-sequence mapping (Section~\ref{sec:background:mapping}), sequence-to-graph mapping follows the \emph{seed-and-extend strategy}. After constructing the graph using a linear reference genome and the associated variations for that genome, the nodes' of the graph are indexed as a pre-processing step. Later, this index is used in the \emph{seeding} step, which aims to find seed matches between the query read and a region of the graph. After optionally clustering or filtering these seed matches with a \emph{filtering} or \emph{chaining} step, \emph{alignment} is performed between all of the remaining seed locations of the graph and the query read. Even though sequence-to-sequence mapping is a well-studied problem, sequence-to-graph mapping is a newer problem and there are \damlaII{only} a few existing tools, which are optimized for \damlaII{only short reads or only long reads}, or specialized for a specific use cases. 

} 
 
\chapter{Bottleneck Analysis of the Genome Assembly Pipeline Using Long Reads}
\label{ch3-nanopore}

Due to the repetitive regions in the genome, the short-read length of the most dominant NGS technologies (\eg100-150 bp reads) causes errors and ambiguities for read mapping \cite{treangen2011repetitive,firtina2016genomic}, and poses computational challenges and accuracy problems to \emph{de novo} assembly \cite{alkan2011limitations}. Repetitive sequences are usually longer than the length of a short read and an entire repetitive sequence cannot be spanned by a single short read. Thus, short reads lead to highly-fragmented, incomplete assemblies \cite{lu2016oxford,alkan2011limitations,magi2017nanopore}. However, a long read can span an entire repetitive sequence and enable continuous and complete assemblies.The demand for sequencing technologies that can produce longer reads has resulted in the emergence of even newer alternative sequencing technologies.

Nanopore sequencing technology \cite{clarke2009continuous} is one example of such technologies that can produce long read lengths. Nanopore sequencing is an emerging single-molecule DNA sequencing technology, which exhibits many attractive qualities, and in time, it could potentially surpass current sequencing technologies. Nanopore sequencing promises high sequencing throughput, low cost, and longer read length, and it does \emph{not} require an amplification step before the sequencing process \cite{marx2015nanopores, branton2008potential, laver2015assessing, ip2015minion}.

Using biological nanopores for DNA sequencing was first proposed in the 1990s \cite{kasianowicz1996characterization}, but the first nanopore sequencing device, MinION \cite{minionwebpage}, was only recently (in May 2014) made commercially available by Oxford Nanopore Technologies (ONT). MinION is an inexpensive, pocket-sized, portable, high-throughput sequencing apparatus that produces data in real-time. These properties enable new potential applications of genome sequencing, such as rapid surveillance of Ebola, Zika or other epidemics \cite{quick2016real}, near-patient testing \cite{quick2014reference}, and other applications that require real-time data analysis. In addition, the MinION technology has two major advantages. First, it is capable of generating ultra-long reads (\eg 882 kilobase pairs or longer \cite{jain2018nanopore,lomanultralong}). MinION's long reads greatly simplify the genome assembly process by decreasing the computational requirements \cite{lu2016oxford,madoui2015genome}. Second, it is small and portable. MinION is named as the first DNA sequencing device used in outer space to help the detection of life elsewhere in the universe with the help of its size and portability \cite{nasawebpage}. With the help of continuous updates to the MinION device and the nanopore chemistry, the first nanopore human reference genome was generated by using only MinION devices \cite{jain2018nanopore}.\par
Nanopores are suitable for sequencing because they:

\begin{itemize}
\item Do not require any labeling of the DNA or nucleotide for detection during sequencing,
\item Rely on the electronic or chemical structure of the different nucleotides for identification,
\item Allow sequencing very long reads, and
\item Provide portability, low cost, and high throughput.
\end{itemize}

Despite all these advantageous characteristics, nanopore sequencing has one major drawback: high error rates. In May 2016, ONT released a new version of MinION with a new nanopore chemistry called R9 \cite{r9webpage}, to provide higher accuracy and higher speed, which replaced the previous version R7. Although the R9 chemistry improves the data accuracy, the improvements are not enough for cutting-edge applications. Thus, nanopore sequence analysis tools have a critical role to overcome high error rates and to take better advantage of the technology. Also, \emph{faster} tools are critically needed to 1) take better advantage of the real-time data production capability of MinION and 2) enable real-time data analysis.

Our goal in this work is to comprehensively analyze current publicly-available tools for nanopore sequence analysis
to understand their advantages, disadvantages, and bottlenecks. It is important to understand where the current tools do \emph{not} perform well, to develop better tools. To this end, we analyze the tools associated with the multiple steps in the genome assembly pipeline using nanopore sequence data in terms of accuracy, speed, memory efficiency, and scalability.

\section{Steps and Tools} \label{sec:steps-and-tools}

\subsection{Basecalling}\label{sec:basecalling}
When a strand of DNA passes through the nanopore (which is called the \emph{translocation} of the strand through the nanopore), it causes drops in the electric current passing between the walls of the pore. The amount of change in the current depends on the type of base passing through the pore. Basecalling, the initial step of the entire pipeline, translates the raw signal output of the nanopore sequencer into bases (A, C, G, T) to generate DNA reads. Most of the current basecallers divide the raw current signal into discrete blocks, which are called \emph{events}. After event-detection, each event is decoded into a most-likely set of bases. In the ideal case, each consecutive event should differ by one base. However, in practice, this is not the case because of the non-stable speed of the translocation. Also, determining the correct length of the homopolymers (\ie repeating stretches of one kind of base, \eg AAAAAAA) is challenging. Both of these problems make \emph{deletions} the dominant error of nanopore sequencing \cite{clivebrownwebpage,de2017sequencer}. Thus, basecalling is the most important step of the pipeline that plays a critical role in decreasing the error rate.

In this work, we analyze five state-of-the-art basecalling tools (Table~\ref{table:step1}). For a detailed comparison of these and other basecallers (including Albacore \cite{albacorenews}, which is not freely available,  and Chiron \cite{teng2018chiron}), we refer the reader to an ongoing basecaller comparison study \cite{basecallercomparisongithub}. Note that this ongoing study does \emph{not} capture the accuracy and performance of the entire genome assembly pipeline using nanopore sequence data.

\begin{table}[h!]
\small
\caption{State-of-the-art nanopore basecalling tools.}\label{table:step1}
\begin{tabularx}{\textwidth}{p{2cm} p{2cm} p{8cm} p{2cm}}
\hline
Tool & Strategy & Multi-threading Support & Reference\\ 
\hline
Metrichor & RNN & (cloud-based) & \cite{metrichorwebpage}\\
Nanonet & RNN & with \texttt{-jobs} parameter & \cite{nanonetwebpage}\\
Scrappie & RNN & with \texttt{export OMP\_NUM\_THREADS} command & \cite{scrappiewebpage}\\
Nanocall & HMM & with \texttt{--threads} parameter & \cite{david2016nanocall,nanocallgithub}\\
DeepNano & RNN & no support; split dataset and run it in parallel & \cite{bovza2017deepnano,deepnanogithub}\\
\hline
\end{tabularx}
\end{table}

\subsubsection*{Metrichor}
Metrichor \cite{metrichorwebpage} is ONT's cloud-based basecaller, and its source code is not publicly available. Before the R9 update, Metrichor was using Hidden Markov Models (HMM) \cite{eddy1996hidden} for basecalling \cite{r9webpage}. After the R9 update, it started using recurrent neural networks (RNN) \cite{schuster1997bidirectional, pearlmutter2008learning} for basecalling \cite{r9webpage}.

\subsubsection*{Nanonet}
Nanonet \cite{nanonetwebpage} has also been developed by ONT, and it is available on Github. Since Metrichor requires an Internet connection and its source code is not available, Nanonet is an offline and open-source alternative for Metrichor. Nanonet is implemented in Python. It also uses RNN for basecalling \cite{nanonetwebpage}. The tool supports multi-threading by sharing the computation needed to call each single read between concurrent threads. In other words, only one read is called at a time.

\subsubsection*{Scrappie}
Scrappie \cite{scrappiewebpage} is the newest proprietary basecaller developed by ONT. It is named as the first basecaller that explicitly addresses basecalling errors in homopolymer regions. In order to determine the correct length of homopolymers, Scrappie performs transducer-based basecalling \cite{clivebrownwebpage}. For versions R9.4 and R9.5, Scrappie can perform basecalling with the raw current signal, without requiring event detection. It is a C-based local basecaller and is still under development \cite{clivebrownwebpage}.

\subsubsection*{Nanocall}
Nanocall \cite{david2016nanocall} uses Hidden Markov Models for basecalling, and it is independently developed by a research group. It was released before the R9 update when Metrichor was also using an HMM-based approach for basecalling, to provide the first offline and open-source alternative for Metrichor. However, after the R9 update, when Metrichor started to perform basecalling with a more powerful RNN-based approach, Nanocall's accuracy fell short of Metrichor's accuracy \cite{nanocallgithub}. Thus, although Nanocall supports R9 and upper versions of nanopore data, its usefulness is limited \cite{nanocallgithub}. Nanocall is a C++-based command-line tool. It supports multi-threading where each thread performs basecalling for \emph{different} groups of raw reads.

\subsubsection*{DeepNano}
DeepNano \cite{bovza2017deepnano} is also independently developed by a research group before the R9 update. It uses an RNN-based approach to perform basecalling. Thus, it is considered to be the first RNN-based basecaller. DeepNano is implemented in Python. It does not have multi-threading support.

\subsection{Read-to-Read Overlap Finding}\label{sec:overlapfinding}
Previous genome assembly methods designed for accurate and short reads (\ie de Bruijn graph (DBG) approach \cite{pevzner2001eulerian,compeau2011apply}) are not suitable for nanopore reads because of the high error rates of the current nanopore sequencing devices \cite{koren2013reducing,de2017sequencer,magi2017nanopore,chu2016innovations}. Instead, overlap-layout-consensus (OLC) algorithms \cite{li2012comparison} are used for nanopore sequencing reads since they perform better with longer, error-prone reads. OLC-based assembly algorithms start with finding the read-to-read overlaps, which is the second step of the pipeline. Read-to-read overlap is defined to be a common sequence between two reads \cite{chu2016innovations}. GraphMap \cite{sovic2016fast} and Minimap \cite{li2016minimap} are the commonly-used state-of-the-art tools for this step (Table~\ref{table:step2}).

\begin{table}[h!]
\small
\caption{State-of-the-art read-to-read overlap finding tools.}\label{table:step2}
\begin{tabularx}{\textwidth}{p{2cm} p{4cm} p{6cm} p{2cm}}
\hline
Tool & Strategy & Multi-threading Support & Reference\\ 
\hline
GraphMap & \textit{k}-mer similarity & with \texttt{--threads} parameter & \cite{sovic2016fast,graphmapgithub}\\
Minimap & minimizer similarity & with \texttt{-t} parameter & \cite{li2016minimap,minimapgithub}\\
\hline
\multicolumn{4}{l}{Note: Both GraphMap and Minimap also have read mapping functionality.}
\end{tabularx}
\end{table}

\subsubsection*{GraphMap}
GraphMap first partitions the entire read dataset into \textit{k}-length substrings (i.e. \textit{k}-mers), and then creates a hash table. GraphMap uses gapped \textit{k}-mers, \ie k-mers that can contain insertions or deletions (indels) \cite{sovic2016fast, burkhardt2003better}. In the hash table, for each \textit{k}-mer entry, three pieces of information are stored: 1) {k}-mer string, 2) the index of the read, and 3) the position in the read where the corresponding \textit{k}-mer comes from. GraphMap detects the overlaps by finding the \textit{k}-mer similarity between any two given reads. Due to this design, GraphMap is a highly sensitive and accurate tool for error-prone long reads. It is a command-line tool written in C++. GraphMap is used for both 1) read-to-read overlap finding with the \texttt{graphmap owler} command and 2) read mapping with the \texttt{graphmap align} command. 

\subsubsection*{Minimap}
Minimap also partitions the entire read dataset into \textit{k}-mers, but instead of creating a hash table for the full set of \textit{k}-mers, it finds the minimum representative set of \textit{k}-mers, called \emph{minimizers}, and creates a hash table with only these minimizers. Minimap finds the overlaps between two reads by finding minimizer similarity. The goals of using minimizers are to 1) reduce the storage requirement of the tool by storing fewer \textit{k}-mers and 2) accelerate the overlap finding process by reducing the search span. Minimap also sorts \textit{k}-mers for cache efficiency. Minimap is fast and cache-efficient, and it does not lose any sensitivity by storing minimizers since the chosen minimizers can represent the whole set of \textit{k}-mers. Minimap is a command-line tool written in C. Like GraphMap, it can both 1) find overlaps between two read sets and 2) map a set of reads to a full genome.

\subsection{Genome Assembly}\label{sec:assembly}
After finding the read-to-read overlaps, OLC-based assembly algorithms generate an \emph{overlap graph}. Genome assembly is performed by traversing this graph, producing the layout of the reads and then constructing a draft assembly. Canu \cite{koren2017canu} and Miniasm \cite{li2016minimap} are the commonly-used error-prone long-read assemblers (Table~\ref{table:step3}).

\begin{table}[h!]
\small
\caption{State-of-the-art assembly tools.}\label{table:step3}
\begin{tabularx}{\textwidth}{p{2cm} p{5cm} p{5cm} p{2cm}}
\hline
Tool & Strategy & Multi-threading Support & Reference\\ 
\hline
Canu & OLC with error correction & auto configuration & \cite{koren2017canu,canugithub}\\
Miniasm & OLC without error correction & no support & \cite{li2016minimap,miniasmgithub}\\
\hline
\end{tabularx}
\end{table}

\subsubsection*{Canu}
Canu performs error-correction as the initial step of its own pipeline. It finds the overlaps of the raw uncorrected reads and uses them for the error-correction. The purpose of error-correction is to improve the accuracy of the bases in the reads \cite{koren2017canu,canu-webpage}. After the error-correction step, Canu finds overlaps between corrected reads and constructs a draft assembly after an additional trimming step. However, error-correction is a computationally expensive step. In its own pipeline, Canu implements its own read-to-read overlap finding tool such that the users do \emph{not} need to perform that step explicitly before running Canu. Most of the steps in the Canu pipeline are multi-threaded. Canu detects the resources that are available in the computer before starting its pipeline and automatically assigns number of threads, number of processes and amount of memory based on the available resources and the assembled genome's estimated size.

\subsubsection*{Miniasm}
Miniasm skips the error-correction step since it is computationally expensive. It constructs a draft assembly from the uncorrected read overlaps computed in the previous step. Although Miniasm lowers computational cost and thus accelerates and simplifies assembly by doing so, the accuracy of the draft assembly depends directly on the accuracy of the uncorrected basecalled reads. Thus, further polishing may be necessary for these draft assemblies. Miniasm does not support multi-threading.

\subsection{Read Mapping and Polishing}\label{sec:polish}
In order to increase the accuracy of the assembly, especially for the rapid assembly methods like Miniasm, which do not have the error-correction step, further polishing may be required. Polishing, \ie post-assembly error-correction, improves the accuracy of the draft assembly by mapping the reads to the assembly and changing the assembly to increase local similarity with the reads \cite{loman2015complete,vaser2017fast,de2017sequencer}. The first step of polishing is mapping the basecalled reads to the generated draft assembly from the previous step. One of the most commonly-used long read mappers for nanopore data is BWA-MEM \cite{li2013aligning}. Read-to-read overlap finding tools, GraphMap and Minimap (Section~\ref{sec:overlapfinding}), can also be used for this step, since they also have a read mapping mode (Table~\ref{table:step4}).\par
After aligning the basecalled reads to the draft assembly, the final polishing of the assembly can be performed with Nanopolish \cite{loman2015complete} or Racon \cite{vaser2017fast} (Table~\ref{table:step5}).

\begin{table}[h!]
\small
\caption{State-of-the-art read mapping tools.}\label{table:step4}
\begin{tabularx}{\textwidth}{p{2cm} p{5cm} p{5cm} p{2cm}}
\hline
Tool & Strategy & Multi-threading Support & Reference\\ 
\hline
BWA-MEM & Burrows-Wheeler Transform & with \texttt{-t} parameter & \cite{li2013aligning,bwagithub}\\
GraphMap & \textit{k}-mer similarity & with \texttt{--threads} parameter & \cite{sovic2016fast,graphmapgithub}\\
Minimap & minimizer similarity & with \texttt{-t} parameter & \cite{li2016minimap,minimapgithub}\\
\hline
\end{tabularx}
\end{table}

\begin{table}[h!]
\small
\caption{State-of-the-art polishing tools.}\label{table:step5}
\begin{tabularx}{\textwidth}{p{2cm} p{4.9cm} p{5.1cm} p{2cm}}
\hline
Tool & Strategy & Multi-threading Support & Reference\\ 
\hline
Nanopolish & Hidden Markov Model & with \texttt{--threads}, \texttt{-P} parameters & \cite{loman2015complete,nanopolishgithub}\\
Racon & Partial order alignment graph & with \texttt{--threads} parameter & \cite{vaser2017fast,racongithub}\\
\hline
\end{tabularx}
\end{table}

\subsubsection*{Nanopolish}
Nanopolish uses the raw signal data of reads along with the mappings from the previous step to improve the assembly base quality by evaluating and maximizing the probabilities for each base with a Hidden Markov Model-based approach \cite{loman2015complete}. It can increase the accuracy of the draft assembly by correcting the homopolymer-rich parts of the genome. Although this approach can increase the accuracy significantly, it is computationally expensive, and thus time consuming. Nanopolish developers recommend BWA-MEM as the read mapper before running Nanopolish \cite{nanopolishgithub}.

\subsubsection*{Racon}
Racon constructs partial order alignment graphs \cite{lee2002multiple,vaser2017fast} in order to find a consensus sequence between the reads and the draft assembly. After dividing the sequence into segments, Racon tries to find the best alignment to increase the accuracy of the draft assembly. Racon is a fast polishing tool, but it does not promise a high increase in accuracy as Nanopolish promises. However, multiple iterations of Racon runs or a combination of Racon and Nanopolish runs can improve accuracy significantly. Racon developers recommend Minimap as the read mapper to use before running Racon, since Minimap is both fast and sensitive \cite{vaser2017fast}.
 
\section{Experimental Methodology} \label{sec:nanopore-methodology}

\subsection{Dataset}
In this work, we use \textit{Escherichia coli} genome data as the test case, sequenced using the MinION with an R9 flowcell \cite{datawebpage}.\par
MinION sequencing has two types of workflows. In the 1D workflow, only the template strand of the double-stranded DNA is sequenced. In contrast, in the 2D workflow, with the help of a hairpin ligation, both the template and complement strands pass through the pore and are sequenced. After the release of R9 chemistry, 1D data became very usable in contrast to previous chemistries. Thus, we perform the analysis of the tools on 1D data.\par
MinION outputs one file in the \emph{fast5} format for each read. The fast5 file format is a hierarchical data format, capable of storing both raw signal data and basecalled data returned by Metrichor. This dataset includes 164,472 reads, \ie fast5 files. Since all these files include both raw signal data and basecalled reads, we can use this dataset for both 1) using the local basecallers to convert raw signal data into the basecalled reads and 2) using the already basecalled reads by Metrichor.

\subsection{Evaluation Systems}
In this work, for accuracy and performance evaluations of different tools, we use three separate systems with different specifications. We use the first computer in the first part of the analysis, accuracy analysis. We use the second and third computers in the second part of the analysis, performance analysis, to compare the scalability of the analyzed tools in the two machines with different specifications (Table~\ref{table:specification}).

\begin{table*}[t!]
\footnotesize
\caption{Specifications of evaluation systems.}\label{table:specification}
\begin{threeparttable}
\begin{tabularx}{\textwidth}{p{1.4cm} p{3.0cm} p{3.0cm} p{3.0cm} p{3.0cm}}
\hline
Name & Model & \specialcel{CPU\\Specifications} & \specialcel{Main Memory\\Specifications} & \specialcel{NUMA\tnote{*}\\Specifications}\\ 
\hline
System 1 & \specialcel{40-core Intel\textsuperscript \textregistered Xeon\textsuperscript \textregistered\\E5-2630 v4 CPU\\@ 2.20GHz} & \specialcel{20 physical cores\\2 threads per core\\40 logical cores with\\hyper-threading\tnote{**}} &  \specialcel{128GB DDR4\\2 channels,\\2 ranks/channel\\Speed: 2400MHz} & 2 NUMA nodes, each with 10 physical cores, 64GB of memory and an 25MB of last level cache (LLC)\\
& & & & \\
\specialcel{System 2\\(\emph{desktop})} & \specialcel{8-core Intel\textsuperscript \textregistered Core\\i7-2600 CPU\\@ 3.40GHz} & \specialcel{4 physical cores\\2 threads per core\\8 logical cores with\\hyper-threading\tnote{**}} & \specialcel{16GB DDR3\\2 channels,\\2 ranks/channel\\Speed: 1333MHz} & 1 NUMA node, with 4 physical cores, 16GB of memory and an 8MB of LLC\\
& & & & \\
\specialcel{System 3\\(\emph{big-mem})} & \specialcel{80-core Intel\textsuperscript \textregistered Xeon\textsuperscript \textregistered\\E7-4850 CPU\\@ 2.00GHz} & \specialcel{40 physical cores\\2 threads per core\\80 logical cores with\\hyper-threading\tnote{**}} & \specialcel{1TB DDR3\\8 channels,\\4 ranks/channel\\Speed: 1066MHz} & 4 NUMA nodes, each with 10 physical cores, 256GB of memory and an 24MB of LLC\\
\hline
\end{tabularx}
\begin{tablenotes}
\item[*] NUMA (Non-Uniform Memory Access) is a computer memory design, where a processor accesses its local memory faster (\ie with lower latency) than a non-local memory (\ie memory local to another processor in another NUMA node). A NUMA node is composed of the local memory and the CPU cores (See Observation 6 in Section~\ref{sec:basecallresults} for detail).
\item[**] Hyper-threading is Intel's simultaneous multithreading (SMT) implementation (See Observation 5 in Section~\ref{sec:basecallresults} for detail).
\end{tablenotes}
\end{threeparttable}
\end{table*}

We choose the first system for evaluation since it has a larger memory capacity than a usual server and, with the help of a large number of cores, the tasks can be parallelized easily in order to get the output data quickly. We choose the second system, called \emph{desktop}, since it represents a commonly-used desktop server. We choose the third system, called \emph{big-mem}, because of its large memory capacity. This \emph{big-mem} system can be useful for those who would like to get results more quickly.

\subsection{Accuracy Metrics}
We compare each draft assembly generated after the assembly step and each improved assembly generated after the polishing step with the reference genome, by using the \texttt{dnadiff} command under the MUMmer package \cite{mummergithub}. We use six metrics to measure accuracy, as defined in Table~\ref{table:accuracymetrics}: 1) number of bases in the assembly, 2) number of contigs, 3) average identity, 4) coverage, 5) number of mismatches, and 6) number of indels.

\subsection{Performance Metrics}
We analyze the performance of each tool by running the associated command-line of each tool with the \texttt{/usr/bin/time -v} command. We use four metrics to quantify performance as defined in Table~\ref{table:perfmetrics}: 1) wall clock time, 2) CPU time, 3) peak memory usage, and 4) parallel speedup.

\begin{table*}[h!]
\footnotesize
\caption{Accuracy metrics.}\label{table:accuracymetrics}
\begin{tabularx}{\textwidth}{p{3.4cm} p{7.4cm} p{3.3cm}}
\hline
Metric Name & Definition & Preferred Values\\ 
\hline
Number of bases & Total number of bases in the assembly & $\simeq$ Length of reference genome\\
Number of contigs & Total number of segments in the assembly & Lower ($\simeq$ 1)\\
Average identity & Percentage similarity between the assembly and the reference genome & Higher ($\simeq$ 100\%)\\
Coverage & Ratio of the number of aligned bases in the reference genome to the length of reference genome & Higher ($\simeq$ 100\%)\\
Number of mismatches & Total number of single-base differences between the assembly and the reference genome & Lower ($\simeq$ 0)\\
Number of indels & Total number of insertions and deletions between the assembly and the reference genome & Lower ($\simeq$ 0)\\
\hline
\end{tabularx}
\end{table*}

\begin{table*}[h!]
\footnotesize
\caption{Performance metrics.}\label{table:perfmetrics}
\begin{threeparttable}
\begin{tabularx}{\textwidth}{p{3.4cm} p{7.4cm} p{3.3cm}}
\hline
Metric Name & Definition & Preferred Values\\ 
\hline
Wall clock time & Elapsed time from the start of a program to the end & Lower\\
CPU time & Total amount of time the CPU spends in user mode (\ie to run the program's code) and kernel mode (\ie to execute system calls made by the program)\tnote{*}& Lower\\
Peak memory usage & Maximum amount of memory used by a program during its whole lifetime & Lower\\
Parallel speedup & Ratio of the time to run a program with 1 thread to the time to run it with N threads & Higher\\
\hline
\end{tabularx}
\begin{tablenotes}
\item[*] If wall clock time < CPU time for a specific program, it means that the program runs in parallel. 
\end{tablenotes}
\end{threeparttable}
\end{table*}

\section{Results and Analysis} \label{sec:results-analysis}

In this section, we present our results obtained by analyzing the performance of different tools for each step in the genome assembly pipeline using nanopore sequence data in terms of accuracy and performance, using all the metrics we provide in Table~\ref{table:accuracymetrics} and Table~\ref{table:perfmetrics}. Additionally, Table~\ref{table:overalltools} shows the tool version, the executed command, and the output of each analyzed tool. We divide our analysis into three main parts.

\begin{table*}[h!]
\footnotesize
\caption{Versions, commands to execute, and outputs for each analyzed tool.}\label{table:overalltools}
\begin{threeparttable}
\begin{tabularx}{\textwidth}{p{3.4cm} p{7.2cm} Z}
\hline
& Command\tnote{*} & Output\\
\hline
\multicolumn{3}{l}{\textcolor{mypink}{\textbf{Basecalling Tools}}}\\
\hline
Nanonet--v2.0 & \texttt{nanonetcall fast5\_dir/ -{}-jobs N -{}-chemistry r9} & \texttt{reads.fasta}\\
Scrappie--v1.0.1& \specialcel{\texttt{(1)export OMP\_NUM\_THREADS=N}\\\texttt{(2)scrappie events -{}-segmentation}\\ \texttt{\textcolor{white}{(2)}Segment\_Linear:split\_hairpin}\\\texttt{\textcolor{white}{(2)}fast5\_dir/ ...}} & \specialcel{\\\\\texttt{reads.fasta}}\\
Nanocall--v0.7.4& \texttt{nanocall -t N fast5\_dir/} & \texttt{reads.fasta}\\
DeepNano--e8a621e & \texttt{python basecall.py --directory fast5\_dir/ -{}-chemistry r9} & \texttt{reads.fasta}\\
\hline
\multicolumn{3}{l}{\textcolor{myblue}{\textbf{Read-to-Read Overlap Finding Tools}}}\\
\hline
GraphMap--v0.5.2 & \texttt{graphmap owler -L paf -t N -r reads.fasta -d reads.fasta} & \texttt{overlaps.paf}\\
Minimap--v0.2 & \texttt{minimap -Sw5 -L100 -m0 -tN reads.fasta reads.fasta} & \texttt{overlaps.paf}\\
\hline
\multicolumn{3}{l}{\textcolor{mygreen}{\textbf{Assembly Finding Tools}}}\\
\hline
Canu--v1.6 & \texttt{canu -p ecoli -d canu-ecoli genomeSize=4.6m -nanopore-raw reads.fasta} & \texttt{draft.fasta}\\
Miniasm--v0.2 & \texttt{miniasm -f reads.fasta overlaps.paf} & \texttt{draft.gfa --> draft.fasta}\\
\hline
\multicolumn{3}{l}{\textcolor{myorange}{\textbf{Read Mapping Tools}}}\\
\hline
BWA-MEM--0.7.15 & \specialcel{\texttt{(1)bwa index draft.fasta}\\ \texttt{(2)bwa mem -x ont2d -t N}\\\texttt{\textcolor{white}{(2)}draft.fasta reads.fasta}} & \specialcel{\\\texttt{mappings.sam -->}\\\texttt{\textcolor{white}{ --> }mappings.bam}}\\
Minimap--v0.2 & \texttt{minimap -tN draft.fasta reads.fasta} & \texttt{mappings.paf}\\
\hline
\multicolumn{3}{l}{\textcolor{mypurple}{\textbf{Polishing Tools}}}\\
\hline
Nanopolish--v0.7.1 & \specialcel{\texttt{(1)python nanopolish\_makerange.py}\\\texttt{\textcolor{white}{(2)}draft.fasta | parallel -P M}\\ \texttt{(2)nanopolish variants -{}-consensus}\\\texttt{\textcolor{white}{(2)}polished.\verb|{1}|.fa -w \verb|{1}| -r reads.fasta}\\\texttt{\textcolor{white}{(2)}-b mappings.bam -g draft.fasta}\\\texttt{\textcolor{white}{(2)}-t N}\\\texttt{(3)python nanopolish\_merge.py}\\\texttt{\textcolor{white}{(3)}polished.*.fa}} & \specialcel{\\\\\\\\\\\\\texttt{polished.fasta}}\\
Racon--v0.5.0 & \texttt{racon (--sam) --bq -1 -t N reads.fastq mappings.paf/(mappings.sam) draft.fasta} & \specialcel{\\\\\texttt{polished.fasta}}\\
\hline
\end{tabularx}
\begin{tablenotes}
\item[*] N corresponds to the number of threads and M corresponds to the number of parallel jobs.
\end{tablenotes}
\end{threeparttable}
\end{table*}

In the first part of our analysis, we examine the first three steps of the pipeline (\emph{cf.} Figure~\ref{fig:pipeline}). To this end, we first execute each basecalling tool (\ie one of Nanonet, Scrappie, Nanocall or DeepNano). Since Metrichor is a cloud-based tool and its source code is not available, we cannot execute Metrichor and get the performance metrics for it. After recording the performance metrics for each basecaller run, we execute either GraphMap or Minimap followed by Miniasm, or Canu itself, and record the performance metrics for each run. We obtain a draft assembly for each combination of these basecalling, read-to-read overlap finding and assembly tools. For each draft assembly, we assess its accuracy by comparing the resulting draft assembly with the existing reference genome. We show the accuracy results in Table~\ref{table:accuracyfirst3}. We show the performance results in Table~\ref{table:perffirst3}. We will refer to these tables in Sections~\ref{sec:basecallresults} --~\ref{sec:assemblyresults}.

In the second part of our analysis, we examine the last two steps of the pipeline (\emph{cf.} Figure~\ref{fig:pipeline}). To this end, for each obtained draft assembly, we execute each possible combination of different read mappers (\ie BWA-MEM or Minimap) and different polishers (\ie Nanopolish or Racon), and record the performance metrics for each step (\ie read mapping and polishing). We obtain a polished assembly after each run, and assess its accuracy by comparing it with the existing reference genome. For these two analyses, we use the first system, which has 40 logical cores, and execute each tool using 40 threads, which is the possible maximum number of threads for that particular machine. We show the accuracy results in Table~\ref{table:accuracylast2}. We show the performance results in Table~\ref{table:perflast2}. We will refer to these tables in Section~\ref{sec:polishresults}.

In the third part of our analysis, we assess the scalability of all of the tools that have multi-threading support. For this purpose, we use the second and third systems to compare the scalability of these tools on two different system configurations. For each tool, we change the number of threads and observe the corresponding change in speed, memory usage, and parallel speedup. These results are depicted in Figures~\ref{fig:basecallplot} --~\ref{fig:nanopolishplot}, and we will refer to them throughout Sections~\ref{sec:basecallresults} --~\ref{sec:polishresults}.\par
Sections~\ref{sec:basecallresults} --~\ref{sec:polishresults} describe the major observations we make for each of the five steps of the pipeline (\emph{cf.} Figure~\ref{fig:pipeline}) based on our extensive evaluation results.

\begin{table*}[t!]
\scriptsize
\caption{Accuracy analysis results for the first three steps of the pipeline.}\label{table:accuracyfirst3}
\scalebox{0.95}{
\begin{tabular}{l | r r r r r r}
\hline
& \specialcell{\#\\Bases} & \specialcell{\#\\Contigs} & \specialcell{Identity\\(\%)} & \specialcell{Coverage\\(\%)} & \specialcell{\#\\Mismatches} & \specialcell{\#\\Indels}\\ 
\hline
\begin{tabular}{p{0.2cm} | p{1.2cm} l p{1.3cm} r p{1.2cm}}1 & \textcolor{mypink}{Metrichor} & + & --- & + & \textcolor{mygreen}{Canu}\end{tabular} & 4,609,499 & 1 & 98.05 & 99.92 & 12,378 & 76,990\\
\begin{tabular}{p{0.2cm} | p{1.2cm} l p{1.3cm} r p{1.2cm}}2 & \textcolor{mypink}{Metrichor} & + & \textcolor{myblue}{Minimap} & + & \textcolor{mygreen}{Miniasm}\end{tabular} & 4,402,675 & 1 & 87.71 & 94.85 & 249,096 & 372,704\\
\begin{tabular}{p{0.2cm} | p{1.2cm} l p{1.3cm} r p{1.2cm}}3 & \textcolor{mypink}{Metrichor} & + & \textcolor{myblue}{GraphMap} & + & \textcolor{mygreen}{Miniasm}\end{tabular}  & 4,500,155 & 2 & 86.22 & 96.95 & 237,747 & 360,199\\
\hline
\begin{tabular}{p{0.2cm} | p{1.2cm} l p{1.3cm} r p{1.2cm}}4 & \textcolor{mypink}{Nanonet} & + & --- & + & \textcolor{mygreen}{Canu}\end{tabular} & 4,581,728 & 1 & 97.92 & 99.97 & 11,971 & 83,248\\
\begin{tabular}{p{0.2cm} | p{1.2cm} l p{1.3cm} r p{1.2cm}}5 & \textcolor{mypink}{Nanonet} & + & \textcolor{myblue}{Minimap} & + & \textcolor{mygreen}{Miniasm}\end{tabular} & 4,350,175 & 1 & 85.50 & 92.76 & 237,518 & 394,852\\
\begin{tabular}{p{0.2cm} | p{1.2cm} l p{1.3cm} r p{1.2cm}}6 & \textcolor{mypink}{Nanonet} & + & \textcolor{myblue}{GraphMap} & + & \textcolor{mygreen}{Miniasm}\end{tabular} & 4,272,545 & 1 & 85.36 & 91.16 & 232,748 & 389,968\\
\hline
\begin{tabular}{p{0.2cm} | p{1.2cm} l p{1.3cm} r p{1.2cm}}7 & \textcolor{mypink}{Scrappie} & + & --- & + & \textcolor{mygreen}{Canu}\end{tabular}& 4,614,149 & 1 & 98.46 & 99.90 & 6,777 & 63,597\\
\begin{tabular}{p{0.2cm} | p{1.2cm} l p{1.3cm} r p{1.2cm}}8 & \textcolor{mypink}{Scrappie} & + & \textcolor{myblue}{Minimap} & + & \textcolor{mygreen}{Miniasm}\end{tabular} & 4,877,399 & 8 & 86.94 & 90.04 & 184,669 & 363,025\\
\begin{tabular}{p{0.2cm} | p{1.2cm} l p{1.3cm} r p{1.2cm}}9 & \textcolor{mypink}{Scrappie} & + & \textcolor{myblue}{GraphMap} & + & \textcolor{mygreen}{Miniasm}\end{tabular} & 4,368,417 & 1 & 86.78 & 89.86 & 189,192 & 372,245\\
\hline
\begin{tabular}{p{0.2cm} | p{1.2cm} l p{1.3cm} r p{1.2cm}}10& \textcolor{mypink}{Nanocall} & + & --- & + & \textcolor{mygreen}{Canu}\end{tabular} & 1,299,808 & 86 & 93.33 & 28.93 & 21,985 & 61,217\\
\begin{tabular}{p{0.2cm} | p{1.2cm} l p{1.3cm} r p{1.2cm}}11 & \textcolor{mypink}{Nanocall} & + & \textcolor{myblue}{Minimap} & + & \textcolor{mygreen}{Miniasm}\end{tabular} & 4,492,964 & 5 & 80.52 & 42.92 & 177,589 & 221,092\\
\begin{tabular}{p{0.2cm} | p{1.2cm} l p{1.3cm} r p{1.2cm}}12 & \textcolor{mypink}{Nanocall} & + & \textcolor{myblue}{GraphMap} & + & \textcolor{mygreen}{Miniasm}\end{tabular} & 4,429,390 & 3 & 80.51 & 41.32 & 171,455 & 213,435\\
\hline
\begin{tabular}{p{0.2cm} | p{1.2cm} l p{1.3cm} r p{1.2cm}}13 & \textcolor{mypink}{DeepNano} & + & --- & + & \textcolor{mygreen}{Canu}\end{tabular} & 7,151,596 & 106 & 92.75 & 99.16 & 38,803 & 211,551\\
\begin{tabular}{p{0.2cm} | p{1.2cm} l p{1.3cm} r p{1.2cm}}14 & \textcolor{mypink}{DeepNano} & + & \textcolor{myblue}{Minimap} & + & \textcolor{mygreen}{Miniasm}\end{tabular} & 4,252,525 & 1 & 82.38 & 65.00 & 199,122 & 335,761\\
\begin{tabular}{p{0.2cm} | p{1.2cm} l p{1.3cm} r p{1.2cm}}15 & \textcolor{mypink}{DeepNano} & + & \textcolor{myblue}{GraphMap} & + & \textcolor{mygreen}{Miniasm}\end{tabular}  & 4,251,548 & 1 & 82.39 & 64.92 & 197,914 & 335,064\\
\hline
\end{tabular}
}
\end{table*}

\subsection{Basecalling Tools}\label{sec:basecallresults}
As we discuss in Section~\ref{sec:basecalling}, ONT's basecallers Metrichor, Nanonet and Scrappie, and another basecaller developed by Boza \textit{et al}. (2017), DeepNano, use Recurrent Neural Networks (RNNs) for basecalling whereas Nanocall developed by David \etal (2016) uses Hidden Markov Models (HMM) for basecalling.

\subsubsection*{Accuracy}
Using RNNs is a more powerful basecalling approach than using HMMs since an RNN 1) does not make any assumptions about sequence length \cite{sutskever2014sequence} and 2) is not affected by the repeats in the sequence \cite{sutskever2014sequence,david2016nanocall,bovza2017deepnano}. However, it is still challenging to determine the correct length of the homopolymers even with an RNN.\par
In order to compare the accuracy of the analyzed basecallers, we group the accuracy results by each basecalling tool and compare them according to the defined accuracy metrics.\par
According to this analysis and the accuracy results shown in Table~\ref{table:accuracyfirst3}, we make the following key observation.\par
\textbf{Observation 1: }\textit{The pipelines that start with Metrichor, Nanonet, or Scrappie as the basecaller have similar identity and coverage trends among all of the evaluated scenarios (\ie tool combinations for the first three steps), but Scrappie has a lower number of mismatches and indels. However, Nanocall and DeepNano cannot reach these three basecallers' accuracies: they have lower identity and lower coverage.}\par
Since Nanonet is the local version of Metrichor, Nanonet and Metrichor's similar accuracy trends are expected. In addition to the power of the RNN-based approach, Scrappie tries to solve the homopolymer basecalling problem. Although Scrappie is in an early stage of development, it leads to a smaller number of indels than Metrichor or Nanonet. Nanocall's poor accuracy results are due to the simple HMM-based approach it uses. Although DeepNano performs better than Nanocall with the help of its RNN-based approach, it results in a higher number of indels and a lower coverage of the reference genome.

\begin{table*}[t!]
\scriptsize
\caption{Performance analysis results for the first three steps of the pipeline.}\label{table:perffirst3}
\scalebox{0.8}{
\begin{threeparttable}
\begin{tabular}{l |r r r| r r r | r r r}
\hline
& \multicolumn{3}{c|}{\textcolor{mypink}{\specialcellll{\textbf{Step 1:}\\Basecaller}}} &  \multicolumn{3}{c|}{\textcolor{myblue}{\specialcellll{\textbf{Step 2:}\\Read-to-Read Overlap Finder}}} & \multicolumn{3}{c}{\textcolor{mygreen}{\specialcellll{\textbf{Step 3:}\\Assembly}}}\\
\cline{2-10}
& \specialcell{Wall\\Clock\\Time\\(h:m:s)} & \specialcell{CPU\\Time\\(h:m:s)} & \specialcell{Memory\\Usage\\(GB)} & \specialcell{Wall\\Clock\\Time\\(h:m:s)} & \specialcell{CPU\\Time\\(h:m:s)} & \specialcell{Memory\\Usage\\(GB)} & \specialcell{Wall\\Clock\\Time\\(h:m:s)} & \specialcell{CPU\\Time\\(h:m:s)} & \specialcell{Memory\\Usage\\(GB)}\\ 
\hline
\begin{tabular}{p{0.15cm} | p{1cm} p{0.2cm} p{1.1cm} p{0.2cm} p{0.8cm}}1 & \textcolor{mypink}{Metrichor} & + & --- & + & \textcolor{mygreen}{Canu}\end{tabular} & \multirow{3}{*}{---\tnote{*}} & \multirow{3}{*}{---\tnote{*}} & \multirow{3}{*}{---\tnote{*}} & --- & --- & --- & 44:12:31 & 502:18:56 & 5.76\\
\begin{tabular}{p{0.15cm} | p{1cm} p{0.2cm} p{1.1cm} p{0.2cm} p{0.8cm}}2 & \textcolor{mypink}{Metrichor} & + & \textcolor{myblue}{Minimap} & + & \textcolor{mygreen}{Miniasm}\end{tabular} & & & & 2:15 & 41:37 & 12.30 & 1:09 & 1:09 & 1.96\\
\begin{tabular}{p{0.15cm} | p{1cm} p{0.2cm} p{1.1cm} p{0.2cm} p{0.8cm}}3 & \textcolor{mypink}{Metrichor} & + & \textcolor{myblue}{GraphMap} & + & \textcolor{mygreen}{Miniasm}\end{tabular} & & & & 6:14 & 1:52:57 & 56.58 & 1:05 & 1:05 & 1.82\\
\hline
\begin{tabular}{p{0.15cm} | p{1cm} p{0.2cm} p{1.1cm} p{0.2cm} p{0.8cm}}4 & \textcolor{mypink}{Nanonet} & + & --- & + & \textcolor{mygreen}{Canu}\end{tabular} & \multirow{3}{*}{17:52:42} & \multirow{3}{*}{714:21:45} & \multirow{3}{*}{1.89} & --- & --- & --- & 11:32:40 & 129:07:16 & 5.27\\
\begin{tabular}{p{0.15cm} | p{1cm} p{0.2cm} p{1.1cm} p{0.2cm} p{0.8cm}}5 & \textcolor{mypink}{Nanonet} & + & \textcolor{myblue}{Minimap} & + & \textcolor{mygreen}{Miniasm}\end{tabular} & & & & 1:13 & 18:55 & 9.45 & 33 & 33 & 0.69\\
\begin{tabular}{p{0.15cm} | p{1cm} p{0.2cm} p{1.1cm} p{0.2cm} p{0.8cm}}6 & \textcolor{mypink}{Nanonet} & + & \textcolor{myblue}{GraphMap} & + & \textcolor{mygreen}{Miniasm}\end{tabular} & & & & 3:18 & 48:27 & 29.16 & 32 & 32 & 0.65\\
\hline
\begin{tabular}{p{0.15cm} | p{1cm} p{0.2cm} p{1.1cm} p{0.2cm} p{0.8cm}}7 & \textcolor{mypink}{Scrappie} & + &--- & + & \textcolor{mygreen}{Canu}\end{tabular} & \multirow{3}{*}{3:11:41} & \multirow{3}{*}{126:19:06} & \multirow{3}{*}{13.36} & --- & --- & --- & 33:47:41 & 385:51:23 & 5.75\\
\begin{tabular}{p{0.15cm} | p{1cm} p{0.2cm} p{1.1cm} p{0.2cm} p{0.8cm}}8 & \textcolor{mypink}{Scrappie} & + & \textcolor{myblue}{Minimap} & + & \textcolor{mygreen}{Miniasm}\end{tabular} & & & & 2:52 & 1:10:26 & 12.40 & 1:29 & 1:29 & 1.98\\
\begin{tabular}{p{0.15cm} | p{1cm} p{0.2cm} p{1.1cm} p{0.2cm} p{0.8cm}}9 & \textcolor{mypink}{Scrappie} & + & \textcolor{myblue}{GraphMap} & + & \textcolor{mygreen}{Miniasm}\end{tabular} & & & & 7:26 & 2:16:02 & 38.31 & 1:23 & 1:23 & 1.87\\
\hline
\begin{tabular}{p{0.15cm} | p{1cm} p{0.2cm} p{1.1cm} p{0.2cm} p{0.8cm}}10 & \textcolor{mypink}{Nanocall} & + & --- & + & \textcolor{mygreen}{Canu}\end{tabular} & \multirow{3}{*}{47:04:53} & \multirow{3}{*}{1857:37:56} & \multirow{3}{*}{37.73} & --- & --- & --- & 1:35:23 & 27:58:29 & 3.77\\
\begin{tabular}{p{0.15cm} | p{1cm} p{0.2cm} p{1.1cm} p{0.2cm} p{0.8cm}}11 & \textcolor{mypink}{Nanocall} & + & \textcolor{myblue}{Minimap} & + & \textcolor{mygreen}{Miniasm}\end{tabular} & & & & 1:15 & 16:08 & 12.19 & 20 & 20 & 0.47\\
\begin{tabular}{p{0.15cm} | p{1cm} p{0.2cm} p{1.1cm} p{0.2cm} p{0.8cm}}12 & \textcolor{mypink}{Nanocall} & + & \textcolor{myblue}{GraphMap} & + & \textcolor{mygreen}{Miniasm}\end{tabular} & & & & 5:14 & 1:09:04 & 56.78 & 16 & 16 & 0.30\\
\hline
\begin{tabular}{p{0.15cm} | p{1cm} p{0.2cm} p{1.1cm} p{0.2cm} p{0.8cm}}13 & \textcolor{mypink}{DeepNano} & + & --- & + & \textcolor{mygreen}{Canu}\end{tabular} & \multirow{3}{*}{23:54:34} & \multirow{3}{*}{811:14:29} & \multirow{3}{*}{8.38} & --- & --- & --- & 1:15:48 & 17:31:07 & 3.61\\
\begin{tabular}{p{0.15cm} | p{1cm} p{0.2cm} p{1.1cm} p{0.2cm} p{0.8cm}}14 & \textcolor{mypink}{DeepNano} & + & \textcolor{myblue}{Minimap} & + & \textcolor{mygreen}{Miniasm}\end{tabular} & & & & 1:50 & 24:30 & 11.71 & 1:03 & 1:03 & 1.31\\
\begin{tabular}{p{0.15cm} | p{1cm} p{0.2cm} p{1.1cm} p{0.2cm} p{0.8cm}}15 & \textcolor{mypink}{DeepNano} & + & \textcolor{myblue}{GraphMap} & + & \textcolor{mygreen}{Miniasm}\end{tabular} & & & & 5:18 & 1:17:06 & 54.64 & 58 & 58 & 1.10\\
\hline
\end{tabular}
\begin{tablenotes}
\item[*] We cannot get the performance metrics for Metrichor since its source code is not available for us to run the tool by ourselves.
\end{tablenotes}
\end{threeparttable}
}
\end{table*}

\subsubsection*{Performance}
RNN and HMM are computationally-intensive algorithms. In HMM-based basecalling, the Viterbi algorithm \cite{forney1973viterbi} is used for decoding. The Viterbi algorithm is a sequential technique and its computation cannot currently be parallelized with multithreading. However, in RNN-based basecalling, multiple threads can work on different sections of the neural network and thus RNN computation can be parallelized with multithreading.\par
In order to measure and compare the performance of the selected basecallers, we first compare the recorded wall clock time, CPU time and memory usage metrics of each scenario for the first step of the pipeline. Based on the results provided in Table~\ref{table:perffirst3}, we make the following key observation.\par
\textbf{Observation 2: }\textit{RNN-based Nanonet and DeepNano are 2.6x and 2.3x faster than HMM-based Nanocall, respectively. Although Scrappie is also an RNN-based tool, it is 5.7x faster than Nanonet due to its C implementation as opposed to Nanonet's Python implementation.}\par
\begin{figure*}[h!]
\centering
\includegraphics[width=\columnwidth,keepaspectratio]{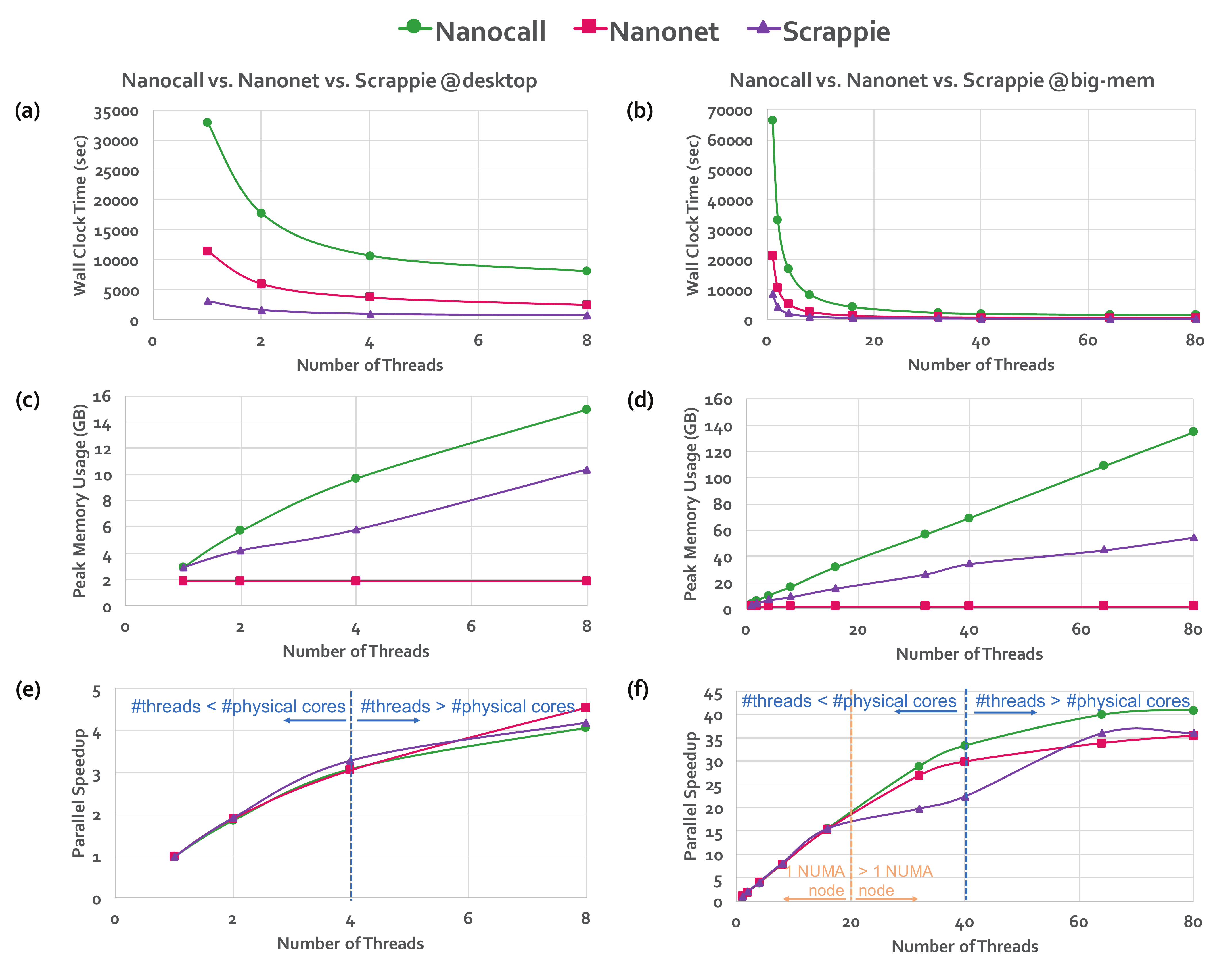}
\caption[Scalability results of Nanocall, Nanonet and Scrappie.]{Scalability results of Nanocall, Nanonet and Scrappie. Wall clock time (a, b), peak memory usage (c, d), and parallel speedup (e, f) results obtained on the \textit{desktop} and \textit{big-mem} systems. The left column (a, c, e) shows the results from the \textit{desktop} system and the right column (b, d, f) shows the results from the \textit{big-mem} system.}\label{fig:basecallplot}
\end{figure*}

For a deeper understanding of these tools' advantages, disadvantages and bottlenecks, we also perform a scalability analysis for each basecaller by running it on the \textit{desktop} server and the \textit{big-mem} server separately, with 1, 2, 4, 8 (maximum for the \textit{desktop} server), 16, 32, 40, 64 and 80 (maximum for the \textit{big-mem} server) threads, and measuring the performance metrics for each configuration. Metrichor and DeepNano are not included in this analysis because Metrichor is a cloud-based tool and its source code is not available for us to change its number of threads, and DeepNano does not support multi-threading. Figure~\ref{fig:basecallplot} shows the speed, memory usage and parallel speedup results of our evaluations. We make four observations.

\textbf{Observation 3: }\textit{When we compare \textit{desktop}'s and \textit{big-mem}'s single thread performance, we observe that \textit{desktop} is approximately 2x faster than \textit{big-mem} (cf. Figure~\ref{fig:basecallplot}a and~\ref{fig:basecallplot}b).}\par
This is mainly because of \textit{desktop}'s higher CPU frequency (see Table 6). It is an indication that all of these three tools are computationally expensive. Larger memory capacity or larger Last-Level Cache (LLC) capacity of \textit{big-mem} cannot make up for the higher CPU speed in \textit{desktop} when there is only one thread.\par
\textbf{Observation 4: }\textit{Scrappie and Nanocall have a linear increase in memory usage when number of threads increases. In contrast, Nanonet has a constant memory usage for all evaluated thread units (cf. Figure~\ref{fig:basecallplot}c and~\ref{fig:basecallplot}d)}.\par
In Scrappie and Nanocall, each thread performs the basecalling for different groups of raw reads. Thus, each thread allocates its own memory space for the corresponding data. This causes the linear increase in memory usage when the level of parallelism increases. In Nanonet, all of the threads share the computation of each read, and thus memory usage is not affected by the amount of thread parallelism.\par
\textbf{Observation 5: }\textit{When the number of threads exceeds the number of physical cores, the simultaneous multithreading overhead prevents continued linear speedup of Nanonet, Scrappie and Nanocall (cf. Figure~\ref{fig:basecallplot}e and~\ref{fig:basecallplot}f).}\par
Simultaneous multithreading (SMT) (\ie running more than one thread per physical core \cite{marr2002hyper,magro2002hyper,tuck2003initial,tullsen1995simultaneous,eggers1997simultaneous,tullsen1996exploiting,yamamoto1995increasing,hirata1992elementary}), or more specifically Intel's hyper-threading (\ie since we use Intel's hyper-threading enabled machines (see Table~\ref{table:specification})) helps to decrease the total runtime but it does \emph{not} provide a linear speedup with the number of threads because of the CPU-intensive workload of Scrappie, Nanocall and Nanonet. If the threads executed are CPU-bound and do not wait for the memory or I/O requests, hyper-threading does not provide linear speedup due to the contention it causes in the shared resources for the computation. This phenomenon has been analyzed extensively in other application domains \cite{marr2002hyper,magro2002hyper,tuck2003initial}.\par
\textbf{Observation 6: }\textit{Data sharing between threads degrades the parallel speedup of Nanonet when cores from multiple NUMA nodes take role in the computation (cf. Figure~\ref{fig:basecallplot}f).}\par
In Nanonet, data is shared between threads and each thread performs different computations on the same data. There are 4 NUMA nodes in \textit{big-mem} (\emph{cf.} Table~\ref{table:specification}), and when data is shared between multiple NUMA nodes, this negatively affects the speedup of Nanonet because accessing the data located in another node (\ie non-local memory) requires longer latency than accessing the data located in local memory. When multiple NUMA nodes start taking role in the computation, Nanocall performs better in terms of scalability since it does \emph{not} require data sharing between different threads.\par
\textbf{Summary.} Based on the observations we make about the analyzed basecalling tools, we conclude that the choice of the tool for this step plays an important role to overcome the high error rates of nanopore sequencing technology. Basecalling with Recurrent Neural Networks (\eg Metrichor, Nanonet, Scrappie) provides higher accuracy and higher speed than basecalling with Hidden Markov Models, and the newest basecaller of ONT, Scrappie, also has the potential to overcome the homopolymer basecalling problem. 

\subsection{Read-to-Read Overlap Finding Tools}\label{sec:overlapresults}
As we discuss in Section~\ref{sec:overlapfinding}, GraphMap and Minimap are the commonly-used tools for this step. GraphMap finds the overlaps using \textit{k}-mer similarity, whereas Minimap finds them by using minimizers instead of the full set of \textit{k}-mers. 

\subsubsection*{Accuracy}
As done in GraphMap, finding the overlaps with the help of full set of \textit{k}-mers is a highly-sensitive and accurate approach. However, it is also resource-intensive. For this reason, instead of the full set of \textit{k}-mers, Minimap uses a minimum representative set of \textit{k}-mers, minimizers, as an alternative approach for finding the overlaps.\par
In order to compare the accuracy of these two approaches, we categorize the results in Table~\ref{table:accuracyfirst3} based on read-to-read overlap finding tools. In other words, we look at the rows with the same basecaller (\ie red-labeled tools) and same assembler (\ie green-labeled tools) but different read-to-read overlap finder (\ie blue-labeled tools). After that, we compare them according to the defined accuracy metrics. We make the following major observation.\par
\textbf{Observation 7: }\textit{Pipelines with GraphMap or Minimap end up with similar values for identity, coverage, number of indels and mismatches. Thus, either of these read-to-read overlap finding tools can be used in the second step of the nanopore sequencing assembly pipeline to achieve similar accuracy.}\par
Minimap and GraphMap do not have a significantly different effect on the accuracy of the generated draft assemblies. This is because Minimap does not lose any sensitivity by storing minimizers instead of the full set of \textit{k}-mers. 

\subsubsection*{Performance}
In order to compare the performance of GraphMap and Minimap, we categorize the results in Table~\ref{table:perffirst3} based on read-to-read overlap finding tools, in a similar way we describe the results in Table~\ref{table:accuracyfirst3} for the accuracy analysis. We also perform a scalability analysis for each of these tools by running them on the \textit{big-mem} server with 1, 2, 4, 8, 16, 32, 40, 64 and 80 threads, and measuring the performance metrics. Because of the high memory usage of GraphMap, data necessary for the tool does not fit in the memory of the \textit{desktop} server and the GraphMap job exits due to a bad memory allocation exception. Thus, we could not perform the scalability analysis of GraphMap in the \textit{desktop} server.\par
\begin{figure}[h!]
\centering
\includegraphics[width=11cm,keepaspectratio]{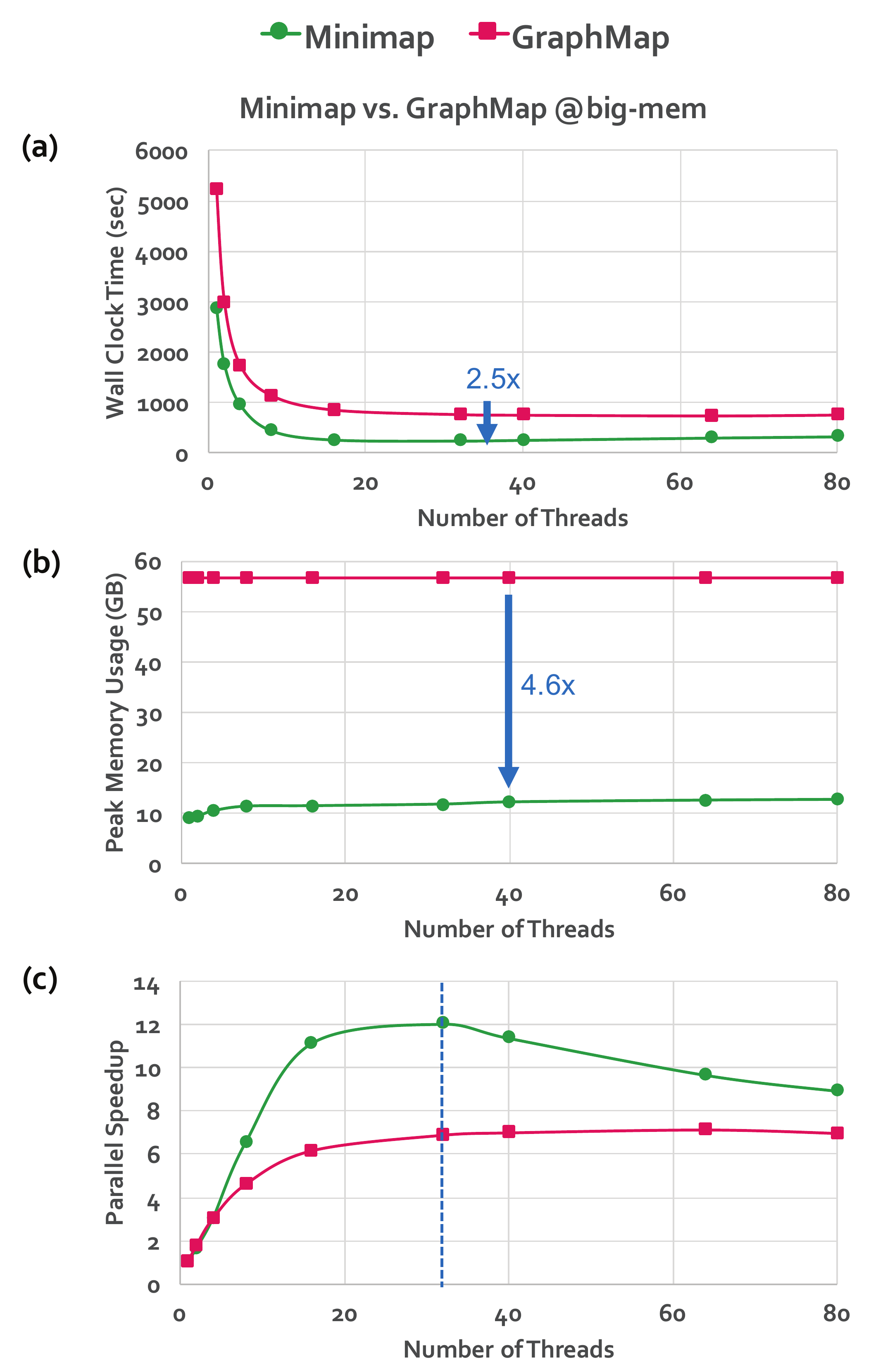}
\vspace{6pt}
\caption[Scalability results of Minimap and GraphMap.]{Scalability results of Minimap and GraphMap. Wall clock time (a), peak memory usage (b), and parallel speedup (c) results obtained on the \textit{big-mem} system.} \label{fig:overlapplot}
\end{figure}

Figure~\ref{fig:overlapplot} depicts the speed, memory usage and parallel speedup results of the scalability analysis for GraphMap and Minimap. We make the following three observations according to the results from Table~\ref{table:perffirst3} and Figure~\ref{fig:overlapplot}.\par
\textbf{Observation 8: }\textit{The memory usage of both GraphMap and Minimap is dependent on the hash table size but independent of number of threads. Minimap requires 4.6x less memory than GraphMap, on average.}\par

This is mainly because Minimap stores only minimizers instead of all \textit{k}-mers. Storing the full set of \textit{k}-mers in GraphMap requires a larger hash table, and thus higher memory usage than Minimap. The high amount of memory requirement causes GraphMap to not run on our \emph{desktop} system for none of the selected number of thread units.

\textbf{Observation 9: }\textit{Minimap is 2.5x faster than GraphMap, on average, across different scenarios in Table~\ref{table:perffirst3}.}\par
Since GraphMap stores all \textit{k}-mers, GraphMap needs to scan its very large dataset while finding the overlaps between two reads. However, in Minimap, the size of dataset that needs to be scanned is greatly shrunk by storing minimizers, as we describe in Observation 8. Thus, Minimap performs much less computation, leading to its 2.5x speedup. Another indication of the different memory usage and its effect on the speed of computation is the Last-Level Cache (LLC) miss rates of these two tools. The LLC miss rate of Minimap is 36\% whereas the LLC miss rate of GraphMap is 55\%. Since the size of data needed by GraphMap is much larger than the LLC size, GraphMap experiences LLC misses more frequently. As a result, GraphMap stalls for longer, waiting for data accesses from main memory, which negatively affects the speed of the tool.\par
\textbf{Observation 10: }\textit{Minimap is more scalable than GraphMap. However, after 32 threads, there is a decrease in the parallel speedup of Minimap (cf. Figure~\ref{fig:overlapplot}c).}\par
Because of its lower computational workload and lower memory usage, we find that Minimap is more scalable than GraphMap. However, in Minimap, threads that finish their work wait for the other active threads to finish their workloads, before starting new work, in order to prevent higher memory usage. Because of this, when the number of threads reaches a high number (\ie 32 in Figure~\ref{fig:overlapplot}c), synchronization overhead greatly increases, causing the parallel speedup to reduce. GraphMap does not suffer from such a synchronization bottleneck and hence does not experience a decrease in speedup. However, GraphMap's speedup saturates when the number of threads reaches a high number due to data sharing between threads.\par
\textbf{Summary.} According to the observations we make about GraphMap and Minimap, we conclude that storing minimizers instead of all \textit{k}-mers, as done by Minimap, does not affect the overall accuracy of the first three steps of the pipeline. Moreover, by storing minimizers, Minimap has a much lower memory usage and thus much higher performance than GraphMap.

\subsection{Assembly Tools}\label{sec:assemblyresults}
As we discuss in Section~\ref{sec:assembly}, Canu and Miniasm are the commonly-used tools for this step.\footnote{In addition, we attempted to evaluate MECAT \cite{xiao2017mecat}, another assembler. We were unable to draw any meaningful conclusions from MECAT, as its memory usage exceeds the 1TB available in our \textit{big-mem} system.}

\subsubsection*{Accuracy}
In order to compare the accuracy of these two tools, we categorize the results in Table~\ref{table:accuracyfirst3} based on assembly tools. We make the following observation.\par
\textbf{Observation 11: }\textit{Canu provides higher accuracy than Miniasm, with the help of the error-correction step that is present in its own pipeline.}

\subsubsection*{Performance}
In order to compare the performance of Canu and Miniasm, we categorize the results in Table~\ref{table:perffirst3} based on assembly tools, in a way similar to what we did in Table~\ref{table:accuracyfirst3} for the accuracy analysis. We could not perform a scalability analysis for these tools since Canu has an auto-configuration mechanism for each sub-step of its own pipeline, which does not allow us to change the number of threads, and Miniasm does not support multi-threading. We make the following observation according to the results in Table~\ref{table:perffirst3}.\par
\textbf{Observation 12: }\textit{Canu is much more computationally intensive and greatly (\ie by 1096.3x) slower than Miniasm, because of its very expensive error-correction step.}\par
\textbf{Summary.} According to the observations we make about Canu and Miniasm, there is a tradeoff between accuracy and performance when deciding on the appropriate tool for this step. Canu produces highly accurate assemblies but it is resource intensive and slow. In contrast, Miniasm is a very fast assembler but it cannot produce as accurate draft assemblies as Canu. We suggest that Miniasm can potentially be used for fast initial analysis and then further polishing can be applied in the next step in order to produce higher-quality assemblies.

\subsection{Read Mapping and Polishing Tools}\label{sec:polishresults}

As we discuss in Section~\ref{sec:polish}, further polishing may be required for improving the accuracy of the low-quality draft assemblies. For this purpose, after aligning the reads to the generated draft assembly with BWA-MEM or Minimap,
one can use Nanopolish or Racon to perform polishing and obtain improved assemblies (\ie consensus sequences).

Nanopolish accepts mappings only in \textit{Sequence Alignment/Map (SAM)} format \cite{li2009sequence} and it works only with draft assemblies generated with the Metrichor-basecalled reads. On the other hand, Racon accepts both \textit{Pairwise Mapping format (PAF)} mappings \cite{li2016minimap} and \textit{SAM}-format mappings, but it requires the input reads and draft assembly files to be in \textit{fastq} format \cite{cock2009sanger}, which includes quality scores. However, by using the \texttt{-bq -1} parameter, it is possible to disable the filtering used in Racon, which requires quality scores. Since our basecalled reads are in \textit{fasta} format \cite{pearson1988improved}, in our experiments, we convert these \textit{fasta} files into \textit{fastq} files and disable the filtering with the corresponding parameter.

BWA-MEM generates mappings in \textit{SAM} format whereas Minimap generates mappings in \textit{PAF} format. Since Nanopolish requires \textit{SAM} format input, we generate the mappings only with BWA-MEM and use them for Nanopolish polishing, in our analysis. On the other hand, since Racon accepts both formats, we generate the mappings and the overlaps with both BWA-MEM and Minimap, respectively, and use them for Racon polishing, in our analysis.

\subsubsection*{Accuracy}
Table~\ref{table:accuracylast2} presents the accuracy metrics results for Nanopolish (\ie Rows 1-3) and Racon (\ie Rows 4-23) pipelines. Based on these results, we make two observations.

\begin{table*}[t!]
\scriptsize
\caption{Accuracy analysis results for the full pipeline with a focus on the last two steps.}\label{table:accuracylast2}
\scalebox{0.76}{
\begin{tabular}{l | r r r r r r}
\hline
& \specialcell{\#\\Bases} & \specialcell{\#\\Contigs} & \specialcell{Identity\\(\%)} & \specialcell{Coverage\\(\%)} & \specialcell{\#\\Mismatches} & \specialcell{\#\\Indels}\\ 
\hline
\begin{tabular}{p{0.15cm} | p{1.1cm} l p{1.1cm} l p{0.8cm} l p{1.6cm} r p{1.2cm}}1 & \textcolor{mypink}{Metrichor} & + & --- & + & \textcolor{mygreen}{Canu} & + &  \textcolor{myorange}{BWA-MEM} & + & \textcolor{mypurple}{Nanopolish}\end{tabular} & 4,683,072 & 1 & 99.48 & 99.93 & 8,198 & 15,581\\
\begin{tabular}{p{0.15cm} | p{1.1cm} l p{1.1cm} l p{0.8cm} l p{1.6cm} r p{1.2cm}}2 & \textcolor{mypink}{Metrichor} & + & \textcolor{myblue}{Minimap} & + & \textcolor{mygreen}{Miniasm} & + & \textcolor{myorange}{BWA-MEM} & + & \textcolor{mypurple}{Nanopolish}\end{tabular} & 4,540,352 & 1 & 92.33 & 96.31 & 162,884 & 182,965\\
\begin{tabular}{p{0.15cm} | p{1.1cm} l p{1.1cm} l p{0.8cm} l p{1.6cm} r p{1.2cm}}3 & \textcolor{mypink}{Metrichor} & + & \textcolor{myblue}{GraphMap} & + & \textcolor{mygreen}{Miniasm} & + & \textcolor{myorange}{BWA-MEM} & + & \textcolor{mypurple}{Nanopolish}\end{tabular} & 4,637,916 & 2 & 92.38 & 95.80 & 159,206 & 180,603\\
\hline
\hline
\begin{tabular}{p{0.15cm} | p{1.1cm} l p{1.1cm} l p{0.8cm} l p{1.6cm} r p{1.2cm}}4 & \textcolor{mypink}{Metrichor} & + & --- & + & \textcolor{mygreen}{Canu} & + &  \textcolor{myorange}{BWA-MEM} & + & \textcolor{mypurple}{Racon}\end{tabular} & 4,650,502 & 1 & 98.46 & 100.00 & 18,036 & 51,842\\
\begin{tabular}{p{0.15cm} | p{1.1cm} l p{1.1cm} l p{0.8cm} l p{1.6cm} r p{1.2cm}}5 & \textcolor{mypink}{Metrichor} & + & --- & + & \textcolor{mygreen}{Canu} & + & \textcolor{myorange}{Minimap} & + & \textcolor{mypurple}{Racon}\\\end{tabular} & 4,648,710 & 1 & 98.45 & 100.00 & 17,906 & 52,168\\
\begin{tabular}{p{0.15cm} | p{1.1cm} l p{1.1cm} l p{0.8cm} l p{1.6cm} r p{1.2cm}}6 & \textcolor{mypink}{Metrichor} & + & \textcolor{myblue}{Minimap} & + & \textcolor{mygreen}{Miniasm} & + & \textcolor{myorange}{BWA-MEM} & + & \textcolor{mypurple}{Racon}\end{tabular} & 4,598,267 & 1 & 97.70 & 99.91 & 24,014 & 82,906\\
\begin{tabular}{p{0.15cm} | p{1.1cm} l p{1.1cm} l p{0.8cm} l p{1.6cm} r p{1.2cm}}7 & \textcolor{mypink}{Metrichor} & + & \textcolor{myblue}{Minimap} & + & \textcolor{mygreen}{Miniasm} & + & \textcolor{myorange}{Minimap} & + & \textcolor{mypurple}{Racon}\end{tabular} & 4,600,109 & 1 & 97.78 & 100.00 & 23,339 & 79,721\\
\hline
\begin{tabular}{p{0.15cm} | p{1.1cm} l p{1.1cm} l p{0.8cm} l p{1.6cm} r p{1.2cm}}8 & \textcolor{mypink}{Nanonet} & + & --- & + & \textcolor{mygreen}{Canu} & + &  \textcolor{myorange}{BWA-MEM} & + & \textcolor{mypurple}{Racon}\end{tabular} & 4,622,285 & 1 & 98.48 & 100.00 & 16,872 & 52,509\\
\begin{tabular}{p{0.15cm} | p{1.1cm} l p{1.1cm} l p{0.8cm} l p{1.6cm} r p{1.2cm}}9 & \textcolor{mypink}{Nanonet} & + & --- & + & \textcolor{mygreen}{Canu} & + & \textcolor{myorange}{Minimap} & + & \textcolor{mypurple}{Racon}\\\end{tabular} & 4,620,597 & 1 & 98.49 & 100.00 & 16,874 & 52,232\\
\begin{tabular}{p{0.15cm} | p{1.1cm} l p{1.1cm} l p{0.8cm} l p{1.6cm} r p{1.2cm}}10 & \textcolor{mypink}{Nanonet} & + & \textcolor{myblue}{Minimap} & + & \textcolor{mygreen}{Miniasm} & + & \textcolor{myorange}{BWA-MEM} & + & \textcolor{mypurple}{Racon}\end{tabular} & 4,593,402 & 1 & 98.01 & 99.97 & 20,322 & 72,284\\
\begin{tabular}{p{0.15cm} | p{1.1cm} l p{1.1cm} l p{0.8cm} l p{1.6cm} r p{1.2cm}}11 & \textcolor{mypink}{Nanonet} & + & \textcolor{myblue}{Minimap} & + & \textcolor{mygreen}{Miniasm} & + & \textcolor{myorange}{Minimap} & + & \textcolor{mypurple}{Racon}\end{tabular} & 4,592,907 & 1 & 98.04 & 100.00 & 20,170 & 70,705\\
\hline
\begin{tabular}{p{0.15cm} | p{1.1cm} l p{1.1cm} l p{0.8cm} l p{1.6cm} r p{1.2cm}}12 & \textcolor{mypink}{Scrappie} & + & --- & + & \textcolor{mygreen}{Canu} & + &  \textcolor{myorange}{BWA-MEM} & + & \textcolor{mypurple}{Racon}\end{tabular} & 4,673,871 & 1 & 98.40 & 99.98 & 13,583 & 60,612\\
\begin{tabular}{p{0.15cm} | p{1.1cm} l p{1.1cm} l p{0.8cm} l p{1.6cm} r p{1.2cm}}13 & \textcolor{mypink}{Scrappie} & + & --- & + &\textcolor{mygreen}{Canu} & + & \textcolor{myorange}{Minimap} & + & \textcolor{mypurple}{Racon}\\\end{tabular} & 4,673,606 & 1 & 98.40 & 99.98 & 13,798 & 60,423\\
\begin{tabular}{p{0.15cm} | p{1.1cm} l p{1.1cm} l p{0.8cm} l p{1.6cm} r p{1.2cm}}14 & \textcolor{mypink}{Scrappie} & + & \textcolor{myblue}{Minimap} & + & \textcolor{mygreen}{Miniasm} & + & \textcolor{myorange}{BWA-MEM} & + & \textcolor{mypurple}{Racon}\end{tabular} & 5,157,041 & 8 & 97.87 & 99.80 & 18,085 & 78,492\\
\begin{tabular}{p{0.15cm} | p{1.1cm} l p{1.1cm} l p{0.8cm} l p{1.6cm} r p{1.2cm}}15 & \textcolor{mypink}{Scrappie} & + & \textcolor{myblue}{Minimap} & + & \textcolor{mygreen}{Miniasm} & + & \textcolor{myorange}{Minimap} & + &\textcolor{mypurple}{Racon}\end{tabular} & 5,156,375 & 8 & 97.87 & 99.94 & 17,922 & 77,807\\
\hline
\begin{tabular}{p{0.15cm} | p{1.1cm} l p{1.1cm} l p{0.8cm} l p{1.6cm} r p{1.2cm}}16 & \textcolor{mypink}{Nanocall} & + & --- & + & \textcolor{mygreen}{Canu} & + &  \textcolor{myorange}{BWA-MEM} & + & \textcolor{mypurple}{Racon}\end{tabular} & 1,383,851 & 86 & 93.49 & 28.82 & 19,057 & 65,244\\
\begin{tabular}{p{0.15cm} | p{1.1cm} l p{1.1cm} l p{0.8cm} l p{1.6cm} r p{1.2cm}}17 & \textcolor{mypink}{Nanocall} & + & --- & + & \textcolor{mygreen}{Canu} & + & \textcolor{myorange}{Minimap} & + & \textcolor{mypurple}{Racon}\\\end{tabular} & 1,367,834 & 86 & 94.43 & 28.74 & 15,610 & 55,275\\
\begin{tabular}{p{0.15cm} | p{1.1cm} l p{1.1cm} l p{0.8cm} l p{1.6cm} r p{1.2cm}}18 & \textcolor{mypink}{Nanocall} & + & \textcolor{myblue}{Minimap} & + & \textcolor{mygreen}{Miniasm} & + & \textcolor{myorange}{BWA-MEM} & + & \textcolor{mypurple}{Racon}\end{tabular} & 4,707,961 & 5 & 90.75 & 97.11 & 91,502 & 347,005\\
\begin{tabular}{p{0.15cm} | p{1.1cm} l p{1.1cm} l p{0.8cm} l p{1.6cm} r p{1.2cm}}19 & \textcolor{mypink}{Nanocall} & + & \textcolor{myblue}{Minimap} & + & \textcolor{mygreen}{Miniasm} & + & \textcolor{myorange}{Minimap} & + & \textcolor{mypurple}{Racon}\end{tabular} & 4,673,069 & 5 & 92.23 & 97.10 & 72,646 & 291,918\\
\hline
\begin{tabular}{p{0.15cm} | p{1.1cm} l p{1.1cm} l p{0.8cm} l p{1.6cm} r p{1.2cm}}20 & \textcolor{mypink}{DeepNano} & + & --- & + & \textcolor{mygreen}{Canu} & + &  \textcolor{myorange}{BWA-MEM} & + & \textcolor{mypurple}{Racon}\end{tabular} & 7,429,290 & 106 & 96.46 & 99.24 & 27,811 & 102,682\\
\begin{tabular}{p{0.15cm} | p{1.1cm} l p{1.1cm} l p{0.8cm} l p{1.6cm} r p{1.2cm}}21 & \textcolor{mypink}{DeepNano} & + & --- & + & \textcolor{mygreen}{Canu} & + & \textcolor{myorange}{Minimap} & + & \textcolor{mypurple}{Racon}\\\end{tabular} & 7,404,454 & 106 & 96.03 & 99.21 & 34,023 & 110,640\\
\begin{tabular}{p{0.15cm} | p{1.1cm} l p{1.1cm} l p{0.8cm} l p{1.6cm} r p{1.2cm}}22 & \textcolor{mypink}{DeepNano} & + & \textcolor{myblue}{Minimap} & + & \textcolor{mygreen}{Miniasm} & + & \textcolor{myorange}{BWA-MEM} & + & \textcolor{mypurple}{Racon}\end{tabular} & 4,566,253 & 1 & 96.76 & 99.86 & 25,791 & 125,386\\
\begin{tabular}{p{0.15cm} | p{1.1cm} l p{1.1cm} l p{0.8cm} l p{1.6cm} r p{1.2cm}}23 & \textcolor{mypink}{DeepNano} & + & \textcolor{myblue}{Minimap} & + & \textcolor{mygreen}{Miniasm} & + & \textcolor{myorange}{Minimap} & + & \textcolor{mypurple}{Racon}\end{tabular} & 4,571,810 & 1 & 96.90 & 99.97 & 24,994 & 119,519\\
\hline
\end{tabular}
}
\end{table*}

\textbf{Observation 13: }\textit{Both Nanopolish and Racon significantly increase the accuracy of the draft assemblies.}\par
For example, Nanopolish increases the identity and coverage of the draft assembly generated with the Metrichor+Minimap+Miniasm pipeline from 87.71\% and 94.85\% (Row 2 of Table~\ref{table:accuracyfirst3}), respectively, to 92.33\% and 96.31\% (Row 2 of Table~\ref{table:accuracylast2}). Similarly, Racon increases them to 97.70\% and 99.91\% (Rows 6--7 of Table~\ref{table:accuracylast2}), respectively.\par
\textbf{Observation 14: }\textit{For Racon, the choice of read mapper does not affect the accuracy of the polishing step.}\par
We observe that using BWA-MEM or Minimap to generate the mappings for Racon results in almost identical accuracy metric results. For example, when we use BWA-MEM before Racon for the draft assembly generated with the Metrichor + Canu pipeline (Row 4 of Table~\ref{table:accuracylast2}), Racon results with 98.46\% identity, 100.00\% coverage, 18,036 mismatches and 51,842 indels. When we use Minimap, instead (Row 5 of Table~\ref{table:accuracylast2}), Racon results with 98.45\% identity, 100.00\% coverage, 17,096 mismatches and 52,168 indels, which is almost identical to the BWA-MEM results.  

\subsubsection*{Performance}
In the first part of the performance analysis for Nanopolish, we divide the draft assemblies into 50kb-segments and polish 4 of these segments in parallel with 10 threads for each segment. For Racon, each draft assembly is polished using 40 threads, but the tool, by default, divides the input sequence into windows of 20kb length. Table~\ref{table:perflast2} presents the performance results for Nanopolish (\ie Rows 1-3) and Racon (\ie Rows 4-23) pipelines. Based on these results, we make the following two observations.

\begin{table*}[t!]
\caption{Performance analysis results for the full pipeline with a focus on the last two steps.}\label{table:perflast2}
\scriptsize
\scalebox{0.76}{
\begin{tabular}{l | r r r | r r r}
\hline
& \multicolumn{3}{c|}{\textcolor{myorange}{\specialcellll{\textbf{Step 4:} Read Mapper}}} &  \multicolumn{3}{c}{\textcolor{mypurple}{\specialcellll{\textbf{Step 5:} Polisher}}}\\
\cline{2-7}
& \specialcell{Wall\\Clock\\Time\\(h:m:s)} & \specialcell{CPU\\Time\\(h:m:s)} & \specialcell{Memory\\Usage\\(GB)} & \specialcell{Wall\\Clock\\Time\\(h:m:s)} & \specialcell{CPU\\Time\\(h:m:s)} & \specialcell{Memory\\Usage\\(GB)}\\ 
\hline
\begin{tabular}{p{0.15cm} | p{1.1cm} l p{1.1cm} l p{0.8cm} l p{1.6cm} r p{1.2cm}}1 & \textcolor{mypink}{Metrichor} & + & --- & + & \textcolor{mygreen}{Canu} & + &  \textcolor{myorange}{BWA-MEM} & + & \textcolor{mypurple}{Nanopolish}\end{tabular} & 24:43 & 15:47:21 & 5.26 & 5:51:00 & 191:18:52 & 13.38\\
\begin{tabular}{p{0.15cm} | p{1.1cm} l p{1.1cm} l p{0.8cm} l p{1.6cm} r p{1.2cm}}2 & \textcolor{mypink}{Metrichor} & + & \textcolor{myblue}{Minimap} & + & \textcolor{mygreen}{Miniasm} & + & \textcolor{myorange}{BWA-MEM} & + & \textcolor{mypurple}{Nanopolish}\end{tabular} & 12:33 & 7:50:54 & 3.75 & 122:52:00 & 4458:36:10 & 31.36\\
\begin{tabular}{p{0.15cm} | p{1.1cm} l p{1.1cm} l p{0.8cm} l p{1.6cm} r p{1.2cm}}3 & \textcolor{mypink}{Metrichor} & + & \textcolor{myblue}{GraphMap} & + & \textcolor{mygreen}{Miniasm} & + & \textcolor{myorange}{BWA-MEM} & + & \textcolor{mypurple}{Nanopolish}\end{tabular} & 12:47 & 7:57:58 & 3.60 & 129:46:00 & 4799:03:51 & 31.31\\
\hline
\hline
\begin{tabular}{p{0.15cm} | p{1.1cm} l p{1.1cm} l p{0.8cm} l p{1.6cm} r p{1.2cm}}4 & \textcolor{mypink}{Metrichor} & + & --- & + & \textcolor{mygreen}{Canu} & + &  \textcolor{myorange}{BWA-MEM} & + & \textcolor{mypurple}{Racon}\end{tabular} & 24:20 & 15:43:40 & 6.60 & 14:44 & 9:09:22 & 8.11\\
\begin{tabular}{p{0.15cm} | p{1.1cm} l p{1.1cm} l p{0.8cm} l p{1.6cm} r p{1.2cm}}5 & \textcolor{mypink}{Metrichor} & + & --- & + & \textcolor{mygreen}{Canu} & + & \textcolor{myorange}{Minimap} & + & \textcolor{mypurple}{Racon}\end{tabular} & 3 & 1:35 & 0.26 & 15:12 & 9:45:33 & 14.55\\
\begin{tabular}{p{0.15cm} | p{1.1cm} l p{1.1cm} l p{0.8cm} l p{1.6cm} r p{1.2cm}}6 & \textcolor{mypink}{Metrichor} & + & \textcolor{myblue}{Minimap} & + & \textcolor{mygreen}{Miniasm} & + & \textcolor{myorange}{BWA-MEM} & + & \textcolor{mypurple}{Racon}\end{tabular} & 12:10 & 7:48:10 & 5.19 & 15:43 & 9:33:39 & 9.98\\
\begin{tabular}{p{0.15cm} | p{1.1cm} l p{1.1cm} l p{0.8cm} l p{1.6cm} r p{1.2cm}}7 & \textcolor{mypink}{Metrichor} & + & \textcolor{myblue}{Minimap} & + & \textcolor{mygreen}{Miniasm} & + & \textcolor{myorange}{Minimap} & + & \textcolor{mypurple}{Racon}\end{tabular} & 3 & 1:24 & 0.26 & 20:28 & 8:57:40 & 18.24\\
\hline
\begin{tabular}{p{0.15cm} | p{1.1cm} l p{1.1cm} l p{0.8cm} l p{1.6cm} r p{1.2cm}}8 & \textcolor{mypink}{Nanonet} & + & --- & + & \textcolor{mygreen}{Canu} & + &  \textcolor{myorange}{BWA-MEM} & + & \textcolor{mypurple}{Racon}\end{tabular} & 9:08 & 5:53:18 & 4.84 & 6:33 & 4:02:10 & 4.47\\
\begin{tabular}{p{0.15cm} | p{1.1cm} l p{1.1cm} l p{0.8cm} l p{1.6cm} r p{1.2cm}}9 & \textcolor{mypink}{Nanonet} & + & --- & + & \textcolor{mygreen}{Canu} & + & \textcolor{myorange}{Minimap} & + & \textcolor{mypurple}{Racon}\end{tabular} & 2 & 54 & 0.26 & 6:45 & 4:17:26 & 7.93\\
\begin{tabular}{p{0.15cm} | p{1.1cm} l p{1.1cm} l p{0.8cm} l p{1.6cm} r p{1.2cm}}10 & \textcolor{mypink}{Nanonet} & + & \textcolor{myblue}{Minimap} & + & \textcolor{mygreen}{Miniasm} & + & \textcolor{myorange}{BWA-MEM} & + & \textcolor{mypurple}{Racon}\end{tabular} & 4:40 & 2:58:02 & 3.88 & 7:08 & 4:19:30 & 5.35\\
\begin{tabular}{p{0.15cm} | p{1.1cm} l p{1.1cm} l p{0.8cm} l p{1.6cm} r p{1.2cm}}11 & \textcolor{mypink}{Nanonet} & + & \textcolor{myblue}{Minimap} & + & \textcolor{mygreen}{Miniasm} & + & \textcolor{myorange}{Minimap} & + & \textcolor{mypurple}{Racon}\end{tabular} & 2 & 46 & 0.26 & 7:01 & 4:18:48 & 9.53\\
\hline
\begin{tabular}{p{0.15cm} | p{1.1cm} l p{1.1cm} l p{0.8cm} l p{1.6cm} r p{1.2cm}}12 & \textcolor{mypink}{Scrappie} & + & --- & + & \textcolor{mygreen}{Canu} & + &  \textcolor{myorange}{BWA-MEM} & + & \textcolor{mypurple}{Racon}\end{tabular} & 33:41 & 21:11:06 & 8.66 & 13:32 & 8:24:44 & 7.58\\
\begin{tabular}{p{0.15cm} | p{1.1cm} l p{1.1cm} l p{0.8cm} l p{1.6cm} r p{1.2cm}}13 & \textcolor{mypink}{Scrappie} & + & --- & + & \textcolor{mygreen}{Canu} & + & \textcolor{myorange}{Minimap} & + & \textcolor{mypurple}{Racon}\end{tabular} & 3 & 1:39 & 0.27 & 18:45 & 7:43:17 & 13.20\\
\begin{tabular}{p{0.15cm} | p{1.1cm} l p{1.1cm} l p{0.8cm} l p{1.6cm} r p{1.2cm}}14 & \textcolor{mypink}{Scrappie} & + & \textcolor{myblue}{Minimap} & + & \textcolor{mygreen}{Miniasm} & + & \textcolor{myorange}{BWA-MEM} & + & \textcolor{mypurple}{Racon}\end{tabular} & 22:41 & 14:31:00 & 6.08 & 14:37 & 8:53:59 & 9.50\\
\begin{tabular}{p{0.15cm} | p{1.1cm} l p{1.1cm} l p{0.8cm} l p{1.6cm} r p{1.2cm}}15 & \textcolor{mypink}{Scrappie} & + & \textcolor{myblue}{Minimap} & + & \textcolor{mygreen}{Miniasm} & + & \textcolor{myorange}{Minimap} & + & \textcolor{mypurple}{Racon}\end{tabular} & 3 & 1:27 & 0.27 & 15:10 & 9:02:45 & 12.72\\
\hline
\begin{tabular}{p{0.15cm} | p{1.1cm} l p{1.1cm} l p{0.8cm} l p{1.6cm} r p{1.2cm}}16 & \textcolor{mypink}{Nanocall} & + & --- & + & \textcolor{mygreen}{Canu} & + &  \textcolor{myorange}{BWA-MEM} & + & \textcolor{mypurple}{Racon}\end{tabular} & 4:52 & 3:01:15 & 3.80 & 11:07 & 3:26:52 & 5.63\\
\begin{tabular}{p{0.15cm} | p{1.1cm} l p{1.1cm} l p{0.8cm} l p{1.6cm} r p{1.2cm}}17 & \textcolor{mypink}{Nanocall} & + & --- & + & \textcolor{mygreen}{Canu} & + & \textcolor{myorange}{Minimap} & + & \textcolor{mypurple}{Racon}\end{tabular} & 3 & 1:16 & 0.22 & 7:28 & 2:50:35 & 3.62\\
\begin{tabular}{p{0.15cm} | p{1.1cm} l p{1.1cm} l p{0.8cm} l p{1.6cm} r p{1.2cm}}18 & \textcolor{mypink}{Nanocall} & + & \textcolor{myblue}{Minimap} & + & \textcolor{mygreen}{Miniasm} & + & \textcolor{myorange}{BWA-MEM} & + & \textcolor{mypurple}{Racon}\end{tabular} & 16:06 & 10:27:20 & 5.06 & 18:56 & 11:32:45 & 11.47\\
\begin{tabular}{p{0.15cm} | p{1.1cm} l p{1.1cm} l p{0.8cm} l p{1.6cm} r p{1.2cm}}19 & \textcolor{mypink}{Nanocall} & + & \textcolor{myblue}{Minimap} & + & \textcolor{mygreen}{Miniasm} & + & \textcolor{myorange}{Minimap} & + & \textcolor{mypurple}{Racon}\end{tabular} & 4 & 1:18 & 0.26 & 11:49 & 7:08:59 & 10.98\\
\hline
\begin{tabular}{p{0.15cm} | p{1.1cm} l p{1.1cm} l p{0.8cm} l p{1.6cm} r p{1.2cm}}20 & \textcolor{mypink}{DeepNano} & + & --- & + & \textcolor{mygreen}{Canu} & + &  \textcolor{myorange}{BWA-MEM} & + & \textcolor{mypurple}{Racon}\end{tabular} & 17:36 & 11:30:20 & 4.43 & 12:48 & 7:13:04 & 8.88\\
\begin{tabular}{p{0.15cm} | p{1.1cm} l p{1.1cm} l p{0.8cm} l p{1.6cm} r p{1.2cm}}21 & \textcolor{mypink}{DeepNano} & + & --- & + & \textcolor{mygreen}{Canu} & + & \textcolor{myorange}{Minimap} & + & \textcolor{mypurple}{Racon}\end{tabular} & 3 & 1:24 & 0.28 & 11:39 & 6:55:01 & 3.73\\
\begin{tabular}{p{0.15cm} | p{1.1cm} l p{1.1cm} l p{0.8cm} l p{1.6cm} r p{1.2cm}}22 & \textcolor{mypink}{DeepNano} & + & \textcolor{myblue}{Minimap} & + & \textcolor{mygreen}{Miniasm} & + & \textcolor{myorange}{BWA-MEM} & + & \textcolor{mypurple}{Racon}\end{tabular} & 8:15 & 5:22:29 & 4.11 & 14:16 & 8:34:32 & 10.30\\
\begin{tabular}{p{0.15cm} | p{1.1cm} l p{1.1cm} l p{0.8cm} l p{1.6cm} r p{1.2cm}}23 & \textcolor{mypink}{DeepNano} & + & \textcolor{myblue}{Minimap} & + & \textcolor{mygreen}{Miniasm} & + & \textcolor{myorange}{Minimap} & + & \textcolor{mypurple}{Racon}\end{tabular} & 3 & 1:10 & 0.26 & \textcolor{white}{xxx:}12:29 & \textcolor{white}{xxx}7:55:32 & 17.11\\
\hline
\end{tabular}
}
\end{table*}

\textbf{Observation 15: }\textit{Nanopolish is computationally much more intensive and thus greatly slower than Racon.}\par
Nanopolish runs take days to complete whereas Racon runs take minutes. This is mainly because Nanopolish works on each base individually, whereas Racon works on the windows. Since each window is much longer (\ie 20kb) than a single base, the computational workload is greatly smaller in Racon. Also, Racon only uses the mappings/overlaps for polishing, whereas Nanopolish uses raw signal data and an HMM-based approach in order to generate the consensus sequence, which is computationally more expensive.\par
\textbf{Observation 16: }\textit{BWA-MEM is computationally more expensive than Minimap.}\par
Although the choice of BWA-MEM and Minimap for the read mapping step does not affect the accuracy of the polishing step, these two tools have a significant difference in performance.

\begin{figure*}[b!]
\centering
\includegraphics[width=\columnwidth,keepaspectratio]{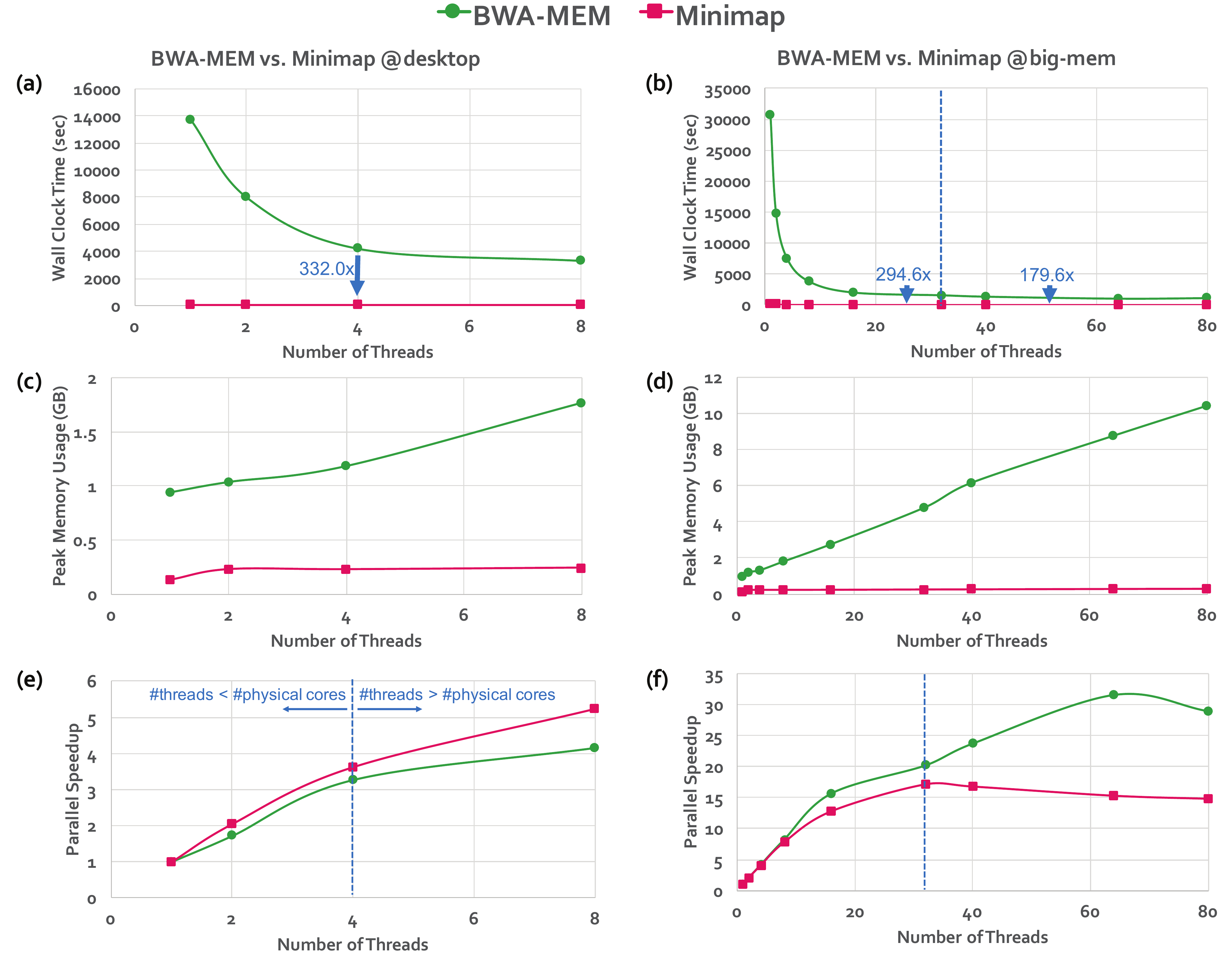}
\caption[Scalability results of BWA-MEM and Minimap.]{Scalability results of BWA-MEM and Minimap. Wall clock time (a, b), peak memory usage (c, d), and parallel speedup (e, f) results obtained on the \textit{desktop} and \textit{big-mem} systems. The left column (a, c, e) shows the results from the \textit{desktop} system and the right column (b, d, f) shows the results from the \textit{big-mem} system.}\label{fig:mapperplot}
\end{figure*}

For a deeper performance analysis of these read mapping and polishing tools, we perform a scalability analysis for each read mapper and each polisher by running them on the \textit{desktop} system and the \textit{big-mem} system separately, with 1, 2, 4, 8 (maximum for \textit{desktop} server), 16, 32, 40, 64 and 80 (maximum for \textit{big-mem} server) threads, and measuring the performance metrics. Figure~\ref{fig:mapperplot} shows the the speed, memory usage and parallel speedup of BWA-MEM and Minimap. We make two observations.

\textbf{Observation 17: }\textit{On both systems, Minimap is greatly faster than BWA-MEM (cf. Figure~\ref{fig:mapperplot}a and~\ref{fig:mapperplot}b). However, when the number of threads reaches high value, Minimap's performance degrades due to the synchronization overhead between its threads (cf. Figure~\ref{fig:mapperplot}f).}\par
On the \textit{desktop} system, Minimap is 332.0x faster than BWA-MEM, on average (see Figure~\ref{fig:mapperplot}a). On the \textit{big-mem} system, Minimap is 294.6x and 179.6x faster than BWA-MEM, on average, when the number of threads is smaller and greater than 32, respectively. This is due to the synchronization overhead that increases with the number of threads used in Minimap (see Observation 10). As we also show in Figure~\ref{fig:mapperplot}f, Minimap's speedup reduces when the number of threads exceeds 32, which is another indication of the synchronization overhead that causes Minimap to slow down.\par
\textbf{Observation 18: }\textit{Minimap's memory usage is independent of the number of threads and stays constant. In contrast, BWA-MEM's memory usage increases linearly with the number of threads (cf. Figure~\ref{fig:mapperplot}c and~\ref{fig:mapperplot}d).}\par
In Minimap, memory usage is dependent on the hash table size and is independent of number of threads (see Observation 8). In contrast, in BWA-MEM, each thread separately performs computation for different groups of reads (as in Scrappie and Nanocall, see Observation 4). This causes the linear increase in memory usage of BWA-MEM when the number of threads increases.\par
\begin{figure}[t!]
\centering
\includegraphics[width=10cm,keepaspectratio]{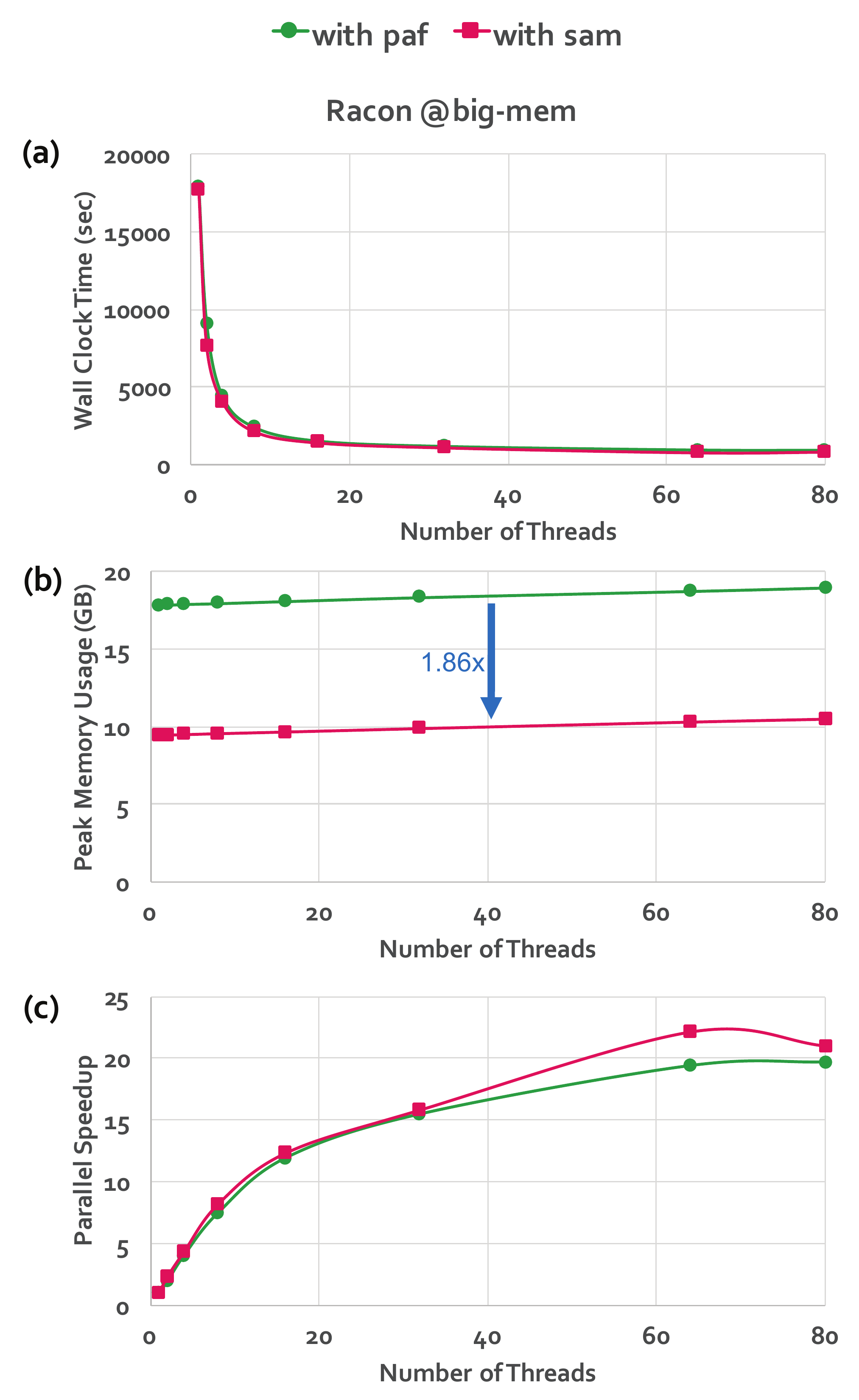}
\caption[Scalability results of Racon.]{Scalability results of Racon. Wall clock time (a), peak memory usage (b), and parallel speedup (c) results obtained on the \textit{big-mem} system.}\label{fig:raconplot}
\end{figure}

Figure~\ref{fig:raconplot} shows the scalability results for Racon on the \textit{big-mem} system. We obtain the results on both of the systems. However, we only show the results for the \textit{big-mem} system since the results for both of the systems are similar. We separately test the tool by using \textit{PAF} mappings and \textit{SAM} mappings. Based on the results, we make the following observation.\par 

\textbf{Observation 19: }\textit{Racon's memory usage is independent of the number of threads for both \textit{PAF} mode and \textit{SAM} mode. However, the memory usage of \textit{PAF} mode is 1.86x higher than the memory usage of \textit{SAM} mode, on average (cf. Figure~\ref{fig:raconplot}b).}\par
The memory usage of Racon depends on the number of mappings received from the fourth step since Racon performs polishing by using these mappings. Racon's memory usage is higher for the PAF mode because the number of mappings stored in the PAF files is greater than the number of mappings stored in the SAM files (\ie 1.4x). However, using \textit{PAF} mappings or \textit{SAM} mappings do not significantly affect the speed (see Figure~\ref{fig:raconplot}a) and the parallel speedup (see Figure~\ref{fig:raconplot}c) of Racon.\par

\begin{figure}[t!]
\centering
\includegraphics[width=10cm,keepaspectratio]{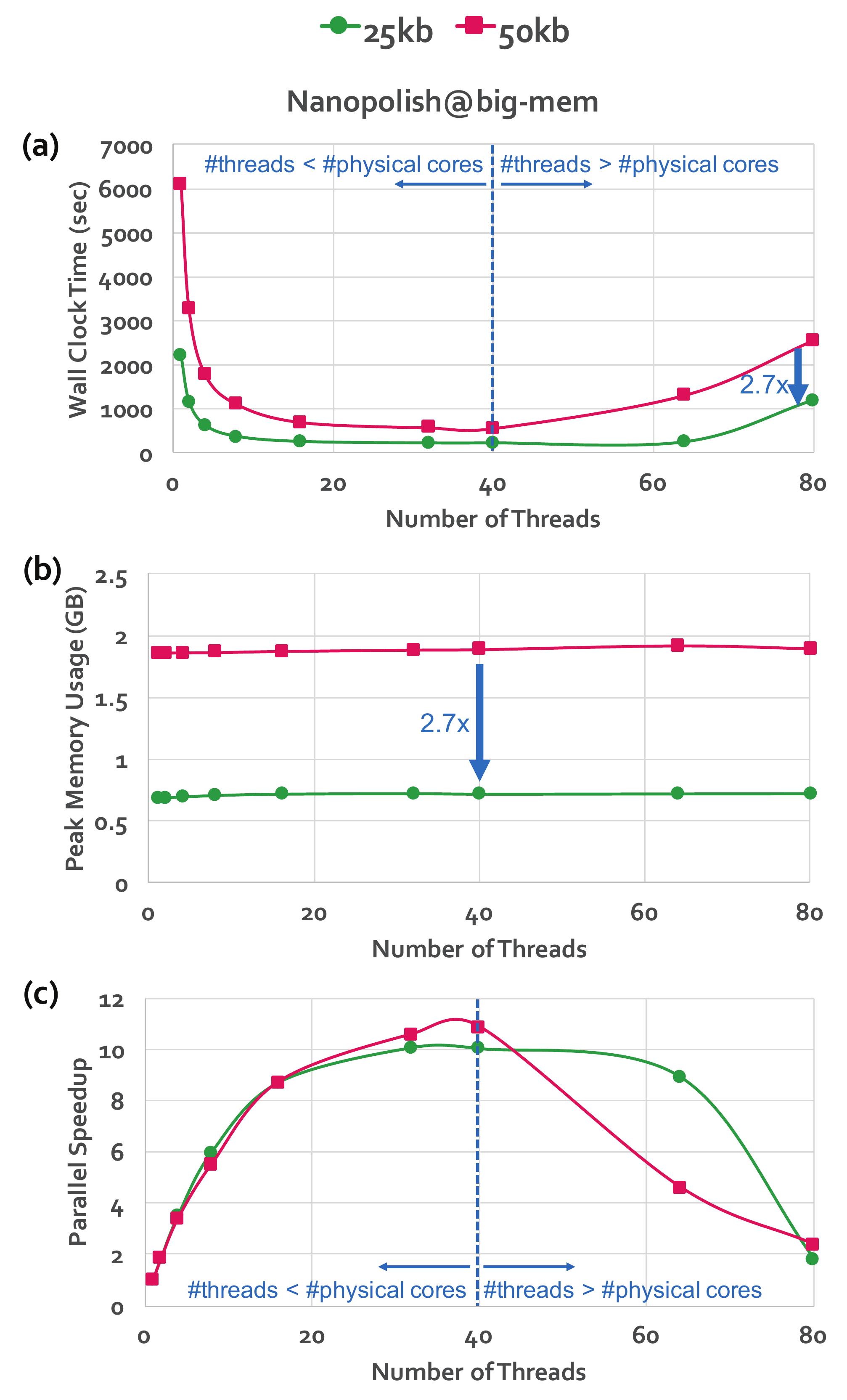}
\caption[Scalability results of Nanopolish.]{Scalability results of Nanopolish. Wall clock time (a), peak memory usage (b), and parallel speedup (c) results obtained on the \textit{big-mem} system.}\label{fig:nanopolishplot}
\end{figure}

Figure~\ref{fig:nanopolishplot} shows the scalability results for Nanopolish. We test the tool by separately using a 25kb and a 50kb segment length to assess the scalability of the tool with respect to the segment length, in addition to the scalability with respect to the number of threads. We measure the performance metrics. We only show the results for the \textit{big-mem} system since the results for both of the systems are similar. Based on the results, we make the following observation.

\textbf{Observation 20: }\textit{Nanopolish's memory usage is independent of the number of threads. However, its memory usage in dependent on the segment length (cf. Figure~\ref{fig:nanopolishplot}b).}\par
The memory usage of Nanopolish is not affected by the number of threads. However, it is dependent on the segment length. Nanopolish uses more memory for longer segments. When the segment length is doubled from 25kb to 50kb, the increase in the memory usage (\ie 2.7x) is greater than 2.0x. This is because the memory usage of Nanopolish depends both on the length of the segment and the number of read mappings that map to this segment. For both of the segments, the memory usage also affects the speed. The Nanopolish run for the 25kb-segment is 2.7x faster than the run for the 50kb-segment (see Figure~\ref{fig:nanopolishplot}a).\par
\textbf{Observation 21: }\textit{Nanopolish's performance greatly degrades when the number of threads exceeds the number of physical cores (cf. Figure~\ref{fig:nanopolishplot}c).}

Hyper-threading causes a slowdown for Nanopolish because of the CPU-intensive workload of Nanopolish and the resulting high contention in the shared resources between the threads executing on the same core, as we discuss in Observation 5.\par
\textbf{Summary.} Based on the observations we make about tools for the optional last two steps of the pipeline, we conclude that further polishing can significantly increase the accuracy of the assemblies. Since BWA-MEM and Nanopolish are more resource-intensive than Minimap and Racon, pipelines with Minimap and Racon can provide a significant speedup compared to the pipelines with BWA-MEM and Nanopolish, while resulting with high-quality consensus sequences. 

\section{Recommendations} \label{sec:recommendations}

\subsection{Recommendations for Tool Users}
Based on the results we have collected and observations we have made for each step of the genome assembly pipeline using nanopore sequence data and the associated tools, we make the following major recommendations for the current and future tool users.

\begin{itemize}
\item ONT's basecalling tools, Metrichor, Nanonet, and Scrappie, are the best choices for the basecalling step in terms of both accuracy and performance. Among these tools, Scrappie is the newest, fastest and most accurate basecaller. Thus, we recommend using Scrappie for the basecalling step (See analysis in Section~\ref{sec:basecallresults}).
\item For the read-to-read overlap finding step, Minimap is faster than GraphMap, and it requires low memory. Also, it has similar accuracy to GraphMap. Thus, we recommend Minimap for the read-to-read overlap finding step (See analysis in Section~\ref{sec:overlapresults}).
\item For the assembly step, if execution time is not an important concern, we recommend using Canu since it produces much more accurate assemblies. However, for a fast initial analysis, we recommend using Miniasm since it is fast and its accuracy can be increased with an additional polishing step. If Miniasm is used for assembly, we definitely recommend further polishing to increase the accuracy of the final assembly (See analysis in Section~\ref{sec:assemblyresults}). Even though polishing takes a similar amount of time if we use Miniasm or Canu, the accuracy improvements are much \emph{smaller} for a genome assembled using Canu. We hope that future work can improve the performance of polishing when the assembled genome already has high accuracy, to reduce the execution time of the overall assembly pipeline.
\item For the polishing step, we recommend using Racon since it is much faster than Nanopolish. Racon also produces highly-accurate assemblies (See analysis in Section~\ref{sec:polishresults}).
\item In the future, laptops may become a popular platform for running genome assembly tools, as the portability of a laptop makes it a good fit for in-field analysis. Compared to the desktop and server platforms that we use to test our pipelines, a laptop has even greater memory constraints and lower computational power, and we must factor in limited battery life when evaluating the tools. Based on the scalability studies we perform using our desktop and server platforms, we would likely recommend using Minimap followed by Miniasm for the assembly step, and Minimap followed by Racon for the polishing step, when performing assembly on a laptop. These three tools use relatively low amounts of memory, and execute quickly, which we expect would make the tools a good fit for the various constraints of a laptop. Despite their low memory usage and fast execution, our recommended pipeline can produce high-quality assemblies that are suitable for fast initial in-field analyses. We leave it to future work to quantitatively study the genome assembly pipeline using nanopore sequence data on laptops and other mobile devices.
\end{itemize}

\subsection{Recommendations for Tool Developers}
Based on our analyses, we make the following recommendations for the tool developers.

\begin{itemize}
\item The choice of language to implement the tool plays a crucial role regarding the overall performance of the tool. For example, although the basecallers Scrappie and Nanonet belong to the same family (\ie they both use the more accurate RNNs for basecalling), Scrappie is significantly faster than Nanonet since Scrappie is implemented in C whereas Nanonet is implemented in Python (See analysis in Section~\ref{sec:basecallresults}).
\item Memory usage is an important factor that greatly affects the performance and the usability of the tool. While developing new tools or improving the current ones, the developers should be aware of the memory hierarchy. Data structure choices that can minimize the memory requirements and cache-efficient algorithms have a positive impact on the overall performance of the tools. Keeping memory usage in check with the number of threads can enable not only a usable (\ie runnable on machines with relatively small memories) tool but also a fast one. For example, we find that GraphMap cannot even run with a single-thread in our \emph{desktop} machine due to excessively high memory usage (See analyses in Sections~\ref{sec:basecallresults}--~\ref{sec:polishresults}).
\item Scalability of the tool with the number of cores/threads is an important requirement. It is important to make the tool efficiently parallelized to decrease the overall runtime. Design choices should be made wisely while considering the possible overheads that parallelization can add. For example, we find that the parallel speedup of Minimap reduces when the number of threads reaches a high number due to a large increase in the overhead of synchronization between threads (See analyses in Sections~\ref{sec:basecallresults}--~\ref{sec:polishresults}).
\item Since parallelizing the tool can increase the memory usage, dividing the input data into batches, or limiting the memory usage of each thread, or dividing the computation instead of dividing the dataset between simultaneous threads can prevent large increases in memory usage, while providing performance benefits from parallelization. For example, in Nanonet, all of the threads share the computation of each read, and thus memory usage is not affected by the amount of thread parallelism. As a result, Nanonet's usability is not limited to machines with relatively larger memories (See analyses in Sections~\ref{sec:basecallresults}--~\ref{sec:polishresults}).
\end{itemize}

\section{Summary} \label{sec:nanopore-conclusion}
We analyze the multiple steps and the associated state-of-the-art tools in the genome assembly pipeline using nanopore sequence data
in terms of accuracy, speed, memory efficiency and scalability. We make four major conclusions based on our experimental analyses of the whole pipeline. First, the basecalling tools with higher accuracy and performance, like Scrappie, can overcome the major drawback of nanopore sequencing technology, \ie high error rates. Second, the read-to-read overlap finding tools, Minimap and GraphMap, perform similarly in terms of accuracy. However, Minimap performs better than GraphMap in terms of speed and memory usage by storing only minimizers instead of all \textit{k}-mers, and GraphMap is not scalable when running on machines with relatively small memories. Third, the fast but less accurate assembler Miniasm can be used for a very fast initial assembly, and further polishing can be applied on top of it to increase the accuracy of the final assembly. Fourth, a state-of-the-art polishing tool, Racon, generates high-quality consensus sequences while providing a significant speedup over another polishing tool, Nanopolish.\par
We hope and believe that our observations and analyses will guide researchers and practitioners to make conscious and effective choices while deciding between different tools for each step of the genome assembly pipeline using long reads. We also hope that the bottlenecks or the effects of design choices we have found and exposed can help developers in building new tools or improving the current ones. Based on our analysis and recommendations, we also show that we need high-performance, memory-efficient, low-power, and scalable designs for genome sequence analysis in order to exploit the advantages that genome sequencing provides.

\chapter{GenASM: A High-Performance, Low-Power Approximate String Matching Acceleration Framework for Genome Sequence Analysis} 
\label{ch4-genasm}

Read mapping is one of the first key steps in genome sequence analysis. For both short and long reads, \emph{multiple} steps of \revII{read mapping} must account for the \revonur{sequencing errors, \revII{and \sgii{for} the differences} caused by genetic mutations and variations}. These \revIII{errors and} \revII{differences} take the form of base insertions, deletions, and/or substitutions\revonur{~\cite{navarro2001guided,waterman1976some,smith1981identification,wu1992fast,myers1999fast,ukkonen1985algorithms}}.
As a result, \revII{read mapping} must perform \emph{approximate} (or \emph{fuzzy}) \emph{string matching} (ASM). Several algorithms exist for ASM, but state-of-the-art \revII{read mapping} tools typically make use of an expensive dynamic programming based algorithm\revonur{~\cite{smith1981identification,levenshtein1966binary,needleman1970general}} that scales quadratically in both execution time and required storage. This ASM algorithm has been shown to be the major bottleneck in \revII{read mapping}~\cite{alser2017gatekeeper,turakhia2018darwin,fujiki2018genax,alser2020accelerating,ham2020genesis,nag2019gencache,huangfu2019medal}. Unfortunately, as sequencing technologies advance, the growth in the rate that sequencing devices generate reads is far outpacing the corresponding growth in computational power~\cite{check2014technology, alser2020accelerating}, placing greater pressure on the ASM bottleneck.
Beyond \revII{read mapping}, ASM is a key technique for \revII{other bioinformatics problems such as whole genome alignment (WGA)~\cite{delcher1999alignment,kurtz2004versatile,bray2003avid,hohl2002efficient,schwartz2003human,brudno2003lagan,dewey2019whole,darwin-wga,lin2020gsalign,marccais2018mummer4,li2018minimap2} and multiple sequence alignment (MSA)~\cite{sankoff1975minimal,carrillo1988multiple,paten2011cactus,higgins1988clustal,lipman1989tool,notredame2000t,lee2002multiple,notredame2002recent,edgar2006multiple}, where two or more whole genomes, or regions of multiple genomes (from the same or different species), are compared to determine their similarity 
for predicting evolutionary relationships or finding common regions (e.g., genes).}
Thus, there is a \revII{pressing} need to develop \revII{techniques} for \revII{genome sequence analysis} that \sgrev{provide} \revonur{fast and efficient} ASM.

In this work, we propose \emph{GenASM}, \sg{an ASM} \revIV{acceleration} framework for \revII{genome sequence analysis}. Our goal is to design a fast, efficient, and flexible framework for both short and long reads, which \sg{can be used to accelerate} \emph{multiple steps} of the genome sequence analysis pipeline. 
\sg{To avoid implementing more complex hardware for the dynamic programming \revII{based} algorithm\revonur{~\cite{Fei2018, kaplan2020bioseal, turakhia2018darwin, gupta2019rapid, Banerjee2019,jiang2007reconfigurable,rucci2018swifold, chen2014accelerating}, we base GenASM upon}
the \textit{Bitap} algorithm~\cite{baeza1992new, wu1992fast}.
Bitap uses only fast and simple bitwise operations to perform 
\revonur{approximate} string matching, making it amenable to \revonur{efficient} hardware acceleration. To our knowledge, GenASM is the first work that enhances and accelerates Bitap.}

\section{Approximate String Matching (ASM)}
\label{sec:approx-background}

\begin{figure}[h!]
\centering
\includegraphics[width=10.5cm,keepaspectratio]{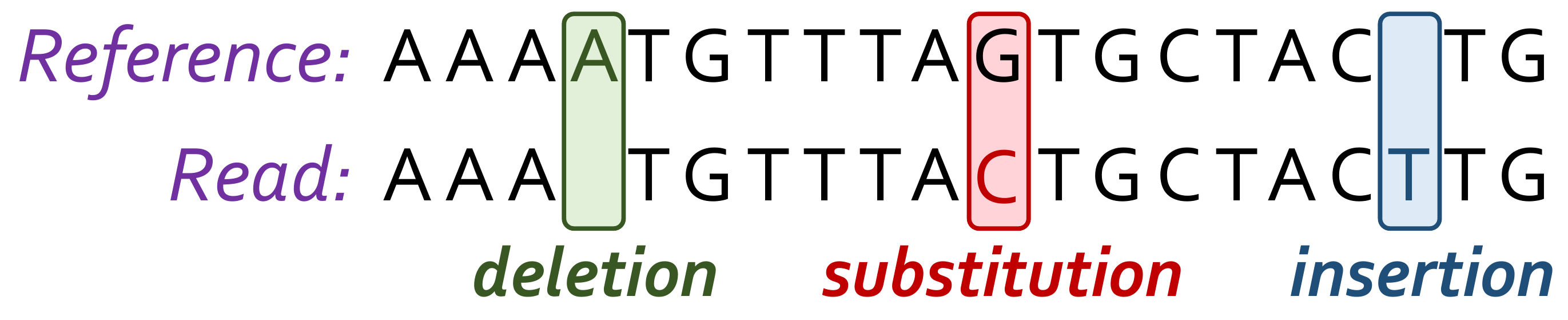}
\caption{Three types of errors (i.e., edits).}\label{fig:edits}
\end{figure}

The goal of approximate string matching \cite{navarro2001guided} is to detect the differences and similarities between \revII{two sequences. \revIII{Given a query read sequence~$Q$=[$q_1$$q_2$\ldots$q_m$], a reference text sequence~$T$=[$t_1$$t_2$\ldots$t_n$] (where $m=|Q|$, $n=|T|$, $n \geq m$),}} and an edit distance threshold $E$, the approximate string matching problem is to identify a set of approximate matches of \revII{$Q$ in $T$} (allowing for at most $E$ differences). 
\revonur{The differences between two sequences of the same species \sgrev{can result from sequencing errors\revIII{~\cite{fox2014accuracy,amarasinghe2020opportunities}} and/or genetic variations\revIII{~\cite{feuk2006structural,alkan2011genome}}}}.
\revII{Reads are prone to sequencing errors, \sgii{which account for about 0.1\% of the length of short reads\revIII{~\cite{glenn2011field,quail2012tale,goodwin2016coming}} and 
\revIII{10--15\%} of the length of} long reads\revIII{~\cite{jain2018nanopore, weirather2017comprehensive,ardui2018single,van2018third}}.


The differences, known as \emph{edits},} can be classified as \textit{substitutions}, \textit{deletions}, \sgrev{or} \textit{insertions} in one or both sequences~\cite{levenshtein1966binary}. Figure~\ref{fig:edits} shows each possible kind of edit. \revII{In ASM, to detect} a deleted character or an inserted \revIII{character, 
we} need to examine all possible \sgrev{\emph{prefixes}} (i.e., substrings that include the first character of the string) \revIII{or \revV{\emph{suffixes}} (i.e., substrings that include the last character of the string)} of the two input sequences, and keep track of the pairs of prefixes \revIII{or suffixes} that provide the \revonur{minimum number of edits}.

Approximate string matching is needed not only to determine the minimum number of edits between two genomic sequences, but also to provide the location and type of each edit. As \revII{two} sequences could have a large number of different possible arrangements of the edit operations and matches (and hence different \revII{\emph{alignments}}), the approximate string matching algorithm usually involves a traceback step. 
\revII{The alignment score is the sum of all edit penalties and match scores along the alignment, as defined by a user-specified scoring function.}
This step finds the \revII{\emph{optimal alignment} as the combination of edit operations to build up the highest alignment score.}

Approximate string matching is typically implemented as a dynamic programming 
\revII{based} algorithm. Existing implementations, such as Levenshtein distance \cite{levenshtein1966binary}, Smith-Waterman \cite{smith1981identification}, and Needleman-Wunsch \cite{needleman1970general}, have quadratic time and space complexity (i.e., $O(m \times n)$ between two sequences with lengths $m$ and $n$). \revII{Therefore, it is desirable to find lower-complexity algorithms for ASM.}

\section{Bitap Algorithm}
\label{sec:background-bitap}

One candidate to replace dynamic programming \revII{based} \revIII{algorithms} for \revIV{ASM} is the \textit{Bitap} algorithm~\cite{baeza1992new, wu1992fast}.
Bitap tackles the problem of computing the minimum edit distance between a \revonur{reference} text (e.g., reference genome) and a query pattern (e.g., read) with a maximum of \textit{k} many errors. When \textit{k} is 0, the algorithm finds the exact matches. 

\begin{figure}[h!]
\centering
\includegraphics[width=\columnwidth,keepaspectratio]{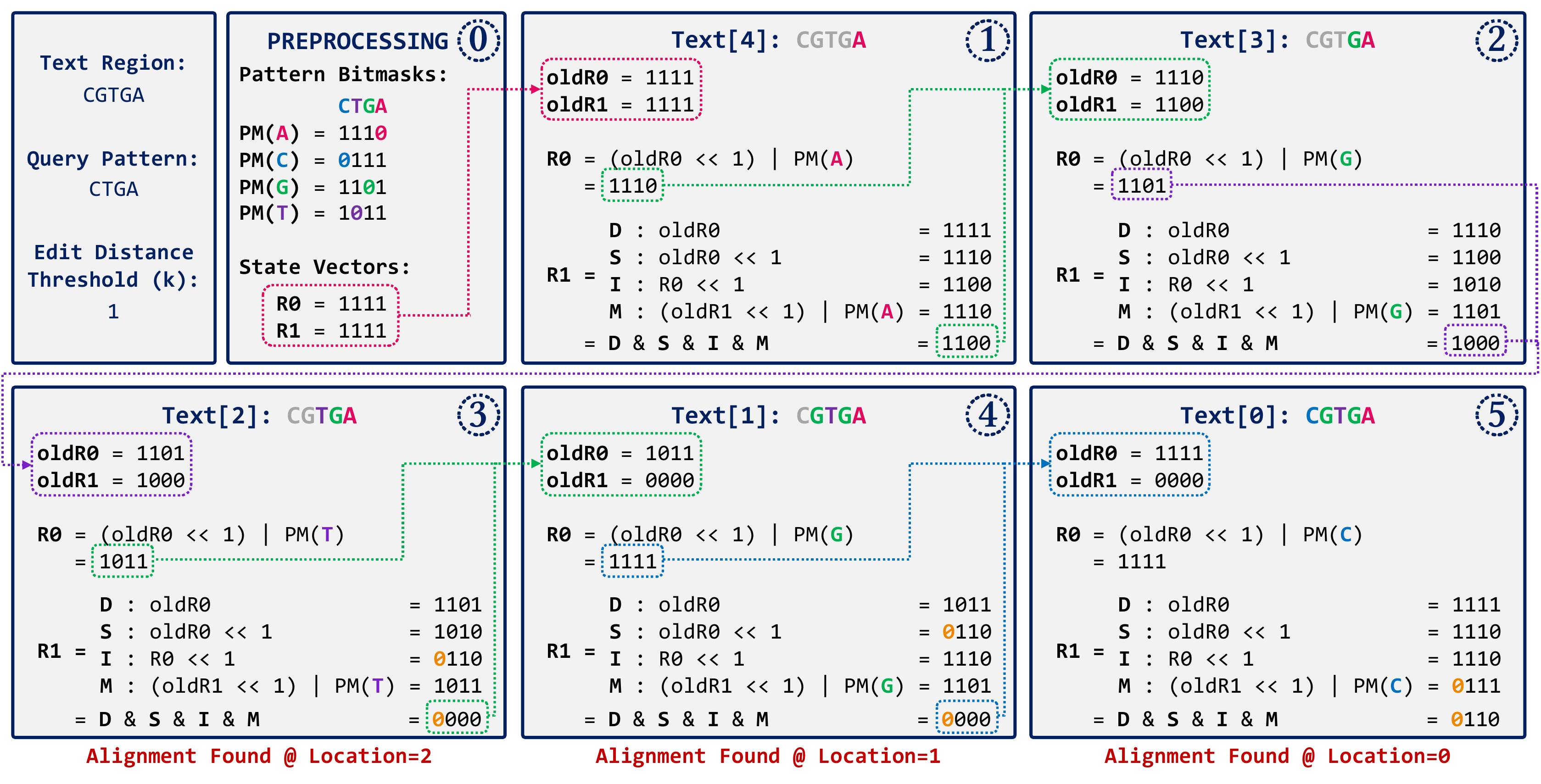}
\caption{\revonur{Example for the Bitap algorithm.}} \label{fig:bitap-dc-example}
\end{figure}

\revonur{
Algorithm~\ref{bitap-search-alg} shows the \textit{Bitap} algorithm and Figure~\ref{fig:bitap-dc-example} shows an example for the execution of the algorithm. The algorithm starts with a pre-processing procedure (\revIII{Line~4} in Algorithm~\ref{bitap-search-alg}; \circlednumberr{0} in Figure~\ref{fig:bitap-dc-example}) that converts the query pattern into $m$-sized pattern bitmasks, \textit{PM}. 
We generate one pattern bitmask for each character in the alphabet. \revII{Since $0$ means match in the Bitap algorithm, we} \revII{set} $PM[a][i]=0$ \revIII{when} $pattern[i] = a$, where $a$ is a character from the alphabet (\revIII{e.g.,} A, C, G, T). 
These pattern bitmasks help us to represent the \revIII{query pattern} in a binary format.
After the bitmasks are prepared for each character, \sgii{every bit} of all \revII{status \sgii{bitvectors} ($R[d]$, where $d$ is in range $[0,k]$)} \sgii{is initialized to 1} (\revIII{Lines~5--6} in Algorithm~\ref{bitap-search-alg}; \circlednumberr{0} in Figure~\ref{fig:bitap-dc-example}). \revII{Each $R[d]$ bitvector at text iteration $i$ holds the partial match information between $text[i:(n-1)]$ \revV{(Line~8)} and the query with maximum of $d$ \revIII{errors. 
Since} at the beginning of the execution there are no matches, we initialize all status bitvectors with 1s.}
The \revII{status} \sgii{bitvectors} of the previous iteration with edit distance $d$ is kept in $oldR[d]$ 
(\revIII{Lines~10--11}) 
to take \revII{partial matches into consideration in the next iterations}.}

\begin{algorithm}[h!]
\fontsize{10}{10}
\caption{Bitap Algorithm}\label{bitap-search-alg}
\textbf{Inputs:} \texttt{text} (reference), \texttt{pattern} (query), \texttt{k} (edit distance threshold)\\
\textbf{Outputs:} \texttt{startLoc} (matching location), \texttt{editDist} (minimum edit distance)
\begin{algorithmic}[1]
    \State $\texttt{n} \gets \texttt{length of \revIII{reference text}}$
    \State $\texttt{m} \gets \texttt{length of \revIII{query pattern}}$
    \Procedure{Pre-Processing}{}
        \State $\texttt{PM} \gets $\texttt{generatePatternBitmaskACGT(pattern)}
        \Comment{\comm{pre-process the pattern}}
        \For{\texttt{d in 0:k }}
            \State $\texttt{R[d]} \gets \texttt{111..111}$ \Comment{\comm{initialize R bitvectors to 1s}}
        \EndFor
    \EndProcedure
    \Procedure{Edit Distance Calculation}{}
        \For{\texttt{i in (n-1):\revV{-1:}0}}
        \Comment{\comm{iterate over each text character}}
            \State $\texttt{curChar} \gets \texttt{text[i]}$
            \For{\texttt{d in 0:k }}
                \State $\texttt{oldR[d]} \gets \texttt{R[d]}$ \Comment{\comm{copy previous iterations' bitvectors as oldR}}
            \EndFor
            \State $\texttt{curPM} \gets \texttt{PM[curChar]}$
            \Comment{\comm{retrieve the pattern bitmask}}
            \State $\texttt{R[0]} \gets \texttt{(oldR[0]}\verb|<<|1)\texttt{ | curPM}$ \Comment{\comm{status bitvector for exact match}}
            \For{\texttt{d in 1:k }}
            \Comment{\comm{iterate over each edit distance}}
                \State $\texttt{deletion \revIII{(D)}} \gets \texttt{oldR[d-1]}$
                \State $\texttt{substitution \revIII{(S)}} \gets \texttt{(oldR[d-1]}\verb|<<|\texttt{1)}$
                \State $\texttt{insertion \revIII{(I)}} \gets \texttt{(R[d-1]}\verb|<<|\texttt{1)}$
                \State $\texttt{match \revIII{(M)}} \gets \texttt{(oldR[d]}\verb|<<|\texttt{1)}\texttt{ | curPM}$
                \State $\texttt{R[d]} \gets \texttt{\revIII{D \& S \& I \& M}}$
                \Comment{\comm{status bitvector for $d$ errors}}
            \EndFor
            \If {\texttt{MSB of R[d] == 0, where 0 $\leq$ d $\leq$ k}}
            \Comment{\comm{check if MSB is 0}}
                \State $\texttt{startLoc} \gets \texttt{i}$
                \Comment{\comm{matching location}}
                \State $\texttt{editDist} \gets \texttt{d}$
                \Comment{\comm{found minimum edit distance}}
            \EndIf
        \EndFor
    \EndProcedure
\end{algorithmic}
\end{algorithm}

\revII{The algorithm examines each text character one by one, one per iteration.} At each text iteration (\circlednumberr{1}--\circlednumberr{5}), the \revII{pattern} bitmask of the current text character ($PM$) is retrieved \revIII{(Line~12)}. \revV{After the status bitvector for exact match is computed ($R[0]$; Line~13), the}
\sgii{status bitvectors for each distance \revV{($R[d]; d=1...k$)} are computed using the rules in \revIII{Lines~15--19}.
For a distance $d$, three intermediate bitvectors for the error cases (one each for deletion, insertion, substitution; D/I/S in Figure~\ref{fig:bitap-dc-example}) are calculated by using $oldR[d-1]$ or $R[d-1]$, since a new error is being added (i.e., the distance is increasing by 1), while the intermediate bitvector \revV{for the} match case (M) is calculated using $oldR[d]$.}
\sgii{For a deletion \revIII{(Line~15)}, we are looking for \revIII{a string match} if the current pattern character is missing, so we copy the partial match information of the previous character ($oldR[d-1]$; consuming a text character) \emph{without} any shifting (\emph{not} consuming a pattern character) to serve as the deletion bitvector \revIII{(labeled as $D$ of $R1$ bitvectors in \circlednumberr{1}--\circlednumberr{5})}.
For a substitution \revIII{(Line~16)}, we are looking for \revIII{a string match} if the current pattern character and \revV{the} current text character do not match, so we take the partial match information of the previous character ($oldR[d-1]$; consuming a text character) and shift it left by one (consuming a pattern character) before saving it as the substitution bitvector \revIII{(labeled as $S$ of $R1$ bitvectors in \circlednumberr{1}--\circlednumberr{5})}.
For an insertion \revIII{(Line~17)}, we are looking for \revIII{a string match} if the current text character is missing, so we copy the partial match information of the \emph{current} character ($R[d-1]$; \emph{not} consuming a text character) and shift it left by one (consuming a pattern character) before saving it as the insertion bitvector \revIII{(labeled as $I$ of $R1$ bitvectors in \circlednumberr{1}--\circlednumberr{5})}.
For a match \revIII{(Line~18)}, we are looking for \revIII{a string match} only if the current pattern character matches the current text character, so we take the partial match information of the previous character ($oldR[d]$; consuming a text character but \emph{not} increasing the edit distance), shift it left by one (consuming a pattern character), \revII{and perform an OR operation with the pattern bitmask of the current text character ($curPM$; comparing the text character and the pattern character)} before saving the result as the match bitvector \revIII{(labeled as $R0$ bitvectors and $M$ of $R1$ bitvectors in \circlednumberr{1}--\circlednumberr{5})}.}

%
%
%

\revII{After computing all four \sgii{intermediate} bitvectors,
in order to take all possible partial matches into consideration, we perform an \sgii{AND operation} \revIII{(Line~19)} with these four bitvectors to \revIII{preserve} all 0s \revIII{that exist in any of them 
(i.e., all potential locations for a string match with 
an edit distance of $d$ up to this point)}. \sgii{We save the ANDed result as the} $R[d]$ status bitvector for the current iteration. \sgii{This process is repeated for each potential edit distance value from 0 to $k$.}}
\sgii{If} \revII{the most significant bit of \sgii{the} $R[d]$ bitvector becomes 0 \revIII{(Lines~20--22)}, \revIV{then there} is a match starting at position $i$ of the text with an edit distance $d$ \revIII{(as shown in \circlednumberr{3}--\circlednumberr{5}).}}
The traversal of the text \sgii{then} continues until all possible text positions are \sgii{examined.}

\section{Motivation and Goals} \label{sec:motivation}

\sgrev{Although the Bitap algorithm is highly suitable for hardware acceleration due to the simple nature of its bitwise operations, \revIV{we find that} it has five limitations that hinder its \revIII{applicability and} efficient hardware acceleration for \revIII{genome analysis}. In this section, we discuss each of these limitations. In order to overcome these limitations and design an effective and efficient accelerator, we find that we need to both \revIII{(1)~modify and extend} the Bitap algorithm and \revIII{(2)~develop} specialized hardware that can exploit the new opportunities that our algorithmic modifications provide.}

\revonur{

\vspace{-1pt}
\subsection{Limitations of Bitap on Existing Systems}
\label{sec:motivation:limitations}
\vspace{-3pt}

\textbf{\revII{No Support for Long Reads.}}
\sgrev{In state-of-the-art implementations of Bitap}, the query length is limited by the word size of the machine running \sgrev{the algorithm.  This is due to (1)~the fact that the bitvector length must be equal to the query length, and (2)~the need to perform bitwise operations on the bitvectors. By limiting the bitvector length to a word, each bitwise operation can be done using a single CPU instruction.
Unfortunately, the lack of multi-word queries prevents these implementations from working for long reads, whose lengths are on the order of thousands \revIII{to millions} of base pairs
(which require thousands of \revII{bits} 
to store).}

\textbf{Data Dependency Between Iterations.} 
As we show in Section~\ref{sec:background-bitap}, the computed bitvectors at each \revIII{text iteration} (i.e., R[d]) \revIII{of the Bitap algorithm} depend on the bitvectors computed in the previous \revIII{text} iteration (i.e., oldR[d-1] and oldR[d]; \revIII{Lines~11, 13, 15, 16, and 18 of Algorithm~\ref{bitap-search-alg})}. Furthermore, for each \revIII{text} character, there is an inner loop that iterates for the maximum edit distance number of iterations \revIII{(Line~14)}. The bitvectors computed in each of these inner iterations (i.e., R[d]) are also dependent on the previous inner iteration's computed bitvectors (i.e., R[d-1]; \revIII{Line~17})}. \sgrev{This two-level data dependency forces the \revII{consecutive} iterations to take place sequentially.}

\textbf{\revII{No Support for Traceback.}}
\revII{Although the baseline Bitap algorithm can find possible matching locations of each query read within the reference text, this covers only the first step of approximate string matching required for genome sequence analysis. Since there could be multiple different alignments between the read and the reference, \revIII{the} traceback \revIII{operation~\cite{myers1988optimal,gotoh1986alignment,gotoh1982improved,miller1988sequence,waterman1984efficient,altschul1986optimal,fickett1984fast,smith1981identification,waterman1976some,ukkonen1985algorithms}} is needed to find the \emph{optimal alignment}, which is the alignment with the minimum edit distance (or with the highest score based on a user-defined scoring function). However, Bitap does not include any such support for optimal alignment identification.}

\textbf{Limited Compute Parallelism.} 
\revII{Even \revIII{after} we solve the algorithmic limitations of Bitap, we find that we cannot extract significant performance benefits with just algorithmic enhancements alone. For example,}
\sgrev{while Bitap iterates over each character of the input text sequentially \revIII{(Line~8)},} we can enable \emph{text-level parallelism} \sgrev{to improve its performance (Section~\ref{sec:bitap-search}).} 
However, the \sgrev{achievable} level of parallelism is limited by the number of compute units in existing systems. For example, our studies show that Bitap is bottlenecked by computation on CPUs, since the working set fits within the private caches but the limited number of cores prevents the further speedup of the algorithm.

\textbf{Limited Memory Bandwidth.} 
\sgrev{  
We would expect that a GPU, which has thousands of compute units, can overcome the limited compute parallelism issues that CPUs experience.} 
However, \sgrev{we find that a GPU implementation of the \revIII{Bitap} algorithm suffers from the limited amount of memory bandwidth available for each GPU thread. Even} when we run a CUDA implementation of the baseline Bitap algorithm~\cite{li2011fast}, \sgrev{whose bandwidth requirements are significantly lower than our modified algorithm, the limited \revIII{memory} bandwidth bottlenecks the algorithm's performance.}
\sgrev{We find that the bottleneck is exacerbated after the number of threads per block reaches 32, as Bitap} becomes shared cache-bound (i.e., on-GPU L2 cache-bound). The small number of registers becomes insufficient to hold the intermediate data required for Bitap execution. Furthermore, when the working set of a thread does not fit within the private memory of the thread, destructive interference between threads while accessing the shared memory creates bottlenecks in the algorithm on GPUs.
\sgrev{We expect these issues to worsen when we implement traceback, which requires significantly higher bandwidth than Bitap.}

\vspace{-1pt}
\subsection{Our Goal}
\label{sec:motivation:genasm}
\vspace{-1pt}

\revII{Our goal in this work is to overcome these limitations and use Bitap in a fast, efficient, and flexible \revII{ASM} framework for both short and long reads. We find that this \revIII{goal} cannot be achieved by modifying only the algorithm or only the hardware. \revIII{We design} \emph{GenASM}, the first ASM \revIV{acceleration} framework for \revII{genome sequence analysis}. Through careful modification and co-design of the enhanced Bitap algorithm and hardware, \revIII{GenASM aims to} successfully replace the expensive dynamic programming \revII{based} algorithm used for ASM in genomics with the efficient bitwise-operation-based Bitap algorithm, which can accelerate \emph{multiple steps} of genome sequence analysis.}

\section{GenASM: A High-Level Overview}
\label{sec:bitmac-overall}

\begin{figure}[h!]
\centering
\vspace{17pt}
\includegraphics[width=\columnwidth,keepaspectratio]{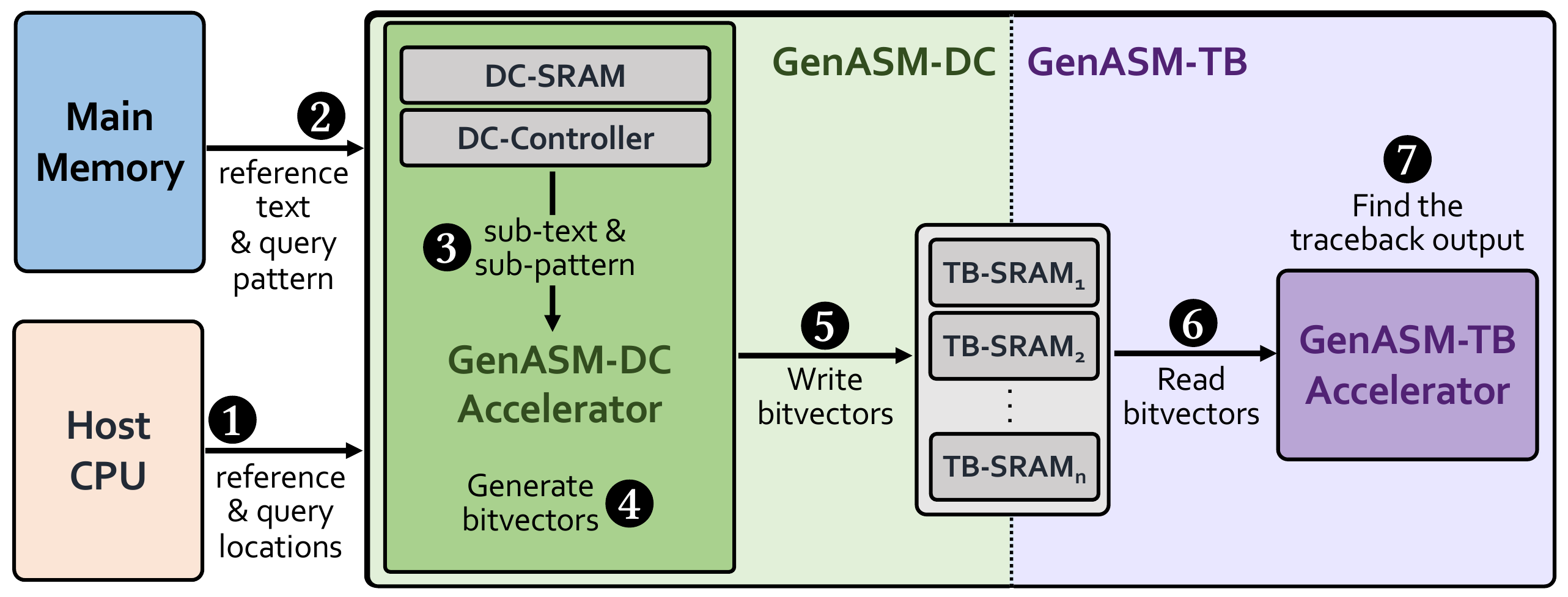}
\caption{\revII{Overview of GenASM.}} \label{fig:bitmac-pipeline}
\vspace{17pt}
\end{figure}

\revonur{In GenASM, we \emph{co-design} our modified Bitap algorithm for distance calculation (DC) and our new Bitap-compatible traceback (TB) algorithm with an area- and power-efficient hardware accelerator.
GenASM consists of two components, as shown in Figure~\ref{fig:bitmac-pipeline}:
(1)~GenASM-DC (Section~\ref{sec:bitap-search}), which for each read generates the bitvectors
and \revonur{performs the minimum edit distance calculation (DC)}; and
(2)~GenASM-TB (Section~\ref{sec:bitap-traceback}), which uses the bitvectors to perform traceback \revonur{(TB)} and find the optimal alignment. GenASM is a flexible framework \revII{that} can be used for different use cases \revIII{(Section~\ref{sec:bitmac-framework}}).}

\sgii{GenASM execution starts when} the host CPU issues \sgii{a} task to GenASM with the reference and the query sequences' locations (\circlednumber{1} in Figure~\ref{fig:bitmac-pipeline}).
GenASM-\hl{DC reads the corresponding reference \revV{text} region and the query \revV{pattern} from the memory.  GenASM-DC then writes these to its dedicated SRAM, which we call DC-SRAM (\circlednumber{2}). After that, GenASM-DC divides the reference \revIII{text (e.g., reference genome) and query pattern (e.g., read)} into multiple overlapping windows \revIII{(\circlednumber{3})}, and for each \emph{sub-text} (i.e., the portion of the reference \revIII{text} in one window) and \emph{sub-pattern} (i.e., the portion of the \revIII{query pattern} in one window), GenASM-DC searches for the sub-pattern within the sub-text and generates the bitvectors (\circlednumber{4}). Each processing element (PE) of GenASM-DC writes the generated bitvectors to its own dedicated SRAM, which we call TB-SRAM (\circlednumber{5}). Once GenASM-DC completes its search for the current window, GenASM-TB starts reading the stored bitvectors from TB-SRAMs (\circlednumber{6}) and generates the window's traceback output (\circlednumber{7}).} Once GenASM-TB generates this output, \rev{GenASM} computes the next window and repeats \sgii{Steps \circlednumber{3}--\circlednumber{7}} until all windows are completed.

Our hardware accelerators are designed to maximize parallelism and minimize memory footprint. Our modified GenASM-DC algorithm is highly parallelizable, and performs only simple and regular bitwise operations, so we implement \revonur{the GenASM-DC accelerator} as a systolic array based accelerator. GenASM-TB \revonur{accelerator} requires simple logic operations to perform the TB-SRAM accesses and the required control flow to complete the traceback operation. Both of our hardware accelerators are highly efficient in terms of area and power. 
We discuss \revII{them} in detail in Section~\ref{sec:bitmac-hw}.



\section{GenASM-DC Algorithm} \label{sec:bitap-search}

\begin{figure}[t!]
\centering
\includegraphics[width=0.9\columnwidth,keepaspectratio]{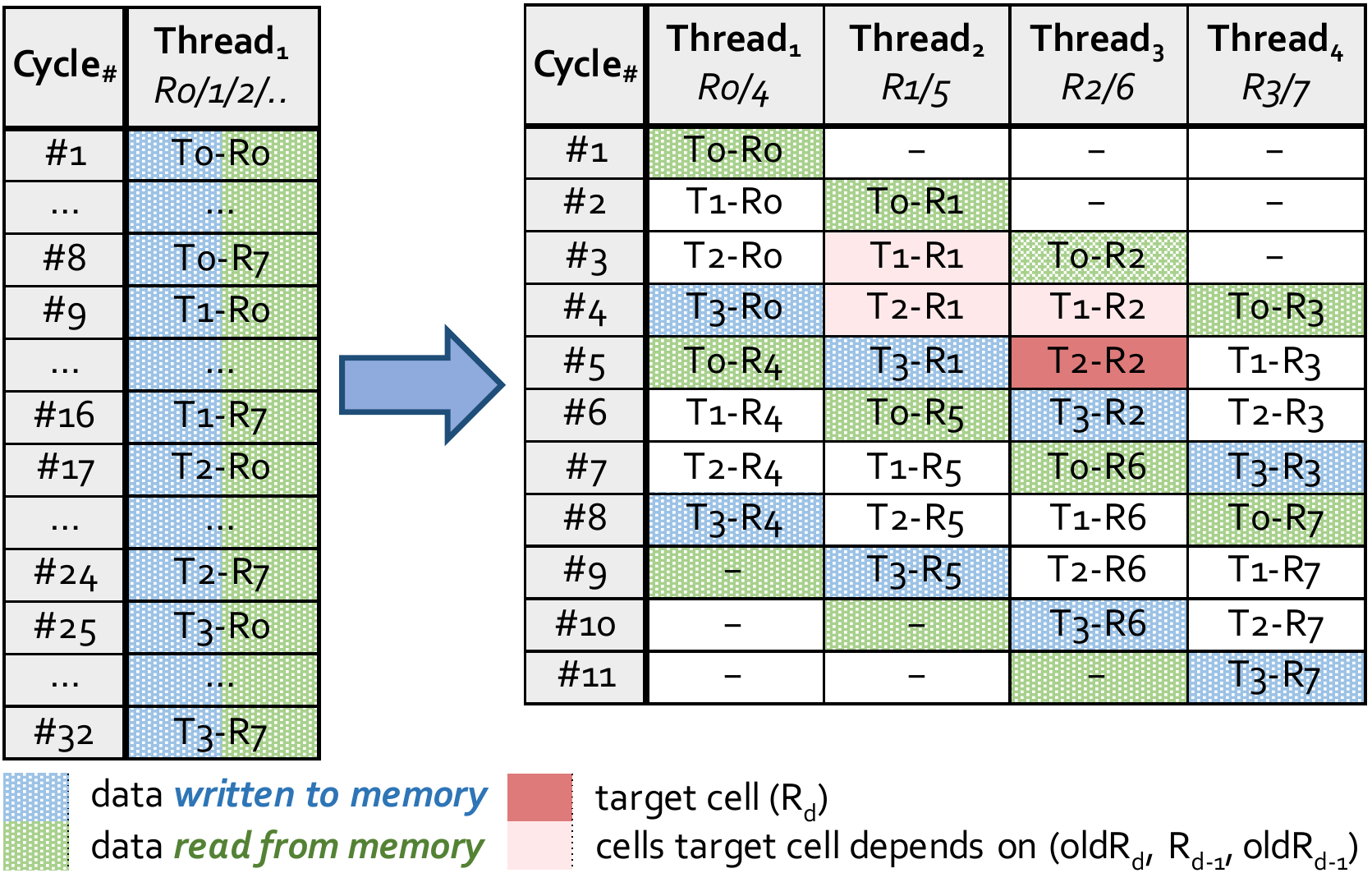}
\caption{Loop unrolling in GenASM-DC.}
\label{fig:data_depend}
\end{figure}

\rev{
We modify the baseline Bitap algorithm (Section~\ref{sec:background-bitap}) to (1)~enable efficient alignment of \revII{long reads}, (2)~remove the data dependency between the iterations, and (3)~provide parallelism for the large \revII{number} of iterations. 

\textbf{Long Read Support.}
\sgrev{The GenASM-DC algorithm \revII{overcomes} the word-length limit of Bitap \revIII{(Section~\ref{sec:motivation:limitations})} by} storing the bitvectors in multiple words when the query is longer than the word size. Although this modification leads to additional computation when performing shifts, it helps GenASM to support both short and long reads.
\sgrev{When shifting word~$i$ of a multi-word bitvector, the bit shifted out \revIII{(MSB)} of word~$i-1$ needs to be stored separately before performing the shift on word $i-1$. Then, that saved bit needs to be loaded as the least significant bit (LSB) of word~$i$ when the shift occurs.}  
\revIII{This causes the complexity of the algorithm to be $\lceil \frac{m}{w} \rceil \times n \times k$, where $m$ is the query length, $w$ is the word size, $n$ is the text length, and $k$ is the edit distance.}

\textbf{Loop Dependency Removal.}
In order to solve \revonur{the two-level data dependency limitation of the baseline Bitap algorithm (Section~\ref{sec:motivation:limitations}), GenASM-DC performs loop unrolling and enables computing non-neighbor (i.e., independent) bitvectors in parallel. Figure~\ref{fig:data_depend} shows an example for unrolling with four threads for text characters T0--T3 and status bitvectors R0--R7.}
\revII{\sg{For the iteration where $R[d]$ represents T2--R2} 
(i.e., \sgii{the} target cell shaded \sgii{in} dark red), $R[d-1]$ refers to T2--R1, $oldR[d-1]$ refers to T1--R1, and $oldR[d]$ refers to T1--R2 (i.e., \revIII{cells T2--R2 is dependent on}, shaded \revIII{in} light red). Based on this example, T2--R2 depends on T1--R2, T2--R1, and T1--R1, but it does not depend on T3--R1, T1--R3, or T0--R4. Thus, these independent bitvectors can be computed in parallel without waiting for \sgii{one another}.}

\textbf{Text-Level Parallelism.}
\sgrev{In addition to the parallelism enabled \revII{by} removing the loop dependencies, we enable GenASM-DC algorithm to exploit text-level parallelism. This parallelism is enabled by} dividing the text into overlapping sub-texts and searching the \revIII{query} in each of these sub-texts in parallel. 
The overlap 
ensures that we do not miss any possible match that may fall around the edges of a sub-text. To guarantee this, the overlap \sgrev{needs to} be of length $m+k$, where $m$ is the \revIII{query} length and $k$ is the edit distance threshold. 

}

\section{G\lowercase{en}ASM-TB Algorithm} \label{sec:bitap-traceback}

After finding the matching location of the text and the edit distance with GenASM-DC, our new traceback\revIII{~\cite{myers1988optimal,gotoh1986alignment,gotoh1982improved,miller1988sequence,waterman1984efficient,altschul1986optimal,fickett1984fast,smith1981identification,waterman1976some,ukkonen1985algorithms}} algorithm, GenASM-TB, finds the sequence of matches, substitutions, insertions and deletions, along with their positions \revV{(i.e., CIGAR string)} for the matched \revIV{region (i.e., the text region that starts from the location reported by GenASM-DC and has a length of $m+k$)}, \revIII{and reports the optimal alignment}. \revIII{Traceback execution (1)~starts from the first character of the matched region between the reference text and query pattern, (2)~examines \revIV{each character and decides} which of the four operations should be picked \revIV{in} each iteration, and (3)~ends when we reach the last character of the matched region.}
GenASM-TB uses the \revonur{intermediate bitvectors generated \revII{and saved in} each iteration of the GenASM-DC algorithm (i.e., match, substitution, deletion and insertion bitvectors generated in \revIII{Lines~15--18} in Algorithm~\ref{bitap-search-alg})}. After a \sgii{value 0} is found at the MSB of one of the $R[d]$ bitvectors \revII{(i.e., \sgii{a} \revIII{string} match is found with $d$ errors)}, 
GenASM-TB \sg{walks through the bitvectors back to the LSB, following a chain of 0s (which indicate matches at each location) and} 
reverting the bitwise operations. \revonur{\sgii{At each position, based} on which of the \revII{four} bitvectors \sgii{holds a value 0} \revII{in} each iteration (starting with \sgii{an MSB with a 0 and ending with an LSB with a 0}), the sequence of matches, substitutions, insertions and deletions (i.e., traceback output) is found for each position of the \revII{corresponding alignment found by GenASM-DC.}} Unlike GenASM-DC, GenASM-TB has an irregular \revonur{control flow} within the stored intermediate bitvectors, which depends on the \revIV{text and \revV{the} pattern.}

\revonur{Algorithm~\ref{bitap-traceback-alg} shows the \textit{GenASM-TB} algorithm and Figure~\ref{fig:bitap-traceback-alg} shows an example for the execution of the algorithm \revIII{for each of the alignments found in \circlednumberr{3}--\circlednumberr{5} of Figure~\ref{fig:bitap-dc-example}}.}
\revonur{In Figure~\ref{fig:bitap-traceback-alg}, <$x,y,z$> stands for \texttt{patternI}, \texttt{textI} and \texttt{curError}, respectively \revIII{(Lines~6--8 in Algorithm~\ref{bitap-traceback-alg})}. \texttt{patternI} represents the position of \sgii{a 0 currently being processed} within a given bitvector (i.e., pattern index), \texttt{textI} represents the outer loop iteration index (i.e., text index; $i$ in Algorithm~\ref{bitap-search-alg}), and \texttt{curError} represents the inner loop iteration index (i.e., number of remaining errors; $d$ in Algorithm~\ref{bitap-search-alg}). 

{
\begin{algorithm}[h!]
\fontsize{10}{10}
\revIII{\textbf{Inputs:} \texttt{text} (reference), \texttt{n}, \texttt{pattern} (query), \texttt{m}, \texttt{W} (window size), \texttt{O} (overlap size)\\
\textbf{Output:} \texttt{CIGAR} (complete traceback output)}
\caption{GenASM-TB Algorithm}\label{bitap-traceback-alg}
\begin{algorithmic}[1]
    \State \revIII{$\texttt{<curPattern,curText>} \gets$ \texttt{<0,0>}}
    \Comment{\comm{start positions of sub-pattern and sub-text}}
    \While{\revIII{\texttt{(curPattern < m) \& (curText < n)}}}
        
        \State \revIII{$\texttt{sub-pattern} \gets \texttt{pattern[curPattern:(curPattern+W)]}$}
        \State \revIII{$\texttt{sub-text} \gets \texttt{text[curText:(curText+W)]}$}
        
        \State \revIII{$\texttt{intermediate bitvectors} \gets \texttt{GenASM-DC(sub-pattern,sub-text,W)}$}

        \State $\texttt{patternI} \gets \texttt{W-1}$ 
        \Comment{\comm{pattern index (position of 0 being processed)}}
        \State $\texttt{textI} \gets \texttt{0}$
        \Comment{\comm{text index}}
        \State $\texttt{curError} \gets \texttt{editDist from GenASM-DC}$
        \Comment{\comm{number of remaining errors}}
        \State $\revIII{\texttt{<patternConsumed,textConsumed>} \gets \texttt{<0,0>}}$
    
        \State $\texttt{prev} \gets \texttt{""}$
        \Comment{\comm{output of previous TB iteration}}
        \While{\texttt{textConsumed<(W-O) \& patternConsumed<(W-O)}}
            \State $\texttt{status} \gets \texttt{0}$
            \If{\texttt{ins[textI][curError][patternI]=0 \& prev='I'}}
                \State $\texttt{status} \gets \texttt{3; add "I" to CIGAR;}$
                \Comment{\comm{insertion-extend}}
            \ElsIf{\texttt{del[textI][curError][patternI]=0 \& prev='D'}}
                \State $\texttt{status} \gets \texttt{4; add "D" to CIGAR;}$
                \Comment{\comm{deletion-extend}}
            \ElsIf {\texttt{match[textI][curError][patternI]=0}}
                \State $\texttt{status} \gets \texttt{1; add "M" to CIGAR; } \texttt{prev} \gets \texttt{"M"}$
                \Comment{\comm{match}}
            \ElsIf{\texttt{subs[textI][curError][patternI]=0}}
                \State $\texttt{status} \gets \texttt{2; add "S" to CIGAR; } \texttt{prev} \gets \texttt{"S"}$
                \Comment{\comm{substitution}}
            \ElsIf{\texttt{ins[textI][curError][patternI]=0}}
                \State $\texttt{status} \gets \texttt{3; add "I" to CIGAR; }
                \texttt{prev} \gets \texttt{"I"}$
                \Comment{\comm{insertion-open}}
            \ElsIf {\texttt{del[textI][curError][patternI]=0}}
                \State $\texttt{status} \gets \texttt{4; add "D" to CIGAR; }
                \texttt{prev} \gets \texttt{"D"}$
                \Comment{\comm{deletion-open}}
            \EndIf
            \If {\texttt{(status > 1)}}         
                \State $\texttt{curError-{}-}$ 
                \Comment{\comm{S, D, or I}}
            \EndIf
            \If {\texttt{(status > 0) \&\& (status != 3)}}  
                \State $\texttt{textI++; textConsumed++}$
                \Comment{\comm{M, S, or D}}
            \EndIf
            \If {\texttt{(status > 0) \&\& (status != 4)}} 
                \State $\texttt{patternI-{}-; patternConsumed++}$
                \Comment{\comm{M, S, or I}}
            \EndIf
        \EndWhile
        \State $\revIII{\texttt{curPattern} \gets \texttt{curPattern+patternConsumed}}$
        \State $\revIII{\texttt{curText} \gets \texttt{curText+textConsumed}}$
    \EndWhile
\end{algorithmic}
\end{algorithm}
}

\begin{figure}[h!]
\centering
\includegraphics[width=13cm,keepaspectratio]{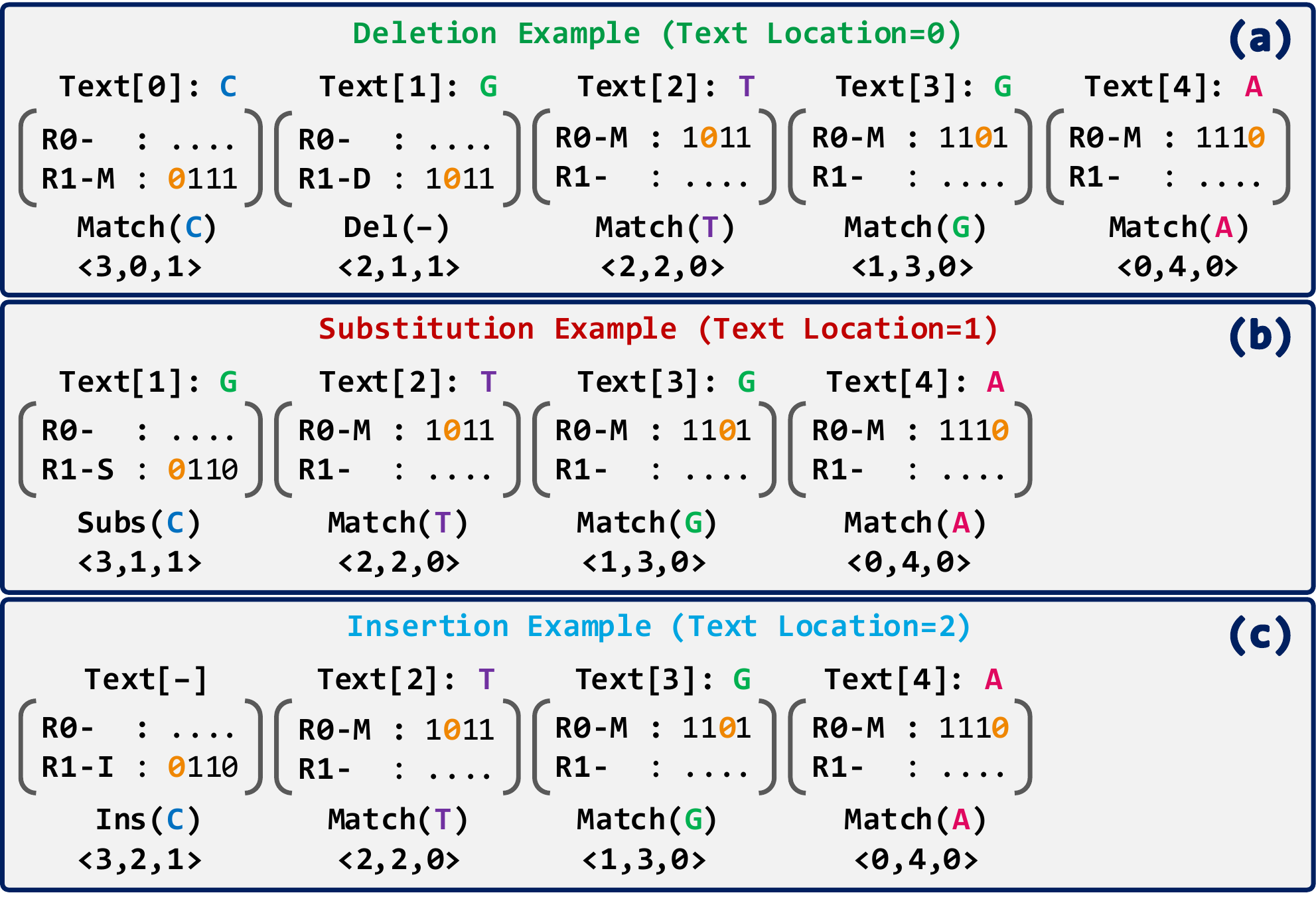}
\caption{Traceback \revII{example} with GenASM-TB algorithm.} \label{fig:bitap-traceback-alg}
\end{figure}

When we find a $0$ at \revIII{\texttt{\small{match[textI][curError][patternI]}}} (i.e., a \emph{match \revV{(M)}} is found for the current position; \revIII{Line~17}), one character \revIII{each} from both text and query is consumed, but the number of remaining errors \revII{stays the} same. Thus, the pointer moves to the next text character (as the text character \revV{is consumed}), and \sgii{the 0 currently being processed} \revII{(highlighted with orange color in Figure~\ref{fig:bitap-traceback-alg})} is right-shifted by one (as the query character is \revV{also} consumed). In other words, \texttt{textI} is incremented \revIII{(Line~28)}, \texttt{patternI} is decremented \revIII{(Line~30)}, but \texttt{curError} remains \revV{the} same. Thus, <$x,y,z$> becomes <$x-1,y+1,z$> after we find a match. For example, in Figure~\ref{fig:bitap-traceback-alg}\revII{a, for Text[0]}, we have <$3,0,1$> for the indices, and after \sgii{the match} is found, at the next position (Text[1]), we have <$2,1,1$>.

When we find a $0$ at \revIII{\texttt{\small{subs[textI][curError][patternI]}}} (i.e., a \emph{substitution \revV{(S)}} is found for the current position; \revIII{Line~19)}, one character \revII{each} from both text and query is consumed, and the number of remaining errors is decremented \revIII{(Line~26}). Thus, <$x,y,z$> becomes <$x-1,y+1,z-1$> after we find a substitution \revII{(e.g., \sgii{Text[\revIII{1}] in} Figure~\ref{fig:bitap-traceback-alg}b)}.

When we find a $0$ at \revIII{\texttt{\small{ins[textI][curError][patternI]}}} (i.e., an \emph{insertion \revV{(I)}} is found for the current position; \revV{Lines~13 and 21)}, the inserted character does not appear \revV{in the 
text}, and only a character from the pattern is consumed. \sgii{The 0 currently being processed} is right-shifted by one, but the text pointer remains the same, and the number of remaining errors is decremented. Thus, <$x,y,z$> becomes <$x-1,y,z-1$> after we find an insertion \revII{(e.g., \sgii{Text[\revIII{--}] in} Figure~\ref{fig:bitap-traceback-alg}c)}.
    
When we find a $0$ at \revIII{\texttt{\small{del[textI][curError][patternI]}}} (i.e., a \emph{deletion \revV{(D)}} is found for the current position; \revV{Lines~15 and 23)}, the deleted character does not appear \revV{in the 
pattern}, and only a character from the text is consumed. \sgii{The 0 currently being processed} is not right-shifted, but the pointer moves to the next text character, and the number of remaining errors is also decremented. Thus, <$x,y,z$> becomes <$x,y+1,z-1$> after we find an insertion \revII{(e.g., \sgii{Text[1] in} Figure~\ref{fig:bitap-traceback-alg}a)}.
}

\textbf{\revII{Divide-and-Conquer Approach.}} Since GenASM-DC stores all of the intermediate bitvectors, 
in the worst case, the length of the text region that the query pattern maps \revonur{to} can be $m+k$, assuming all of the errors are deletions from the pattern. Since we need to store all of the bitvectors for $m+k$ characters, and compute \revV{$4 \times k$} many bitvectors within each text iteration (each $m$ bits long), for long reads with high error rates, the memory requirement becomes \textasciitilde \revV{80GB}, \revonur{when m is 10,000 and k is 1,500}. 

In order to decrease the memory footprint of the algorithm, we follow two key ideas. First, we apply a divide-and-conquer approach \rev{(similar to the tiling approach of Darwin's alignment accelerator, GACT~\cite{turakhia2018darwin})}. Instead of storing all of the bitvectors for $m+k$ text characters, we divide the text and \revIII{pattern} into overlapping windows \revonur{(i.e., sub-text and \revIII{sub-pattern; Lines~3--4 in Algorithm~\ref{bitap-traceback-alg})}} and perform the traceback computation \revII{for each window}. After all of the windows' partial traceback outputs are generated, we merge them to find the complete traceback \revV{output}. This approach helps us to decrease the memory footprint \revonur{from \revV{$((m+k) \times 4 \times k \times m)$} \revIII{bits} to \revV{$(W \times 4 \times W \times W)$ \revIII{bits}}}, where $W$ is the window size. This divide-and-conquer approach also helps us to reduce the complexity of the bitvector generation step (Section~\ref{sec:bitap-search}) from \revIII{$\lceil \frac{m}{w} \rceil \times n \times k$ to $\lceil \frac{W}{w} \rceil \times W \times W$}. 
Second, instead of storing all 4 bitvectors (i.e., match, substitution, insertion, deletion) separately, we only need to store bitvectors for match, insertion, and deletion, as the substitution bitvector can be obtained easily by left-shifting the deletion bitvector by 1 \revII{(\revIII{Line~16} in Algorithm~\ref{bitap-search-alg})}.
This modification helps us to decrease the required write bandwidth and the memory footprint 
to \revV{$(W \times 3 \times W \times W)$} \revIII{bits}.

GenASM-TB restricts the number of consumed characters from the text or the pattern to \texttt{W-O} \revonur{(\revIII{Line~11} in Algorithm~\ref{bitap-traceback-alg})} to ensure that consecutive windows share $O$ characters \revII{(i.e., overlap size)}, and thus, the traceback output can be generated accurately. 
\revIII{The sub-text and the sub-pattern corresponding to each window are found using the number of consumed text characters (\texttt{textConsumed}) and the number of consumed pattern characters (\texttt{patternConsumed}) in the previous window (Lines~31--32 in Algorithm~\ref{bitap-traceback-alg})}.


\textbf{Partial Support for Complex Scoring Schemes.} 
We extend the GenASM-TB algorithm to provide \revonur{partial} support \revIII{(Section~\ref{sec:results-accuracy})} for
non-unit costs for different edits and \revonur{the} affine gap penalty model\revonur{~\cite{gotoh1982improved,miller1988sequence,waterman1984efficient,altschul1986optimal}}. 
\revIII{By changing the order in which different traceback cases are checked in \revIII{Lines~13--24} in Algorithm~\ref{bitap-traceback-alg}, we can support different types of scoring schemes. For example, in order} \sgii{to mimic \revIII{the behavior of the} affine gap penalty \revIII{model}}, we check whether the \revonur{traceback output} that has been chosen for the previous position \revII{(i.e., \texttt{prev}) is an insertion or a deletion}. If the previous edit is a gap (insertion or deletion), and 
there is a $0$ at the current position of the insertion or deletion bitvector \revIII{(Lines~13 and 15 in Algorithm~\ref{bitap-traceback-alg})}, then we prioritize extending this previously opened gap, and choose \revII{insertion-extend or deletion-extend} as the current position's \revonur{traceback output}, depending on the type of the previous gap. \revIV{As another example}, in order to mimic the behavior of non-unit costs for different edits, we can simply sort three error cases \revII{(substitution, insertion-open, deletion-open)} from the lowest penalty to the highest penalty. \revonur{If substitutions have a lower penalty than gap openings, the order shown in Algorithm~\ref{bitap-traceback-alg} should remain the same. However, if substitutions have a greater penalty than gap openings, we should check for the substitution case after checking the insertion-open and deletion-open cases \revIII{(i.e., Lines~19--20 should come after Line~24 in Algorithm~\ref{bitap-traceback-alg})}.}

\section{G\lowercase{en}ASM Hardware Design} \label{sec:bitmac-hw}


\textbf{GenASM-DC Hardware.} We implement GenASM-DC as a linear cyclic systolic array\revIII{~\cite{kung1978systolic,kung1982systolic}} based accelerator.
The accelerator is optimized to reduce both the memory bandwidth and the memory footprint. Feedback logic enabling cyclic systolic behavior allows us to fix the required number of memory ports\revIII{~\cite{kung1982systolic}} and to reduce memory footprint. 

\begin{figure*}[t!]
\centering
\includegraphics[width=\columnwidth,keepaspectratio]{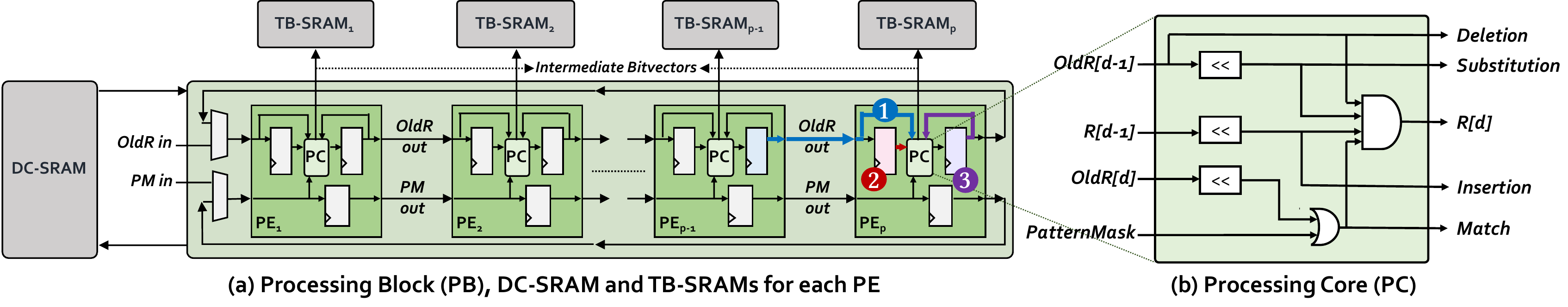}
\caption{\revonur{Hardware design of GenASM-DC.}}
\label{fig:bitmac-dc-pb}
\end{figure*}

A GenASM-DC accelerator consists of a processing block (PB; Figure~\ref{fig:bitmac-dc-pb}a) along with \revonur{a} control and memory management logic.
A PB consists of multiple processing elements (PEs). Each PE contains a single processing core (PC; Figure~\ref{fig:bitmac-dc-pb}b) and flip-flop-based storage logic. The PC is the primary compute unit, and implements
\revIII{Lines 15--19 of Algorithm~\ref{bitap-search-alg}} to perform the approximate string matching for a $w$-bit query pattern.
The number of PEs in a PB is based on compute, area, memory bandwidth and power requirements. This block also implements the logic to load data from outside of the array \revIII{(i.e., DC-SRAM; Figure~\ref{fig:bitmac-dc-pb}a)} or internally for cyclic operations.

GenASM-DC uses two types of \revonur{SRAM buffers (Figure~\ref{fig:bitmac-dc-pb}a)}: (1)~DC-SRAM, \revonur{which stores the reference text, the pattern bitmasks for the query read, and the intermediate data generated from PEs (i.e., $oldR$ values and MSBs required for shifts; \revIII{Section~\ref{sec:bitap-search}});}
and (2)~TB-SRAM, \revonur{which stores} the intermediate bitvectors \revonur{from GenASM-DC for later use by} GenASM-TB. For a 64-PE configuration with 64~bits of processing per PE, \revIII{and} \revonur{for the case where we have a long (10Kbp) read\footnote{\revIII{Although we use 10Kbp-long reads in our analysis (Section~\ref{sec:methodology:datasets}), GenASM does \emph{not} have any limitation on the length of \revIV{reads as a result of} our divide-and-conquer approach (Section~\ref{sec:bitap-traceback}).}} with a high error rate (15\%) and a corresponding text region of 11.5Kbp}, GenASM-DC requires a total of \revonur{8KB DC-SRAM storage.} 
For each PE, we have a dedicated \revIII{TB-SRAM, 
which} stores the match, insertion and deletion bitvectors generated by the corresponding PE. For the same configuration of GenASM-DC, 
each PE requires a total of 1.5KB TB-SRAM storage, with a single R/W port. In each cycle, 192 bits of data (24B) is written to each TB-SRAM by each PE.

\sgii{When each thread (i.e., each column) in Figure~\ref{fig:data_depend} is mapped to a PE, GenASM-DC coordinates the data dependencies across DC iterations, with the help of two flip-flops in each PE. For example, T2--R2 in Figure~\ref{fig:data_depend} is generated by $PE_x$ in $Cycle_y$, and is mapped to $R[d]$. In order to generate T2--R2, T2--R1 (which maps to $R[d-1]$) needs to be generated by $PE_{x-1}$ in $Cycle_{y-1}$ (\circlednumber{1} in Figure~\ref{fig:bitmac-dc-pb}), T1--R1 (which maps to $oldR[d-1]$) needs to be generated by $PE_{x-1}$ in $Cycle_{y-2}$ (\circlednumber{2}), and T1--R2 (which maps to $oldR[d]$) needs to be} generated by $PE_{x}$ in $Cycle_{y-1}$ (\circlednumber{3}), where $x$ is the PE index and $y$ is the cycle index. 
With this dependency-aware mapping, regardless of the number of instantiated PEs, \revV{we can successfully limit DC-SRAM traffic for a single PB 
to only one read and one write per cycle.}


\textbf{GenASM-TB Hardware.}
After GenASM-DC finishes writing all of the \revV{intermediate} bitvectors to TB-SRAMs, GenASM-TB \revIII{reads} them by following an irregular control flow, which depends on the \revV{text and the pattern} to find the optimal alignment \revIII{(by \revIV{implementing} Algorithm~\ref{bitap-traceback-alg})}.

\begin{figure}[b!] 
\centering
\includegraphics[width=12cm,keepaspectratio]{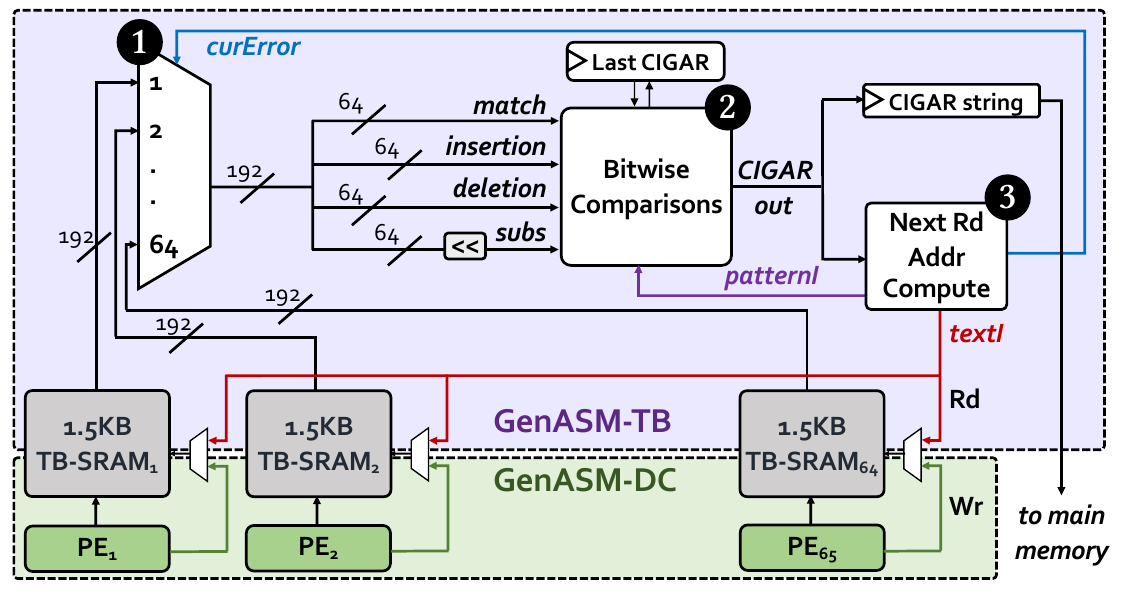}
\caption{\revonur{Hardware design of GenASM-TB.
}} \label{fig:bitmac-tb-hw-overall}
\end{figure}

\revonur{In our GenASM configuration, where we have 64 PEs and 64~bits per PE in a GenASM-DC accelerator, and the window size ($W$) is 64 (Section~\ref{sec:bitap-traceback}), we have one 1.5KB TB-SRAM (which fits \revIII{our} 24B/cycle $\times$ 64 cycles/window \revIII{output storage requirement}) for each of the 64 PEs.}
As Figure~\ref{fig:bitmac-tb-hw-overall} shows, 
a single GenASM-TB accelerator is connected to all of these 64 TB-SRAMs (96KB, in total). 
In each GenASM-TB cycle, we read from only one TB-SRAM. \revIII{\texttt{curError} provides the index of the TB-SRAM that we read from; \texttt{textI} provides the starting index within this TB-SRAM, which we read the next set of bitvectors from; and \texttt{patternI} provides the position of the 0 being processed (Algorithm~\ref{bitap-traceback-alg}).}

We implement the GenASM-TB hardware using very simple logic (Figure~\ref{fig:bitmac-tb-hw-overall}), which \revIII{\circlednumber{1}~reads} the bitvectors from one of the TB-SRAMs using the computed address, \revIII{\circlednumber{2}~performs} the required \revIII{bitwise comparisons} to find the CIGAR character for the current position, and \revIII{\circlednumber{3}~computes} the next TB-SRAM address to read the new set of bitvectors. After GenASM-TB finds the complete CIGAR string, it writes the output to main memory and completes its \revIII{ execution. 
}

\textbf{Overall System.}
\label{sec:overall-system}
\revIII{We design our system to take advantage of 
modern 3D-stacked memory systems~\cite{ghose2019demystifying, kim2016ramulator}, 
such as the Hybrid Memory Cube (HMC)~\cite{hmc}
or High-Bandwidth Memory (HBM)~\cite{hbm, lee2016simultaneous}.
Such memories are made up of multiple layers of DRAM arrays that
are stacked vertically in a single package.
These layers} are connected via high-bandwidth links called
\emph{through-silicon vias} (TSVs) 
\revIII{that provide lower-latency and more energy-efficient data access to the layers than the external DRAM I/O pins~\cite{davis2005demystifying,lee2016simultaneous}.
Memories such as HMC and HBM include a dedicated \emph{logic layer}
that connects to the TSVs and allows processing elements to
be implemented in memory to exploit the efficient data access.
Due to thermal and area constraints, only simple processing elements that execute low-complexity operations (e.g., bitwise logic, simple arithmetic, simple cores) can be included in the logic layer~\cite{boroumand2018google,drumond2017mondrian,tetris,ahn2015pim,ahn2015scalable,hsieh2016accelerating,hsieh2016transparent,mutlu2019processing,boroumand2019conda,pattnaik2016scheduling,kim2018grim}.}

\revIII{We decide to implement GenASM in the logic layer of 3D-stacked memory, for two reasons.}
First, we can exploit the natural subdivision within 3D-stacked memory (e.g., vaults in HMC\revIII{~\cite{hmc}, pseudo-channels in HBM~\cite{hbm}}) to efficiently enable parallelism across multiple GenASM accelerators. This subdivision allows accelerators to work in parallel without interfering with each other. Second, we can reduce the power consumed for DRAM accesses by reducing off-chip data movement across the memory channel\revIII{~\cite{mutlu2019processing}}.
\revIII{Both} of our hardware accelerators are highly efficient in terms of area and power (Section~\ref{sec:results:area-power}), and can fit within the 
\revIII{logic layer's constraints}.

\revIII{To illustrate how GenASM takes advantage of 3D-stacked memory, we discuss an example implementation of GenASM inside the logic layer of a 16GB HMC with 32~vaults~\cite{hmc}.}
Within each vault, the logic layer contains a GenASM-DC accelerator, its associated DC-SRAM (8KB), a GenASM-TB accelerator, and TB-SRAMs (64$\times$1.5KB). Since we have small SRAM buffers for both DC and TB to exploit locality, GenASM accesses the memory and utilizes the memory bandwidth only to read the reference and the query sequences. \revmicro{One GenASM accelerator at each vault requires 105--142 MB/s bandwidth, thus the total bandwidth requirement of all 32 GenASM accelerators is 3.3--4.4 GB/s \revIII{(which is much less than peak bandwidth provided by modern 3D-stacked memories)}. 
}

\section{G\lowercase{en}ASM Framework} 
\label{sec:bitmac-framework}
\vspace{-2pt}

\rev{
We demonstrate the efficiency and flexibility of the GenASM acceleration framework by describing three use cases of approximate string matching in genome sequence analysis: (1)~read alignment step of short and long read mapping, (2)~pre-alignment filtering for short reads, and (3)~edit distance calculation between \revIII{any two 
sequences}. We believe \revonurcan{the} GenASM framework can be useful for many other use cases, and we discuss some of them briefly in Section~\ref{sec:bitmac-framework-other}.

\textbf{Read Alignment of Short and Long Reads.}\label{sec:bitmac-framework-aln}
As we explain in Section~\ref{sec:background:mapping}, read alignment is the last step of short and long read mapping. In read alignment, all of the \revV{remaining} \revonur{candidate mapping \revIII{regions of the reference genome} and the \revIII{query reads} are aligned,} in order to identify the mapping that yields \revonur{either the lowest total number of errors (if using edit distance based scoring) or the highest score (if using \revV{a user-defined scoring function}).}
Thus, read alignment can be a use case for approximate string matching, \revonurcan{since} errors (i.e., substitutions, insertions, deletions) should be \revonur{taken into account} when aligning the sequences. As part of read alignment, we also need to generate the traceback output for the best alignment between \revonur{the reference \revIII{region and the read.}}

For read alignment, \revonurcan{the} whole GenASM pipeline, \revonurcan{as} explained in Section~\ref{sec:bitmac-overall}, should be executed, including the traceback step. In general, read alignment requires more complex scoring schemes, where different types of edits have non-unit costs. Thus, GenASM-TB should be configured based on the given cost \revonurcan{of} each type of edit (Section~\ref{sec:bitap-traceback}). As GenASM framework can work with arbitrary length sequences, we can use it to accelerate both short read and long read alignment.

\textbf{Pre-Alignment Filtering for Short Reads.}\label{sec:bitmac-framework-filter}
In the pre-alignment filtering step of short read mapping, the candidate \revII{mapping} locations, reported by the seeding step, are further filtered by using different mechanisms. 
\sgii{Although the regions of the reference at these candidate mapping locations share common seeds with query reads,}
they are not necessarily \emph{similar} sequences. 
To avoid examining dissimilar sequences at the \revonurcan{downstream} computationally-expensive read alignment step, \revonurcan{a} pre-alignment filter estimates the edit distance between \revonur{every read} and \revIII{the regions of the reference at each read's candidate mapping locations}, and \revonurcan{uses} this estimation to quickly decide whether or not read alignment is needed. If the sequences are dissimilar enough, significant amount of time is saved by avoiding the expensive alignment step\revonurcan{~\cite{alser2017gatekeeper, alser2019sneakysnake, Alser2019, Xin2013, Xin2015}}.

\revonur{
In pre-alignment filtering, since we only need to estimate \revIII{(approximately)} the edit distance and check whether it is above a user-defined threshold, GenASM-DC can be used as a pre-alignment filter. As GenASM-DC is very efficient when we have shorter sequences and a low error threshold (due to \revonur{the} $O(m \times n \times k)$ complexity of the underlying Bitap algorithm, where $m$ is the query length, $n$ is the reference length, and $k$ is the number of allowed errors), GenASM framework can \revonur{efficiently} accelerate the pre-alignment filtering step of \revIII{especially} short read mapping.\footnote{\revIII{Although we believe that GenASM can also be used as a pre-alignment filter for long reads, we leave the evaluation of this use case for future work.}}
}

\vspace{1pt}
\textbf{Edit Distance Calculation.}\label{sec:bitmac-framework-edc}
Edit distance, also called Levenshtein distance~\cite{levenshtein1966binary}, is the minimum number of edits (i.e., substitutions, insertions and deletions) required to convert one sequence to another. Edit distance calculation is one of the fundamental operations in genomics to measure the similarity or distance between two sequences~\cite{vsovsic2017edlib}. As we explain in Section~\ref{sec:background-bitap}, the Bitap algorithm, which is the underlying algorithm of GenASM-DC, is \revonur{originally} designed for edit distance calculation. Thus, GenASM framework can accelerate edit distance calculation between any \revonurcan{two} arbitrary-length genomic sequences.

Although GenASM-DC can find the edit distance by itself and \revIII{traceback 
is} optional for this use case, DC-TB interaction is required \revonurcan{in our accelerator} \revonur{to exploit} the efficient divide-and-conquer approach GenASM follows. \revonur{Thus,} 
GenASM-DC and GenASM-TB work together to find the minimum edit distance in a fast and memory-efficient way, but the traceback output \revonur{is not generated or reported by default (though it can optionally be enabled)}.
}
\section{Evaluation Methodology} \label{sec:methodology}

\textbf{Area and Power Analysis.}
We synthesize \revonur{and place \& route} the GenASM-DC and GenASM-TB accelerator datapaths using \revonur{the} Synopsys Design Compiler~\cite{synopsysdc} with a typical 28nm \revonurcan{low-power} process, with memories generated using an \revonurcan{industry-grade} SRAM compiler, to analyze the accelerators' area and power.
Our synthesis targets \revIV{post-routing} timing closure at \hl{1GHz} clock frequency.
We then use \revonur{an in-house cycle-accurate} simulator parameterized with the synthesis and memory estimations to drive the performance and \revonur{power} analysis.



We evaluate a 16GB HMC-like 3D-stacked DRAM architecture, with 32 vaults~\cite{hmc} and 256GB/s of internal bandwidth~\cite{boroumand2018google,hmc},
and a clock frequency of 1.25GHz~\cite{hmc}. 
The amount of available area in the logic layer for GenASM is around 3.5--4.4 mm\textsuperscript{2} per vault~\cite{drumond2017mondrian,boroumand2018google}. The power budget of our PIM logic per vault is 312mW~\cite{drumond2017mondrian}.

\textbf{Performance Model.}
We build a \revonur{spreadsheet-based} analytical model \revonur{for GenASM-DC and GenASM-TB, which} considers reference \revV{genome} (i.e., text) length, query \revonur{read} (i.e., pattern) length, maximum edit distance, window size, \revV{hardware design} parameters (number of PEs, bit width of \revonurcan{each} PE) and number of vaults as input parameters and projects compute cycles, DRAM \revonur{read/write bandwidth, SRAM read/write} bandwidth, and memory footprint. \revonur{We verify the analytically-estimated} cycle counts for various PE configurations with the cycle counts collected from \revonurcan{our} RTL simulations.  


\revonur{\textbf{Read Alignment Comparisons.}}
For the read alignment \revonur{use case}, we compare GenASM with the read alignment steps of two \revonurcan{commonly-used} state-of-the-art read mappers: Minimap2~\cite{li2018minimap2} and BWA-MEM~\cite{li2013aligning}, running on \revonur{an} Intel\textsuperscript{\textregistered} Xeon\textsuperscript{\textregistered} Gold 6126 CPU\revonur{~\cite{intel_cpu}} operating at 2.60GHz, with 64GB DDR4 memory. Software baselines are run with a single thread and with 12 threads.
We measure the execution time and power consumption of 
the alignment steps in Minimap2 and BWA-MEM.  We measure the individual power consumed by each tool using Intel's PCM power utility~\cite{intelpcm}. 

\revmicro{We also compare GenASM with \revII{a} state-of-the-art GPU-accelerated short read alignment tool, GASAL2~\cite{gasal2}.  We run GASAL2 on an Nvidia Titan V GPU\revonur{~\cite{nvidia_titan}} with 12GB HBM2 memory~\cite{hbm}. To fully utilize the GPU, we configure the number of alignments per batch based on the GPU's number of multiprocessors and the maximum number of threads per multiprocessor, as described in the GASAL2 paper~\cite{gasal2}. To better analyze the high parallelism that \revonur{the GPU} provides, we replicate our datasets to obtain datasets with 100K, 1M and 10M reference-read pairs for short reads. We run the datasets with GASAL2, and collect kernel time and average power consumption using \emph{nvprof}~\cite{nvprof}.}

\par

We also compare GenASM with two state-of-the-art \revonur{hardware-based} alignment accelerators, GACT of Darwin \cite{turakhia2018darwin} and SillaX of GenAx \cite{fujiki2018genax}. 
We synthesize and execute the open-source RTL for GACT~\cite{darwingithub}.
We estimate the performance of SillaX using data reported by the original work~\cite{fujiki2018genax}.

We \sgii{analyze the alignment accuracy of GenASM} by comparing the alignment outputs \sgii{(i.e., alignment score, edit distance, and CIGAR string)} of GenASM with the alignment outputs of BWA-MEM \revII{and Minimap2, for short reads and long reads, respectively.}
We obtain the BWA-MEM and Minimap2 alignments by running the tools with their default settings.

\revonur{\textbf{Pre-Alignment Filtering Comparisons.} We} compare 
GenASM with Shouji~\cite{Alser2019}, which is \revonur{the state-of-the-art} FPGA-based pre-alignment filter for short reads. \revonur{For execution time and filtering accuracy \revonurcan{analyses}, we use data reported by the original work~\cite{Alser2019}. For power analysis, we report the total power consumption of Shouji using the power analysis tool in Xilinx Vivado~\cite{vivado}, after synthesizing and implementing the open-source FPGA design of Shouji~\cite{shoujigithub}.}

\revonur{\textbf{Edit Distance Calculation Comparisons.} We compare}
GenASM with the state-of-the-art \revonur{software-based} read alignment library, Edlib~\cite{vsovsic2017edlib}, \revonur{running on an Intel\textsuperscript{\textregistered} Xeon\textsuperscript{\textregistered} Gold 6126 CPU~\cite{intel_cpu} operating at 2.60GHz, with 64GB DDR4 memory.} Edlib uses \revonur{the} Myers' bitvector algorithm~\cite{myers1999fast} to find the edit distance between two sequences. We use the default global \revonur{Needleman-Wunsch (NW)~\cite{needleman1970general}} mode of Edlib to perform our comparisons. \revonur{We measure the power consumed by Edlib using Intel's PCM power utility~\cite{intelpcm}.}

\revonur{We also compare GenASM with ASAP~\cite{Banerjee2019}, which is the state-of-the-art FPGA-based} \revonurcan{accelerator for computing the edit distance between two short reads.} \revonur{We estimate the performance of ASAP using data reported by the original work~\cite{Banerjee2019}.}

\textbf{Datasets.}
\label{sec:methodology:datasets}
For the read alignment use case, we evaluate GenASM using the latest major release of the human genome assembly, GRCh38 \cite{ncbi38genome}. We use the 1--22, X, and Y chromosomes by filtering the unmapped contigs, unlocalized contigs, and mitochondrial genome.
Genome characters are encoded into 2-bit patterns
(A = 00, C = 01, G = 10, T = 11). 
With this encoding, the reference genome uses \SI{715}{\mega\byte} of memory.

We generate four sets of long reads \revV{(i.e., PacBio and ONT datasets)} using PBSIM~\cite{ono2012pbsim} and three sets of short reads \revV{(i.e., Illumina datasets)} using Mason~\cite{holtgrewe2010mason}.
For the PacBio datasets, we use the default error profile for the continuous long reads (CLR) in PBSIM. For the ONT datasets, we modify the settings to match the error profile of ONT reads sequenced using R9.0 chemistry~\cite{jain2017minion}. Both datasets have 240,000 reads of length 10Kbp, \revonur{each simulated with 10\% and 15\% error rates}. The Illumina datasets have \hl{200,000 reads of length 100bp, 150bp, and 250bp}, \revonur{each simulated with a 5\% error rate}.

For the pre-alignment filtering use case, we use two datasets that Shouji~\cite{Alser2019} provides as test cases: \revIII{reference-read pairs (1)~of length 100bp with an edit distance threshold of 5, and (2)~of length 250bp with an edit distance threshold of 15.}

For the edit distance calculation use case, we use the publicly-available dataset that Edlib~\cite{vsovsic2017edlib} \revV{provides. 
The} dataset includes two real DNA sequences, which are 100Kbp and 1Mbp in \revonur{length, and artificially-mutated versions of the original DNA sequences with measures of similarity ranging between 60\%--99\%. Evaluating this set of sequences with varying values of similarity and length enables us to demonstrate how these parameters affect performance.}

\section{Results} \label{sec:results}

\subsection{Area and Power Analysis}\label{sec:results:area-power}

Table \ref{table:area} shows the area and power breakdown of each component in GenASM, \revonur{and the total area overhead and power consumption of (1)~a single GenASM accelerator (in 1 vault) and (2)~32 GenASM accelerators (in 32 vaults). Both GenASM-DC and GenASM-TB \revonur{operate} at 1GHz.}

\revonur{The area} overhead of \revonur{one} GenASM accelerator is \SI{0.334}{\milli\meter\squared}, and \revonur{the} power consumption of \revonur{one} GenASM accelerator, including the SRAM power, is \SI{101}{\milli\watt}. \revonur{When we compare GenASM with a \revonur{single core of a} modern Intel\textsuperscript{\textregistered} Xeon\textsuperscript{\textregistered} Gold 6126 CPU\revonur{~\cite{intel_cpu}} (which we conservatively estimate to use \SI{10.4}{\watt}~\cite{intel_cpu} and \SI{32.2}{\milli\meter\squared}~\cite{skylake-diearea} per core), we find that GenASM is significantly more efficient in terms of both area and power consumption.}
As we have one GenASM accelerator \revonur{per vault}, 
the total area overhead of \revonur{GenASM in all 32 vaults is} \SI{10.69}{\milli\meter\squared}. 
Similarly, the total power consumption of \revonur{32 GenASM accelerators is} \SI{3.23}{\watt}.

\begin{table}[h!]
\small
\centering
\caption{Area and power breakdown of GenASM.}
\label{table:area}
\includegraphics[width=13cm,keepaspectratio]{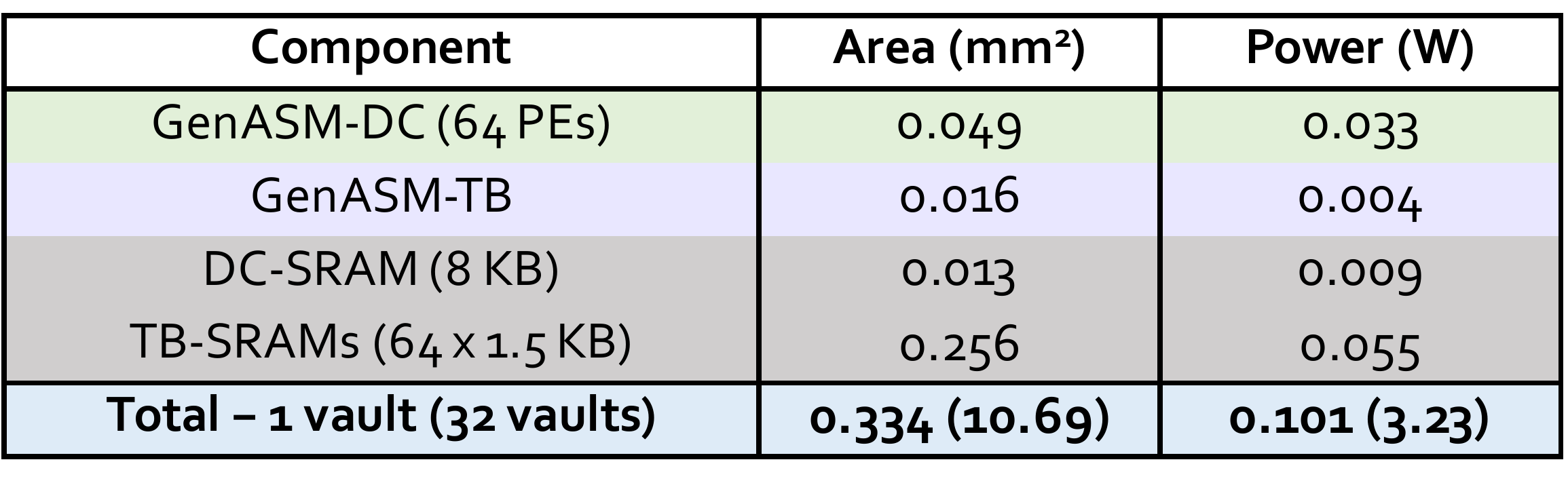}
\end{table}



\subsection{Use Case 1: Read Alignment}\label{sec:results-alignment}

\begin{figure}[t!]
\centering
\includegraphics[width=\columnwidth,keepaspectratio]{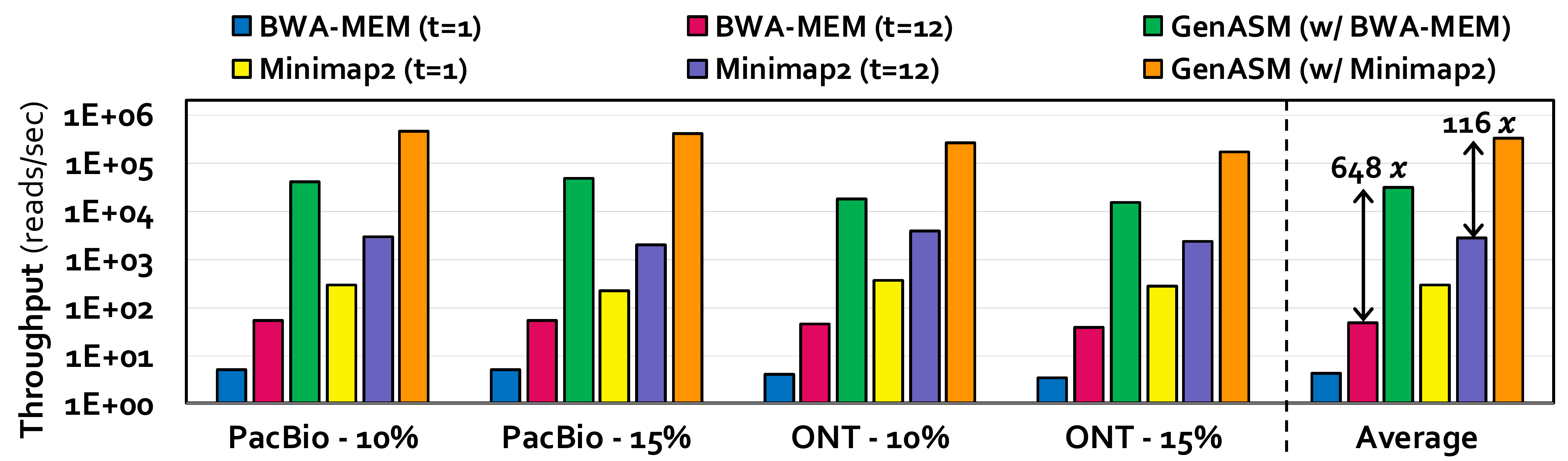}
\caption{\revonur{Throughput comparison of GenASM and the alignment steps of BWA-MEM and Minimap2 for long reads.}} \label{fig:throughput-result-long}
\end{figure}

\textbf{Software Baselines (CPU).} Figure~\ref{fig:throughput-result-long} shows the read alignment throughput (reads/sec) of GenASM and the alignment steps 
of BWA-MEM and Minimap2, when aligning long noisy PacBio and ONT reads against the human reference genome. 
When comparing with BWA-MEM, we run GenASM with the candidate locations reported by BWA-MEM's filtering step. Similarly, when comparing with Minimap2, we run GenASM with the candidate locations reported by Minimap2's filtering step. GenASM's throughput is determined by the throughput of the execution of GenASM-DC and GenASM-TB with window size ($W$) \revIII{of} 64 and overlap size ($O$) \revIII{of} 24.

As Figure~\ref{fig:throughput-result-long} shows, GenASM \revonur{provides (1)}~\hl{$7173\times$} and \hl{$648\times$} throughput improvement over the alignment step of BWA-MEM for its \revIII{single-thread} and \revIII{12-thread} execution, respectively, \revonur{and (2)~}\hl{$1126\times$} and \hl{$116\times$} throughput improvement over the alignment step of Minimap2 for its \revIII{single-thread} and \revIII{12-thread} execution, respectively. 

Based on our power analysis with long reads, we find that power consumption of BWA-MEM's alignment step is \SI{58.6}{\watt} and \SI{109.5}{\watt}, and power consumption of Minimap2's read alignment step is \SI{59.8}{\watt} and \SI{118.9}{\watt} for their \revIII{single-thread} and \revIII{12-thread} executions, respectively. \revonur{GenASM consumes only 3.23W, and thus reduces the power consumption of} the alignment steps of BWA-MEM and Minimap2 by $18\times$ and $19\times$ for single-thread execution, and by $34\times$ and $37\times$ for 12-thread execution, respectively. 

\begin{figure}[t!]
\centering
\includegraphics[width=\columnwidth,keepaspectratio]{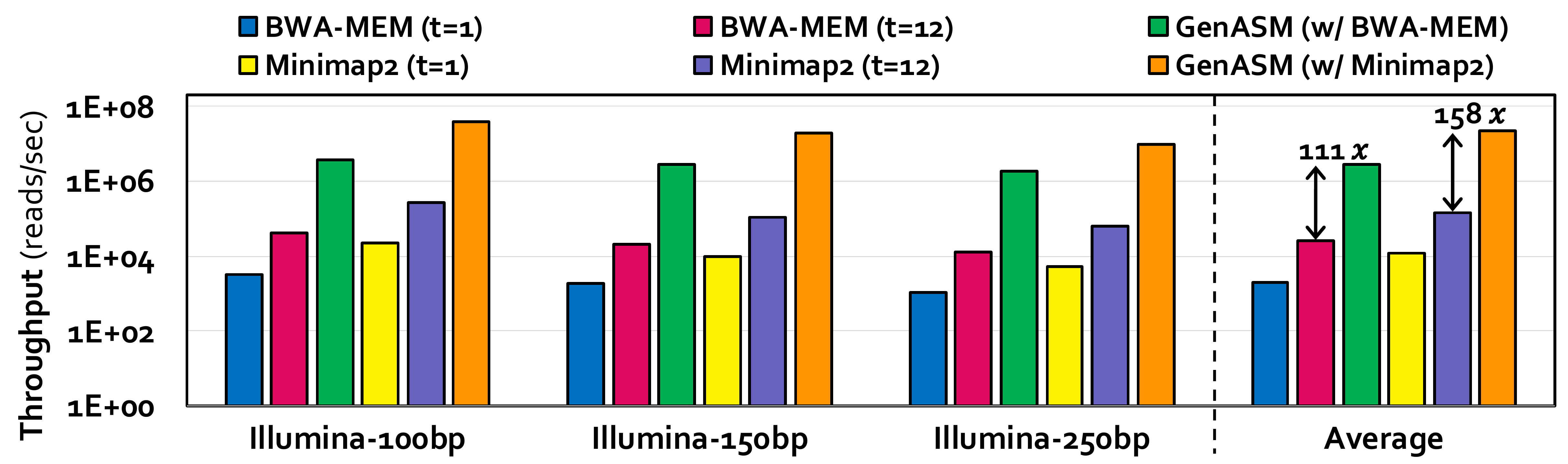}
\caption{Throughput comparison of GenASM and the alignment steps of BWA-MEM and Minimap2 for short reads.} \label{fig:throughput-result-short}
\end{figure}

Figure~\ref{fig:throughput-result-short} \revonur{compares} the read alignment throughput (reads/sec) of GenASM \revonur{with that of} the alignment steps \revonur{of BWA-MEM and Minimap2, when aligning short} Illumina reads against the human reference genome. 
\revonur{GenASM provides (1)~}\hl{$1390\times$} and \hl{$111\times$} throughput improvement over the alignment step of BWA-MEM for its \revIII{single-thread} and \revIII{12-thread} execution, respectively, \revonur{and (2)~}\hl{$1839\times$} and \hl{$158\times$} throughput improvement over the alignment step of Minimap2 for its \revIII{single-thread} and \revIII{12-thread} execution. 

Based on our power analysis with short reads, we find that GenASM reduces the power consumption over the alignment steps of BWA-MEM and Minimap2 by $16\times$ and $18\times$ for \revIII{single-thread execution}, and by $33\times$ and $31\times$ for \revIII{12-thread} \revonur{execution}, respectively. 

\begin{figure}[h!]
\centering
\includegraphics[width=\columnwidth,keepaspectratio]{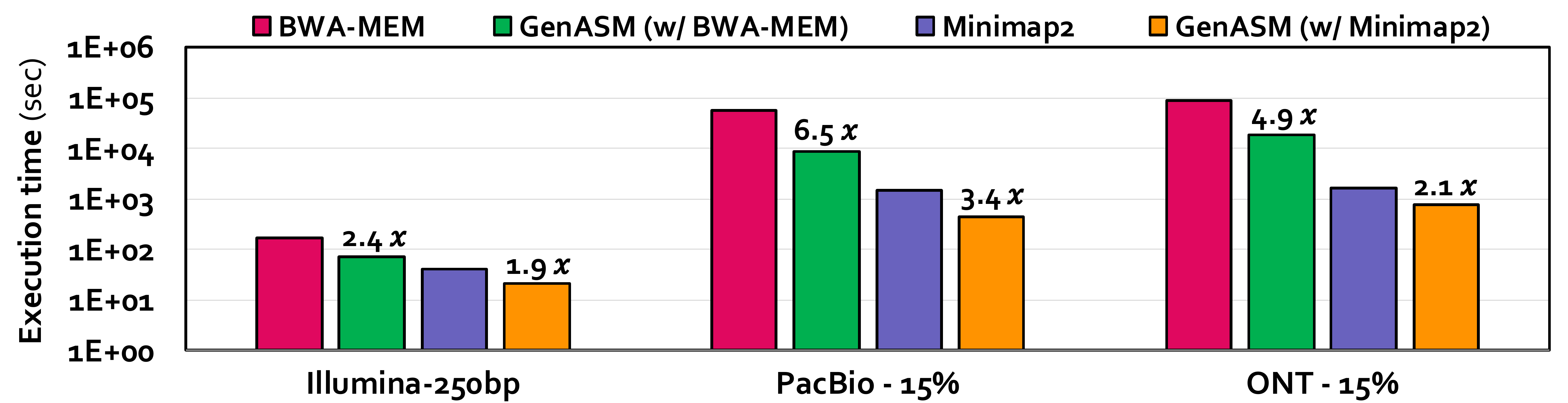}
\caption{\revIII{Total execution time of \revonur{the entire} BWA-MEM and Minimap2 pipelines with and without GenASM.}} \label{fig:fullpipeline-mixed}
\end{figure}

Figure~\ref{fig:fullpipeline-mixed} shows the total execution time of \revonur{the entire} BWA-MEM and Minimap2 pipelines, 
\revonur{along with the total execution time when the alignment steps of each pipeline are replaced by GenASM,}
for the three representative input datasets.
\revIV{As Figure~\ref{fig:fullpipeline-mixed} shows, GenASM provides (1)~\hl{$2.4\times$} and \hl{$1.9\times$} speedup for Illumina reads (250bp); (2)~\hl{$6.5\times$} and \hl{$3.4\times$} speedup for PacBio reads (15\%); and (3)~\hl{$4.9\times$} and \hl{$2.1\times$} speedup for ONT reads (15\%), over the \revonur{entire} pipeline executions of BWA-MEM and Minimap2, respectively.}

\textbf{Software Baselines (GPU).}
We compare GenASM with \revonur{the state-of-the-art GPU aligner,} GASAL2~\cite{gasal2}, \revonur{using three datasets of varying size (100K, 1M, and 10M reference-read pairs)}. \revIV{Based on our analysis,} we \revonur{make three findings.
First, for} 100bp Illumina reads, GenASM provides \revonur{9.9$\times$, 9.2$\times$, and 8.5$\times$} speedup over GASAL2, while reducing the power consumption by \revonur{15.6$\times$, 17.3$\times$ and 17.6$\times$} for 100K, 1M, and 10M datasets, respectively.
\revonur{Second, for} 150bp Illumina reads, GenASM provides \revonur{15.8$\times$, 13.1$\times$, and 13.4$\times$} speedup over GASAL2, while reducing the power consumption by \revonur{15.4$\times$, 18.0$\times$, and 18.7$\times$} for 100K, 1M, and 10M datasets, respectively.
\revonur{Third, for} 250bp Illumina reads, GenASM provides 21.5$\times$, 20.6$\times$, and 21.1$\times$ speedup over GASAL2, while reducing the power consumption by 16.8$\times$, 20.2$\times$, and 20.6$\times$ for 100K, 1M, and 10M datasets, respectively.
We conclude that GenASM provides significant performance benefits and energy efficiency over GPU aligners for short \revIII{reads.



\textbf{Hardware Baselines.}} We compare GenASM with two state-of-the-art hardware accelerators for read alignment: GACT \revonur{(from Darwin~\cite{turakhia2018darwin})} and SillaX \revonur{(from GenAx~\cite{fujiki2018genax})}.

Darwin is a hardware accelerator designed for \emph{long} read alignment~\cite{turakhia2018darwin}. Darwin contains components that accelerate both the filtering (D-SOFT) and alignment (GACT) steps of read mapping. 
The open-source RTL code available for the GACT accelerator of Darwin allows us to estimate the throughput, area and power consumption of GACT and compare it with GenASM for read alignment. In Darwin, GACT logic and the associated 128KB SRAM are responsible for filling the dynamic programming matrix, generating the traceback pointers and finding the maximum score. 
Thus, we believe that it is fair to compare the power consumption and the area of the GACT logic and GenASM logic, along with their associated SRAMs.

In order to have an iso-bandwidth comparison with Darwin's GACT, we compare \revIII{only} a single array of GACT and a single set of GenASM-DC and GenASM-TB, because (1)~GenASM utilizes the high \revonur{memory} bandwidth that PIM provides \revonur{only} to parallelize many sets of GenASM-DC and GenASM-TB, and a single set of GenASM-DC and GenASM-TB does \emph{not} require high bandwidth, and (2)~all internal data of both GenASM and Darwin \revIII{is} provided by local SRAMs. We synthesize both designs \revonur{(i.e., GenASM and GACT)} at \revonur{an iso-PVT (process, voltage, temperature) corner}, with \revonur{the} same number of PEs, and with their optimum parameters.

\begin{figure}[h!]
\centering
\includegraphics[width=\columnwidth,keepaspectratio]{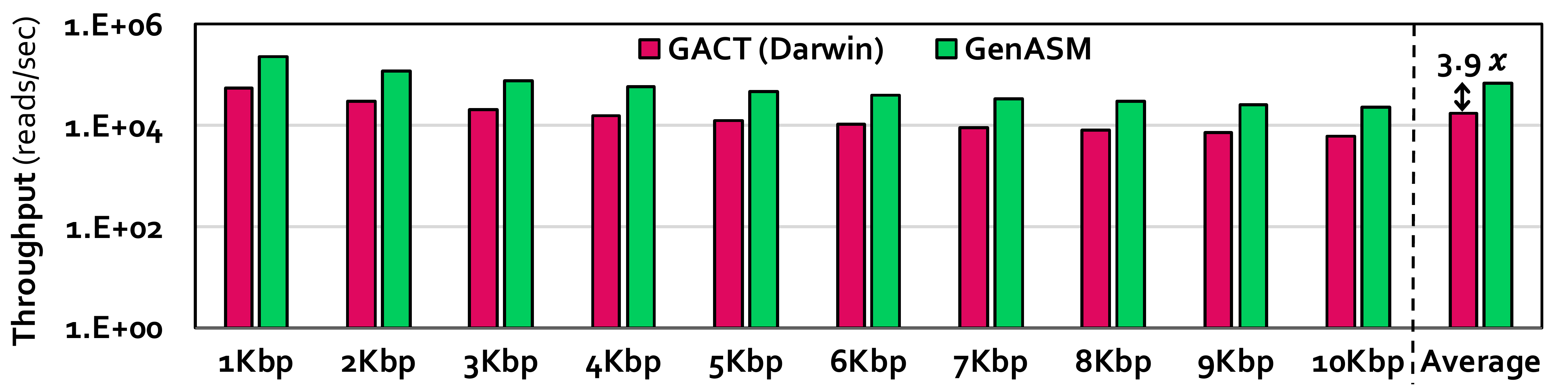}
\caption{Throughput comparison of GenASM and GACT \revonur{from} Darwin for long reads.} \label{fig:darwin-long}
\end{figure}

\revII{As Figure~\ref{fig:darwin-long} shows}, for a single GACT array with 64 PEs at 1GHz, the throughput of GACT decreases from 55,556 to 6,289 alignments per second when the sequence length increases from 1Kbp to 10Kbp, while consuming \SI{277.7}{\milli\watt} of power. In comparison, 
\revIV{for} a single GenASM accelerator at 1GHz (with a 64-PE configuration), the throughput decreases from \hl{236,686} to \hl{23,669} alignments per second when the sequence length increases from 1Kbp to 10Kbp, while consuming \revonur{\SI{101}{\milli\watt}} of power. \revonur{This shows that, on average,} GenASM \revII{provides $3.9\times$} better throughput than GACT, while \revIII{consuming} $2.7\times$ \revIII{less power}. Thus, GenASM provides \revII{$10.5\times$} better throughput per unit power \revonur{for long reads} when compared to GACT.

\begin{figure}[h!]
\centering
\includegraphics[width=\columnwidth,keepaspectratio]{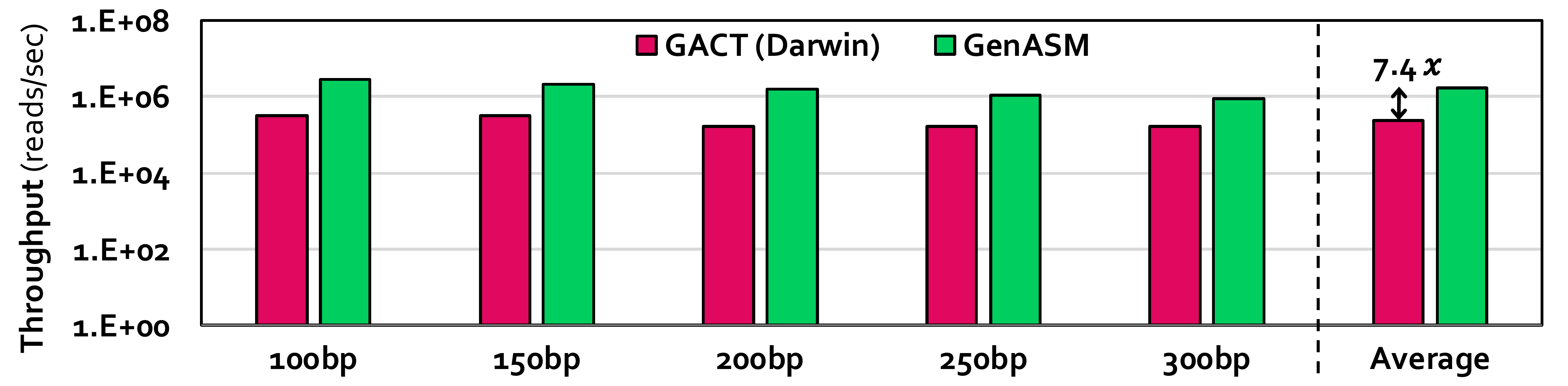}
\caption{Throughput comparison of GenASM and GACT \revonur{from} Darwin for short reads.} \label{fig:darwin-short}
\end{figure}

\revII{As Figure~\ref{fig:darwin-short} shows,} we also compare the throughput of GenASM and GACT for short read alignment \revII{(i.e., 100--300bp reads)}. We find that GenASM performs \revII{7.4$\times$} better than GACT when 
aligning short reads, \revII{on average}. \revonur{Thus, GenASM provides \revII{$20.0\times$} better throughput per unit power for short reads when compared to GACT.}

We compare the required area for the GACT logic with 128KB of SRAM and the required area for the GenASM logic (GenASM-DC and GenASM-TB) with 8KB of DC-SRAM and 96KB of TB-SRAMs, at 28nm. We find that \revonur{GenASM requires $1.7\times$ less area than 
GACT. Thus, GenASM provides \revII{$6.6\times$ and $12.6\times$} better throughput per unit area for long reads and for short reads, respectively, when compared to GACT.}

The main difference between GenASM and GACT is the underlying algorithms. GenASM uses \revonur{our} modified 
Bitap algorithm, which requires only simple and fast bitwise operations. On the other hand, GACT uses the complex and computationally more expensive \revV{dynamic programming based algorithm} for alignment. This is the main reason why GenASM is more efficient than GACT of Darwin. 
\revonur{GenAx} is a hardware accelerator designed for \emph{short} read alignment~\cite{fujiki2018genax}. Similar to Darwin, \revonur{GenAx accelerates} both the filtering and alignment steps of read mapping. Unlike GenAx, whose design is optimized only for short reads, GenASM is more robust and works with \emph{both} short and long reads.
While we are unable to reimplement GenAx, the throughput analysis of SillaX \revonur{(the alignment accelerator of GenAx)} provided by the original work~\cite{fujiki2018genax} allows us to \revonur{provide a} performance comparison between GenASM and SillaX for short read alignment. 

We compare SillaX with GenASM at their optimal operating frequencies (2GHz for SillaX, 1GHz for GenASM), and find that GenASM provides $1.9\times$ higher throughput for \revIII{short 
reads} (101bp) than SillaX (whose approximate throughput is 50M alignments per second). 
\revII{Using} the area and power numbers \revII{reported} for the \revIII{computation} logic of SillaX, we find that GenASM requires 63\% less logic area (\SI{2.08}{\milli\meter\squared} vs.\ \SI{5.64}{\milli\meter\squared}) and 82\% less logic power \revIII{(\SI{1.18}{\watt} vs.\ \SI{6.6}{\watt}). 

In} order to compare the total area of SillaX and GenASM, we perform a CACTI-based analysis~\cite{wilton1996cacti} for \revonur{the SillaX SRAM (\SI{2.02}{\mega\byte})}. \revonur{We find that the SillaX SRAM consumes an area of} \SI{3.47}{\milli\meter\squared}, \revonur{resulting in a total area of \SI{9.11}{\milli\meter\squared} for SillaX}. Although GenASM \revonur{(\SI{10.69}{\milli\meter\squared})} requires \revonur{17\% more total area than SillaX,} 
we find that GenASM provides $1.6\times$ better throughput per unit area for short reads than \revonur{SillaX}.

\textbf{Accuracy Analysis.} \label{sec:results-accuracy}
We compare the traceback outputs of GenASM and \revIII{(1)~BWA-MEM} for short reads, \revIII{(2)~Minimap2} for long reads, to assess the accuracy and correctness of GenASM-TB. We find that the optimum $(W,O)$ setting (i.e., window size and overlap size) for the GenASM-TB algorithm in terms of performance and accuracy is $W=64$ and $O=24$. With this setting, GenASM \revIII{completes the alignment of all reads in each dataset, and increasing} the window size does \emph{not} change the alignment output.

For short reads, we use the default scoring setting of BWA-MEM (i.e., match=+1, substitution=-4, gap opening=-6, and gap extension=-1). 
\revonur{For 96.6\% of the short reads, GenASM \revIV{finds} an alignment whose score is equal to the score of the alignment reported by BWA-MEM. This \revV{fraction increases to 99.7\% 
when we consider scores that are within $\pm4.5\%$ of the \revIV{scores reported} by BWA-MEM.}}

For long reads, we use the default scoring setting of Minimap2 (i.e., match=+2, substitution=-4, gap opening=-4, and gap extension=-2). 
\revonur{For 99.6\% of the long reads with a 10\% error rate, GenASM \revV{finds} an alignment whose score is within $\pm0.4\%$ of the score of the alignment reported by Minimap2. For 99.7\% of the long reads with a 15\% error rate, GenASM \revV{finds} an alignment whose score is within $\pm0.7\%$ of the score of the alignment reported by Minimap2.}

\revmicro{
There are two reasons for the \revonur{difference between the alignment scores reported by GenASM and the scores reported by the baseline tools.} First, GenASM performs \revonur{traceback} for the alignment with the minimum edit distance. However, the \revonur{baseline can report an alignment that has a higher number of edits but a lower score than the alignment reported by GenASM}, when more complex scoring schemes are used. Second, during the TB stage, GenASM follows a fixed order at each iteration when picking between substitutions, insertions, or deletions (based on the penalty of each error type). \revonur{While} we pick the error type with the lowest possible cost at the current iteration, another error type with a higher \revonur{initial} cost may lead to a better (i.e., lower-cost) alignment in later iterations, which cannot be known beforehand.\footnote{\revIII{We can add support for different orderings by adding more configurability to the GenASM-TB accelerator, which we leave for future work.}}}

Although GenASM is optimized for \revonur{unit-cost based scoring (i.e., edit distance) and \revIII{currently} provides only \revV{partial} support} for more complex scoring schemes, we show that GenASM framework can still serve as a fast, memory- and power-efficient, \revIII{and quite accurate} alternative for read alignment.


\subsection{Use Case 2: Pre-Alignment Filtering} \label{sec:results-filtering}

\rev{
We compare GenASM with the state-of-the-art FPGA-based pre-alignment filter for short reads, Shouji~\cite{Alser2019}, \revIV{using} two datasets provided in~\cite{Alser2019}. When we compare 
Shouji (with maximum filtering units) and GenASM for the dataset with 100bp sequences, we find that GenASM provides $3.7\times$ speedup over Shouji, \revonur{while reducing power consumption by $1.7\times$}. When we perform the same analysis with 250bp sequences, we find that \revonur{GenASM does not provide speedup over Shouji, but reduces power consumption by $1.6\times$}.

In pre-alignment filtering for short reads, 
\revII{only} GenASM-DC is executed \revIII{(Section~\ref{sec:bitmac-framework-filter})}. The complexity \revonur{of} GenASM-DC 
is \revII{$O(n\times m \times k)$} whereas the complexity of Shouji is \revII{$O(m\times k)$}, where $n$ is the text length, $m$ is the read length, and $k$ is the edit distance threshold. \revIII{Going from the 100bp dataset to the 250bp dataset}, all these three parameters increase linearly. Thus, \revonur{the} speedup of GenASM over Shouji \revonur{for pre-alignment filtering} decreases for datasets with longer reads.

\revII{To analyze filtering accuracy}, \revonur{we use Edlib~\cite{vsovsic2017edlib} to generate the ground truth edit distance value for each sequence pair in the datasets (similar to Shouji)}}. 
We evaluate the accuracy of GenASM as a pre-alignment filter by computing its false accept rate and false reject rate \revonur{(as defined in \cite{Alser2019})}.

The false accept rate\revonur{~\cite{Alser2019}} is the ratio of the number of dissimilar sequences that are falsely accepted by the filter \revIII{(as similar)} and the total number of dissimilar sequences that are rejected by the ground truth. The goal is to minimize the false accept rate to maximize the number of dissimilar sequences that are eliminated \revIII{by the filter}. For the 100bp dataset 
with \revonur{an} edit distance threshold of 5, Shouji has \revonur{a 4\% false accept rate, whereas GenASM has a false accept rate of only 0.02\%}. 
For the 250bp dataset with \revonur{an} edit distance threshold of 15, Shouji has \revonur{a 17\% false accept rate, whereas GenASM has a false accept rate of only 0.002\%.} 
Thus, GenASM provides a very low rate of \revonur{falsely-accepted} dissimilar sequences, and significantly improves the accuracy of pre-alignment filtering compared to Shouji. 

\revonur{While Shouji approximates the edit distance, GenASM calculates the actual distance. Although calculation requires more computation than approximation, a computed distance results \revIII{in} a near-zero (0.002\%) false accept rate.\footnote{\revIV{The reason for the non-zero false accept rate of GenASM is \revIV{that} when there is a deletion in the first character of the query, 
GenASM does \emph{not} count this as an edit, and skips this extra character of the text when computing the edit distance. Since GenASM reports an edit distance that is one lower than the edit distance reported by the ground truth, if GenASM's reported edit distance is below the threshold but the ground truth's is not, GenASM leads to a false accept.}}
Thus, GenASM filters more false-positive locations out, leaving fewer candidate locations for the expensive alignment step to process. This \revIII{greatly} reduces the combined execution time of filtering and alignment. Thus, even though GenASM does not provide any speedup over Shouji when filtering the 250bp sequences, its lower false accept rate makes it a better option for this step of the pipeline with greater overall benefits.} 


The false reject rate\revonur{~\cite{Alser2019}} is the ratio of the number of similar sequences that are rejected by the filter \revIII{(as dissimilar)} and the total number of similar sequences that are accepted by the ground truth. The false reject rate should always be equal to 0\%. We observe that GenASM always \revonur{provides a 0\% false reject rate, and thus does} not filter out similar sequence pairs, \revIII{as does Shouji}.

\subsection{Use Case 3: Edit Distance Calculation}\label{sec:results-edit}

\begin{figure}[h!]
\centering
\includegraphics[width=\columnwidth,keepaspectratio]{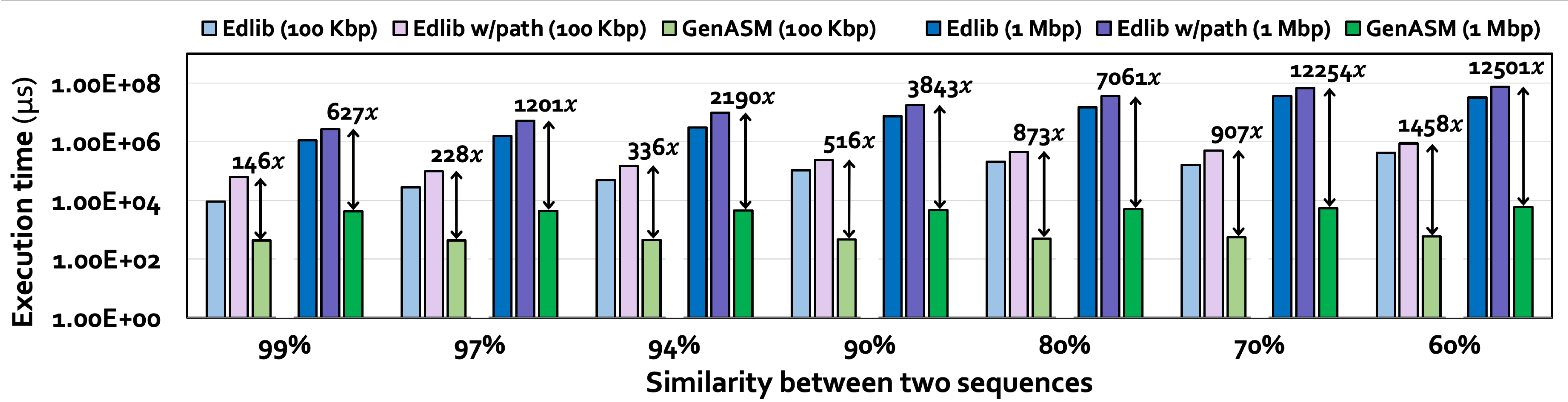}
\caption{Execution time comparison of GenASM and Edlib for edit distance calculation.} \label{fig:execution-result-edlib}
\end{figure}

We compare GenASM with the state-of-the-art edit distance calculation library, Edlib~\cite{vsovsic2017edlib}. Figure~\ref{fig:execution-result-edlib} compares the execution time of Edlib (with and without finding the traceback \revonur{output}) and GenASM when finding the edit distance between \revonur{two sequences of length 100Kbp, and also two sequences of length 1Mbp}, which have similarity ranging from 60\% to 99\% (Section~\ref{sec:methodology:datasets}). 
\revonur{Since Edlib is a \revIII{single-thread} edit distance calculation tool, for a fair comparison, we compare the throughput of only \revonur{one} GenASM accelerator \revIII{(i.e., in one vault)} with \revIII{a single-thread} execution of the Edlib tool.}

As Figure~\ref{fig:execution-result-edlib} shows, when performing edit distance calculation between two 100Kbp sequences, GenASM \revIII{provides 22--716$\times$ and 146--1458$\times$ speedup over Edlib execution without and with traceback, respectively}. GenASM has the same execution time for both of the cases. When the sequence length increases from 100Kbp to 1Mbp, the execution time of GenASM increases linearly \revIII{(since $W$ is constant, but $m+k$ increases linearly)}. However, due to its quadratic complexity, Edlib cannot scale linearly. Thus, for the edit distance calculation of 1Mbp sequences, GenASM \revIII{provides 262--5413$\times$ and 627--12501$\times$ speedup over Edlib execution without and with traceback, respectively.}

Although both \revonur{the} GenASM algorithm and Edlib’s underlying Myers’ algorithm~\cite{myers1999fast} use bitwise operations only for edit distance calculation and exploit bit-level parallelism, the main \revIII{advantages of the GenASM algorithm come} from \revIV{(1)~the} divide-and-conquer approach we follow for efficient support for longer sequences, \revIII{and \revIV{(2)~our} efficient co-design \revV{of the GenASM algorithm} with the GenASM hardware accelerator}. 

\revonur{Based on our power analysis, we find that
power consumption of Edlib is \SI{55.3}{\watt} and \SI{58.8}{\watt} when finding the edit distance between two 100Kbp sequences and two 1Mbp sequences, respectively. \revII{Thus,} GenASM reduces power consumption by $548\times$ and $582\times$ over Edlib, respectively.}

\revonur{
We also compare GenASM with ASAP~\cite{Banerjee2019}, the state-of-the-art FPGA-based accelerator for edit distance calculation. While we are unable to reimplement ASAP, the execution time and power consumption analysis of ASAP provided in~\cite{Banerjee2019} allows us to provide a comparison between GenASM and ASAP. ASAP is optimized for shorter sequences and reports execution time \revIII{only} for sequences of length 64bp--320bp~\cite{Banerjee2019}. Based on\revV{~\cite{Banerjee2019},} 
the execution time of one ASAP accelerator increases from \SI{6.8}{\micro\second} to \SI{18.8}{\micro\second} when the sequence length increases from 64bp to 320bp, while consuming \SI{6.8}{\watt} of power. In comparison, we report that the execution time of one GenASM accelerator increases from \SI{0.017}{\micro\second} to \SI{2.025}{\micro\second} when the sequence length increases from 64bp to 320bp, while consuming \SI{0.101}{\watt} of power. This shows that GenASM provides 9.3--400$\times$ speedup over ASAP, while \revIII{consuming $67\times$ less power}.
}

\subsection{Sources of Improvement in GenASM} 

\revonur{GenASM's} performance improvements come from our \revonur{algorithm/hardware} co-design, \revIII{i.e.,} both from our modified algorithm and our co-designed architecture for this algorithm. The sources of the large improvements in GenASM are (1)~the very \revIII{simple} computations it performs; (2)~the divide-and-conquer approach we follow, which makes our design efficient for both short and long reads despite their different error profiles; and (3)~the very high degree of parallelism obtained with the help of specialized compute units, dedicated SRAMs for both GenASM-DC and GenASM-TB, and the vault-level parallelism provided by \revonur{processing in the logic layer of 3D-stacked memory}. 

\textbf{Algorithm-Level.} \revII{
Our divide-and-conquer approach allows us to decrease the execution time of GenASM-DC from \revII{$(\frac{m \times(m+k)\times k}{P{\times}w})$} cycles to $((\frac{W{\times}W{\times}min(W,k)}{P{\times}w}){\times}\frac{m+k}{W-O})$ cycles, where $m$ is the pattern size, $k$ is the \revV{edit distance threshold}, $P$ is the number of PEs that GenASM-DC has (i.e., 64), $w$ is the number of bits processed by each PE (i.e., 64), $W$ is the window size (i.e., 64), and $O$ is the overlap size between windows (i.e., 24). Although the total GenASM-TB execution time does \emph{not} change ($(m+k)$ cycles vs. $((W-O)\times\frac{m+k}{W-O})$ cycles),
our divide-and-conquer approach helps us decrease the GenASM-DC execution time by $3662\times$ for long reads, and by $1.6-3.9\times$ for short reads. 
}


\textbf{Hardware-Level.} GenASM-DC's systolic-array-based design removes the data dependency limitation of the underlying Bitap algorithm, and provides $64\times$ parallelism by performing 64 iterations of the GenASM-DC algorithm \revV{in parallel}. 
Our hardware accelerator for GenASM-TB makes use of specialized per-PE TB-SRAMs, which eliminates the otherwise very high \revonur{memory} bandwidth consumption of traceback and enables efficient execution.

\textbf{Technology-Level.} With the help of \revonur{3D-stacked memory's} vault-level parallelism, we can obtain $32\times$ parallelism by performing 32 alignments \revIII{in parallel in} different vaults.

\section{\rev{Other Use Cases of GenASM}}\label{sec:bitmac-framework-other}

\revIII{We have quantitatively evaluated three use cases of approximate string matching for genome sequence analysis (Section~\ref{sec:results}). We discuss \revonur{four} other potential use cases of GenASM, whose evaluation we leave for future work.}

\textbf{Read\revIV{-to-Read Overlap} \revIII{Finding Step of de} Novo Assembly.}
\sg{\emph{De novo} assembly\revonur{~\cite{chaisson2015genetic}} is an alternate genome sequencing approach that assembles an entire DNA sequence without the use of a reference genome.}
The first step of \textit{de novo} assembly is \revIII{to find} read-to-read overlaps since the reference genome does not exist\revIII{~\cite{cali2017nanopore}}.
\revonur{Pairwise read alignment (i.e., read-to-read alignment)} is the last step of read-to-read overlap finding\revonur{~\cite{pevzner2001eulerian,li2018minimap2}}. As sequencing devices can introduce errors to the reads, read alignment in overlap finding also needs to take these errors into account. \sg{GenASM can be used for the \revII{pairwise read} alignment step \revII{of} overlap finding.}

\textbf{Hash-Table Based Indexing.}
In the indexing step of read mapping, the reference genome is indexed and stored as a hash table, \revIII{whose keys are all possible fixed-length substrings (i.e., seeds) and whose values are \revIV{the locations} of these seeds in the reference genome.} 
\revII{This index structure is queried in the seeding step to find \revIV{the} candidate matching 
locations of query reads. As we need to find \revIV{the locations} of each seed in the reference text \revIII{to form the index structure}, GenASM can be \revIV{used 
to} generate \revIV{the hash-table based index}.

\textbf{Whole Genome Alignment.}
Whole genome alignment~\revII{\cite{dewey2019whole, paten2011cactus}} is the method of aligning two genomes (from \sg{the same or different} species) for predicting evolutionary \revIII{or familial} relationships between \revIII{these genomes. In whole genome alignment, we need to align two very long sequences. Since GenASM can operate on arbitrary-length sequences \revIV{as a result} of our divide-and-conquer approach, whole genome alignment can be accelerated using the GenASM framework.}}


\revonur{
\textbf{Generic Text Search.} Although GenASM-DC is optimized for genomic sequences (i.e., DNA sequences), which are composed of only 4 characters (i.e., A, C, G and T), GenASM-DC can be extended to support larger alphabets, thus enabling generic text search. \revV{When generating the pattern bitmasks during the pre-processing step, the only change that is required is to generate bitmasks for the \revIII{entire} alphabet, instead of for only four characters.} There is no change required to the edit distance calculation step.

As special cases of general text search, the alphabet can be defined as RNA bases (i.e., A, C, G, U) for RNA sequences or as amino acids (i.e., A, R, N, D, C, Q, E, G, H, I, L, K, M, F, P, S, T, W, Y, V) for protein sequences. This enables GenASM to be used \revIII{for 
RNA} sequence alignment or protein sequence alignment~\cite{haque2009pairwise,smith1981identification,needleman1970general,lipman1985rapid,altschul1990basic,altschul1997gapped,kent2002blat,needleman1970general,notredame2000t,higgins1988clustal,thompson1994clustalw,edgar2004muscle,zou2015halign}. 

}


\section{Related Work} \label{sec:related-work}

To our knowledge, this is the first approximate string matching \revIII{acceleration} framework that enhances and accelerates the Bitap algorithm, and \revIII{demonstrates the effectiveness of the framework for multiple use cases in genome sequence analysis.}
Many previous works have attempted to improve (in software or in hardware) the performance of a \emph{single} step of the genome sequence analysis pipeline. Recent acceleration works tend to follow one of two key \revIII{directions\revonur{~\cite{alser2020accelerating}}.



The} first approach is to build pre-alignment filters that use heuristics to first check the differences between two genomic sequences before using the computationally-expensive approximate string matching algorithms. 
Examples of such filters are the Adjacency Filter \cite{Xin2013} that is implemented for standard CPUs, SHD \cite{Xin2015} that uses SIMD-capable CPUs, and GRIM-Filter \cite{kim2018grim} that is built in 3D-stacked memory. Many works also exploit the large amounts of parallelism offered by FPGA architectures for pre-alignment filtering,
such as GateKeeper~\cite{alser2017gatekeeper}, MAGNET~\cite{alser2017magnet}, Shouji~\cite{Alser2019}, and \revonur{SneakySnake~\cite{alser2019sneakysnake}}. 
A recent work, GenCache~\cite{nag2019gencache}, proposes an in-cache accelerator to improve the filtering (i.e., seeding) mechanism of GenAx (for short reads) by using in-cache operations~\cite{aga2017compute} and software modifications.

The second approach is to use hardware accelerators for the computationally-expensive read alignment step. 
\revonur{Examples of such hardware accelerators are RADAR~\cite{huangfu2018radar}, FindeR~\cite{zokaee2019finder}, and AligneR~\cite{zokaee2018aligner}, which make use of ReRAM based designs for faster FM-index search, or RAPID~\cite{gupta2019rapid} and BioSEAL~\cite{kaplan2020bioseal}, which target dynamic programming acceleration \revonur{with} \revIII{processing-in-memory}.
\revIII{Other read alignment acceleration works include} SIMD-capable CPUs~\cite{Daily2016}, multicore CPUs~\cite{Georganas2015, Liu2015}, and specialized hardware accelerators such as GPUs (e.g., GSWABE~\cite{Liu2015}, CUDASW++ 3.0~\cite{Liu2013}), FPGAs (e.g., FPGASW~\cite{Fei2018}, ASAP~\cite{Banerjee2019}), or ASICs (e.g., Darwin~\cite{turakhia2018darwin} and GenAx~\cite{fujiki2018genax})}. 

\rev{In contrast to GenASM, all of these prior works focus on accelerating \revonur{only} a single \revV{use case in genome sequence analysis}, whereas GenASM \revonur{is capable of accelerating at least} three different use cases (i.e., read alignment, pre-alignment filtering, edit distance calculation) where approximate string matching is required.}


\section{Summary} \label{sec:conclusion}

We propose GenASM, an approximate string matching \revIV{(ASM)} acceleration framework for genome sequence analysis \hl{built upon} \revonur{our modified and enhanced} Bitap algorithm. GenASM performs bitvector-based 
\revIV{ASM}, which can accelerate multiple steps of genome sequence analysis. 
We co-design our \revonur{highly-parallel,} scalable and memory-efficient algorithms with low-power and area-efficient hardware accelerators. 
We evaluate GenASM for three different use cases of ASM in genome sequence analysis \revIV{for both short and long reads}: read alignment, pre-alignment filtering, and edit distance calculation. 
We show that GenASM is significantly faster and more power- and area-efficient
than state-of-the-art software and hardware tools for each of these use cases.

\chapter{BitMAc: FPGA-Based Near-Memory Acceleration of Bitvector-Based Sequence Alignment} 
\label{ch5-bitmac} 

\damlaII{Modern data-intensive applications demand high computation capabilities with strict power constraints. Unfortunately, such applications suffer from a significant waste of both execution cycles and energy in current computing systems due to the costly data movement between the computation units and the memory units~\cite{singh2021nero,mutlu2019processing,boroumand2018google,singh2021fpga,oliveira2021pimbench,mutlu2021primer_pim,ghose2019processing,upmem2021,hajinazar2021simdram,fernandez2020natsa,singh2019napel}. GenASM-based sequence alignment (Chapter~\ref{ch4-genasm}) is an example for such applications. }

Recent FPGAs couple a reconfigurable fabric with high-bandwidth memory (HBM) to enable more efficient data movement and improve overall performance and energy efficiency. This trend is an example of a paradigm shift to \emph{near-memory computing}. In this work, we propose BitMAc, where we leverage such an FPGA with high-bandwidth memory (HBM) for presenting an FPGA-based prototype for our GenASM accelerators. In BitMAc, we map GenASM on an FPGA with a state-of-the-art 3D-stacked memory (HBM2), where HBM2 offers high memory bandwidth and FPGA resources offer high parallelism by instantiating multiple copies of the GenASM accelerators. We exploit intra-level parallelism by instantiating multiple processing elements (PEs) for the DC execution, and inter-level parallelism by running multiple independent GenASM executions in parallel. We show that due to the simplicity of the GenASM algorithms, BitMAc is a low-cost and scalable solution for bitvector-based sequence alignment, with its high energy-efficiency and low resource requirements.

\section{Near-Memory Computing with Modern FPGAs} \label{sec:fpga}

FPGAs are one of the most commonly used form of reconfigurable hardware engines today, and their computational capabilities are greatly increasing every generation due to increased number of transistors on the FPGA chips. Besides these increasing computational capabilities, modern FPGAs provide (1)~an advanced technology node of 7-14nm FinFET~\cite{intelfinfet,xilinxfinfet} that offers higher performance, (2)~different types of on-chip memories (e.g., M20Ks for Intel/Altera chips or UltraRAM (URAM) for Xilinx chips that offer large on-chip memory next to the logic, and (3)~the integration of high-bandwidth memory (HBM) on the same package with an FPGA that allows us to implement our accelerator logic much closer to the memory with an order of magnitude more bandwidth than traditional DDR4-based FPGA boards~\damlaII{\cite{singh2021fpga,singh2021nero}}. Thus, modern FPGA architectures can deliver unprecedented levels of integration and compute capability due to these new advances and features, which provide an opportunity to largely alleviate the memory bottleneck of real-world data-intensive applications.

One example of such modern FPGAs is Intel's Stratix 10 MX device~\cite{stratix10mx}. Intel Stratix 10 MX integrates 3D-stacked High-Bandwidth DRAM Memory (HBM2) alongside a high-performance monolithic 14 nm FPGA fabric die. The fabric die contains 2,100K logic elements (LEs), 94.5 Mbits of embedded eSRAM blocks (each with 47.25 Mbit), over 134 Mbits of embedded M20K memory blocks (each with 20 Kbit), and hard memory controllers.

\begin{figure}[h!]
 \centering
 \includegraphics[width=13cm,keepaspectratio]{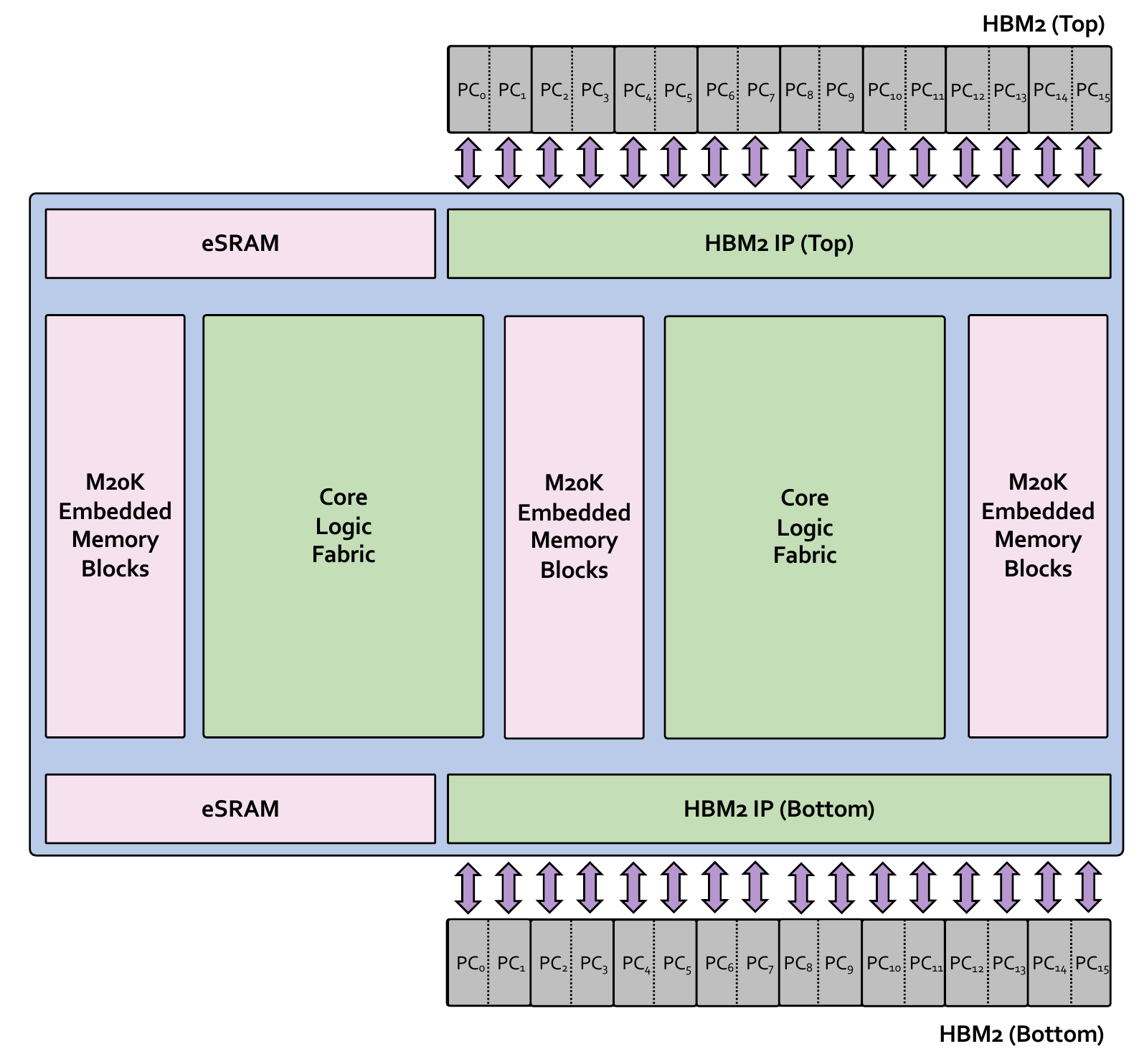}
 \caption{High-level schematic of the Intel Stratix 10 MX device.}
 \label{fig:system-fpga-stratix}
\end{figure}

In Figure~\ref{fig:system-fpga-stratix}, we show a high-level schematic of the Intel Stratix 10 MX device. The FPGA fabric die is connected to two HBM stacks, each of which has 8GB of capacity and 8 independent channels or 16 independent pseudo-channels~\cite{hbm}. Since there are two HBM stacks per each Stratix 10 MX device, in total we have 16GB of memory capacity and 32 independent pseudo-channels. Each HBM2 channel supports a 128-bit DDR data bus, thus, providing two independent 64-bit data bus for each pseudo-channel. For each pseudo-channel, the accesses happen with the burst length of 4, thus at each access, a pseudo-channel reads/writes 32B of data.

The HBM2 IP includes a soft logic adaptor implemented in core logic to efficiently interface user logic to the HBM2 controller. The user interface to the HBM2 controller is maintained through the AXI4 protocol~\cite{ambaaxi4}. The HBM2 IP provides 16 AXI interfaces for each HBM2 controller, with one AXI interface available per HBM2 pseudo-channel. Thus, each AXI interface supports a 256-bit wide write data and 256-bit wide read data interface to/from the HBM2 controller.

\section{BitMAc Implementation} \label{sec:bitmac-design}

\begin{figure}[h!]
 \centering
 \includegraphics[width=14cm,keepaspectratio]{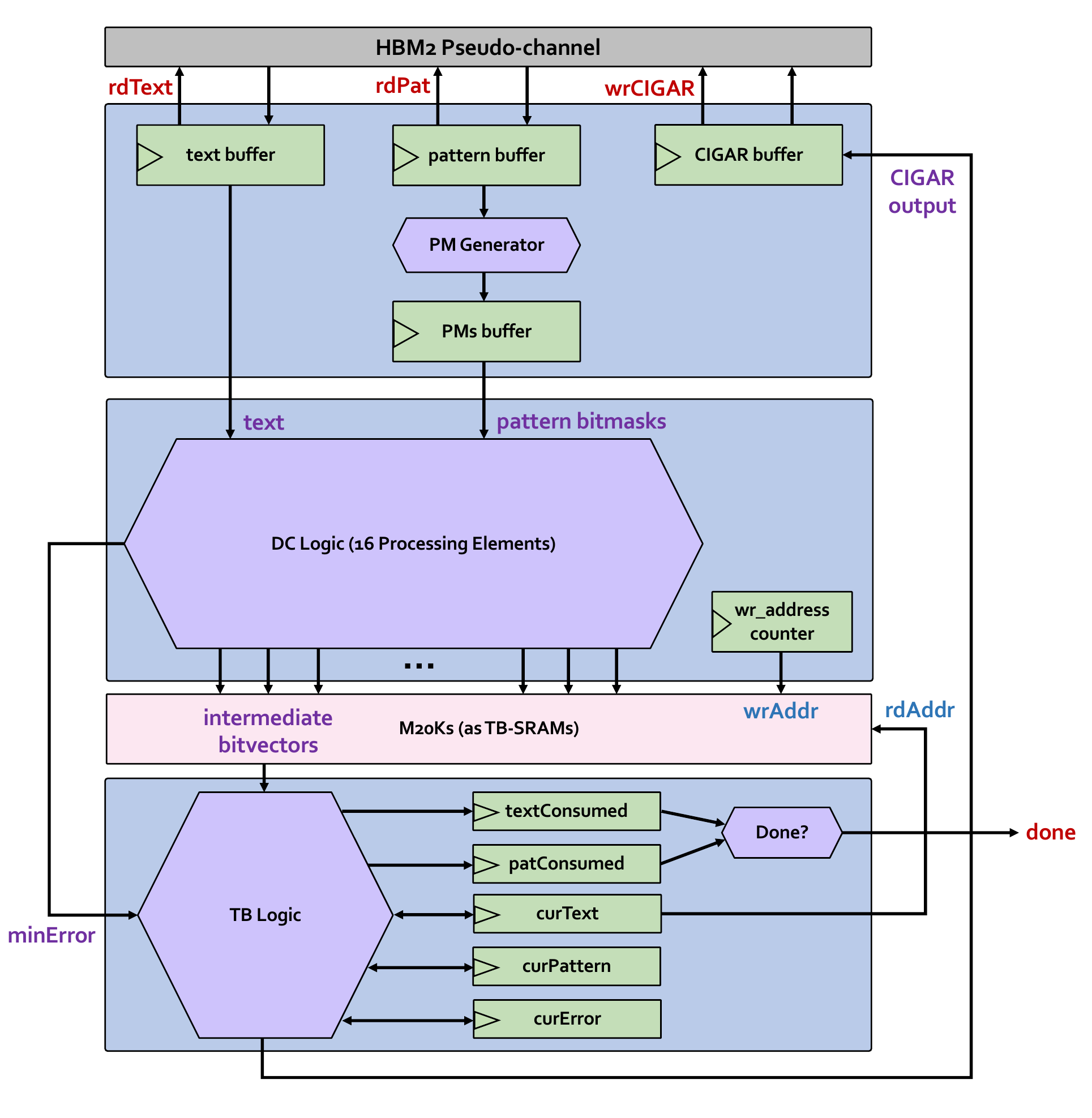}
 \caption{The overview of a BitMAc accelerator attached to one pseudo-channel of an Intel Stratix 10 MX device.}
 \label{fig:bitmac-fpga-mapping}
\end{figure}

In BitMAc, we map our GenASM accelerators and their associated SRAMs (See Section~\ref{sec:bitmac-hw}) to the Intel Stratix 10 MX device. As we show in Figure~\ref{fig:bitmac-fpga-mapping}, we map the DC and TB datapaths along with the HBM2 interface to the FPGA core logic fabric. Due to their larger total capacity, we map TB-SRAMs to the M20Ks. On the other hand, since DC-SRAM is required only to store the input text and pattern, we store them into registers and thus, map them to the core logic fabric as well.

\subsection{Mapping TB-SRAMs to M20Ks}\label{sec:bitmac-m20k-mapping}

In our GenASM design, we set the window size (W) as 64 and at each cycle, we store 3 out of the 4 intermediate bitvectors (i.e., deletion, insertion, and match bitvectors since substitution can be obtained using the deletion bitvector). However, for a better mapping, here in our BitMAc design, to match the width of the data port of M20Ks (40 bits) and to decrease the amount of data written to M20Ks, we set the window size as 60 and we store only 2 out of the 4 intermediate bitvectors (i.e., deletion and match) since substitution bitvector can be obtained using the deletion bitvector and if none of these 3 bitvectors contain the 0-of-interest at the position-of-interest, it means that insertion bitvector has it, thus it is not required to explicitly store the fourth bitvector. These changes enable us to decrease the amount of data generated by each processing element of the DC accelerator to be 120 bits (instead of the original 192 bits), which is 3 times of the width of an M20K data bus.

Since we have 6847 many M20Ks in our FPGA and since we need 3 M20K ports for each processing element, within our budget, we can instantiate 2282 many processing elements (PEs). To be able to instantiate multiple BitMAc accelerators to exploit inter-level parallelism, we set the number of PEs as 16, based on our empirical analysis. This enables us to set the error threshold ($k$) for each window of the BitMAc execution as 15 since for each error value ($0...k$), we have one dedicated PE (Figure~\ref{fig:data_depend}). With this configuration, we can instantiate maximum of 142 BitMAc accelerators on our FPGA.

\subsection{Mapping DC and TB Datapaths to the FPGA Logic}\label{sec:bitmac-logic-mapping}

We map the GenASM-DC and GenASM-TB datapaths to the FPGA logic without any modifications. We also map our FSM design, which controls the memory accesses, DC and TB executions to the FPGA logic. Different than our GenASM design, we also implement the memory components required to store the inputs and outputs of the system as registers, instead of the DC-SRAM in the GenASM design. The reason of this design choice is to be able to dedicate all of the M20Ks solely for storing the intermediate bitvectors. 

\subsection{Mapping of the Main Memory to the HBM2 Stacks}\label{sec:bitmac-memory-mapping}

In order to exploit the high-bandwidth that the HBM2 stacks provide, we store our inputs (text-pattern pairs) and output (CIGAR string) in the HBM2 memory. Since we have 32 pseudo-channels in total, we can attach 4 BitMAc accelerators to each of these pseudo-channels not to exceed our 142 accelerator limit (See Section~\ref{sec:bitmac-m20k-mapping}). 

Since at each burst access of a pseudo-channel, we can read/write 32B of data, we can read one pair of a 60bp-length text (120 bits with a 2-bit implementation) and a 60bp-length pattern (120 bits with a 2-bit implementation) at each burst. For a better memory alignment, we can also set each of these sequences as 64bp (256 bits in total). Similarly, for the CIGAR output, we can have maximum of 256 bits. Thus, in memory, we can reserve 64B memory space for each pair of text and pattern (16B text + 16B pattern + 32B CIGAR output). This also enables us to (1)~have a single memory address counter for each pseudo-channel, and (2)~have 1 memory access for reading and 1 memory access for writing data from/to HBM2 for each window's execution.
\section{Evaluation} \label{sec:bitmac-evaluation}

\subsection{Methodology} \label{sec:bitmac-evaluation-methodology}

We implement our DC and TB accelerator datapaths using SystemVerilog and incorporate the M20Ks and the HBM2 interface for both top and bottom HBM2 stacks using M20K IP and HBM2 IP of Intel Quartus Prime~\cite{intelquartus}, respectively. The complete BitMAc design has 4 BitMAc accelerators connected to each pseudo-channel (128 in total), where each BitMAc accelerator contains a DC accelerator with 16 PEs, a TB accelerator, an FSM, and 13.2KB of M20Ks. We synthesize \revonur{and place \& route} the complete BitMAc design clocked at 200 MHz, and report the resource utilization and power analysis results based on the compilation flow that Intel Quartus Prime provides. We also modify the \revonur{spreadsheet-based} analytical model of GenASM based on our new BitMAc design, and report our performance results based on this analytical model.  

We perform our BitMAc evaluation for the read alignment \revonur{use case} (See Section~\ref{sec:bitmac-framework-aln}). Similar to GenASM, we compare BitMAc with (1)~the read alignment steps of Minimap2~\cite{li2018minimap2} and BWA-MEM~\cite{li2013aligning} as the CPU-based baselines, and (2)~GASAL2~\cite{gasal2} as the GPU-based baseline. \damla{We also compare BitMAc with GenASM to show the comparison for the FPGA- and ASIC-based implementations, respectively.} You can refer to Section~\ref{sec:methodology} for more details on the baselines and the datasets used.

\subsection{Power Analysis} \label{sec:bitmac-evaluation-power}

Table~\ref{table:bitmac-fpga-power} shows the dynamic and total on-chip power dissipation of different configurations of our BitMAc design on an Intel Stratix 10 MX device. We show that the total power dissipation of a single BitMAc accelerator is \SI{6}{\watt}. 
We also show that for 32 BitMAc accelerators (one BitMAc accelerator per pseudo-channel), the total power dissipation is \SI{17.2}{\watt}, and for 128 BitMAc accelerators (four BitMAc accelerators per pseudo-channel; our complete design), the total power dissipation is \SI{48.9}{\watt}. 

\begin{table}[h!]
\centering
\caption{On-chip power dissipation of the BitMAc design.}
\label{table:bitmac-fpga-power}
\includegraphics[width=\columnwidth,keepaspectratio]{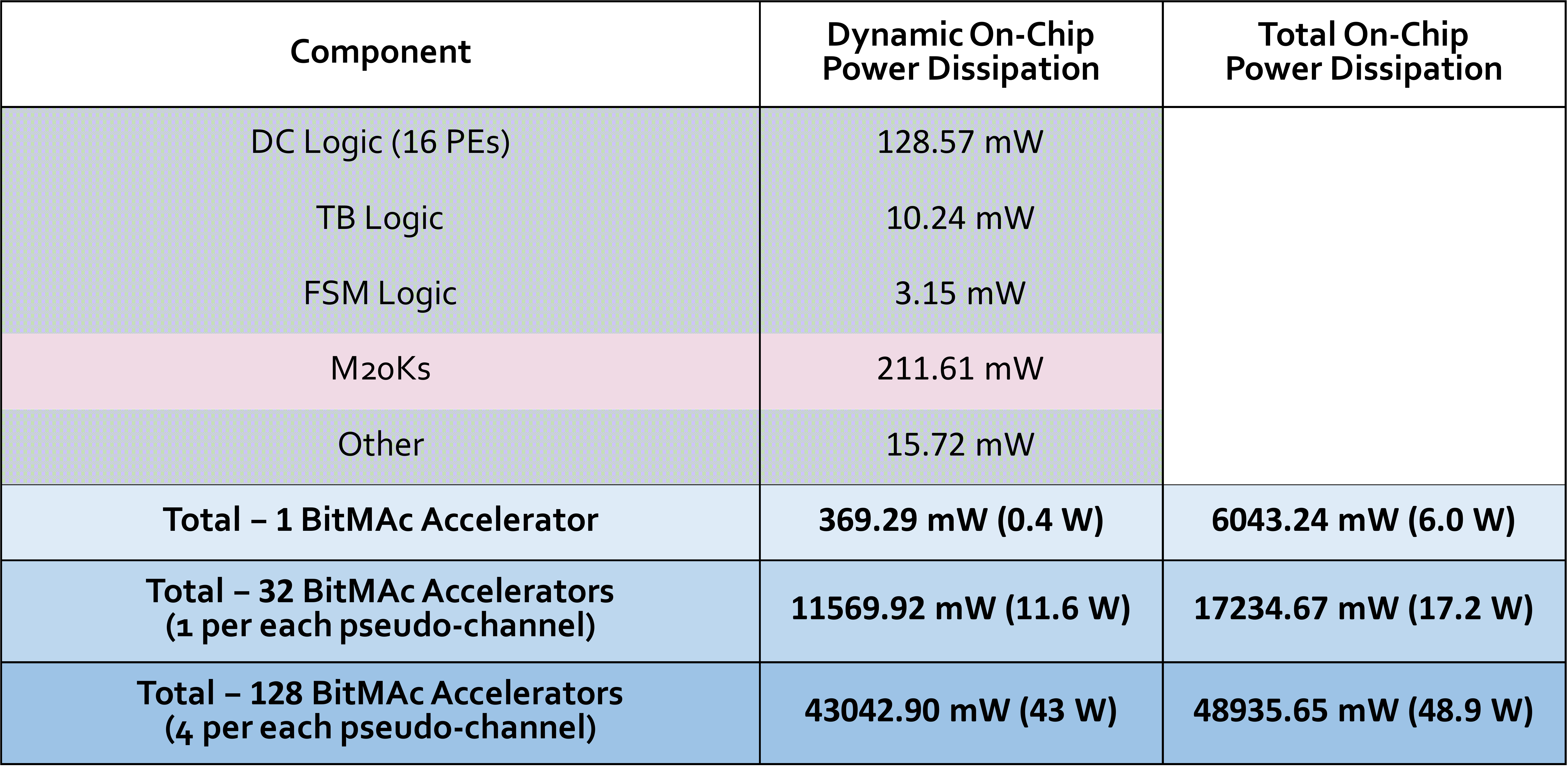}
\end{table}

When we look at the power dissipation analysis by block type for our complete BitMAc design, we find that 59\% of the total dissipation accounts for the M20Ks, 9\% accounts for the combination cells, 7\% accounts for the register cells, 13\% accounts for the clock network, and 12\% accounts for the static power dissipation. Thus, M20Ks are the main contributors to the total on-chip power dissipation.

\subsection{Performance Analysis} \label{sec:bitmac-evaluation-performance}

\begin{figure}[h!]
\centering
\includegraphics[width=\columnwidth,keepaspectratio]{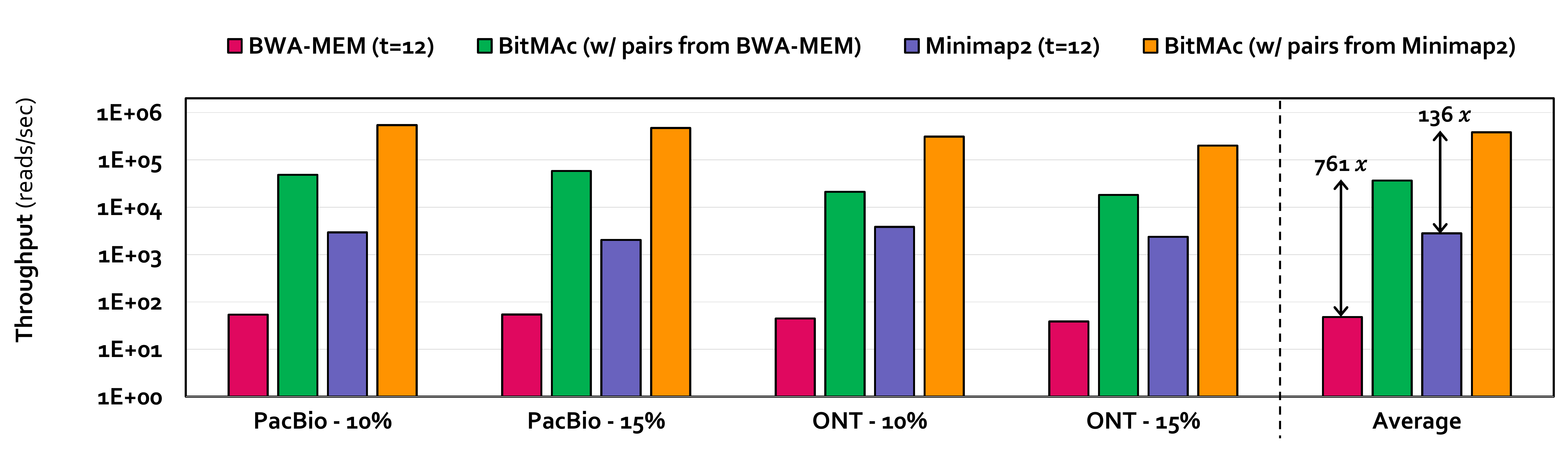}
\caption{\revonur{Throughput comparison of BitMAc and the alignment steps of BWA-MEM and Minimap2 for long reads.}} \label{fig:bitmac-throughput-result-long}
\end{figure}

\textbf{CPU-based Baselines.} 
Figure~\ref{fig:bitmac-throughput-result-long} shows the read alignment throughput (reads/sec) of BitMAc and the alignment steps of BWA-MEM and Minimap2, when aligning long noisy PacBio and ONT reads against the human reference genome. 
As Figure~\ref{fig:bitmac-throughput-result-long} shows, BitMAc provides $761\times$ and $136\times$ throughput improvement over the alignment steps of BWA-MEM and Minimap2 for their \revIII{12-thread} execution, respectively, while reducing the power consumption by $1.9\times$ and $2.0\times$ for 12-thread execution. 

\begin{figure}[h!]
\centering
\includegraphics[width=\columnwidth,keepaspectratio]{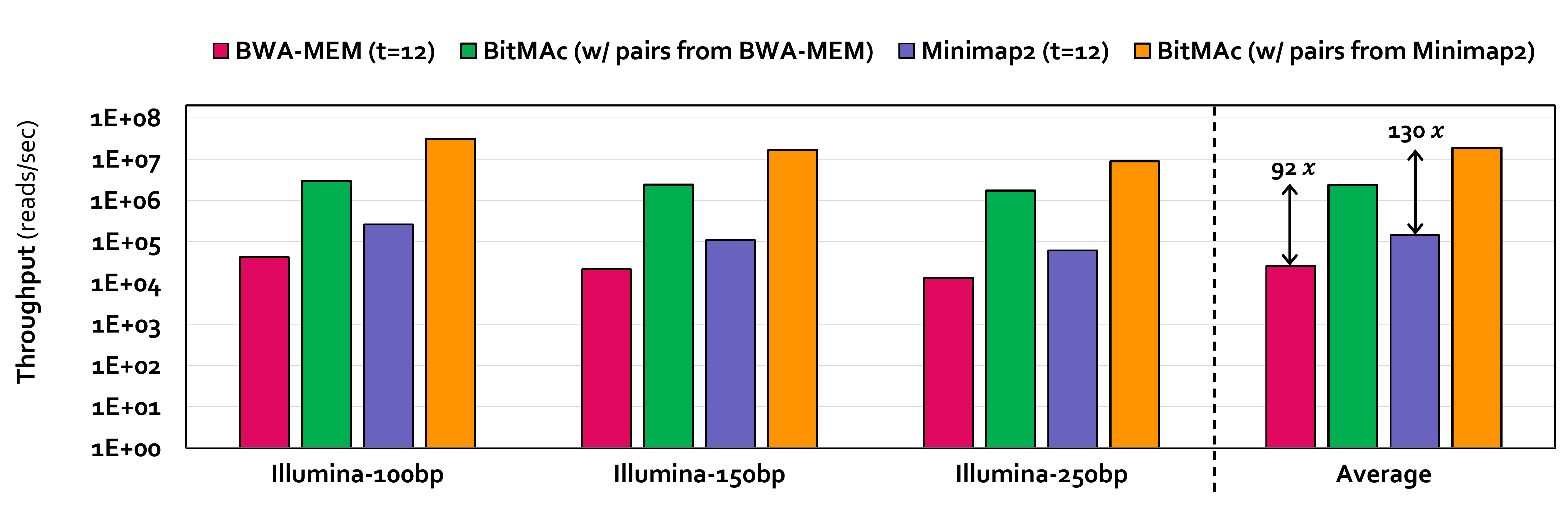}
\caption{Throughput comparison of BitMAc and the alignment steps of BWA-MEM and Minimap2 for short reads.} \label{fig:bitmac-throughput-result-short}
\end{figure}

Figure~\ref{fig:bitmac-throughput-result-short} \revonur{compares} the read alignment throughput (reads/sec) of BitMAc \revonur{with that of} the alignment steps \revonur{of BWA-MEM and Minimap2, when aligning short} Illumina reads against the human reference genome. 
We find that BitMAc provides $92\times$ and $130\times$ throughput improvement over the alignment steps of BWA-MEM and Minimap2 for their \revIII{12-thread} execution, respectively, while reducing the power consumption by $2.2\times$ and $2.0\times$ for 12-thread execution. 

\textbf{GPU-based Baseline.}
We compare BitMAc with \revonur{the state-of-the-art GPU aligner,} GASAL2~\cite{gasal2}, \revonur{using three datasets of varying size (100K, 1M, and 10M reference-read pairs)}. \revIV{Based on our analysis,} we \revonur{make three findings.
First, for} 100bp Illumina reads, BitMAc provides \revonur{$7.9\times$, $7.3\times$, and 6.8$\times$} speedup over GASAL2, while reducing the power consumption by \revonur{2.9\%, 12.3\% and 13.7\%} for 100K, 1M, and 10M datasets, respectively.
\revonur{Second, for} 150bp Illumina reads, BitMAc provides \revonur{$13.2\times$, $10.9\times$, and $11.1\times$} speedup over GASAL2, while reducing the power consumption by \revonur{1.5\%, 15.6\% and 19.1\%} for 100K, 1M, and 10M datasets, respectively.
\revonur{Third, for} 250bp Illumina reads, BitMAc provides $19.4\times$, $18.7\times$, and $19.0\times$ speedup over GASAL2, while reducing the power consumption by \revonur{9.9\%, 25.2\% and 26.6\%} for 100K, 1M, and 10M datasets, respectively.

\subsection{FPGA Resource Utilization} \label{sec:bitmac-evaluation-resourceutil}

We list the resource utilization of different configurations of our BitMAc design on an Intel Stratix 10 MX device in Table~\ref{table:bitmac-resourceutil}. We make four key observations. 
First, we find that for the complete BitMAc design (128 BitMAc accelerators), the DC logic corresponds to 73\% of the logic utilization, the TB logic corresponds to 23\% of the logic utilization, the FSM logic corresponds to 1\% of the logic utilization, and the memory interface logic corresponds to 3\% of the logic utilization.
Second, we find that there is a linear scaling of the FPGA resources with the number of instances we include in our design. Third, we observe that the logic utilization (64\%) is lower than the on-chip memory utilization (90\%). Fourth, due to high amount of data that needs to be stored for the TB execution on M20Ks, we are bottlenecked by the amount of on-chip memory (i.e., M20Ks) we have. Because of this limitation, we also cannot saturate the high bandwidth that multiple HBM2 stacks on the FPGA provide.
Thus, in order to scale further and fully exploit the high-bandwidth that HBM2 stacks provide, we need (1)~algorithm-level modifications to decrease the amount of data that need to be stored in M20Ks, and (2)~newer FPGA chips that provide a higher amount of on-chip memory capacity.

\begin{table}[h!]
\centering
\caption{FPGA resource utilization of the BitMAc design.}
\label{table:bitmac-resourceutil}
\includegraphics[width=\columnwidth,keepaspectratio]{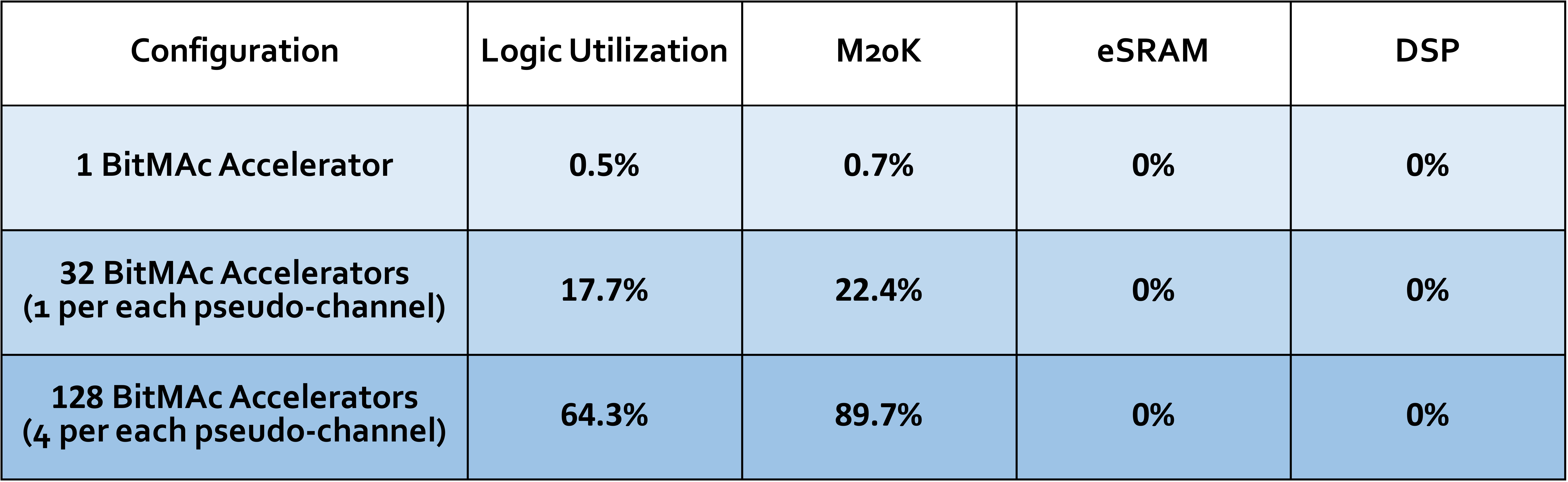}
\end{table}

\damla{
\subsection{FPGA-based BitMAc vs. ASIC-based GenASM} \label{sec:bitmac-genasm-comparison}

We also compare BitMAc with GenASM~(Chapter~\ref{ch4-genasm}) to show the comparison for the FPGA- and ASIC-based implementations, respectively. For GenASM, we consider the configuration explained in Section~\ref{sec:methodology} and for BitMAc, we consider the configuration explained in Section~\ref{sec:bitmac-evaluation-methodology}. GenASM operates at 1GHz while having 64 PEs per accelerator and 32 accelerators in total in its complete design. On the other hand, BitMAc operates at 200MHz while having 16 PEs per accelerator and 128 accelerators in total in its complete design.

Based on our analysis, we find that BitMAc provides $1.18\times$ throughput improvement over GenASM for both short and long reads with the help of our algorithmic and hardware design changes for a more efficient FPGA mapping. However, due to the cost of reconfigurability of the FPGAs and dedicating the whole chip for the BitMAc design, a single FPGA-based BitMAc accelerator increases the power consumption by $4\times$ compared to a single ASIC-based GenASM accelerator when we only take the dynamic on-chip power dissipation into consideration. On the other hand, when we take the total on-chip power dissipation into consideration, BitMAc increases the power consumption by $59\times$.
}
\section{Summary} \label{sec:bitmac-discussion}

We propose BitMAc, where we leverage a modern FPGA with high-bandwidth memory (HBM) for presenting an FPGA-based prototype for our GenASM accelerators. In BitMAc, we map GenASM on Stratix 10 MX FPGA with a state-of-the-art 3D-stacked memory (HBM2), where HBM2 offers high memory bandwidth and FPGA resources offer high parallelism by instantiating multiple copies of the GenASM accelerators. 

After re-modifying the GenASM algorithms for a better mapping to existing FPGA resources, we show that BitMAc provides 64\% logic utilization and 90\% on-chip memory utilization, while having \SI{48.9}{\watt} of total power consumption. We compare BitMAc with state-of-the-art CPU-based and GPU-based read alignment tools. For long reads, BitMAc provides $761\times$ and 136$\times$ speedup over the alignment steps of the state-of-the-art read mappers, BWA-MEM and Minimap2, respectively, while reducing power consumption by $1.9\times$ and $2.0\times$. For short reads, BitMAc provides 92$\times$ and 130$\times$ speedup over the alignment steps of BWA-MEM and Minimap2, respectively, while reducing power consumption by $2.2\times$ and $2.0\times$. We also show that BitMAc provides significant speedup compared to the GPU-based baseline, GASAL2. Thus, BitMAc is a low-cost and scalable FPGA-based solution for bitvector-based sequence alignment.
 
\chapter{\mech: A Hardware Acceleration Framework for Sequence-to-Graph Mapping}\label{ch6-segram} 

\damla{

\sgh{An emerging problem with using a single reference genome for an entire species is the DNA variation that exists from organism to organism within a population (known as \emph{genetic diversity}), which results from DNA mutations over time. 
The use of a single reference genome can bias the \zulal{mapping} 
process and downstream analysis towards the DNA composition and variations present in the reference organism, \zulal{because}
(1)~the organism whose DNA is being constructed \zulal{may have} 
a different set of variations, and
(2)~the reference organism's variations might be uncommon among most organisms in the population~\cite{paten2017genome}.
As a result, the reconstructed DNA may not be a faithful reproduction of the original sequence.
Combined with errors that can be introduced during genome sequencing (with error rates as high as 5--10\% for long reads with thousands of base pairs~\cite{jain2018nanopore, weirather2017comprehensive,ardui2018single,van2018third}), reference bias can lead to significant inaccuracies during mapping.}
\sghi{This can create issues for a wide range of genomic studies, from identifying mutations that lead to cancer, to tracking mutating variants of viruses such as SARS-CoV-2~\cite{computonics2020hackathon}, where the details of the variation from the reference are critical to our understanding~\cite{ambler2019gengraph}.}

\sgh{An emerging technique to overcome reference bias is the use of graph-based representations of a species' genome, known as \emph{genome graphs}~\cite{pevzner2001eulerian, paten2017genome}.
A genome graph represents the reference genome \emph{and} known genetic variations in the population as a graph-based data structure.
A node represents one or more base pairs, and edges connect the base pairs in a node to all of the possible base pairs that come next in the sequence, with multiple outgoing edges from a node capture genetic variation.
Genome graphs are growing in popularity for a number of applications, such as variation calling~\cite{garrison2018variation}, genome assembly~\cite{compeau2011apply, pevzner2001eulerian, zerbino2008velvet, simpson2009abyss}, error correction~\cite{salmela2014lordec}, and multiple sequence alignment~\cite{paten2011cactus, lee2002multiple}}.



\sgh{For genome sequence analysis, instead of mapping an organism's reads to the linear DNA sequence of a single reference organism (known as sequence-to-sequence mapping), \emph{sequence-to-graph mapping} captures the inherent genetic diversity among a population and can result in significantly more accurate 
\zulal{read mapping}~\cite{garrison2018variation}.
Like sequence-to-sequence mapping, sequence-to-graph mapping}
follows the \emph{seed-and-extend strategy}~\cite{rautiainen2020graphaligner}. 
\sgh{The first time the graph is constructed, its nodes are indexed for fast lookup. During mapping,}
this \sgh{pre-processed} index is used in the \emph{seeding} step, which aims to find \sgh{potential} seed matches between the query read and a region of the graph. After optionally clustering or \sgh{\emph{filtering} the potential matches,}
\emph{alignment} is performed between all of the remaining seed locations of the graph and the query read.


Due to its nascent nature, only a few software tools exist for graph-based genome sequence analysis~\cite{nurk2021complete, rautiainen2020graphaligner, garrison2018variation, rakocevic2019fast, kim2019graph}.
Given the additional complexities and overheads of processing a genome graph instead of a linear reference genome, graph-based analysis exacerbates analysis bottlenecks such as read-to-reference mapping.
\textbf{Our goal} is to design high-performance, scalable, power- and area-efficient hardware accelerators that alleviate bottlenecks in both the seeding and alignment steps of sequence-to-graph mapping, with support for both short (e.g., Illumina) and long (e.g., PacBio, ONT) read sequencing technologies.

In this work, we propose \mech, a hardware acceleration framework for sequence-to-graph mapping and alignment.
For seeding, we base \mech on a memory-efficient minimizer-based seeding algorithm, and for alignment, we develop a new bitvector-based, highly-parallel sequence-to-graph alignment algorithm. We \emph{co-design} both of our algorithms with high-performance, area- and power-efficient hardware accelerators. \mech consists of two components: (1)~\ms, which provides hardware support to 
execute our minimizer-based seeding algorithm \zulal{efficiently}, and (2)~\ba, which provides hardware support to 
execute our bitvector-based sequence-to-graph alignment algorithm \zulal{effectively}. \damla{To our knowledge, \mech is the \emph{first} hardware acceleration framework for sequence-to-graph mapping, \ms is the \emph{first} hardware accelerator for minimizer-based seeding, and \ba is the \emph{first} hardware accelerator for sequence-to-graph alignment.}

\section{Minimizer-Based Indexing \& Seeding}
In many applications of string comparisons for sequence analysis, the first step is to find the set of \emph{seeds} to represent each sequence. Seeds are chosen from the set of \emph{k-mers}, which are exact matching subsequences of length \emph{k} between the query sequence and reference~\cite{roberts2004reducing}. Considering ultra-long reads being produced by recent sequencing machines, the size of the k-mer set can be enormous depending on \emph{k}, making it hard to store and process. 

One possible approach for reducing storage requirements is to apply fixed k-mer sampling methods~\cite{liu2019fast}. Another approach divides the sequence into windows with a predefined size and selects a k-mer from each window according to a scoring mechanism as representatives. These unique k-mers, called \emph{minimizers}~\cite{roberts2004reducing, schleimer2003winnowing}, ensure that two different sequences are represented with the same seed if they contain a long enough common subsequence. Since minimizer-based seeding eliminates some possible k-mers, sensitivity and speed can vary depending on the window size and scoring~\cite{jain2020weighted}.

\section{Sequence-to-Graph Alignment}
The goal of aligning a sequence to a graph is finding the path on the graph that yields the sequence's highest alignment score~\cite{jain2019accelerating}. Sequence-to-graph alignment algorithm with quadratic time complexity was first formulated by Navarro~\cite{navarro2001guided} and it traverses the DP matrix row by row instead of the original column-wise fashion of linear alignment. After calculating the terms on a row, the algorithm makes searches on the graph and propagates the values to the next row. There are many efforts for optimizing or accelerating the dynamic programming for linear alignment. However,  obtaining efficient solutions for sequence-to-graph alignment demands attention with the growing trend in genome graphs. 
}

\section{Motivation and Goal} \label{sec:gengraph-motivation}

\damla{
As shown in GenASM (Chapter~\ref{ch4-genasm}) and other prior works~\damlaII{\cite{alser2017gatekeeper,turakhia2018darwin,fujiki2018genax,kim2020geniehd,alser2019sneakysnake,goyal2017ultra,fujiki2020seedex,bingol2021gatekeeper,nag2019gencache,kim2018grim,lavenier2016dna,kaplan2018rassa,kaplan2020bioseal}}, sequence-to-sequence mapping is one of the major bottlenecks of the genome sequence analysis pipeline and need to be accelerated using specialized hardware. Since a graph-based representation of the genome is more complex than the linear representation, sequence-to-graph mapping places greater pressure on this bottleneck. Thus, there is a pressing need to develop techniques that provide fast, efficient, and low-cost sequence-to-graph mapping, which support both short reads (e.g., Illumina reads) and long reads (e.g., PacBio and ONT reads).

\damlaI{When we analyze the state-of-the-art sequence-to-graph mapping tool, GraphAligner~\cite{rautiainen2020graphaligner}, in order to reveal the performance bottlenecks, we make three key observations. (1)~With respect to memory accesses, GraphAligner is 86.7\% L1 data cache bound, i.e., GraphAligner requires improvements to the L1 data cache (e.g., lower access latency) in order to improve its performance. (2)~The bandwidth utilization of GraphAligner is less than 3\% (high number of L1 data cache hits due to high temporal locality) which shows that DRAM bandwidth is not the bottleneck of the performance of GraphAligner. (3) When we perform a scalability analysis by running GraphAligner with 10, 20, and 40 threads, we observe that it scales sublinearly (i.e., its parallel efficiency does not exceed 0.4), indicating that only increasing the number of threads does not realize the full benefits of parallelism, motivating the need for efficient acceleration.}

Even though there are several hardware accelerators designed \sgh{to alleviate bottlenecks in} several steps of linear read mapping (e.g., pre-alignment filtering~\cite{kim2018grim,alser2019sneakysnake}, sequence-to-sequence alignment~\cite{turakhia2018darwin,fujiki2018genax,cali2020genasm}), none of these designs can be employed directly for the sequence-to-graph mapping problem. \damla{This is because linear sequence mapping is a special case of sequence-to-graph mapping, where all nodes have only one edge, and hence the corresponding accelerators are limited to this special case, but unsuitable for the general problem, where we also need to consider multiple edges that a node can have (i.e., hops).} However, with the growing importance and usage of genome graphs, it is crucial to have efficient designs for sequence-to-graph mapping, which are tuned to work with both short and long reads. 

In this work, our goal is to design a high-performance, memory-efficient, and low-power hardware acceleration framework for sequence-to-graph mapping and alignment. To this end, we propose \mech, the first hardware acceleration framework for sequence-to-graph mapping and alignment. For an efficient and general-purpose acceleration, \mech aims to accelerate both seeding and sequence-to-graph alignment steps of the sequence-to-graph mapping pipeline, which are optimized for both short and long reads. We base \mech upon a minimizer-based seeding algorithm and we propose a novel bitvector-based algorithm to perform approximate string matching between a read and \damla{a graph-based reference}. \damla{To our knowledge, \mech proposes the first hardware accelerator for minimizer-based seeding and the first hardware accelerator for sequence-to-graph alignment.}
}
\section{Overview of \mech} \label{sec:background}

\damla{

In \mech, we \emph{co-design} our minimizer-based seeding algorithm and bitvector-based sequence-to-graph alignment algorithm with highly parallel, low power, and area efficient accelerators. \mech consists of two main components: (1)~\ms~(MS), which finds the minimizers for a given query read and fetches the candidate seed locations for the selected minimizers; and (2)~\ba~(BA), which for each candidate seed, aligns the query read for the subgraph surrounding the seed and finds the optimal alignment.

Before \mech execution starts, as the pre-processing steps, each chromosome's graph structure is generated, then each graph's nodes are indexed, and both the resulting graph and hash table index are pre-loaded to the main memory.

\begin{figure}[b!]
\centering
\includegraphics[width=\columnwidth,keepaspectratio]{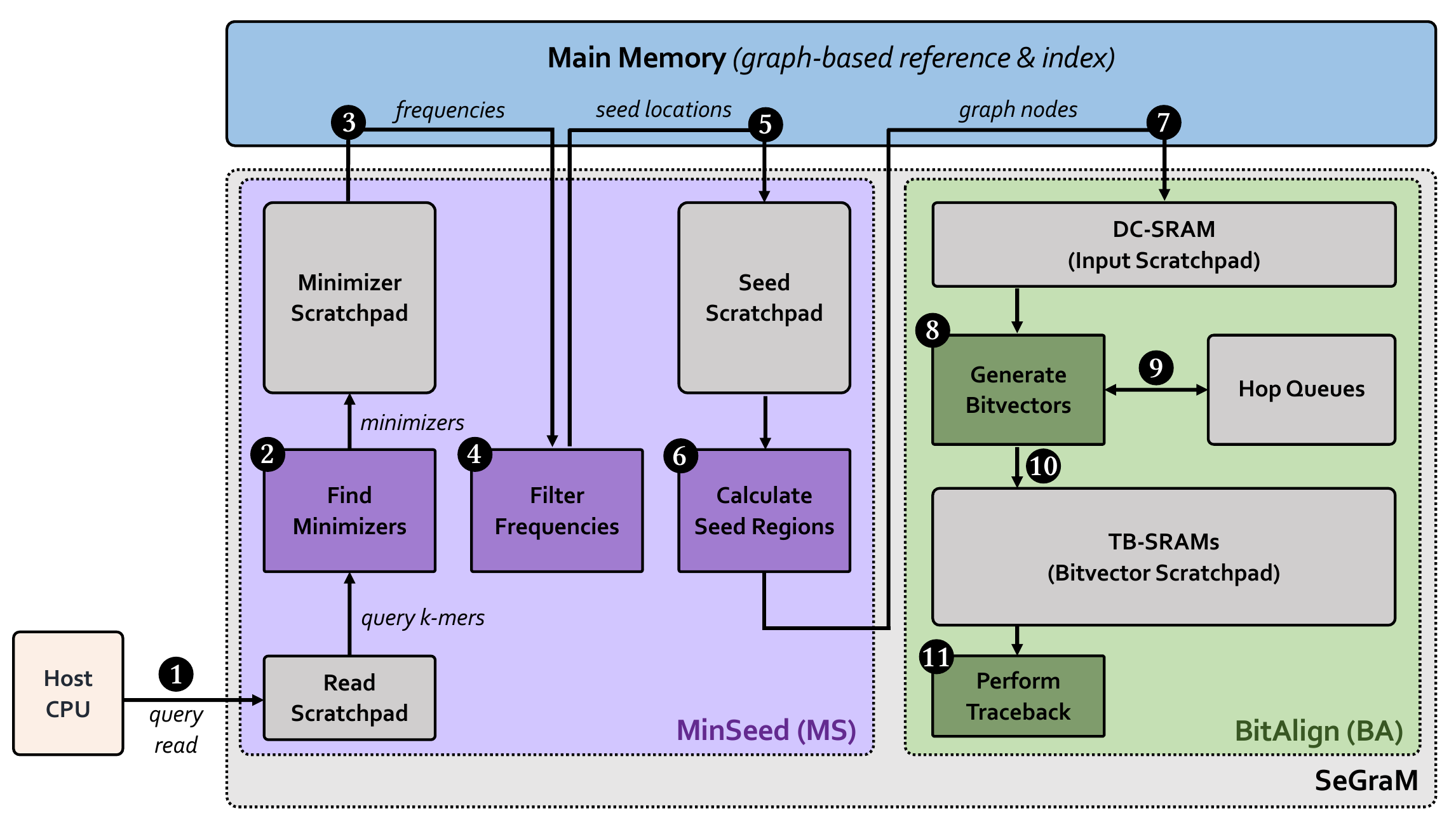}
\caption{Overview of \mech.} 
\label{fig:gengraph-pipeline}
\end{figure}

\mech execution starts when the query read is streamed from the host CPU and \ms writes it to the \emph{read scratchpad} (\circlednumber{1}). Using all of the \emph{k-mers} of the query read, \ms finds the minimizers and writes them to the \emph{minimizer scratchpad} (\circlednumber{2}). For each minimizer, \ms fetches its frequency from the hash table at the main memory (\circlednumber{3}) and filters out the minimizers whose frequency is above an user-defined threshold (\circlednumber{4}). Next, \ms fetches the seed locations of the remaining minimizers from the main memory, and writes them to the \emph{seed scratchpad} (\circlednumber{5}). Finally, \ms calculates the candidate reference region for each seed (\circlednumber{6}), fetches the graph nodes from the memory for each candidate region and writes them to the \emph{input scratchpad} (\circlednumber{7}). Once \ms sends the subgraph corresponding to the candidate reference region along with the query read, \ba execution starts with generating the bitvectors (\circlednumber{8}) required for the distance calculation step of the alignment (i.e., seed-extension). While generating these bitvectors, \ba writes them to the \emph{hop queues} in order to handle the hops required for the graph alignment (\circlednumber{9}), and also, to the \emph{bitvector scratchpad} (\circlednumber{10}). Once \ba finishes generating and writing all the bitvectors, it starts reading them back from the \emph{bitvector scratchpad}, performs the traceback operation, and finds the optimal alignment between the subgraph and the query read (\circlednumber{11}).

}
\section{Pre-Processing for \mech} \label{sec:preprocessing}

\damla{

\mech mechanism requires two pre-processing steps before starting its execution: (1)~generating the graph-based reference using a linear reference genome (i.e., as a FASTA file~\cite{fasta}) and its associated variations (i.e., as VCF file(s)~\cite{vcf}), and topologically sorting this graph~\cite{kahn1962topological}; and (2)~indexing the nodes of this generated graph and generating the hash table-based index. 

\textbf{Graph-based reference generation.}
As the first pre-processing step, we generate the graph-based reference from the input FASTA file and VCF file(s) \damla{using} the vg toolkit's~\cite{garrison2018variation} \texttt{vg construct} command. We generate one graph for each chromosome, and we set the maximum sequence length of each node as 16Kbp. For the alignment step of sequence-to-graph mapping, we need to make sure the nodes of our graphs are topologically sorted. Thus, as our next step, we sort our graphs using the \texttt{vg ids -s} command. Then, we convert our VG-formatted graphs to GFA-formatted~\cite{gfa} graphs using the \texttt{vg view} command since GFA is easier to work with for the later steps of the pre-processing. 

As we show in Figure~\ref{fig:gengraph-graphstructure}, in order to store the graph-based reference, we generate three separate table structures \damla{by using the GFA-formatted graphs}: (1)~the nodes table, which stores the nodes of the graphs as the <node id> as the key and the <node's sequence's length, starting address at the sequences table, node's number of outgoing edges, starting address at the edges table> as the value; (2)~the sequences table, which stores the associated sequences of each node; and (3)~the edges table, which stores the associated outgoing nodes (by node ID) of each node.

\begin{figure}[h!]
\centering
\includegraphics[width=0.9\columnwidth,keepaspectratio]{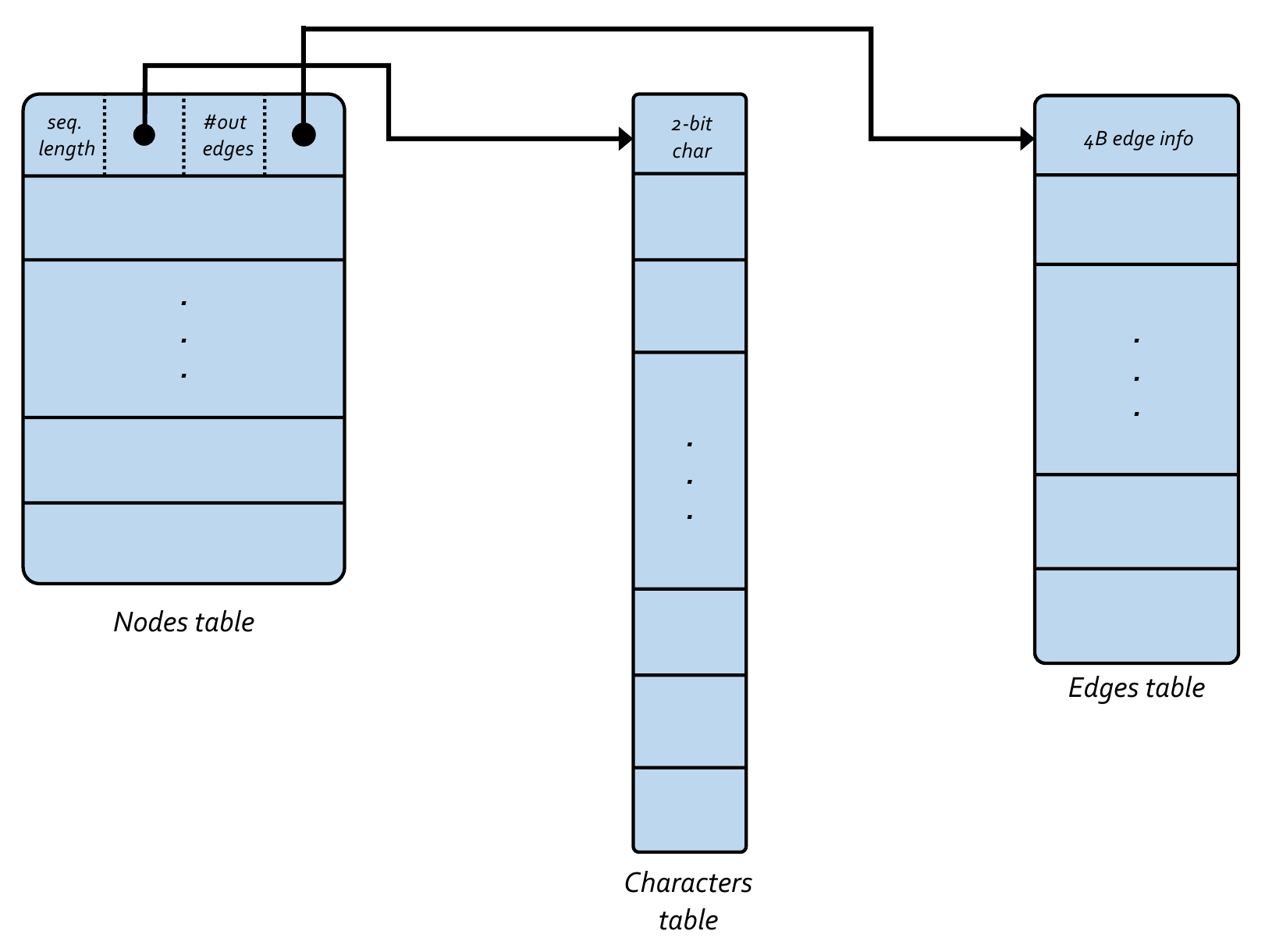}
\caption{Memory layout of the graph-based reference structure. 
} \label{fig:gengraph-graphstructure}
\end{figure}

We use the stats (i.e., number of nodes, number of edges, and total sequence length) for each chromosome's associated graph to determine the data sizes of our graph structure.
Based on our analysis, we find that each node in the nodes table requires 32B per entry and the total size of the nodes table is \emph{\#nodes} $*$ \emph{32B}. \damla{Since we can store characters in the sequences table using a 2-bit representation (since we only have 4 possible characters: A (00), C (01), G (10), and T (11)),} the total size is \emph{total sequence length} $*$ \emph{2bit}. We also find that each entry in the edges table requires 4B, thus the total size of the edges table is \emph{\#edges} $*$ \emph{4B}. \damlaI{In total, it takes 1.4 GB to store the graph structures for 24 chromosomes of the human genome.}

\textbf{Hash table-based index generation.}
As the second pre-processing step, we generate the hash table-based index for each of the generated graphs (one for each chromosome). Different than the traditional linear read mapping, in sequence-to-graph mapping's indexing step, the nodes of the graph structure are indexed and stored in the hash table. As we explain in Section~\ref{sec:minseed_algo}, since \mech performs minimizer-based seeding, we also use minimizers~\cite{roberts2004reducing,li2016minimap,li2018minimap2} as the keys and their exact matching locations in the graphs' nodes as the values of the index while generating the index.

As we show in Figure~\ref{fig:gengraph-indexstructure}, in order to store the hash table-based index, we use a three-level structure. In the first-level of the hash table, similar to Minimap2~\cite{li2018minimap2}, we have \emph{buckets} to decrease the memory footprint of the index. Each entry in this first-level of the index stores <starting address of the minimizers in the corresponding bucket in the second-level, number of minimizers in the corresponding bucket>. In the second-level of the hash table, we have the \emph{minimizers}. Each entry in this second-level of the index stores <hash value of the corresponding minimizer, starting address of the seed locations of the corresponding minimizer in the second-level, number of locations of the corresponding minimizer>. Minimizers are sorted based on their hash values in the second-level of the index. Finally, in the third-level of the hash table, we have the \emph{seed locations}. Each entry in this third-level of the index stores <node ID of the corresponding seed location, relative offset of the corresponding seed location within the node>. Locations are grouped based on their corresponding minimizers and sorted within each group based on their values.

\begin{figure}[h!]
\centering
\includegraphics[width=0.9\columnwidth,keepaspectratio]{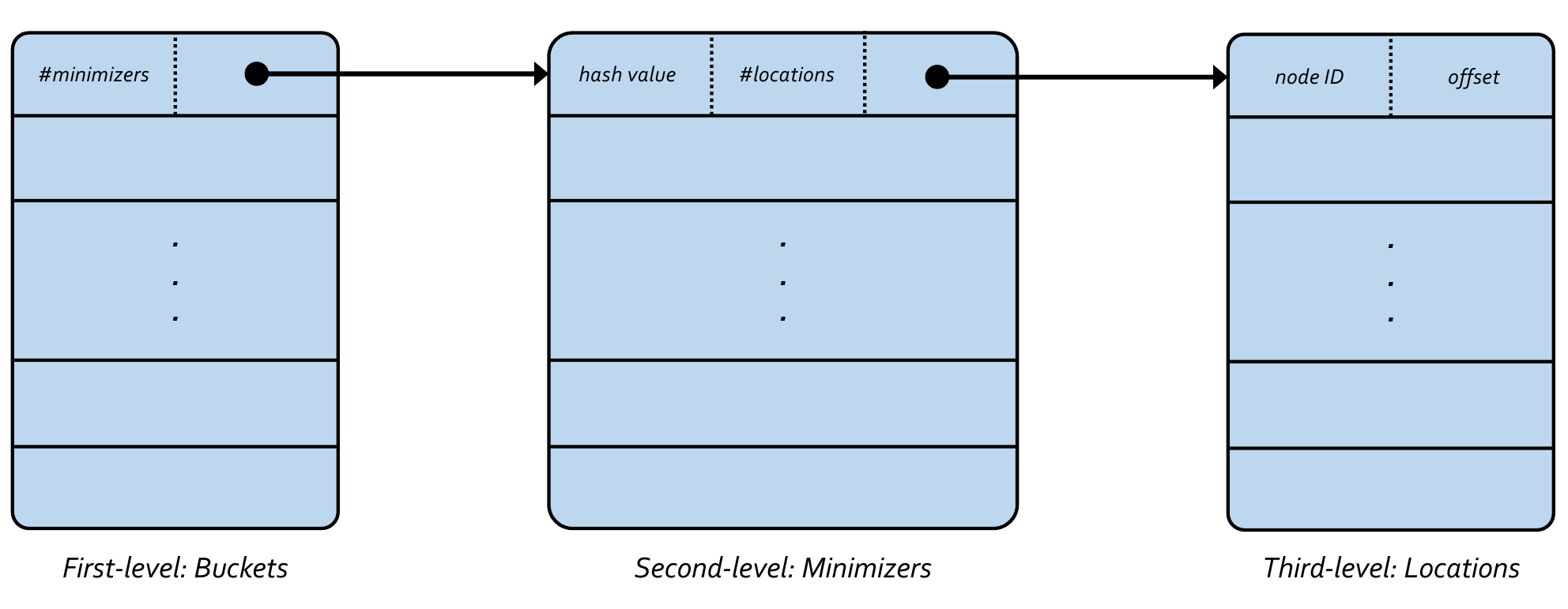}
\caption{Memory layout of the hash table-based index structure.
} \label{fig:gengraph-indexstructure}
\end{figure}

We use the stats (i.e., number of distinct minimizers, total number of locations, maximum number of minimizers per bucket, and maximum number of locations per minimizer) for each graph to determine the data sizes of our index structure. Based on our empirical analysis, we find that $2^{24}$ is the optimum number of buckets in terms of memory footprint and the number of keys per each bucket. We also find that each bucket entry requires 4B of data, thus the total size of the first-level of the index is $2^{24}$ $*$ \emph{4B}. Each minimizer in the second-level of the index requires 12B of data, thus the total size of the second-level of the index is \emph{\#distinct minimizers} $*$ \emph{12B}. Each location in the third-level of the index requires 8B of data, thus the total size of the third-level of the index is \emph{\#total number of locations} $*$ \emph{8B}. \damlaI{In total, it takes 9.8 GB to store the hash table-based index for 24 chromosomes of the human genome.} 

}
\section{MinSeed Algorithm} \label{sec:minseed_algo}

\damla{

We base our seeding algorithm, \ms, upon Minimap2's minimizer-based seeding algorithm (i.e., \texttt{mm\_sketch}). A \emph{minimizer}~\cite{roberts2004reducing,li2016minimap,li2018minimap2} is the smallest \emph{k-mer} in a window of \emph{w} consecutive k-mers. The goals of using minimizers, or \emph{<w,k>}-minimizers, instead of the full set of k-mers, are to decrease the storage requirements of the index by storing fewer number of k-mers, and speeding up the queries that are made to this index. In Figure~\ref{fig:minimizers}, we show an example of how <5,3>-minimizer of a sequence is selected among the full set of k-mers. After finding the 5 adjacent 3-mers, we sort them and select the smallest. In this example, sorting is done simply by the lexicographical order.

\begin{figure}[h!]
\centering
\includegraphics[width=0.75\columnwidth,keepaspectratio]{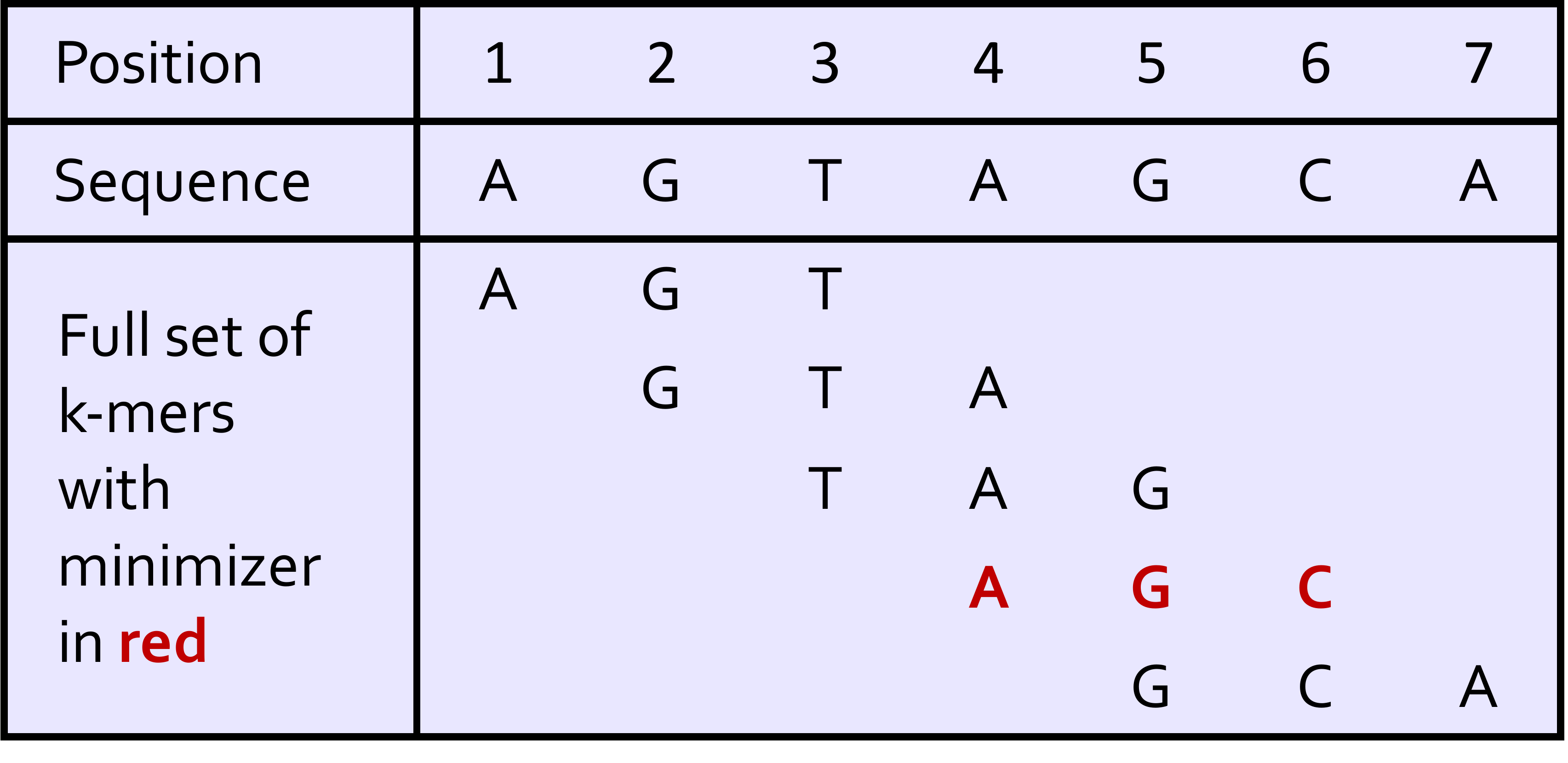}
\caption{Example of finding the minimizers of a given sequence.} \label{fig:minimizers}
\end{figure}

\ms algorithm starts with computing the minimizers of a given query read. 
Even though using two loops such that the outer loop iterates over the query read while the inner loop finds the minimum k-mer within each window is an easy solution for minimizer computation, using a queue that caches the previous minimum k-mers can avoid the inner loop and provide a $O(m)$ complexity algorithm, where $m$ is the length of the query read~\cite{li2016minimap,li2018minimap2,jain2020weighted}.

After finding the minimizers, \ms queries the hash table-based index stored in the memory to fetch the frequencies (i.e., \emph{\#locations}) of each minimizer. 
If the frequency of a minimizer is above the user-defined threshold, then it is discarded. 
If the minimizer is not discarded, then all the seed locations for that minimizer is fetched from the index in the memory.

After fetching all the seed locations, using the node ID and relative offset of the seed locations along with the relative offset of the corresponding minimizer within the query read, the rightmost and leftmost positions (i.e., right-extension and left-extension) of each seed are calculated. \damla{As we show in Figure~\ref{fig:seedregion}, to find the leftmost position of the seed region ($x$), we need the start position of the minimizer within the query read ($a$), the start position of the seed ($c$), and the error rate ($E$). Similarly, to find the rightmost position of the seed region ($y$), we need the end position of the minimizer within the query read ($b$), the end position of the seed ($d$), the query read length ($m$), and the error rate ($E$).}
Finally, for each seed, the subgraph surrounded by these positions is fetched from the memory 
and provided as the output of the \ms algorithm.

\begin{figure}[h!]
\centering
\includegraphics[width=0.85\columnwidth,keepaspectratio]{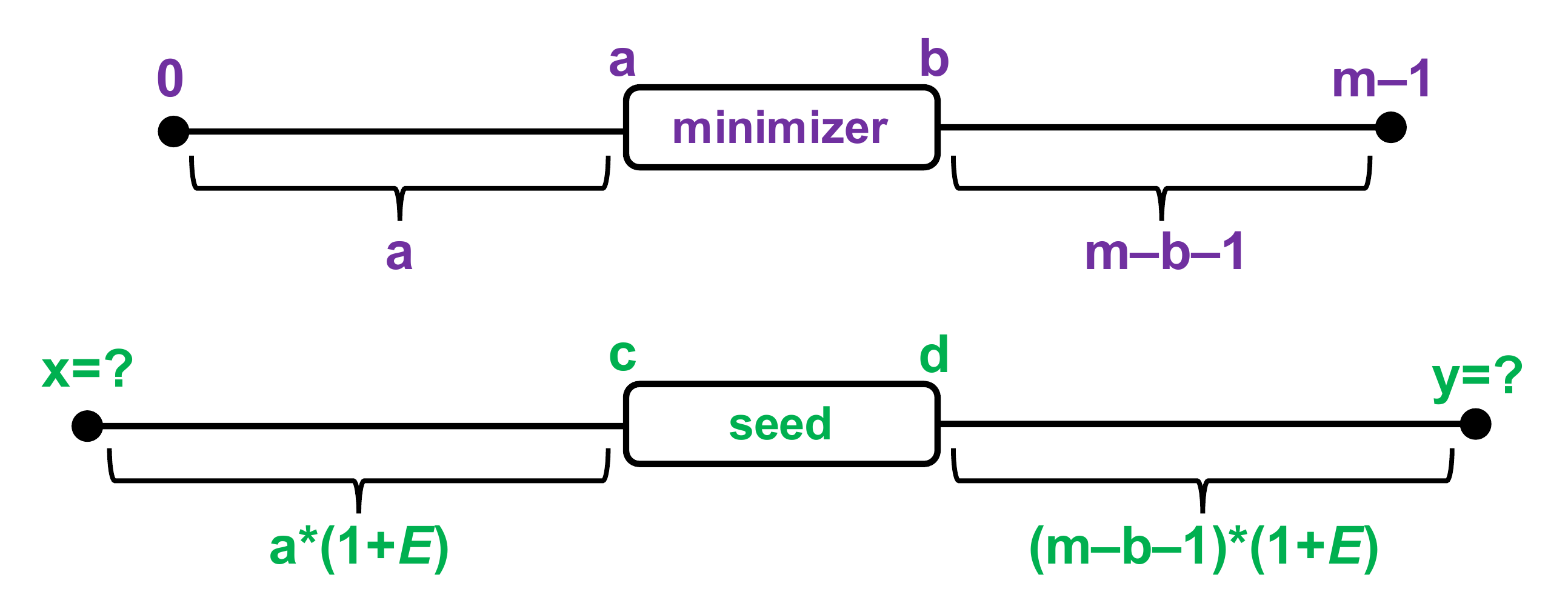}
\caption{Calculations for finding the candidate seed region.} \label{fig:seedregion}
\end{figure}

}

\section{BitAlign Algorithm} \label{sec:bitgraph_algo}

\damla{

After finding the subgraphs, our new sequence-to-graph alignment algorithm, \ba, performs distance calculation and traceback operations between the query read and the subgraph in order to find the \emph{optimal alignment}. In order to provide an efficient, hardware-friendly, and low-cost solution, we \damla{generalize the bitvector-based sequence alignment algorithm, GenASM (Sections~\ref{sec:bitap-search}~and~\ref{sec:bitap-traceback}), for sequence-to-graph alignment and exploit the bit-parallelism the algorithm provides.}

The major difference between sequence-to-sequence alignment and sequence-to-graph alignment is, in sequence-to-sequence alignment, we are only interested in the neighbor (i.e., previous/adjacent) text character whereas in sequence-to-graph alignment, due to possible incoming/outgoing edges from non-neighbor characters, we have to incorporate those edges as well. More formally, in the original GenASM algorithm, edges were implicit: the character at position \texttt{i < n-1} had a single \emph{successor} at position \texttt{i+1}, and iteration \texttt{i} requires the results of \texttt{i+1}. In a graph however, edges must be stored explicitly: every node \texttt{i} has a set of successors \texttt{\{$j_0$,$j_1$,...\}}, and iteration \texttt{i} now requires the results of all \texttt{$j_x$}. By topologically sorting the nodes, we ensure that \texttt{$\forall x. i < j_x$}, and thus the results of all successors are available in iteration \texttt{i}.

Thus, we modify the GenASM algorithm, such that while generating the bitvectors during the distance calculation step, we not only consider the previous text character's bitvectors (\emph{oldR[d] bitvectors}), but also all bitvectors that are corresponding to the \damlaI{outgoing} edges (or \emph{hops}). For example, as we show in Figure~\ref{fig:bitalign-bitvectors}, when generating the bitvectors for the dark blue-shaded node, we need both light blue-shaded nodes' bitvectors. On the other hand, when generating the bitvectors for the dark red-shaded node, we only need the light red-shaded node's bitvectors.

\begin{figure}[h!]
\centering
\vspace{5pt}
\includegraphics[width=\columnwidth,keepaspectratio]{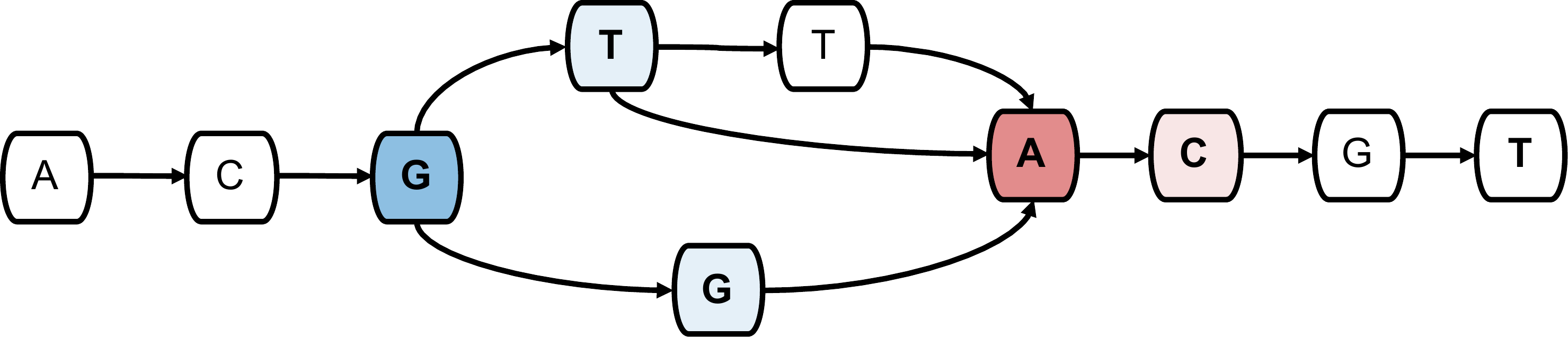}
\caption{Example of the dependency between different nodes of a graph when generating bitvectors.}
\label{fig:bitalign-bitvectors}
\end{figure}

In order to \damla{generalize the GenASM algorithm for} sequence-to-graph alignment, in \ba, we (1)~first linearize the input subgraph (assuming it is topologically sorted), (2)~store the (\emph{R[d] bitvectors}) for all the text iterations (not just for the previous one; \emph{oldR[d]}), and (3)~update how intermediate bitvectors (i.e., match, substitution, deletion, and insertion) are calculated in order to incorporate the hops as well. We show the new \ba algorithm in Algorithm~\ref{bitalign-dc-alg}. When calculating the deletion ($D$), substitution ($S$), and match ($M$) bitvectors, we take the hops into consideration, whereas when calculating the insertion ($I$) bitvector, we do \emph{not} need to.
}

{
\begin{algorithm}[h!]
\fontsize{10}{10}
\caption{BitAlign Algorithm}\label{bitalign-dc-alg}
\textbf{Inputs:} \texttt{graph-nodes} (reference), \texttt{pattern} (query), \texttt{k} (edit distance threshold)\\
\textbf{Outputs:} \texttt{editDist} (minimum edit distance), \texttt{CIGARstr} (traceback output)
    \begin{algorithmic}[1]
        \State $\texttt{n} \gets \texttt{length of \revIII{linearized reference subgraph}}$
        \State $\texttt{m} \gets \texttt{length of \revIII{query pattern}}$
        \State $\texttt{PM} \gets $\texttt{genPatternBitmasks(pattern)}
    \Comment{\comm{pre-process the pattern}}
        \State $\texttt{allR[n][d]} \gets \texttt{111.111}$
        \Comment{\comm{init R[d] bitvectors for all characters}}
        \For{\texttt{i in (n-1):\revV{-1:}0}}
    \Comment{\comm{iterate over each graph node}}
        \State $\texttt{curChar} \gets \texttt{graph-nodes[i].char}$
        \State $\texttt{curPM} \gets \texttt{PM[curChar]}$
        \Comment{\comm{retrieve the pattern bitmask}}
        \State $\texttt{R0} \gets \texttt{111...111}$
        \Comment{\comm{status bitvector for exact match}}
        \For{\texttt{j in graph-nodes[i].successors}}
            \State $\texttt{R0} \gets \texttt{((R[j][0]}\verb|<<|\texttt{1) | curPM) \& R0}$ 
        \EndFor
        \State $\texttt{allR[i][0]} \gets \texttt{R0}$
        \For{\texttt{d in 1:k }}
            \State $\texttt{I} \gets \texttt{(allR[i][d-1]}\verb|<<|\texttt{1)}$
            \Comment{\comm{insertion}}
            \State $\texttt{Rd} \gets \texttt{I}$
            \Comment{\comm{status bitvector for $d$ errors}}
            \For{\texttt{j in graph-nodes[i].successors}}
                \State $\texttt{D} \gets \texttt{allR[j][d-1]}$
                \Comment{\comm{deletion}}
                \State $\texttt{S} \gets \texttt{allR[j][d-1]}\verb|<<|\texttt{1}$
                \Comment{\comm{substitution}}
                \State $\texttt{M} \gets \texttt{(allR[j][d]}\verb|<<|\texttt{1)}\texttt{ | curPM}$
                \Comment{\comm{match}}
                \State $\texttt{Rd} \gets \texttt{D \& S \& M \& Rd}$
            \EndFor
            \State $\texttt{allR[i][d]} \gets \texttt{Rd}$
        \EndFor
    \EndFor
    \State $\texttt{<editDist, CIGAR>} \gets \texttt{traceback(allR, graph-nodes, pattern)}$
    \end{algorithmic}
\end{algorithm}
}

\damla{
For traceback in the style of GenASM, \damla{we need $3(k+1)$ bitvectors to be stored per edge in the graph}. Since the number of edges in the graph can only be bounded very loosely, the potential memory footprint \damla{increases} significantly, which \damla{is} expensive to implement in hardware. We solve this problem by storing only $k+1$ bitvectors per \emph{node}, from which the $3(k+1)$ bitvectors per \emph{edge} can be regenerated on-demand during traceback. While this incurs a minor computational overhead, \damla{this modification helps us to decrease the memory footprint of the algorithm by $3\times$} when the graph is a path, and additional edges incur no memory overhead. Since memory is the main area and power cost of the alignment hardware, this tradeoff is very favorable.
}
\section{\mech Hardware Design} \label{sec:bitgraph_hw}

\damla{

In \mech, we \emph{co-design} our new \ms algorithm for seeding and new \ba algorithm for sequence-to-graph alignment with specialized custom accelerators.

\subsection{\ms Hardware} \label{sec:minseed_hw}

A \ms accelerator consists of: (1)~three computation modules responsible for finding the minimizers from a query read, filtering the frequencies of minimizers if above a threshold, and finding the associated regions of every seed location by calculating the rightmost and leftmost positions; (2)~three scratchpads for storing the query read, its minimizers, and seed locations; and (3)~the memory interface, which handles the frequency, seed location, and subgraph accesses.

As we show in Figure~\ref{fig:minseed-hw}, \ms accelerator gets the query read as the input and finds the subgraphs to align this query as the output. The computation modules are implemented with a simple logic since we require basic logical operations (e.g., comparisons, simple \damla{arithmetic} operations, scratchpad R/W operations). 

\begin{figure}[h!]
\centering
\includegraphics[width=\columnwidth,keepaspectratio]{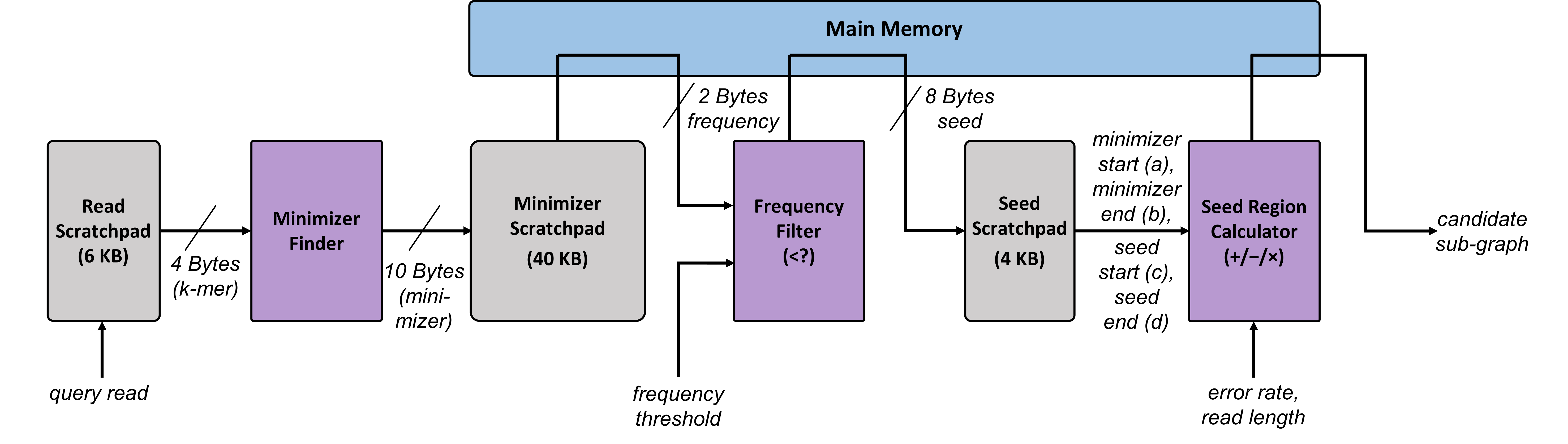}
\caption{Hardware design of \ms.}
\label{fig:minseed-hw}
\end{figure}

\subsection{\ba Hardware} \label{sec:bitalign_hw}

We implement the distance calculation (DC) hardware of \ba as a linear cyclic systolic array-based accelerator. Since we need to incorporate the hops as well, in our new design, we use \emph{hop queue registers} in order to feed the bitvectors of non-neighbor characters/nodes. 

As we show in Figure~\ref{fig:bitalign-hw}, the generated $R[d]$ bitvector from each processing element (PE) is fed to the tail of the hop queue register of the current PE. Each hop queue register then provides the stored bitvectors as the $oldR[d]$ bitvectors to the same PE (required for the match bitvector's calculation) and as the $oldR[d-1]$ bitvectors to the next PE (required for the deletion and substitution bitvectors' calculation) in the next cycle. 

\begin{figure}[t!]
\centering
\includegraphics[width=\columnwidth,keepaspectratio]{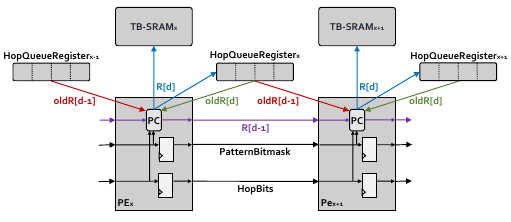}
\caption{Processing element (PE) design of \ba.}
\label{fig:bitalign-hw}
\end{figure}

We implement the successor relation as an adjacency matrix called $hopBits$ (Figure~\ref{fig:bitalign-hopbits}). Based on the $hopBits$ of the current text character, either the actual hop bitvector or all 1s bitvector is used when calculating the match, deletion, and substitution bitvectors of the current PE. $R[d-1]$ bitvector (required for the insertion bitvector's calculation) is directly provided by the previous PE (i.e., not through the hop queue registers). In order to decrease the size of each hop queue register, based on our empirical analysis, we limit the hop length to 12. Thus, each hop queue register contains 12 elements. 

\begin{figure}[b!]
\centering
\includegraphics[width=0.85\columnwidth,keepaspectratio]{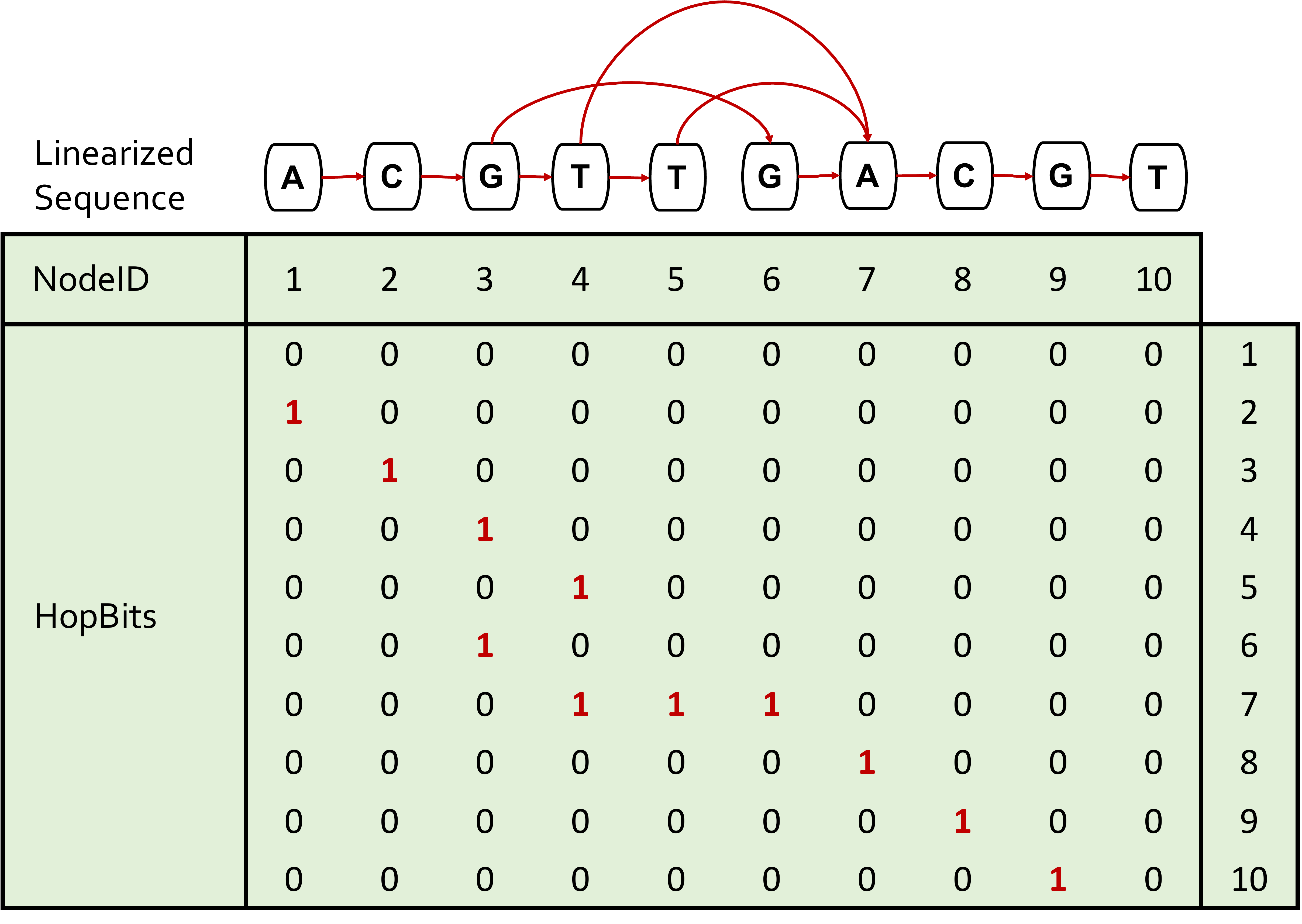}
\caption{Linearized input subgraph and the generated hopBits.} \label{fig:bitalign-hopbits}
\end{figure}

\damla{As we explain in Section~\ref{sec:bitgraph_algo}, in order to decrease the memory footprint of the stored bitvectors required for the traceback (TB) execution, we only store the ANDed version of the intermediate bitvectors ($R[d]$) and re-compute the intermediate bitvectors (i.e., match, substitution, deletion, and insertion) during the TB execution. Thus, each element of the queue register has a length equal to the window size ($W$), instead of $3*W$. Similarly, the size of each TB-SRAM for each PE decreases with this design choice. Another required change to the \ba hardware due to this design choice is for the TB accelerator design. After reading the $R[d]$ bitvectors from the TB-SRAMs and before performing the bitwise comparisons to find the CIGAR character for the current traceback iteration,  we need an additional step that re-generates the intermediate bitvectors using the $R[d]$ bitvectors, required for the TB execution.}

Besides the TB-SRAMs, \ba also requires DC-SRAM to store the linearized reference graph, associated hopBits for each node, and the pattern bitmasks for the query read. For a 128-PE configuration with 128 bits of processing per PE, \ba requires a total of 24KB DC-SRAM storage. Also, each PE requires a total of 2KB TB-SRAM storage, with a single R/W port (128KB, in total). In each cycle, 128 bits of data (16B) is written to each TB-SRAM and to each hop queue register by each PE. 

}
\subsection{Overall System Design} \label{sec:gengraph-overall_hw}

\damla{

Figure~\ref{fig:overall-system} shows the overall design of \mech. 
\mech is connected to a host system.
The host transfers a single query read to \mech{}, which is buffered before being processed. We employ double buffering technique to hide the transfer latency. Our acceleration platform consists of four HBM2E stacks~\cite{hbm}, each with 8 channels. Next to each HBM2E stack, we place one \mech{} module. A channel is exposed to \mech{} as a 256-bit wide port. The theoretical bandwidth available per 3D stack of HBM2E is 307 GB/s~\cite{hbm}. Thus, our complete accelerator design can leverage a peak theoretical bandwidth of 1.2 TB/s. By placing \mech{} in the same package as the four HBM2E stacks, we mimic the configuration of current commercial devices such as GPUs~\cite{v100,a100} and FPGA boards~\cite{vu37p,ad9h7}.
This in-package configuration allows \mech{} to have high-bandwidth memory access without limitations in area and thermal dissipation of other 3D-stacked memory technologies~\cite{hmc}, where accelerators can be placed in the logic layer of the 3D stack. We replicate the content of each HBM2E stack (i.e., graph-based reference and hash table\-based index) among the 4 independent stacks. Within each stack, we distribute the graph and index structures of the 24 chromosomes based on their sizes among the 8 independent channels. 

The main computation pipeline of \mech consists of \ms and \ba modules.  A single \mech consists of 8 \ms modules that exploit data-level parallelism when performing seeding. Each \ms module has exclusive access to one HBM2E channel. This ensures full bandwidth exploitation without contention, and allows us to optimally balance bandwidth and \ms's compute throughput. The \ms module is responsible for finding the minimizers of a given query read and their associated seed locations that are fed to our \ba module. Each \ms module is connected to a single \ba module. The \ba module is responsible for performing alignment between each of the seeds reported by the \ms module and the query read.

\begin{figure}[h!]
\vspace{10pt}
\centering
\includegraphics[width=\columnwidth,keepaspectratio]{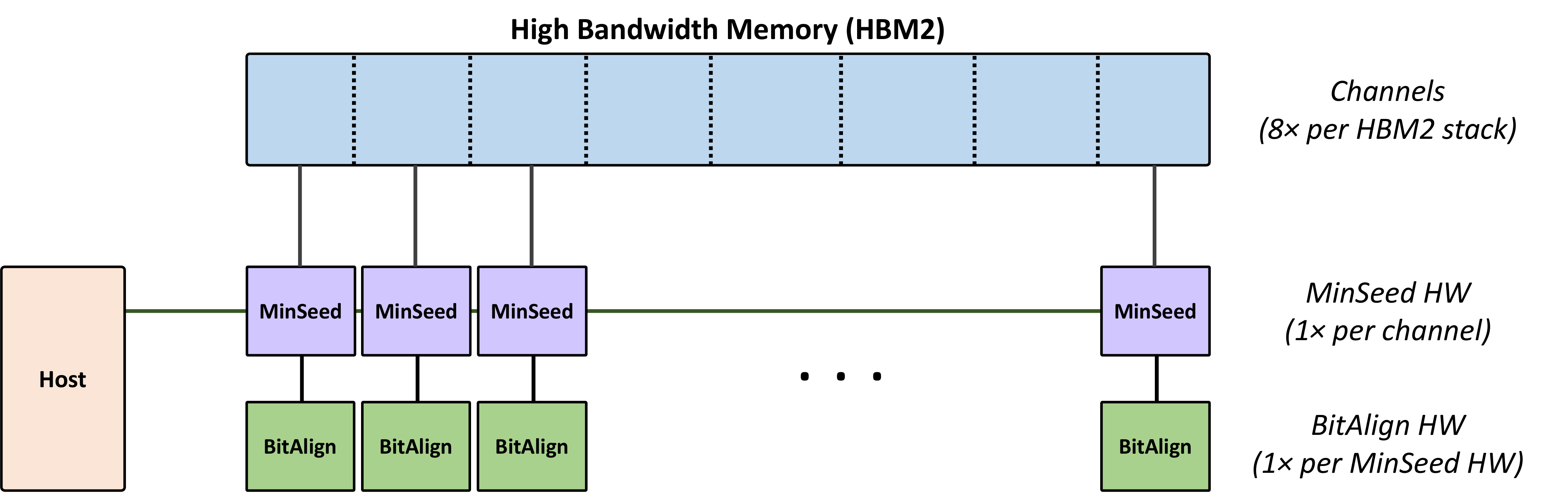}
\caption{Overall system design of \mech.} \label{fig:overall-system}
\vspace{10pt}
\end{figure}

We design the \mech mechanism in a pipelined fashion, such that we can hide the latency of our \ms accelerator when performing seeding while running sequence-to-graph alignment with our \ba accelerator. Thus, while \ba is running, \ms can fetch the next set of minimizers, frequencies, and seeds from the main memory, and write them to their associated scratchpads. In order to enable this, similar to the read scratchpad, we employ double buffering technique for the minimizer and seed scratchpads, not to overwrite the existing minimizers and seeds in those scratchpads that are being executed with the \ba accelerator, in parallel.

}
\section{{\mech} as a Framework} \label{sec:segram-framework}

\damla{

With the help of the flexibility and decomposability of the \mech framework, we can run each accelerator (i.e., \ms and \ba) separately or we can run them together for end-to-end execution. Thus, we describe three use cases of \mech: (1)~end-to-end sequence-to-graph mapping, (2)~sequence-to-graph alignment, and (3)~seeding.

\textbf{End-to-end sequence-to-graph mapping.}
For sequence-to-graph mapping, the whole \mech design should be executed, since both seeding and alignment steps are required, as we explain in Section~\ref{sec:background-genomemapping}. Thus, for this use case, both \ms and \ba should be executed. Also, with the help of the inherited divide-and-conquer approach from the GenASM algorithms, we can use \mech for performing sequence-to-graph mapping for both short and long reads.  

\textbf{Sequence-to-graph alignment.}
Since as an input, \ba requires the subgraph as the reference and the query read as the pattern, it can also act as a sequence-to-graph aligner, without the need of an initial seeding tool/accelerator. Also, since sequence-to-sequence alignment is a special and simpler variant of sequence-to-graph alignment, \ba can also be used for that use case, when the linear input text is represented as a graph, where the nodes only have a single edge.

\textbf{Seeding.}
Similarly, \ms only can be used as the seeding module for both graph-based mapping and linear traditional mapping. \ms is orthogonal to be coupled with any alignment tool or accelerator.

}
\section{Evaluation Methodology} \label{sec:gengraph-methodology}

\damla{

\textbf{Performance, Area and Power Analysis.}
We synthesize and place \& route the \ms and \ba accelerator datapaths using the Synopsys Design Compiler~\cite{synopsysdc} with a typical 28nm low-power process.
Our synthesis targets post-routing timing closure at 1GHz clock frequency.
\damla{We use CACTI~\cite{wilton1996cacti,shivakumar2001cacti} to estimate the area overhead and power consumption of the scratchpad in \ms and \ba.}
\damlaI{Our power analysis for the baseline tools includes the power consumption of the socket but excludes the DRAM power. In our power analysis for SeGraM, we also exclude the HBM power but include power consumption of all logic and scratchpad/SRAM units.}
We then use an in-house cycle-accurate simulator and a spreadsheet-based analytical model parameterized with the synthesis and memory estimations to drive the performance analysis.

\textbf{Baseline Tools.}
We compare \mech with two state-of-the-art sequence-to-graph mappers: vg~\cite{garrison2018variation} and GraphAligner~\cite{rautiainen2020graphaligner}, running on an Intel\textsuperscript{\textregistered} Xeon\textsuperscript{\textregistered} Gold 6126 CPU~\cite{intel_cpu} operating at 2.60GHz, with 64GB DDR4 memory. \damla{We run both tools with 12 threads. }
We measure the execution time and power consumption of 
the baseline tools. We measure the individual power consumed by each tool using Intel's PCM power utility~\cite{intelpcm}. \damla{We also compare \ba with a state-of-the-art sequence-to-graph aligner, PaSGAL~\cite{jain2019accelerating}, and also with three state-of-the-art sequence-to-sequence hardware aligners: Darwin~\cite{turakhia2018darwin}, GenAx~\cite{fujiki2018genax}, and GenASM~\cite{cali2020genasm}. For all these four baselines, we use the numbers reported by their respective papers.} 

\textbf{Datasets.}
\label{sec:gengraph-methodology:datasets}
We evaluate \mech using the latest major release of the human genome assembly, GRCh38 \cite{ncbi38genome}, as the starting reference genome. For the variations, we use 7 VCF files for HG001-007 from the GIAB project (v3.3.2)~\cite{giabvcf}.

As the read datasets, we generate four sets of long reads (i.e., PacBio and ONT datasets) using PBSIM2~\cite{ono2021pbsim2} and three sets of short reads (i.e., Illumina datasets) using Mason~\cite{holtgrewe2010mason}.
For the PacBio and ONT datasets, we have reads of length 10Kbp, each simulated with 5\% and 10\% error rates. The Illumina datasets have reads of length 100bp, 150bp, and 250bp, each simulated with a 1\% error rate.

}

\section{Results} \label{sec:gengraph-results}

\damla{

\subsection{Area and Power Analysis} \label{sec:gengraph-results:area-power}

Table \ref{table:gengraph-areapower} shows the area and power breakdown of the compute (i.e., logic) units \damla{and the memory components (i.e., scratchpads)} in \mech, and the total area overhead and power consumption of (1)~a single \mech accelerator (attached to a single channel), (2)~8 \mech accelerators (in a single stack with 8 channels), and (3)~32 \mech accelerators (in 4 stacks). Our accelerators operate at 1GHz.

\begin{table}[b!]
\centering
\caption{Area and power breakdown of \mech.}
\label{table:gengraph-areapower}
\includegraphics[width=\columnwidth,keepaspectratio]{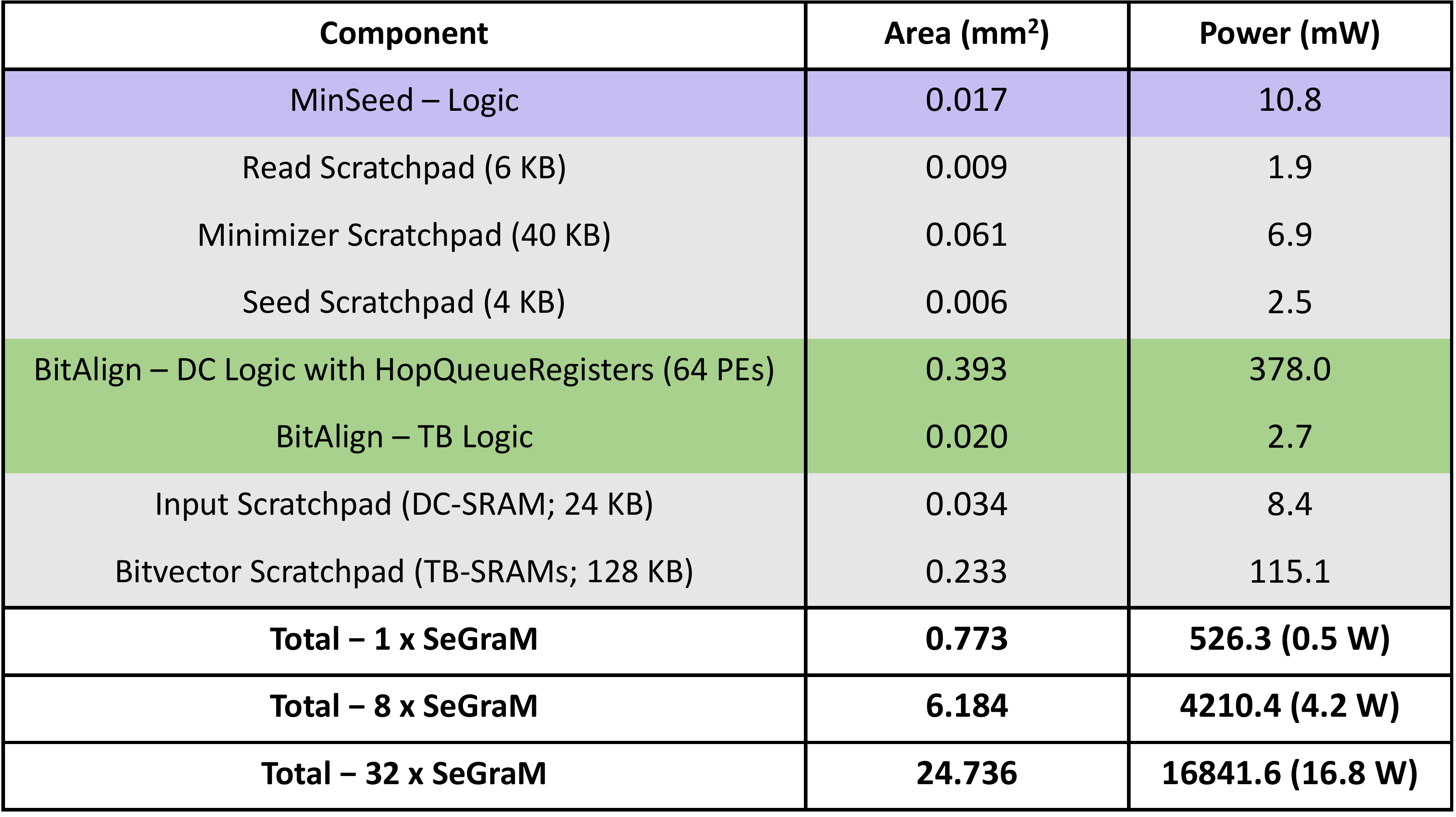}
\end{table}

The area overhead of one \mech accelerator is \SI{0.773}{\milli\meter\squared}, and the power consumption of one \mech accelerator
is \SI{526}{\milli\watt}. We find that the main contributors for the area overhead and power consumption are (1)~hopQueueRegisters since they constitute more than 60\% of the area and power of BitAlign-DC logic, and (2)~the bitvector scratchpads (TB-SRAMs). 
As we have one \mech accelerator per channel, the total area overhead of \mech attached to all 32 channels is \SI{24.7}{\milli\meter\squared}. 
Similarly, the total power consumption of 32 \mech accelerators is \SI{16.8}{\watt}.

\subsection{Analysis of \mech}\label{sec:results-segram}

\damla{
We compare end-to-end execution of \mech with two state-of-the-art sequence-to-graph mapping tools, GraphAligner and vg. We compare both of the tools with \mech for both long and short reads. We measure the execution time and power consumption of the baseline tools for their seeding, filtering/chaining, and alignment steps only (i.e., we do not include the pre-processing steps, which are executed only once).
}

\textbf{Long Read Analysis.} \label{sec:gengraph-results:longreads}
Figure~\ref{fig:gengraph-throughput-result-long-graphaligner} shows the read mapping throughput (reads/sec) of \mech and GraphAligner, when aligning long noisy PacBio and ONT reads against the graph-based representation of the human reference genome. 
We show that, on average, \mech provides $8.8\times$ throughput improvement over GraphAligner's 12-thread execution.

\begin{figure}[h!]
\centering
\includegraphics[width=0.9\columnwidth,keepaspectratio]{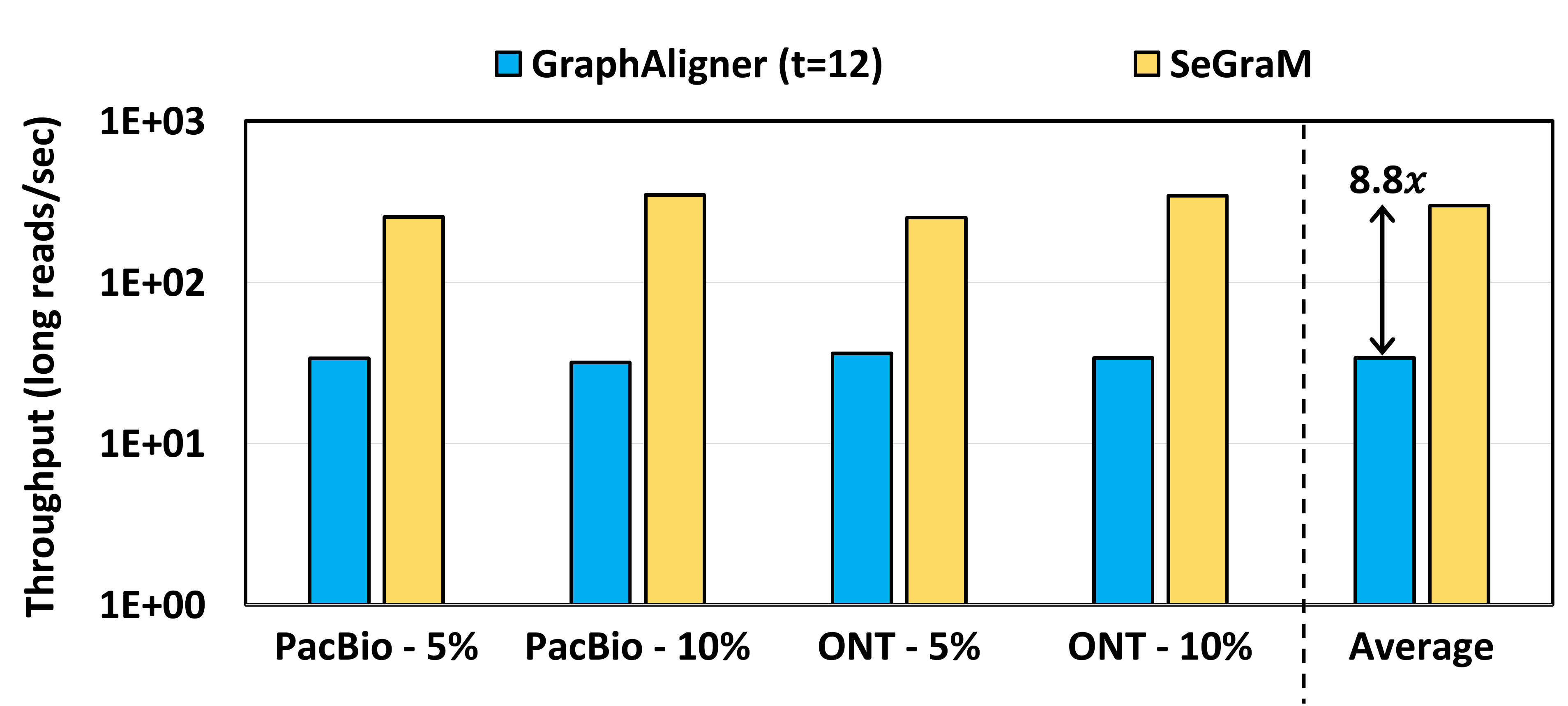}
\caption{\revonur{Throughput comparison of \mech and GraphAligner for long reads.}} \label{fig:gengraph-throughput-result-long-graphaligner}
\end{figure}

Figure~\ref{fig:gengraph-throughput-result-long-vg} shows the read mapping throughput (reads/sec) of \mech and vg, when aligning long noisy PacBio and ONT reads against the graph-based representation of the human reference genome. 
We show that, on average, \mech provides $7.3\times$ throughput improvement over vg's 12-thread execution.

\begin{figure}[h!]
\centering
\includegraphics[width=0.9\columnwidth,keepaspectratio]{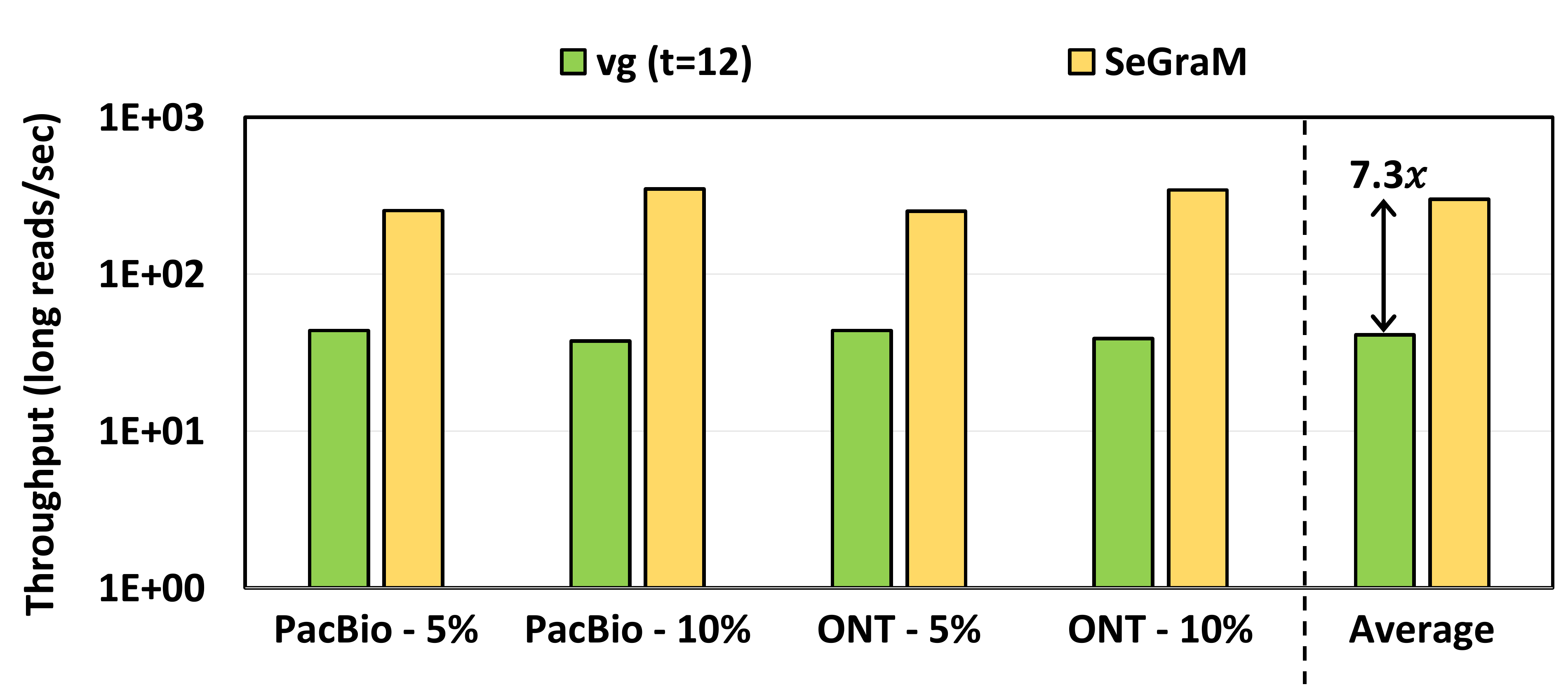}
\caption{\revonur{Throughput comparison of \mech and vg for long reads.}} \label{fig:gengraph-throughput-result-long-vg}
\end{figure}

Based on our power analysis with long reads, we find that power consumption of GraphAligner is \SI{83}{\watt} and power consumption of vg is \SI{109}{\watt} for their 12-thread execution. Thus, \mech reduces the power consumption of GraphAligner and vg by $4.9\times$ and $6.5\times$ over their 12-thread execution. 

\textbf{Short Read Analysis.} \label{sec:gengraph-results:shortreads}
Figure~\ref{fig:gengraph-throughput-result-short-graphaligner} shows the read mapping throughput (reads/sec) of \mech and GraphAligner, when aligning short Illumina reads against the graph-based representation of the human reference genome. 
We show that, on average, \mech provides $168\times$ throughput improvement over GraphAligner's 12-thread execution.

\begin{figure}[h!]
\centering
\includegraphics[width=0.9\columnwidth,keepaspectratio]{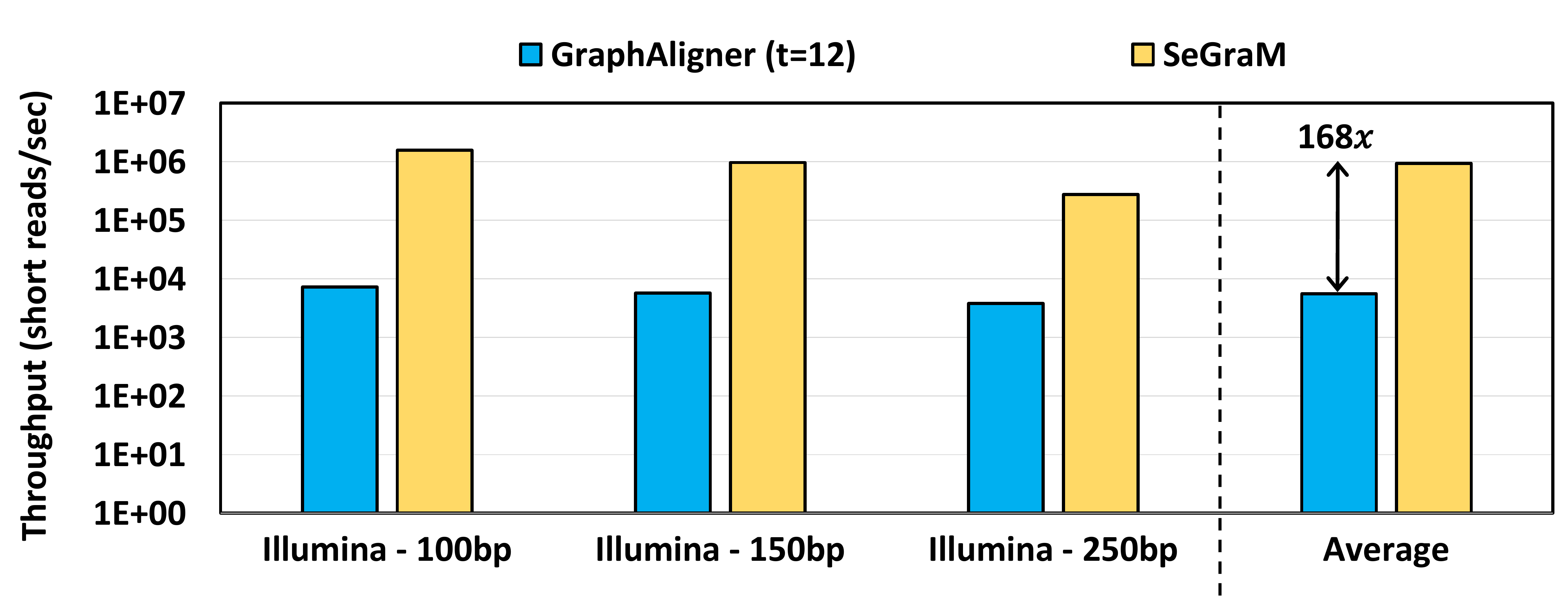}
\caption{\revonur{Throughput comparison of \mech and GraphAligner for short reads.}} \label{fig:gengraph-throughput-result-short-graphaligner}
\end{figure}

Figure~\ref{fig:gengraph-throughput-result-short-vg} shows the read mapping throughput (reads/sec) of \mech and vg, when aligning short Illumina reads against the graph-based representation of the human reference genome. 
We show that, on average, \mech provides $726\times$ throughput improvement over vg's 12-thread execution.

\begin{figure}[h!]
\centering
\includegraphics[width=0.9\columnwidth,keepaspectratio]{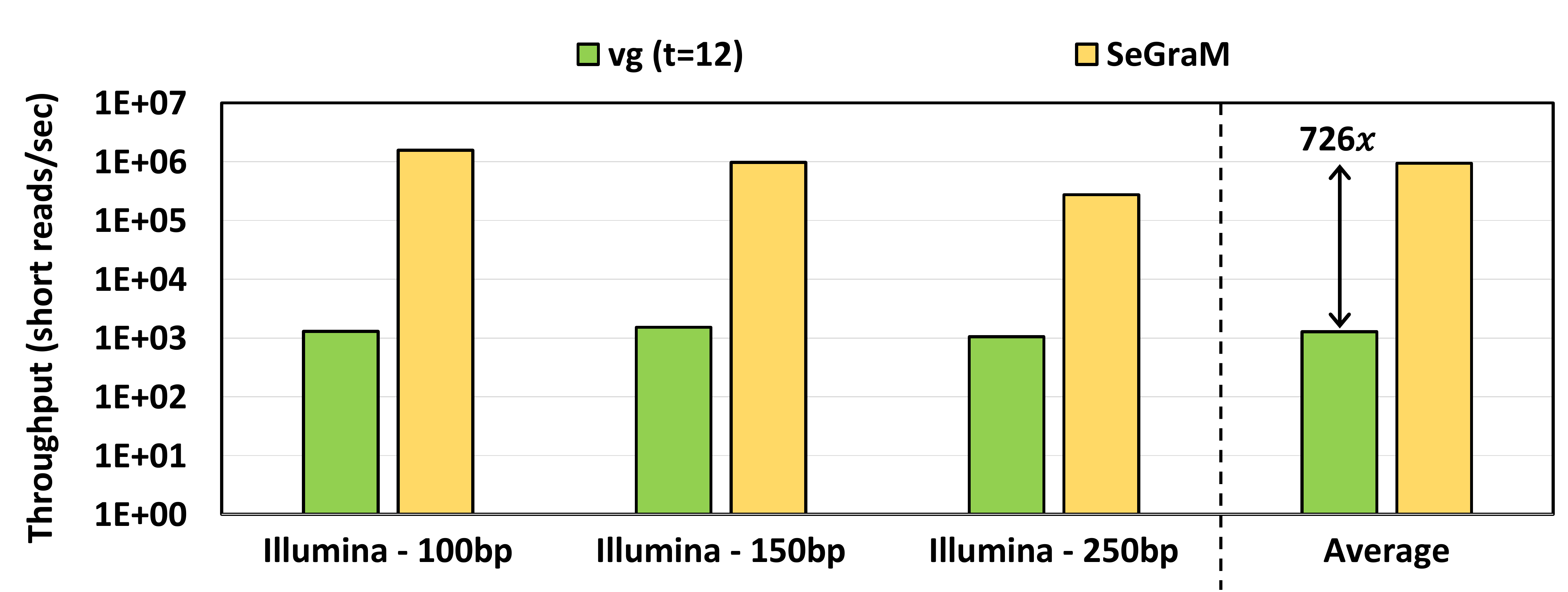}
\caption{\revonur{Throughput comparison of \mech and vg for short reads.}} \label{fig:gengraph-throughput-result-short-vg}
\end{figure}

Based on our power analysis with short reads, we find that power consumption of GraphAligner is \SI{79}{\watt} and power consumption of vg is \SI{83}{\watt} for their 12-thread execution. Thus, \mech reduces the power consumption of GraphAligner and vg by $4.7\times$ and $4.9\times$ over their 12-thread execution.

\damla{
\textbf{Sources of Improvement.} \label{sec:gengraph-results:discussion}
The sources of large performance improvements in \mech are (1)~the efficient and hardware-friendly underlying algorithms for both seeding and sequence-to-graph alignment, (2)~carefully designed dedicated scratchpads based on the empirical data we collect for different data structures (e.g., graphs, minimizers, seeds), (3)~hop queue registers, since they allow us to fetch all the bitvectors for the hops within a single cycle, and (4)~our pipelined overall design, where we can hide the execution latency of \ms, with the \ba execution. Thus, even though we have to increase the sizes of the scratchpads to allow double buffering and add hop queue registers for not increasing the execution time required for sequence-to-graph alignment, these additional area and power overheads help us to provide large increase in throughput for both short and long reads.
}


\subsection{Analysis of \ba}

\textbf{Sequence-to-Graph Alignment.}
As we explain in Section~\ref{sec:segram-framework}, \ba-only can be used for sequence-to-graph alignment, without the need of a preceding seeding tool/accelerator. We compare \ba with the state-of-the-art SIMD-based sequence-to-graph alignment tool, PaSGAL~\cite{jain2019accelerating}. PaSGAL is composed of three main steps: (1)~DP-fwd, where the input graph and query read are aligned using the dynamic programming based graph alignment approach to compute the ending position of the alignment, without running the traceback operation; (2)~DP-rev, where graph and query are aligned in the reverse direction to compute the starting position of the alignment, again without running the traceback operation; and (3)~Traceback, where using the starting and ending positions of the alignment, the corresponding section of the score matrix is re-calculated and traceback is performed to find the optimal alignment. 

Since the input of \ba is the subgraph and the query read, not the complete input graph, we \emph{only} compare \ba with the third step of PaSGAL for a fair comparison. As we show in Figure~\ref{fig:gengraph-throughput-result-pasgal}, based on the results reported in \cite{jain2019accelerating} for the LRC-L1, LRC-L2, MHC1-M1, and MHC1-M2 datasets, \mech provides $41\times$--$539\times$ speedup over the 48-thread AVX512-supported execution of PaSGAL.

\begin{figure}[h!]
\centering
\includegraphics[width=\columnwidth,keepaspectratio]{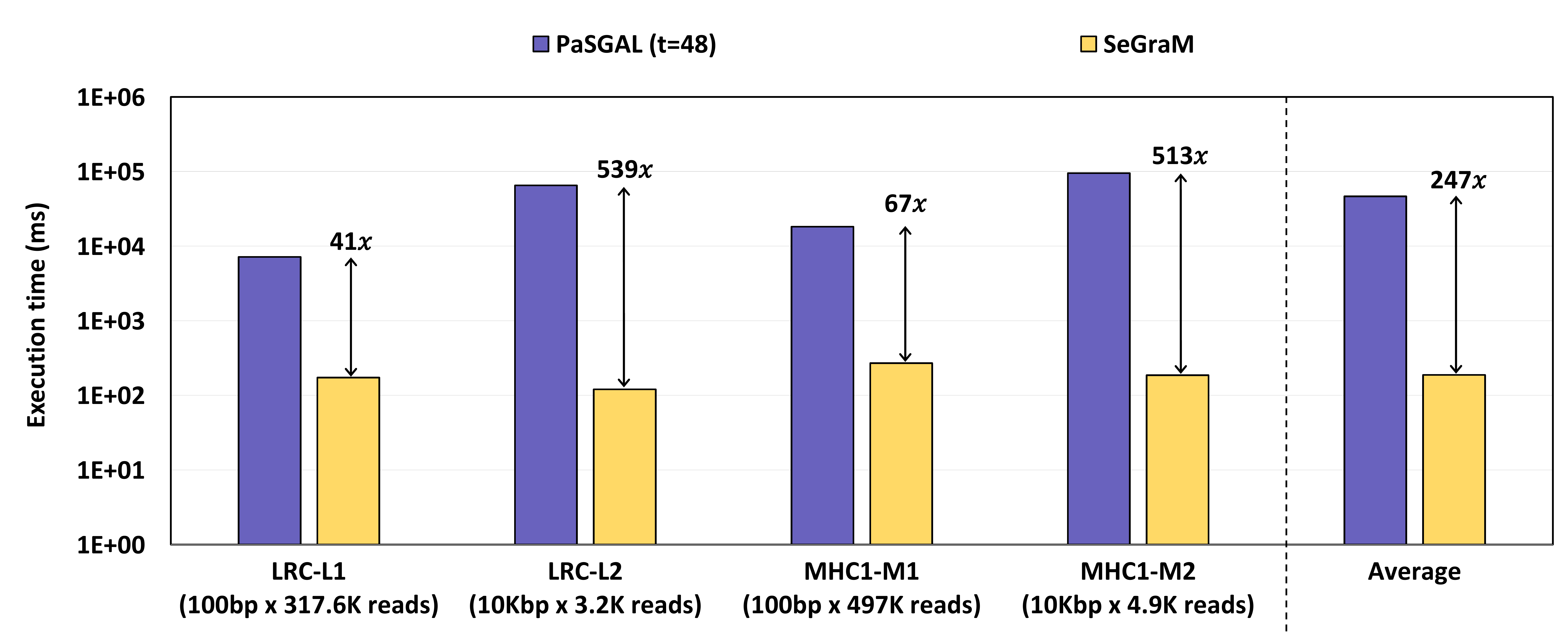}
\caption{\revonur{Performance comparison of \mech and PaSGAL for sequence-to-graph alignment.}} \label{fig:gengraph-throughput-result-pasgal}
\end{figure}

Compared to PaSGAL, \ba shows significant speedup, especially for the long read datasets (i.e., LRC-L2 and MHC1-M2). The reason of this is the inherited divide-and-conquer approach that \ba follows. Instead of aligning the full subgraph and the query read, with the help of the windowing approach, \ba manages to decrease the complexity of sequence-to-graph alignment, and efficiently aligns both short and long reads.

\textbf{Sequence-to-Sequence Alignment.}
Even though sequence-to-sequence alignment tools or accelerators cannot be used for sequence-to-graph alignment since they do \emph{not} consider hops and only consider neighbor text characters, sequence-to-graph alignment tools or accelerators can be used for the traditional sequence-to-sequence alignment problem. Thus, \ba can be used for both sequence-to-sequence alignment and sequence-to-graph alignment. The cost of more functionality in \ba is the extra hop queue registers. However, with the help of these additional memory components, we do \emph{not} sacrifice any performance. 

To show the efficiency of \ba for this special use case of sequence-to-graph alignment (i.e., sequence-to-sequence alignment), we compare \ba with the state-of-the-art hardware accelerators for sequence-to-sequence alignment: GACT of Darwin~\cite{turakhia2018darwin}, SillaX of GenAx~\cite{fujiki2018genax}, and GenASM~\cite{cali2020genasm}. GACT is optimized for long reads, SillaX is optimized for short reads, and GenASM is optimized for both short and long reads. We use the optimum configuration of each accelerator reported in their corresponding papers. 

Based on our analysis with long reads, we find that, on average, \mech provides $4.8\times$ and $1.2\times$ throughput improvement over GACT of Darwin and GenASM, respectively, while having $1.9\times$ and $5.2\times$ higher power consumption, and $1.4\times$ and $2.3\times$ higher area overhead. For short reads, we find that, on average, \mech provides $2.4\times$ and $1.3\times$ throughput improvement over SillaX of GenAx and GenASM, respectively. 

\subsection{Analysis of \ms}

As we explain in Section~\ref{sec:gengraph-overall_hw}, with the help of our pipelined design, \ms execution is not on the critical path of the overall \mech execution. However, since \ms selects the candidate seed locations and sends them to \ba for the final alignment, it plays a critical role on the overall sensitivity of our approach. Since \ms does \emph{not} perform any filtering approach to reduce the number of candidate seed regions that are sent for alignment (except discarding the seeds that have higher frequency than the threshold, which is an optimization that baseline tools already implement), \mech does not decrease the sensitivity of the overall sequence-to-graph mapping execution.

Even though not having a filtering mechanism increases the number of candidate seed regions that are sent for the expensive alignment step, with our highly-efficient \ba accelerator, we alleviate this bottleneck that exists in other existing tools and provide significant improvement over the baseline tools, as we show in Section~\ref{sec:results-segram}. For example, for a long read dataset, GraphAligner decreases the number of seeds extended from 77M to 48K with its filtering/chaining approaches. On the other hand, \ms sends 35M seeds to our \ba accelerator, but still provides higher throughput than GraphAligner. Similarly, for a short read dataset, GraphAligner decreases the number of seeds extended from 828K to 11K, while \ms sends 375K seeds to our \ba accelerator, and still provides much higher throughput. Thus, we show that \mech provides a highly sensitive and also high-performance solution for sequence-to-graph mapping.

}
\section{Related Work} \label{sec:related_work}

\damla{

To our knowledge, we are the first to propose
(1)~a hardware acceleration framework for sequence-to-graph mapping (\mech),  
(2)~a hardware accelerator for minimizer-based seeding (\ms), and 
(3)~a hardware accelerator for sequence-to-graph alignment (\ba).
No prior work has studied hardware for genome graph processing.

\textbf{Software Tools for Sequence-to-Graph Mapping.}
Even though genome graphs gain attention recently, there are only a few tools available specialized for sequence-to-graph mapping or alignment. Examples of sequence-to-graph mapping tools are GraphAligner~\cite{rautiainen2020graphaligner}, vg~\cite{garrison2018variation}, and  HISAT2~\cite{kim2019graph}. There are also some works which focus on alignment only, without an indexing and seeding step, such as PaSGAL~\cite{jain2019accelerating} and abPOA~\cite{gao2020abpoa}. However, these are all software-based tools.

\textbf{Hardware Accelerators for Genome Sequence Analysis.}
Existing hardware accelerators for genome sequence analysis focus on accelerating only the traditional read mapping (i.e., sequence-to-sequence) pipeline, and cannot support genome graphs as their inputs. For example, ERT~\cite{subramaniyan2020accelerating}, NEST~\cite{huangfu2020nest}, SaVI~\cite{laguna2020seed}, and MEDAL~\cite{huangfu2019medal} accelerate the seeding step of sequence-to-sequence mapping. 
Darwin~\cite{turakhia2018darwin}, GenAx~\cite{fujiki2018genax}, GenASM~\cite{cali2020genasm}, and SeedEx~\cite{fujiki2020seedex} accelerate read alignment with only a single reference genome.
These accelerators have no way to track the multiple paths that need to be traversed in a graph, and cannot be easily modified to support multiple path tracking.
\mech builds upon hardware components of GenASM, and incorporates new algorithms and hardware, to efficiently support genome graph mapping and alignment, and can also support sequence-to-sequence mapping (as a graph where each node has only one outgoing edge).

There are also processing-in-memory (PIM) based accelerators for genome sequence analysis, such as GRIM-Filter~\cite{kim2018grim}, RAPID~\cite{gupta2019rapid}, PIM-Aligner~\cite{angizi2020pimaligner}, RADAR~\cite{huangfu2018radar}, FindeR~\cite{zokaee2019finder}, and AligneR~\cite{zokaee2018aligner}. However, similar to read alignment accelerators, they are tuned for a single linear reference only, and cannot support genome graphs. \damlaII{Besides these accelerators mostly focusing on the read mapping steps, there are also other PIM-based accelerators that focus on other steps of the pipeline~\cite{cali2017nanopore}, such as PIM-Assembler~\cite{angizi2020pim} for genome assembly and Helix~\cite{lou2020helix} for nanopore basecalling.}

}
\clearpage
\section{Summary} \label{sec:gengraph-conclusion}

\sgh{

Genome graphs are emerging representations for the DNA of a population, and overcome the biases present in traditional genome sequence analysis, where a single reference genome is used.
Unfortunately, the additional overheads of sequence-to-graph mapping exacerbate the read mapping bottleneck in the genome sequence analysis pipeline.
To alleviate this, we propose \mech, the first acceleration framework for sequence-to-graph mapping and alignment.
For \mech, we co-design algorithms and accelerators for memory-efficient minimizer-based seeding and bitvector-based, highly-parallel sequence-to-graph alignment. 

For sequence-to-graph mapping with long reads, we find that \mech achieves $8.8\times$ and $7.3\times$ speedup over 12-thread execution of state-of-the-art sequence-to-graph mapping tools (GraphAligner and vg, respectively), while reducing power consumption by $4.9\times$ and $6.5\times$. For sequence-to-graph mapping with short reads, we find that \mech achieves $168\times$ and $726\times$ speedup over 12-thread execution of GraphAligner and vg, respectively, while reducing power consumption by $4.7\times$ and $4.9\times$. 
\sgh{For sequence-to-graph alignment,} we
show that \ba provides $41\times$--$539\times$ speedup over PaSGAL, \sgh{a state-of-the-art sequence-to-graph alignment tool}. We conclude that \mech is a high-performance and efficient hardware acceleration framework, which can accelerate multiple steps of the sequence-to-graph mapping pipeline, and of the traditional read mapping pipeline.

}

\chapter{Importance of Accelerating Genome Sequence Analysis} 

We believe the long-term impact of GenASM, BitMAc, SeGraM, and other works that propose hardware acceleration for genome sequence analysis is three-fold:

\damla{
\textbf{Enabling Portable, Fast, and Efficient Genome Sequence Analysis.} 
Recent advances have enabled genome sequencing anywhere in the world with cheap, portable sequencing machines (e.g., ONT’s MinION). Soon, even smaller sequencing devices can enable sequencing using smartphones. Such readily available sequencing technologies can open up several new applications, such as bringing personalized medicine to rural or remote areas, near-patient testing, and rapid infection diagnosis and outbreak tracing \damlaII{(e.g., COVID-19~\cite{wu2020new,harcourt2020isolation,james2020lampore,da2020evolution}, Ebola~\cite{quick2016real,greninger2015rapid}, Zika~\cite{faria2016mobile})}. However, these applications require memory-efficient, low-power, and area-efficient systems to process the generated genome sequence data, as laptops and mobile phones have limited resources (e.g., greater memory constraints, limited battery life). Our approach of co-designing scalable and memory-efficient algorithms with area- and power-efficient hardware accelerators is an important milestone, allowing genome sequence analysis to be performed in highly-resource-constrained environments. Our genomics accelerators can even be implemented in the sequencing machine itself, eliminating expensive sequencer-to-computer data movement and providing a single embedded solution for portable sequencing and sequence analysis. 

\textbf{Rapid Genome Sequence Analysis for Pandemics.} 
Rapid genome sequence analysis plays a critical role during pandemics such as the current COVID-19 (i.e., SARS-CoV-2) crisis \damlaII{in 2020-2021~\cite{fauci2020covid,huang2020clinical,guan2020clinical}}. Rapid analysis can (1)~\damlaII{enable the quick detection of the virus in human DNA samples; (2)~enable the rapid identification of the mutations, sources, and transmission modes of the virus;} (3)~help with the development of new treatments; and (4)~help uncover why some people experience more severe symptoms and higher mortality than others. Given the fast pace at which viruses can proliferate and mutate during a pandemic, there is a need to perform large volumes of viral genomic analysis rapidly and widely, as lost time or limited availability can hinder tracking and harm our ability to control spread and mutations. Today, rapid genome sequence analysis is bottlenecked by the limited computational power and memory bandwidth of existing systems. We believe it is more important than ever to overcome these bottlenecks through the development of high-efficiency, low-cost solutions. Beyond the benefits that our genomics accelerators already yield, we hope that our co-design approach sparks further research from both academia and industry on developing even more powerful and efficient solutions for rapid genome sequence analysis of viruses.

\textbf{Reducing Genomic Accelerator Costs with Multi-Purpose Frameworks.} 
While there is a pressing need for genomic sequence analysis hardware, any fixed-function hardware incurs high per-unit costs, as the non-recurring engineering (NRE) costs can be amortized over only the number of platforms that perform the specific function. To significantly lower NRE costs, we design GenASM and BitMAc to provide substantial benefits for generic \damlaII{approximate string matching} (a widely-used primitive for any text search or error-aware pattern matching), while still optimizing \damlaII{their designs} to maximize benefits for genomic use cases. We believe that such an approach, with flexible frameworks that can serve as general-purpose accelerators but include domain-specific optimizations, opens a promising pathway for a low-cost acceleration of other tasks. For example, other bioinformatics workloads (e.g., a graph processing acceleration framework for genome assembly, a neural network acceleration framework for nanopore basecalling \damlaII{or variant calling~\cite{poplin2018universal}}) can take a similar approach, making what would otherwise be high-cost hardware much cheaper, and addressing key cost concerns in the healthcare industry. \damlaII{We hope that our approach of reusing genome sequence analysis frameworks for general-purpose acceleration will inspire future designers to consider NRE and incorporate general-purpose support into their accelerators.}
}
\chapter{Conclusions and Future Directions} 

\section{Conclusions} 

In this dissertation, we characterize the real-system behavior of the genome sequence analysis pipeline and its associated tools, expose the bottlenecks and tradeoffs of the pipeline and tools, and co-design fast and efficient algorithms along with scalable and energy-efficient customized hardware accelerators for the key pipeline bottlenecks to enable faster genome sequence analysis. Our goals are to (1)~understand where the current tools and algorithms do not perform well in order to develop better tools and algorithms, and (2)~understand the limitations of existing hardware systems when running these tools and algorithms in order to design efficient customized accelerators. Towards this end, we propose four major works.

First, we present the first \damlaIII{experimental analysis of} state-of-the-art tools associated with each step of the genome assembly pipeline using long reads. We analyze the tools in multiple dimensions that are important for both developers and users/practitioners: accuracy, performance, memory usage and scalability. We reveal new bottlenecks and tradeoffs that different combinations of tools and different underlying systems lead to, based on our extensive experimental analyses. We also provide guidelines for both practitioners, such that they can determine the appropriate tools and tool combinations that can satisfy their goals, and tool developers, such that they can make design choices to improve current and future tools. 

Second, we propose GenASM, the first approximate string matching (ASM) acceleration framework for genome sequence analysis. GenASM performs bitvector-based ASM, which can efficiently accelerate multiple steps of genome sequence analysis. 
We modify the underlying ASM algorithm (Bitap~\damlaIII{\cite{baeza1992new, wu1992fast}}) to significantly increase its parallelism and reduce its memory footprint. 
Using this modified algorithm, we design the first hardware accelerator for Bitap. Our hardware accelerator consists of specialized systolic-array-based compute units and on-chip SRAMs that are designed to match the rate of computation with memory capacity and bandwidth, resulting in an efficient design whose performance scales linearly as we increase the number of compute units working in parallel. We demonstrate that GenASM provides significant performance and power benefits for three different use cases in genome sequence analysis. First, GenASM accelerates read alignment for both long reads and short reads. For long reads, GenASM outperforms state-of-the-art software and hardware accelerators by 116$\times$ and \revII{$3.9\times$}, respectively, while reducing power consumption by $37\times$ and 2.7$\times$. For short reads, GenASM outperforms state-of-the-art software and hardware accelerators by $111\times$ and $1.9\times$. Second, GenASM accelerates pre-alignment filtering for short reads, with $3.7\times$ the performance of a state-of-the-art pre-alignment filter, while \revonur{reducing power consumption by \revonur{$1.7\times$}} and significantly improving the filtering accuracy. Third, GenASM accelerates edit distance calculation, with \revIII{22--12501$\times$} \revonur{and 9.3--400$\times$ speedups over the state-of-the-art \revonur{software} library and FPGA-based accelerator, respectively, while \revV{reducing power consumption} by \revonur{548--582$\times$} and \revIII{$67\times$}.} We also briefly discuss \revonur{four other} use cases that can benefit from GenASM.

Third, we propose BitMAc, which is an FPGA-based prototype for GenASM. In BitMAc, we map our GenASM algorithms on Stratix 10 MX FPGA with a state-of-the-art 3D-stacked memory (HBM2), where HBM2 offers high memory bandwidth and FPGA resources offer high parallelism by instantiating multiple copies of the GenASM accelerators. After \damlaIII{modifying} the GenASM algorithms for better mapping to existing FPGA resources, we show that BitMAc provides 64\% logic utilization and 90\% on-chip memory utilization, while having \SI{48.9}{\watt} of total power consumption. We compare BitMAc with state-of-the-art CPU-based and GPU-based read alignment tools. Compared to the alignment steps of the CPU-based read mappers, (1)~for long reads, BitMAc provides $761\times$ and 136$\times$ speedup, while reducing power consumption by $1.9\times$ and $2.0\times$, and (2)~for short reads, BitMAc provides 92$\times$ and 130$\times$ speedup, while reducing power consumption by $2.2\times$ and $2.0\times$. We also show that BitMAc provides significant speedup compared to the GPU-based baseline, while reducing   power consumption.

Fourth, we propose \mech, the first hardware acceleration framework for sequence-to-graph mapping \damla{and alignment}. Reference genomes are conventionally represented as a linear sequence. However, this linear representation of the reference genome \damlaIII{results in} ignoring the variations that exist in a population (i.e., genetic diversity) and introducing biases for the downstream analysis. To address these limitations, recently, graph-based representations of the genomes (i.e., \emph{genome graphs}) have gained attention. As shown in many prior \damlaII{works~\cite{alser2017gatekeeper,turakhia2018darwin,cali2020genasm,fujiki2018genax,kim2020geniehd,alser2019sneakysnake,goyal2017ultra,fujiki2020seedex,bingol2021gatekeeper,nag2019gencache,kim2018grim,lavenier2016dna,kaplan2018rassa,kaplan2020bioseal}}, sequence-to-sequence mapping is one of the major bottlenecks of the genome sequence analysis pipeline and \damlaIII{needs} to be accelerated using specialized hardware. Since graph-representation of the genome is much more complex than the linear representation, sequence-to-graph mapping is placing a greater pressure on this bottleneck. Thus, in this work, our goal is to design a high-performance, scalable, power- and area-efficient hardware accelerator for sequence-to-graph mapping that support both short and long reads. We base \mech on a memory-efficient minimizer-based seeding algorithm and a bitvector-based, highly-parallel sequence-to-graph alignment algorithm. We \emph{co-design} both of our algorithms with high-performance, area- and power-efficient hardware accelerators. \damlaII{\mech consists of two components: (1)~\ms, which provides hardware support to execute our minimizer-based seeding algorithm, and (2)~\ba, which provides hardware support to execute our bitvector-based sequence-to-graph alignment algorithm.} \damla{For sequence-to-graph mapping with long reads, we find that \mech achieves $8.8\times$ and $7.3\times$ speedup over 12-thread execution of state-of-the-art sequence-to-graph mapping tools (GraphAligner~\damlaIII{\cite{rautiainen2020graphaligner}} and vg~\damlaIII{\cite{garrison2018variation}}, respectively), while reducing power consumption by $4.9\times$ and $6.5\times$. For sequence-to-graph mapping with short reads, we find that \mech achieves $168\times$ and $726\times$ speedup over 12-thread execution of GraphAligner and vg, respectively, while reducing power consumption by $4.7\times$ and $4.9\times$. For sequence-to-graph alignment, we show that \ba provides $41\times$--$539\times$ speedup over \damlaII{PaSGAL~\damlaIII{\cite{jain2019accelerating}}, a} state-of-the-art sequence-to-graph alignment tool}.

Overall, we demonstrate that genome sequence analysis can be accelerated by co-designing scalable and energy-efficient customized accelerators along with efficient algorithms for the key steps of genome sequence analysis.
\section{Future Research Directions}

\damla{

This dissertation opens new avenues for genomics research. In this section, we describe several such \damlaII{promising} research directions in which the ideas and approaches in this dissertation can be extended to provide more functionality or to accelerate other key steps of the genome sequence analysis pipeline.

\subsection{Algorithmic Enhancements to GenASM/BitAlign for Broader Functionality} 

As we explain in Chapter~\ref{ch4-genasm}, GenASM is the first work that enhances and accelerates the Bitap algorithm for approximate string matching. We modify Bitap to add efficient support for
long reads and enable parallelism within each ASM operation. We also propose the first Bitap-compatible traceback algorithm. Later, as we explain in Chapter~\ref{ch6-segram}, we further extend the GenASM algorithms and propose \ba, a novel bitvector-based sequence-to-graph alignment algorithm. However, currently, both the GenASM algorithm and the BitAlign algorithm \damlaII{have a limitation}, which affects their accuracy: they only support Levenshtein distance (i.e., edit distance) calculation~\cite{levenshtein1966binary}, where each error (i.e., substitution, insertion or deletion) \damlaII{has the same cost} (i.e., 1). \damlaII{One future work would be to extend} both GenASM and BitAlign to \damlaII{fully support\footnote{\damlaII{GenASM currently offers partial support for non-unit costs for different edits and the affine gap penalty model (Section~\ref{sec:bitap-traceback}).}}} different \damlaII{costs} for each error type and affine gap penalty model, where gap openings and gap extensions are penalized differently. \damlaII{Once the algorithms are modified with this scoring extension, the hardware accelerators need to be modified to support these algorithmic changes. In order to efficiently support the changes, multiple components of the accelerators may need to be modified or more significantly redesigned.}

\subsection{End-to-End Acceleration of the Mapping Pipeline} 

As we explain in Section~\ref{sec:bitmac-framework}, GenASM provides support for the read alignment and pre-alignment filtering steps of the read mapping (i.e., sequence-to-sequence mapping) pipeline (Section~\ref{sec:background:mapping}). Even though GenASM is orthogonal to any indexing and seeding approach, due to Amdahl's Law, it is preferable to accelerate the entire read mapping pipeline rather than its individual steps. Similar to the approaches followed by Illumina's DRAGEN platform~\cite{illuminadragen} and NVIDIA's Parabricks platform~\cite{nvidiaparabricks}, in order to obtain larger amounts of speedup, one future work would be extending our GenASM work such that all of the steps of the read mapping pipeline would be accelerated as a complete hardware accelerator design. This end-to-end design would also help us to reduce the high amount of data movement that takes place while moving data between different compute units that perform different steps of the pipeline.

Similarly, as we explain in Section~\ref{sec:segram-framework}, SeGraM provides support for the seeding and the sequence-to-graph alignment steps of the sequence-to-graph mapping pipeline (Section~\ref{sec:background-genomemapping}). Thus, another \damlaII{promising future work would be to incorporate} the pre-processing steps of sequence-to-graph mapping and also implementing efficient pre-alignment filters as part of our SeGraM design for a more comprehensive and efficient end-to-end design.

\subsection{Bottleneck Analysis and Acceleration of Assembly with Long Reads} 

With the emergence of long read sequencing technologies \damlaII{(ONT~\cite{ontwebsite} and PacBio~\cite{pacbiowebsite})}, \textit{de novo} assembly becomes a promising way of constructing the original genome. When we analyze the genome assembly pipeline using nanopore (ONT) sequence data (Chapter~\ref{ch3-nanopore}), we show that assembly is one of the most computationally-expensive steps of the pipeline. We also show that there is a tradeoff between accuracy and performance when deciding on the appropriate tool for this step. Thus, we believe that there is a need to design an accelerator for generic graph processing algorithms that includes specialized support for the assembly step of the pipeline, which will provide both high performance and high accuracy. 

\damlaII{The expected first step would be to comprehensively analyze} the current state-of-the-art long read assembly tools and revealing the bottlenecks in terms of performance, scalability, and accuracy. This would enable to explore possible algorithmic changes and different acceleration mechanisms (e.g., specialized accelerators, in-memory processing engines, and SIMD architecture) to resolve the bottlenecks. Even though many prior works (e.g.,~\cite{zhong2013medusa,graphp,ahn2015pim,ahn2015scalable,graphpim,song2018graphr,sengupta2015graphreduce,nurvitadhi2014graphgen,gui2019survey,khoram2018accelerating,dai2018graphh,zhang2017boosting,zhang2015efficient,gharaibeh2013efficient,zhou2017tunao,ozdal2016energy,ham2016graphicionado,dai2017foregraph,zhou2016high,dai2016fpgp,attia2014cygraph,delorimier2006graphstep,wang2016gunrock,khorasani2014cusha,wang2019processor}) have proposed hardware accelerators for generic graph processing algorithms, efficiently using these accelerators for genome assembly is an important but unexplored research problem. \damlaII{Thus, exploring state-of-the-art graph processing accelerators and analyzing their suitability for the requirements of the assembly step would be beneficial. A related direction is to co-design an assembly algorithm with an efficient hardware accelerator and evaluate this design with real or simulated long read datasets to assess the performance and accuracy of the approach. Furthermore, exploring how to support different steps of the genome sequence analysis pipeline as well as generic graph processing algorithms with the proposed design would be beneficial.}

}


\section{Final Concluding Remarks}

In this dissertation, we have demonstrated that \emph{genome sequence analysis can be accelerated by co-designing scalable and energy-efficient customized accelerators along with efficient algorithms for the key steps of genome sequence analysis.} 
First, we comprehensively analyze the tools in the genome assembly pipeline for long reads in multiple dimensions (i.e., accuracy, performance, memory usage, and scalability), uncovering bottlenecks and tradeoffs that different combinations of tools and different underlying systems lead to. We show that we need high-performance, memory-efficient, low-power, and scalable designs for genome sequence analysis in order to exploit the advantages that genome sequencing provides. Second, we propose GenASM, an acceleration framework that builds upon bitvector-based approximate string matching (ASM) to accelerate multiple steps of the genome sequence analysis pipeline. We co-design our highly-parallel, scalable and memory-efficient algorithms with low-power and area-efficient hardware accelerators. 
Third, we implement an FPGA-based prototype for GenASM, where state-of-the-art 3D-stacked memory (HBM2) offers high memory bandwidth and FPGA resources offer high parallelism by instantiating multiple copies of the GenASM accelerators. 
Fourth, we propose \mech, the first hardware acceleration framework for sequence-to-graph mapping and alignment. \mech enables the efficient mapping of a sequenced genome to a graph-based reference, providing more comprehensive and accurate genome sequence analysis.
We conclude and hope that this dissertation inspires future work in co-designing algorithms and hardware together to create powerful frameworks that accelerate other genomics workloads and emerging applications.

\newpage
\phantomsection
\addcontentsline{toc}{chapter}{Other Works of the Author}
\section*{Other Works of the Author}

\damla{

Throughout the course of my Ph.D. study, I have worked on several different topics with many fellow graduate students from Carnegie Mellon University, ETH Zurich, and other institutions. In this chapter, I would like to acknowledge these works. 

I have worked on a number of other projects on genomics. 
In collaboration with Jeremie S. Kim, we propose GRIM-Filter~\cite{kim2018grim}, a novel seed location filtering algorithm, which is optimized to exploit 3D-stacked memory systems that integrate computation within a logic layer stacked under memory layers, to perform processing-in-memory (PIM). We also propose AirLift~\cite{kim2019airlift}, a fast and comprehensive technique for remapping alignments from one genome to another. AirLift greatly reduces the time to perform end-to-end \damlaII{BAM-to-BAM~\cite{bam} (i.e., binary alignment/map format; binary version of SAM (sequence alignment/map format~\cite{sam,sam2,li2009sequence})} remapping on a read set from one reference genome to another while maintaining high accuracy and comprehensiveness that is comparable to fully mapping the read set to the new reference.

In collaboration with Can Firtina, we propose Apollo~\cite{firtina2020apollo}, the first machine learning-based
universal technology-independent assembly polishing algorithm. Apollo enables all available reads to contribute to assembly polishing and scales well to polish an assembly of any size (e.g. both small and large genome assemblies) within a single run. Apollo corrects errors in an assembly by using read-to-assembly alignment regardless of the sequencing technology used to generate reads.

In collaboration with Mohammed Alser, we propose a survey on state-of-the-art algorithmic methods and hardware-based acceleration approaches for genome analysis~\cite{alser2020accelerating}. We cover the algorithmic approaches that exploit the structure of the genome as well as the structure of the underlying hardware, and the hardware-based acceleration approaches that exploit specialized microarchitectures or various execution paradigms (e.g., processing inside or near memory).

I have also worked on a number of non-genomics focused projects. In collaboration with Saugata Ghose, we propose an experimental study, where we rigorously analyze the combined DRAM–workload behavior for 9 different DRAM types and 115 modern applications and multiprogrammed workloads~\cite{ghose2019demystifying}. 

In collaboration with Gagandeep Singh, we propose a work, where we leverage an FPGA coupled with high-bandwidth memory (HBM) for improving the pre-alignment filtering step of genome analysis and representative kernels from a weather prediction model~\cite{singh2021fpga}.


}
\newpage
\phantomsection
\addcontentsline{toc}{chapter}{Bibliography}
\begin{singlespace}
\normalsize
\bibliography{main}
\bibliographystyle{plain}
\end{singlespace}

\end{document}